\documentclass{aa}  

\usepackage{graphicx}
\usepackage{placeins}
\usepackage[dvipsnames]{xcolor}
\usepackage{float}
\usepackage{afterpage}
\usepackage{caption}
\usepackage{txfonts}
\newcommand\Tstrut{\rule{0pt}{2.6ex}}         
\newcommand\Bstrut{\rule[-1.3ex]{0pt}{0pt}}   

\begin{document}

   \title{JCMT/SCUBA-2 uncovers an excess of $850\mu$m counts on megaparsec scales around high-redshift quasars}

   \subtitle{Characterization of the overdensities and their alignment with the quasars' Ly$\alpha$ nebulae\thanks{Tables F1 to F40 (the source catalogs) are only available in electronic form
at the CDS via anonymous ftp to cdsarc.cds.unistra.fr (130.79.128.5)
or via https://cdsarc.cds.unistra.fr/cgi-bin/qcat?J/A+A/}\fnmsep\thanks{The reduced images are only available at the CDS via anonymous ftp to cdsarc.u-strasbg.fr (130.79.128.5) or via http://cdsarc.u-strasbg.fr/viz-bin/qcat?J/A+A/}}

   \author{F. Arrigoni Battaia
          \inst{1}
          \and
          A. Obreja\inst{2}
          \and
          C.-C. Chen\inst{3}
          \and
          M. Nowotka\inst{4,5}
          \and
          M. Fumagalli\inst{6,7}
          \and
          J. X. Prochaska\inst{8,9}
          \and
          Y. Yang\inst{10}
          \and
          Z. Cai\inst{11}
          \and
          N. Mu\~noz-Elgueta\inst{1}
          \and
          M. Fossati\inst{6}
          }

   \institute{Max-Planck-Institut fur Astrophysik, Karl-Schwarzschild-Str 1, D-85748 Garching bei M\"unchen, Germany\\
              \email{arrigoni@mpa-garching.mpg.de}
         \and
             Universit\"ats-Sternwarte M\"unchen, Scheinerstraße 1, D-81679 M\"unchen, Germany 
         \and
             Academia Sinica Institute of Astronomy and Astrophysics, No.1, Section 4, Roosevelt Rd., Taipei 10617, Taiwan 
         \and
             Department of Physics, Stanford University, Stanford, CA 94305, USA
         \and
             Kavli Institute for Particle Astrophysics \& Cosmology (KIPAC), Stanford University, Stanford, CA 94305, USA
         \and
            Dipartimento di Fisica G. Occhialini, Universit\`a degli Studi di Milano Bicocca, Piazza della Scienza 3, 20126 Milano, Italy
         \and
            INAF - Osservatorio Astronomico di Trieste, via G. B. Tiepolo 11, 34143 Trieste, Italy   
         \and
            Department of Astronomy \& Astrophysics, UCO/Lick Observatory, University of California, 1156 High Street, Santa Cruz, CA 95064, USA
         \and
            Kavli Institute for the Physics and Mathematics of the Universe (Kavli IPMU), 5-1-5 Kashiwanoha, Kashiwa, 277-8583, Japan
         \and
            Korea Astronomy and Space Science Institute, 776 Daedeokdae-ro, Yuseong-gu, Daejeon 34055, Republic of Korea
         \and
            Department of Astronomy, Tsinghua University, Beijing 100084, China
             }

   \date{}

\abstract{We conducted a systematic survey of the environment of high-redshift quasars at submillimeter wavelengths to unveil and characterize the surrounding distribution of dusty submillimeter galaxies (SMGs). We took sensitive observations with the SCUBA-2 instrument on the \textit{James Clerk Maxwell} Telescope for 3 enormous Lyman-alpha nebulae (ELANe) and 17 quasar fields 
in the redshift range $2<z<4.2$ selected from recent Lyman alpha (Ly$\alpha$) surveys.  
These observations uncovered 523 and 101 sources at 850~$\mu$m and 450~$\mu$m, respectively, with signal-to-noise ratios (S/N) $>4$ or detected in both bands at S/N$>3$. 
We ran self-consistent Monte Carlo simulations to construct 850~$\mu$m number counts and unveil an excess of sources in 75\% of the targeted fields. Overall, regions around ELANe and quasars are overabundant with respect to blank fields by a factor of $3.4\pm0.4$ and $2.5\pm0.2$, respectively (weighted averages). 
Therefore, the excess of submillimeter sources is likely part of the megaparsec-scale environment around these systems. By combining all fields and repeating the count analysis in radial apertures, we find (at high significance, $\gtrsim5\sigma$) a decrease 
in the overdensity factor from $>3$ within $\sim 2$~cMpc to $\sim2$ in the annulus at the edge of the surveyed field ($\sim10$~cMpc), which suggests that the physical extent of the overdensities is larger than our maps. We computed preferred directions for the overdensities of SMGs from the 
positions of the sources and used them to orient and create stacked maps of source densities for the quasars' environment. This stacking unveils an elongated structure reminiscent of a large-scale filament with a scale width of $\approx 3$~cMpc. Finally, the directions of the overdensities 
are roughly aligned with the major axis of the Ly$\alpha$ nebulae, suggesting that the latter trace, on scales of hundreds of kiloparsecs, the central regions of the projected large-scale structure described by the SMGs on megaparsec scales. Confirming member associations of the SMGs is required to further characterize 
their spatial and kinematic distribution around ELANe and quasars.}  

\titlerunning{Excess of 850$\mu$m counts around high-redshift quasars}
\authorrunning{F. Arrigoni Battaia, A. Obreja, C.-C. Chen, et al.}
   \maketitle

%

\section{Introduction}
  
 Quasars, accreting supermassive black holes at the center of galaxies, are one of the most luminous sources throughout cosmic time. Their extraordinary energy budget is expected to affect their host galaxies and the surrounding material out to intergalactic scales (e.g., \citealt{rees88,SilkRees98,Fabian2012,KingPounds2015,Harrison2017}). Current galaxy formation models require this quasar feedback to explain several observables, especially at the massive end of the galaxy population and for intergalactic material in groups and clusters (e.g., \citealt{Croton2006,Dubois2012,Crain2015,Weinberger2017,Costa2022}). 
Bright quasars are relatively rare objects characterized by a strong evolution of their number density with redshift (e.g., \citealt{Shen2020} and references therein). The substantial rise in the number of quasars from cosmic dawn to their peak at $z\sim2$ and the subsequent fall to $z=0$ can be explained in the framework of the hierarchical growth of structures together with a decrease in fuel at late times (e.g., \citealt{ER1988,kh00,Hopkins07}).   
Because of their paucity, bright quasars are usually assumed to pinpoint the location of the highest density peaks in the dark matter (DM) distribution. Indeed, clustering studies infer that high-redshift quasars should inhabit relatively massive DM halos in the currently probed redshift range, $0.5 < z \lesssim 4$ ($M_{\rm DM}\sim 10^{12}-10^{13}$~M$_{\odot}$; e.g., \citealt{white12,Eftekharzadeh2015}). 

To confirm these expectations, several works have tried to directly constrain the density field around quasars by searching for associated unobscured companions (e.g., Lyman alpha emitters; e.g., \citealt{steidel00}) with different techniques and on different scales (from a few hundred kiloparsecs out to megaparsec scales), resulting in heterogeneous results (e.g., \citealt{Willott2005,Kashikawa2007,Trainor2012,Kikuta2017,Mazzucchelli2017,GarciaVergara2019}). Possible causes of this lack of agreement could be related to (i) observational strategies or available instrumentation (i.e., small sample statistics for each study and/or shallow sensitivity) and/or (ii) a physical phenomenon: the galaxies in quasar environments are significantly dustier, resulting in a biased picture when looking only at unobscured tracers (e.g., \citealt{GarciaVergara2020}). 
Recent efforts suggest that both effects could be at work and have started to confirm the expectations from clustering studies and, therefore, the association of quasars with rich environments. For example, \citet{Fossati2021} targeted 27 $z\sim3-4.5$ quasar fields with deep observations ($\sim 4$~hours) using the Multi Unit Spectroscopic Explorer (MUSE; \citealt{Bacon2010}) and found a significant overdensity of Lyman alpha emitters (LAEs) within 300~km~s$^{-1}$ of the quasars' systemic redshift and on scales of a few hundred kiloparsecs. From this overdensity and the kinematics of the LAEs, the authors inferred a DM halo mass in the range $M_{\rm DM}\approx 10^{12}-10^{12.5}$~M$_{\odot}$, which is in line with the aforementioned clustering measurements.
Meanwhile, several studies have reported the presence of companions or overdense environments in submillimeter galaxies (SMGs; e.g., \citealt{Smail1997}) from tens of kiloparsecs up to a few hundred kiloparsecs around $2<z\lesssim 7$ quasars, thanks especially to the Atacama Large Millimeter/submillimeter Array (ALMA; e.g., \citealt{Decarli2017, Trakhtenbrot2017, Venemans2020, Bischetti2021, TC2021, GarciaVergara2022}). Altogether, these works paint a coherent picture that stresses the importance of multiwavelength observations in determining the environment of quasars.

In this framework, submillimeter observations have targeted diverse active galactic nucleus (AGN) samples to constrain the richness of their environment on megaparsec scales. These works discovered an excess of submillimeter source counts around some high-redshift radio galaxies (e.g., \citealt{Stevens2003,Dannerbauer2014}), a few high-redshift quasars (\citealt{Priddey2008}), or IR-luminous AGN selected by the Wide-field Infrared Survey Explorer (WISE) space telescope (\citealt{Jones2017}). For fields with large numbers of detections, it has also been assessed whether the SMGs trace large-scale structures with a preferential alignment with the radio galaxies' jets (\citealt{Stevens2003}). The latest work exploring this aspect targeted the fields of 17 high-redshift radio galaxies and found evidence for the alignment of SMGs with the perpendicular direction with respect to the radio jets (\citealt{Zeballos2018}). This misalignment has been interpreted as a signature that the SMGs trace a large-scale filament feeding the protocluster core. 

Nonetheless, the overdensities are only found in a few fields out of larger parent samples, possibly suggesting that SMGs are not a universal tracer of overdensities around AGN or that they may preferentially coevolve around certain types of quasars (\citealt{Rigby2014,Zeballos2018}). 
To further investigate these hypotheses and extend the types of AGN environments studied, \citet{FAB2018b} targeted one $z\sim2$ enormous Lyman alpha nebula (ELAN; e.g., \citealt{Cai2017a}). ELANe are rare systems characterized by an extreme Lyman alpha (Ly$\alpha$) surface brightness (${\rm SB}_{\rm Ly\alpha}\sim10^{-17}$~erg~s$^{-1}$~cm$^{-2}$~arcsec$^{-2}$) over $>100$~physical kpc, with multiple embedded AGN and active galaxies (e.g., \citealt{hennawi+15,FAB2018,TC2021,FAB2022}), making them good candidates for protoclusters (e.g., \citealt{hennawi+15}). \citet{FAB2018b} reported a factor of 4 overdensity for submillimeter sources on megaparsec scales, providing further evidence of the richness of ELAN environments.
More recently, \citet{Nowotka2022} expanded that survey by targeting three additional ELAN fields and revealed a ubiquitous overabundance of SMGs.  Assuming that all the submillimeter sources are associated with the targeted ELANe, the star formation rate density is found to be similar to that around other quasar samples or in protoclusters at similar redshifts ($z\sim2-3$; e.g., \citealt{Clements2014,Dannerbauer2014,Kato2016}). Therefore, ELANe seem to pinpoint regions where the coevolution of quasars and SMGs is favored.

Larger statistical surveys are needed to confirm these findings and assess whether ELANe inhabit very different environments than other bright quasars characterized by less extreme Ly$\alpha$ nebulae.
In this paper, we report the results obtained by analyzing 20 fields centered on 17 quasars and  3 ELANe.
The paper is structured as follows. Sections~\ref{sec:obs} and \ref{sec:extr_cat} 
describe the sample selection, the observations, their data reduction, and the source extraction, together with the criteria for making our catalogs. Section~\ref{sec:NC}
reports on the calculation of the differential number counts. We highlight our results in Sect.~\ref{sec:results}, including (i) an estimate of the overdensity factors for each field and on average; (ii) a study of the variation in the overdensity as a function of radial aperture; and (iii) a comparison of the overdensity factors with each system's properties (both quasar and Ly$\alpha$ nebula properties). Section~\ref{sec:discussion} discusses (i) the implications of the similarity of overdensities found around all the studied systems; (ii) the presence of a characteristic scale for the overdensities determined by stacking maps of counts; and (iii) the alignment between overdensities and extended Ly$\alpha$ emission in these fields. Finally, we summarize our work in Sect.~\ref{sec:summary}. Throughout this paper, we adopt the cosmological parameters $H_0 = 70$~km~s$^{-1}$~Mpc$^{-1}$, $\Omega_{\rm M} =0.3$, and $\Omega_{\Lambda} =0.7$. In this cosmology, 1~cMpc corresponds to about 0.5~arcmin at the mean redshift of our sample ($z=3.337$).

\section{Observations}
\label{sec:obs}

\subsection{Sample selection}

\begin{table*}
\scriptsize
\caption{Observation log.}
\centering
\begin{tabular}{lccccccccccc}
\hline
\hline
Field   & ID\tablefootmark{a} & RA      & Dec   & $z_{\rm sys}$\tablefootmark{b}        & noise$_{850\mu m}$\tablefootmark{c} & $A_{\rm eff; 850\mu m}$\tablefootmark{d} & noise$_{450\mu m}$\tablefootmark{c} & $A_{\rm eff; 450\mu m}$\tablefootmark{d} & t$_{\rm exp}$\tablefootmark{e} & $\tau$\tablefootmark{f} & Program ID \\      
        &    & (J2000) & (J2000) & & (mJy beam$^{-1}$) & (arcmin$^{2}$) & (mJy beam$^{-1}$) & (arcmin$^{2}$) & (min) &  &          \\
\hline
\multicolumn{12}{c}{ELANe}\\
\hline 
Jackpot\tablefootmark{*}                  & -  & 08:41:58.47      &  +39:21:21.00   &  2.0412 &  0.5  &  126.99  &  3.9  &  123.79 & 423 & 0.04-0.06 & M17BP024, M21AP046 \\ 
MAMMOTH-I         & -  & 14:41:24.46      &  +40:03:09.45   &  2.317  &  0.5  &  126.45  &  3.8  &  125.33 & 423 & 0.02-0.04 & M17BP024, M21AP046 \\ 
Fabulous          & 13 & 10:20:10.00      &  +10:40:02.00   &  3.164  &  0.6  &  127.17  &  4.8  &  123.28 & 402 & 0.04-0.05 & M18AP054, M21AP046 \\ 
\hline
\multicolumn{12}{c}{From QSO MUSEUM sample (Arrigoni Battaia et al. 2019)}\\
\hline
J0525-233\tablefootmark{*}  &  3 & 05:25:06.50   &  -23:38:10.0   &  3.110  &  1.0 & 126.20 & 15.1  & 123.70 & 192 & 0.04     &  M21AP041 \\ 
SDSSJ1209+1138    &  7 & 12:09:18.00     &  +11:38:31.0   &  3.117  &  0.7 & 128.27 & 9.7   & 125.76 & 358 & 0.05-0.06     &  M18BP062, M18AP054\\ 
SDSSJ1025+0452    & 10 & 10:25:09.60     &  +04:52:46.0   &  3.227  &  0.9 & 127.92 & 9.5   & 126.60 & 162 & 0.04-0.05     &  M18AP054\\ 
SDSSJ1557+1540    & 18 & 15:57:43.30     &  +15:40:20.0   &  3.265  &  0.8 & 127.18 & 13.1  & 125.01 & 292 & 0.06-0.07     &  M18BP062, M18AP054  \\ 
SDSSJ1342+1702    & 24 & 13:42:33.20     &  +17:02:46.0   &  3.062  &  0.8 & 126.99 & 13.8  & 126.69 & 261 & 0.04-0.07     &  M18BP062, M18AP054\\ 
Q-0115-30         & 29 & 01:17:34.00     &  -29:46:29.0   &  3.180  &  0.9 & 125.55 & 16.2  & 122.20 & 224 & 0.04     &  M21AP041 \\ 
Q2355+0108        & 46 & 23:58:08.54     &  +01:25:07.2   &  3.385  &  1.1 & 128.27 & 20.5  & 124.66 & 164 & 0.07-0.08     &  M18BP062\\ 
SDSSJ0819+0823    & 50 & 08:19:40.58     &  +08:23:58.0   &  3.197  &  1.0 & 128.22 & 14.4  & 127.81 & 162 & 0.04-0.05     &  M18AP054\\ 
\hline
\multicolumn{12}{c}{From MAGG sample (Fossati et al. 2021)}\\
\hline
J015741-01062957  & -  & 01:57:41.56    &  -01:06:29.6    &  3.5645 & 1.0 & 127.66 & 14.1 & 124.42 & 182 & 0.05-0.06 &  M19BP048\\ 
J020944+051713    & -  & 02:09:44.62    &  +05:17:13.7    &  4.1846 & 1.0 & 127.37 & 13.5 & 124.45 & 182 & 0.05-0.07 &  M19BP048\\ 
J024401-013403    & -  & 02:44:01.84    &  -01:34:03.9    &  4.044  & 0.9 & 127.11 & 10.8 & 124.21 & 211 & 0.05-0.06 &  M19BP048\\ 
J033900-013318\tablefootmark{*}    & -  & 03:39:00.99   &  -01:33:17.6    &  3.204  & 0.9 & 126.99 & 12.5 & 123.88 & 182 & 0.05-0.06 &  M19BP048\\ 
J111113-080401    & -  & 11:11:13.64    &  -08:04:02.5    &  3.930  & 1.0 & 126.82 & 13.8 & 125.48 & 182 & 0.03-0.06 &  M19BP048\\ 
J193957-100241\tablefootmark{*}   & -  & 19:39:57.26    &  -10:02:41.5    &  3.787  & 1.0 & 125.91 & 14.6 & 123.75 & 182 & 0.07      &  M19BP048\\ 
J221527-161133    & -  & 22:15:27.29    &  -16:11:33.0    &  4.000  & 1.0 & 127.31 & 14.7 & 123.24 & 182 & 0.07      &  M19BP048\\ 
J230301-093930    & -  & 23:03:01.45    &  -09:39:30.7    &  3.4774 & 1.0 & 126.68 & 15.9 & 123.60 & 182 & 0.07      &  M19BP048\\ 
\hline
\multicolumn{12}{c}{In both QSO MUSEUM and MAGG samples}\\
\hline
J233446-090812\tablefootmark{*}    & 45 & 23:34:46.40   &  -09:08:12.2    &  3.3261 & 1.0 & 124.06 & 12.6 & 124.19 & 181 & 0.05-0.06 &  M19BP048\\ 
\hline
\end{tabular}
\label{tab:sample}
\tablefoot{
\tablefoottext{a}{ID from the QSO MUSEUM sample.}
\tablefoottext{b}{Systemic redshift of the quasar evaluated from the peak of the C~IV line, correcting for the expected shift as estimated by \citet{Shen2016}. The
intrinsic uncertainty on this correction is $\sim415$~km~s$^{-1}$ and dominate the error badget ($\Delta z\approx0.007$).}
\tablefoottext{c}{Noise level at the center of each map.}
\tablefoottext{d}{Effective area assumed in this work and defined above the 3~times the central noise.}
\tablefoottext{e}{Exposure time on source.}
\tablefoottext{f}{Range of values of the opacity at 225~GHz for each target.}
\tablefoottext{*}{Systems with a radio-loud central or companion quasar.}
}
\end{table*}

\begin{figure}
\centering
\includegraphics[width=0.98\columnwidth]{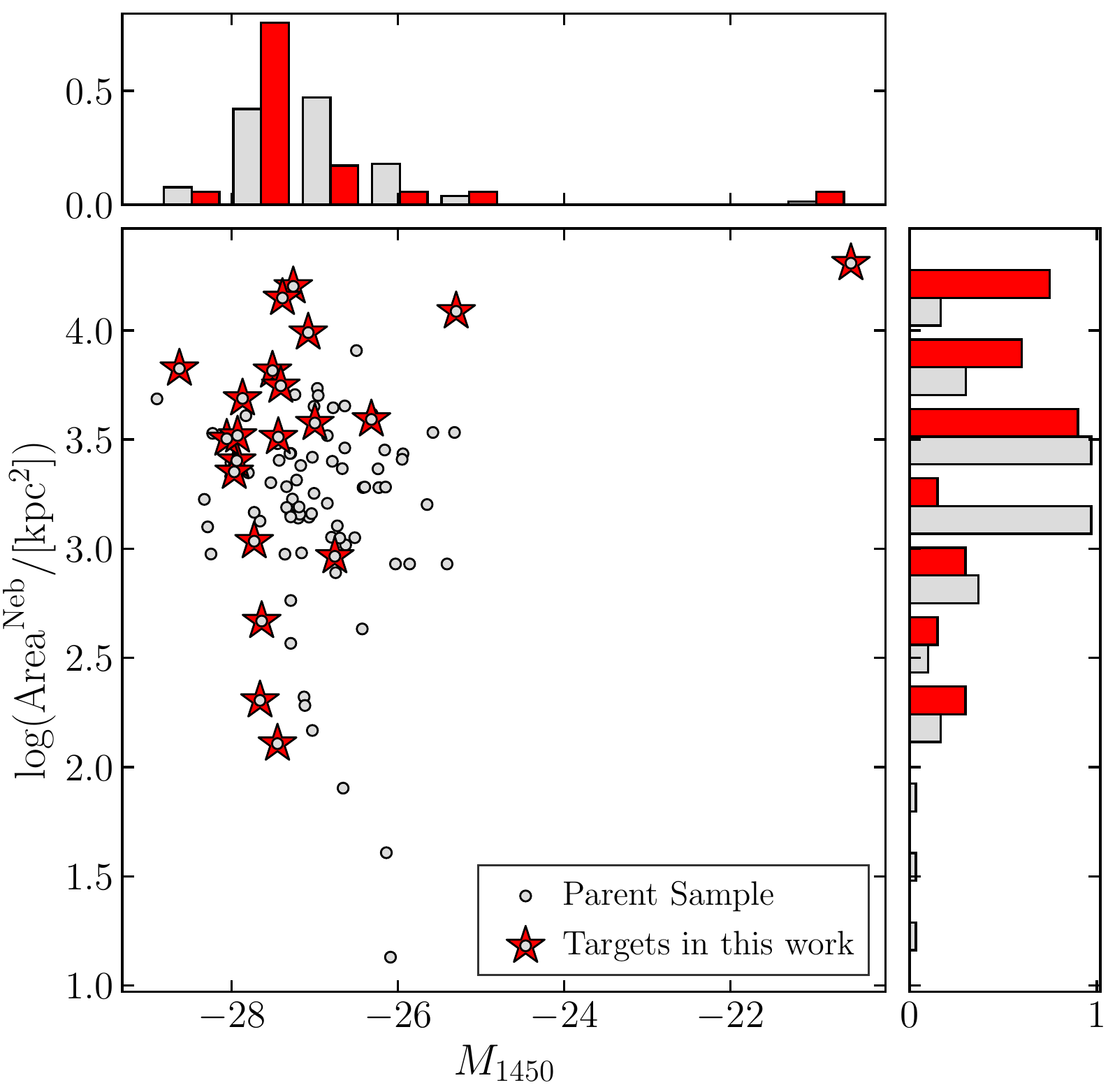}
\caption{Absolute magnitude at 1450~\AA\ $M_{1450}$ of the central AGN versus Ly$\alpha$ nebula areas of the targeted systems (red) in comparison to the parent sample (gray). 
Each histogram on the side is normalized separately.}
\label{fig:sample}
\end{figure}

We report on submillimeter data for 17 quasar fields in the redshift range $z\sim3-4.2$ and for 3 ELAN fields (Jackpot, MAMMOTH-I, Fabulous).  
The latter three objects have been already studied in \citet{Nowotka2022}, but with shorter total exposure times than used here ($\sim$3 versus $\sim$7 hours).
All these systems have information on their Ly$\alpha$ emission on hundreds of kiloparsecs thanks to recent surveys with narrow-band filters or the MUSE instrument:  Fluorescent Lyman-Alpha Survey of cosmic Hydrogen iLlumInated by hIGH-redshifT quasars (FLASHLIGHT; \citealt{FAB2014}), MApping the Most Massive Overdensity Through Hydrogen (MAMMOTH; e.g., \citealt{Cai2016}), Quasar
Snapshot Observations with MUse: Search for Extended Ultraviolet
eMission (QSO MUSEUM; \citealt{FAB2019}), and MUSE Analysis of Gas around Galaxies (MAGG; \citealt{Fossati2021}).
The FLASHLIGHT and QSO MUSEUM surveys targeted so far 26 and 61 relatively bright ($17.0 < i < 20.0$ and $17.4 < i < 19.0$) high-redshift ($2 < z <2.3$ and $3.03 < z < 3.46$) quasars with the aim to uncover surrounding extended Ly$\alpha$ emission and characterize its properties and those of the emitting gas (e.g., \citealt{FAB2016}). The MAMMOTH survey identified overdensities at $z=2-3$ by locating extremely rare, strong intergalactic \ion{H}{I} Ly$\alpha$ absorption systems and absorption groups (\citealt{Cai2017a}). These fields were then observed with narrow-band filters targeting the Ly$\alpha$ line at those redshifts to characterize the population of LAEs. The results of this effort includes the discovery of the MAMMOTH-I ELAN (\citealt{Cai2016}). The MAGG survey targeted 28 relatively bright ($16.5 < r < 20.0$) $z=3.2-4.5$ quasars with the aim to study the halo gas of (i) $z=3-4.5$ star-forming galaxies in the surroundings of optically thick \ion{H}{I} absorbers (\citealt{Lofthouse2020}), (ii) $z=3-4.5$ quasars (\citealt{Fossati2021}), and (iii) $z\sim1$ galaxies (\citealt{Dutta2020}). 
The targets of our work were selected to cover the full range of Ly$\alpha$ nebulae (e.g., area) and quasars (e.g., luminosity) properties of the parent sample, 
after taking into account source visibility.
Five systems have a radio-loud central or companion quasar with unresolved morphology at the resolution of the Very Large Array (VLA) Faint Images of the Radio Sky at Twenty-Centimeters (FIRST) survey ($\sim5$ arcsec; \citealt{Becker1994}) or of 
the NRAO VLA Sky Survey (NVSS; $\sim45$~arcsec; \citealt{Condon1998}). Figure~\ref{fig:sample} shows the absolute magnitude at 1450~\AA\ $M_{1450}$ of the central AGN and 
the areas of the Ly$\alpha$ nebulae around the targeted systems in comparison to the aforementioned parent sample. In Table~\ref{tab:sample} we list the selected sources with their coordinates and redshifts.

\subsection{SCUBA-2/JCMT observations and data reduction}

The observations for the twenty fields in this study (Table~\ref{tab:sample}) were conducted with the Submillimetre Common-User Bolometer Array 2 (SCUBA-2; \citealt{Holland2013}) on the \textit{James Clerk
Maxwell} Telescope (JCMT) during flexible observing under several programs from September 2, 2017, until July 17,  2021 (IDs: M17BP024, M18AP054, M18BP062, M19BP048, M21AP041, M21AP046), under good weather conditions (band 1 and 2, atmospheric opacity at 225~GHz $\tau_{225{\rm GHz}}\leq 0.07$).
For these observations, a Daisy pattern was used to cover an area of $\simeq 13.7\arcmin$ in diameter, which was centered at the location of each known quasar or ELAN in those fields. 
To facilitate the scheduling, the observations were organized into scans/cycles of about 30 minutes. We obtained a total exposure time on source ranging from a minimum of 162 minutes up to $402-423$ minutes for the three ELAN fields. 

For the data reduction, we closely followed the method used in \citet{TC2013a}, \citet{FAB2018b}, and \citet{Nowotka2022}. Briefly, the data were reduced using the Dynamic Iterative Map Maker (DIMM) 
included in the SMURF package from the STARLINK software (\citealt{Jenness2011,Chapin2013}). The standard configuration file dimmconfig\_blank\_field.lis was 
adopted for our science purposes. Data were reduced for each scan and the MOSAIC\_JCMT\_IMAGES recipe in PICARD, the Pipeline for Combining and Analyzing 
Reduced Data (\citealt{Jenness2008}), was used to co-add the reduced scans into the final maps. 

Point source detectability was increased by applying a standard matched filter to the final maps, using the PICARD recipe SCUBA2\_MATCHED\_FILTER. 
For flux calibration, we adopted the recommended flux conversion factors (FCFs) with 10\% upward corrections obtained from the analysis of archival data spanning ten years\footnote{\url{https://www.eaobservatory.org/jcmt/instrumentation/continuum/scuba-2/calibration/}}. These FCFs are date dependent, and we list the relevant values for our work in Appendix~\ref{app:FCFs}. We also generate calibrated maps using the standard FCFs (491 Jy pW$^{-1}$ for 450~$\mu$m and 537 Jy pW$^{-1}$ for 850~$\mu$m) with 10\% upward corrections (\citealt{Dempsey2013}). These values have a difference of only $5-10$\% with respect to the new FCFs. We tested our results against such a change in FCFs and our findings are unchanged. 

\begin{figure*}
\centering
\includegraphics[width=0.78\textwidth]{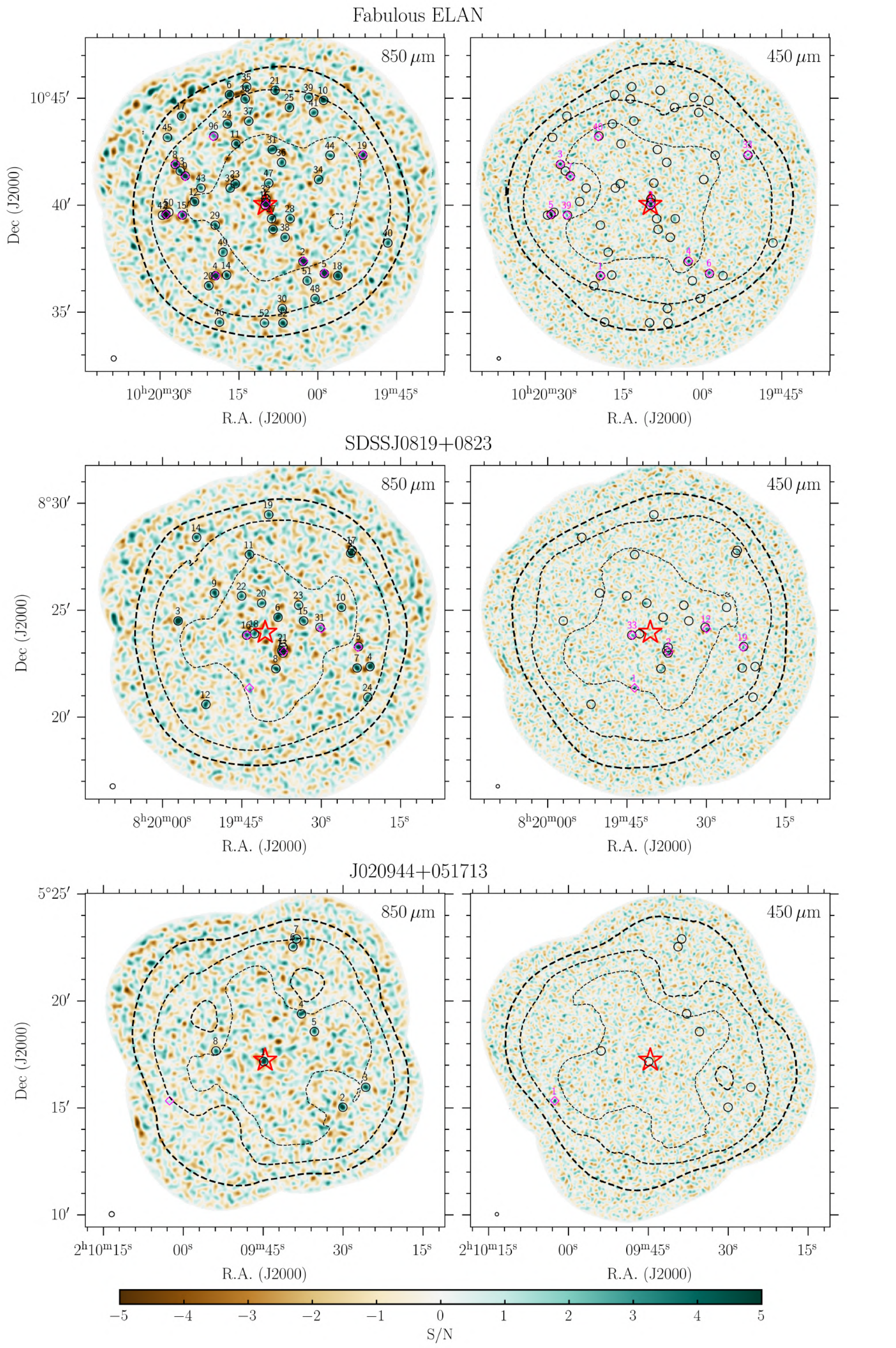}
\caption{SCUBA-2 S/N maps at $850$~$\mu$m (left column) and $450$~$\mu$m (right column) for three example fields: one ELAN (Fabulous), the most overdense quasar field (SDSSJ~0819+0823), and one of the least overdense quasar fields (J020944+051713).
The size of each panel is about $15\arcmin \times 15\arcmin$. The red star in each panel indicates the location of the ELAN and quasars. The $850$ and 450~$\mu$m detections with corresponding ID numbers are highlighted as black circles and magenta diamonds, respectively.  
For each map, we indicate the noise contours (black dashed) for levels of 1.5, 2, and 3 times the central noise, the last of which defines the outer boundary
for source extraction (Table~\ref{tab:sample}). The $850$ and 450~$\mu$m maps for all other fields are shown in Appendix~\ref{app:maps}.
}
\label{fig:Maps1}
\end{figure*}

The final central noise levels for our data are in the range 0.53-1.07~mJy/beam and 3.82-20.50~mJy/beam at 850~$\mu$m and 450~$\mu$m, respectively. In the reminder of this work we focused on the regions of the data characterized by a noise level less than three times the central noise. We refer to this area as the effective area.
Table~\ref{tab:sample} summarizes the observations log, including the map center, the effective area, the central noise, the total exposure time, the opacity, and the program ID for each field and band.

\section{Source extraction and catalogs}
\label{sec:extr_cat}

We extracted the detections from all maps following \citet{FAB2018b} and \citet{Nowotka2022}.
In brief, we focused on the effective area of each map (see Table~\ref{tab:sample}) 
and found all sources with peak S/N~$>3$. This is done by looking recursively for the maximum pixel within the selected region, extracting the position and the information of the peak, 
and subtracting a scaled point spread function (PSF) centered at that position. The process was stopped when the peak S/N went below 3. We used these list of sources as preliminary catalogs, 
and cross-checked the catalogs at 850 and 450~$\mu$m for each field to select counterparts in the other band.
A counterpart is defined as a detected source at 450~$\mu$m laying within the 850~$\mu$m beam. 

In the final catalogs we list every $>4\sigma$ source, but also every $>3\sigma$ source characterized by a $>3\sigma$ counterpart in the other band. Considering all fields, we discovered 523 sources at 850~$\mu$m and 101 sources at 450~$\mu$m. In Tables~F1 
to F40 (available only in electronic form), 
we list the positions, positional error, S/N, fluxes, deboosted fluxes and counterparts for all these sources. 
In the same tables we also list the distances between detections at 850~$\mu$m and their counterparts at 450~$\mu$m. Figure~\ref{fig:Maps1} and Figures~\ref{fig:MapsELAN} to \ref{fig:Maps7} show the final S/N maps at 850 and 450~$\mu$m for each targeted field with the discovered sources overplotted.

\subsection{Reliability of source extraction}
\label{sec:rel}

The aforementioned catalogs could be affected by spurious sources. To quantify them, we use several methods as in \citet{FAB2018b}. First, we inverted the final maps and applied the source extraction algorithm.
The median number of sources per field at $>4\sigma$ at 850 and 450~$\mu$m are found to be six and one, respectively. We stress that the matched filter technique introduce strong negative features in the vicinity of the brighter sources in each field. This results in a larger number of detections in the inverted maps for fields with very bright sources and therefore making this first method not reliable in estimating false detections.
Second, to get a better estimate, we applied the source extraction algorithm to true noise maps, and checked the number of detections with $>4\sigma$. The true noise maps were constructed by using the jackknife resampling technique. In brief, for each field and band, we obtained two maps by co-adding roughly half of the data. The true noise maps were then the result of the subtraction of these two maps. Indeed, any real source in the maps should be subtracted irrespective of its significance, resulting in residual maps that are source-free noise maps.
Importantly, to account for the slight difference in exposure time, the true noise maps were scaled by a factor of 
$\sqrt{t_1\times t_2}/(t_1+t_2)$, where $t_1$ and $t_2$ indicate the exposure time of each pixel from the two maps.
These jackknife maps have a central noise in agreement with the science data to $<4\%$ and $<11\%$ at 850~$\mu$m and 450~$\mu$m, respectively. 
In these maps, the median number of sources at $>4\sigma$ is one and two at $850$ and $450$~$\mu$m, respectively.
We thus expect a similar number of spurious sources in our $>4\sigma$ source catalogs, which corresponds to a median false detection rate of $\sim5\%$ and $\sim45\%$ at $850$ and $450$~$\mu$m, respectively.
Table~\ref{tab:spurious} lists the numbers for each field and band for both methods.

In addition, as done in \cite{FAB2018b}, we estimated the number of spurious detections for the sources with $3<$S/N$<4$ identified as having a counterpart in the other bandpass (lower portion of catalog tables in Appendix~\ref{app:catalogs}). We proceed as follows. First, sources between $3$ and $4\sigma$ were extracted from the jackknife maps and cross-correlated with the detections in the real data at the other wavelength.
An average of 0.2 and 1.0 spurious sources matched a detection in the real maps. This test was then repeated 1000 times using random positions for the spurious sources within the effective area of our observations. The average of spurious sources at $850$ and $450$~$\mu$m was found to be 0.03 and 0.25, stressing that $>3\sigma$ sources selected
in both bandpasses are even more reliable than $>4\sigma$ sources selected in only one bandpass. Table~\ref{tab:spurious} lists the numbers for each field and method.

Altogether, the sources at $450$~$\mu$m without a detection at $850$~$\mu$m are most likely spurious for the shallower observations of the 17 quasar fields. For completeness, we report all the sources in our catalogs. 
Our analysis focuses on the $850$~$\mu$m data and, therefore, our conclusions are not affected by the spurious $450$~$\mu$m sources.

\subsection{Completeness tests}
\label{sec:compl}

For each field and band, we tested at which flux the catalogs can be considered complete by populating the true noise maps discussed in Sect.~\ref{sec:rel} with randomly placed mock sources of a given flux.
We then use the extraction algorithm and considered as recovered sources those detected above 4$\sigma$ and within the beam area. In more detail, sources were injected within the effective area with fluxes from 0.1 to 30.1 mJy (0.1 to 80.1 mJy) with a step of
0.5 mJy (1.0 mJy) for 850 (450) $\mu$m.  
For each step in flux, we iterated the extraction by introducing 1000 sources.
The results of these tests are shown in Fig.~\ref{fig:compl}. As expected, the three ELAN fields (MAMMOTH-1, Jackpot, Fabulous)  have deeper data with the $50\%$ completeness being on average around 3.2 and 24.0 mJy at 850 and 450~$\mu$m, respectively, and
the $80\%$ being on average around 4.0 and 30.4 mJy, respectively. 
The other fields have the average $50\%$ and $80\%$ completeness at 850~$\mu$m around 5.4 and 6.9~mJy, respectively, while the completeness at 450~$\mu$m is much worse ($\lesssim50\%$ at 55~mJy). SDSSJ1209+1138 is the deepest of the quasar fields with the $50\%$ and $80\%$ completeness at 850~$\mu$m around 3.9 and 4.9~mJy, respectively.

\section{Differential number counts}
\label{sec:NC}

In this section we outline the calculation of the differential number counts and the underlying counts model for each field at 850~$\mu$m. Our aim is to ascertain the presence of an excess of sources in these fields with respect to blank reference fields (Sect.~\ref{sec:overd}).
We proceeded following the method of \citet{TC2013a}, already used in the case of ELANe by \citet{FAB2018b} and \citet{Nowotka2022}. 
First, we determine down to which S/N there is a significant excess signal with respect to a pure noise distribution for each field, by inspecting the S/N histograms of the signal maps and the jackknife maps (see Figs.~\ref{fig:SNhistos1}, \ref{fig:SNhistos2}, and \ref{fig:SNhistos3}).
We consider as significant excess a factor of $\geq3\times$ more pixels at a given S/N. This is the case down to a median S/N~$=2.5$ (dashed vertical line in those figures; see Table~\ref{tab:und_models} for the specific values). This positive excess signal is due to real astronomical sources, while the negative excesses are due to the
negative troughs of the matched-filter PSF (\citealt{TC2013a}). This effect can be properly accounted for during sources extraction with a correct PSF.

\begin{figure*}
\centering
\includegraphics[width=1.0\textwidth]{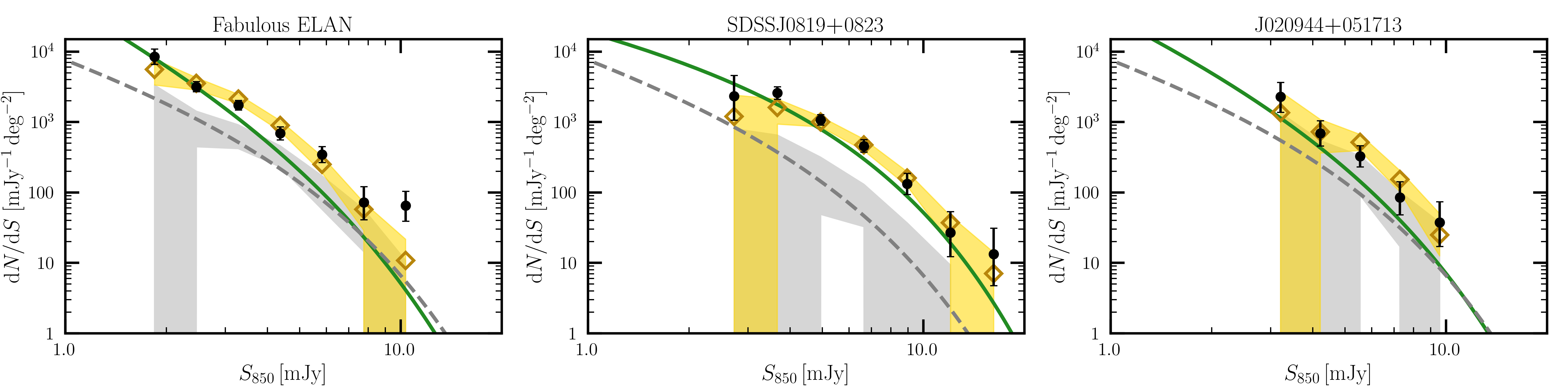}
\caption{Differential number counts recovered from 500 realizations of the underlying number count models determined through Monte Carlo simulations (yellow) and the fiducial model (gray) for three example fields: one ELAN (Fabulous) and two quasar fields, the ones with the largest (SDSSJ~0819+0823) and smallest (J020944+051713) overdensity factors in our sample.
The shaded regions represent the $90\%$ confidence intervals of the counts recovered from the simulated maps. The black points are raw differential number counts as defined in Sect.~\ref{sec:NC}. The underlying count models and the fiducial model (\citealt{Geach2017}) are plotted in green and dashed gray, respectively. The effect of flux boosting can be observed as the shaded regions are lying above their input model. Note that the lower zero limits of the fiducial model realizations at faint fluxes reflect the nonuniform noise structure of our maps, where incompleteness increases with distance from the center of the map. Similar plots for all the other fields are shown in Appendix~\ref{app:diff_counts}.}
\label{fig:counts}
\end{figure*}

As a next step, we extracted sources lowering the detection threshold to the aforementioned smaller value for each field. This can be done as positional information is not relevant for number counts analyses (e.g., \citealt{TC2013b}). In addition, we extracted sources with the same thresholds from the respective jackknife maps.
We then computed the raw number counts in equally spaced logarithmic bins between the minimum and maximum flux densities of the detected sources. 
To do so, we summed the number of sources in each flux bin and divided by the width and detectable area of the bin. The detectable area defines where a source with a certain flux density can be detected above the S/N threshold. The raw number counts were then obtained by subtracting off the spurious counts obtained from the jackknife maps.
The average fraction of spurious sources down to the thresholds considered here is 40\%. 

We have then corrected for the known observational biases that affect the raw counts (e.g., flux boosting, incompleteness, source blending) with the use of Monte Carlo simulations following the exact procedure by \citet{Nowotka2022}. These simulations determine the underlying count model able to correctly match the raw number counts in each field, and have been shown to be consistent to previously published results for both a known blank field, the Chandra Deep Field-North (CDF-N; \citealt{TC2013a}), and an overdense field, MAMMOTH-I (\citealt{FAB2018b}). 
We briefly summarize how the simulations proceed. More details can be found in \citet{Nowotka2022}.

First, to ease comparison with previous works, and in particular with the chosen fiducial blank field count model (\citealt{Geach2017}), we parametrized the underlying count model as a Schechter function of the form

\begin{equation}
\frac{dN}{dS}=\frac{N_0}{S_0}\left(\frac{S}{S_0}\right)^{-\gamma}{\rm exp}\left(-\frac{S}{S_0}\right)    
,\end{equation}
where $N_0$ (units of deg$^{-2}$) is the normalization and $S_0$ (units of mJy) is the characteristic flux.
To avoid degeneracy between $N_0$ and $S_0$, we keep the value of $S_0$ fixed to 2.5, as done in \citet{Geach2017} and \citet{Nowotka2022}.
We fit this function to the raw number counts to obtain the parameters of an initial model used to inject mock sources in the jackknife map at random positions. We injected sources down to 1~mJy, the same lowest flux adopted by the model in \citet{Geach2017}, to which we compare our results to estimate overdensities. 
We constructed in this way 100 simulated maps for each set of model parameters, and obtain the mean recovered counts. A new set of counts was then created by multiplying the raw counts by the ratio between raw counts and mean recovered counts. At this point, the simulation starts a new iteration from the Schechter function fitting. The process was repeated until the recovered counts fall within the $1\sigma$ error of the raw counts. Table~\ref{tab:und_models} lists the parameters of the underlying models found using these simulations.

\begin{table}
\scriptsize
\caption{Best-fit parameters for the underlying count models parametrized as Schechter functions.}
\centering
\begin{tabular}{lccccc}
\hline
\hline
Quasar  & ID\tablefootmark{a} & S/N\tablefootmark{b} &$N_0$                            & $S_0$   & $\gamma$       \\     
        &            & & (deg$^{-2}$)        &  (mJy) &          \\
\hline
Jackpot           & -  & 2.5 & $22200\pm2300$  & $2.50$  & $2.42\pm0.26$  \\ 
MAMMOTH-I         & -  & 2.5 & $20300\pm1700$  & $2.50$  & $2.54\pm0.22$  \\ 
Fabulous          & 13 & 2.5 & $20000\pm3200$  & $2.50$  & $2.42\pm0.39$  \\ 
J~0525-233        &  3 & 2.5 & $33400\pm8500$  & $2.50$  & $1.88\pm0.26$  \\ 
SDSSJ1209+1138    &  7 & 2.6 & $20400\pm3400$  & $2.50$  & $1.85\pm0.24$  \\ 
SDSSJ1025+0452    & 10 & 2.5 & $21600\pm5200$  & $2.50$  & $1.55\pm0.26$  \\ 
SDSSJ1557+1540    & 18 & 2.5 & $29400\pm4500$  & $2.50$  & $1.63\pm0.20$  \\ 
SDSSJ1342+1702    & 24 & 2.5 & $18500\pm6500$  & $2.50$  & $2.28\pm0.46$  \\ 
Q-0115-30         & 29 & 2.5 & $23800\pm7500$  & $2.50$  & $1.72\pm0.36$ \\ 
Q2355+0108        & 46 & 2.7 & $24100\pm4500$  & $2.50$  & $1.37\pm0.64$  \\ 
SDSSJ0819+0823    & 50 & 2.5 & $28200\pm6600$  & $2.50$  & $1.00\pm0.16$  \\ 
J015741-01062957  & -  & 2.5 & $17500\pm7100$  & $2.50$  & $2.13\pm0.53$  \\ 
J020944+051713    & -  & 2.9 & $17300\pm8000$  & $2.50$  & $2.09\pm0.65$  \\ 
J024401-013403    & -  & 2.6 & $19600\pm1800$  & $2.50$  & $2.72\pm0.13$  \\ 
J033900-013318    & -  & 2.5 & $16100\pm3300$  & $2.50$  & $1.00\pm0.18$  \\ 
J111113-080401    & -  & 2.6 & $26900\pm6700$  & $2.50$  & $1.77\pm0.30$  \\ 
J193957-100241    & -  & 2.5 & $27900\pm4800$  & $2.50$  & $1.59\pm0.18$  \\ 
J221527-161133    & -  & 2.5 & $18100\pm4000$  & $2.50$  & $2.37\pm0.42$  \\ 
J230301-093930    & -  & 2.5 & $21500\pm5500$  & $2.50$  & $1.00\pm0.17$ \\ 
J233446-090812    & 45 & 2.7 & $17700\pm5500$  & $2.50$  & $1.00\pm0.18$ \\ 
\hline
\end{tabular}
\tablefoot{
\tablefoottext{a}{ID in QSO MUSEUM.}
\tablefoottext{b}{S/N threshold used for source extraction for computing the number counts.}
}
\label{tab:und_models}
\end{table}

We confirmed that our Monte Carlo simulations properly account for observational biases by 
creating 500 simulated maps for each field  
with mock sources injected into the respective jackknife map following the estimated underlying models. Figures~\ref{fig:counts}, \ref{fig:countsMUSEUM}, and \ref{fig:countsMAGG} show the mean recovered counts together with their 90\% confidence intervals. The simulated counts are in good agreement with the raw counts for all fields. The brightest bins for the MAMMOTH-I ELAN, J0525-2338, and J230301-093930 contain only one or two sources and their counts are not substantially different from the model. These bright sources are the quasars or sources suspected to be associated with these systems given the low probability of a chance detection (\citealt{FAB2018b,Nowotka2022}). Similarly, for the Fabulous ELAN and SDSSJ~1342+1702, we applied the convergence criteria but neglecting the last bin, which clearly deviates from a Schechter due to the presence of a few bright sources likely associated with the systems (see, e.g., \citealt{FAB2022} for the Fabulous ELAN).

Finally, to test whether this procedure is able to assess the presence of an overdensity, we create 500 additional simulated maps, but this time injecting sources according to the fiducial blank field model from \citet{Geach2017} with the parameters $N_0=7180\pm1220$~deg$^{-2}$,$S_0=2.5$~mJy, and $\gamma=1.5\pm0.4$. Figure~\ref{fig:counts}, \ref{fig:countsELAN}, \ref{fig:countsMUSEUM}, and \ref{fig:countsMAGG} show that the realizations of the blank field model (gray shaded region) produce recovered counts that are lower than the raw counts in most fields, highlighting the presence of overdensities. We compute overdensity factors in Sect.~\ref{sec:overd}.

\section{Results}
\label{sec:results}

\subsection{Overdensity factors}
\label{sec:overd}

Our analysis of the differential number counts and the identification of the underlying models point out that most of the targeted fields are overdense in comparison to the reference blank field model by \citet{Geach2017}. In this section we quantify the overdensity factor for each field following the two approaches presented in \citet{Nowotka2022}.

First, the raw counts in each flux bin are corrected for observational biases by dividing them by the ratios of the mean recovered counts and the obtained underlying model. The blank field model is then fit to this corrected counts with only the normalization, $N_0$, as a free parameter. The overdensity factors are then estimated as the ratio between the obtained normalization and the blank field model normalization. In Table~\ref{tab:ov_factors} we list the obtained overdensity factors together with their errors taking into account both Poisson noise and uncertainties from the simulations. We find a weighted average overdensity factor for ELAN fields of $1.9\pm0.2$, which agrees within 1$\sigma$ of the value found in \citet{Nowotka2022} using the same method. For the overall quasar fields we instead obtain a weighted average overdensity factor of $1.8\pm0.1$, with the QSO MUSEUM sample showing higher values with respect to the MAGG sample, $2.3\pm0.2$ and $1.6\pm0.1$, respectively.

\begin{table}
\scriptsize
\caption{Overdensity factors estimated for each field.}
\centering
\begin{tabular}{lcccc}
\hline
\hline
Quasar  & ID\tablefootmark{a} & $\delta_{N_0}$\tablefootmark{b} & $\delta_{\rm cumul}$\tablefootmark{c}         \\     
\hline
Jackpot           & -  &  $1.9\pm0.4$   &   $3.4\pm0.6$  \\ 
MAMMOTH-I         & -  &  $2.3\pm0.4$   &   $3.9\pm0.7$  \\ 
Fabulous          & 13 &  $1.7\pm0.3$   &   $3.0\pm0.6$  \\ 
J0525-233         &  3 &  $3.2\pm0.6$   &   $4.0\pm1.6$  \\ 
SDSSJ1209+1138    &  7 &  $2.1\pm0.4$   &   $2.7\pm0.7$  \\ 
SDSSJ1025+0452    & 10 &  $2.7\pm0.6$   &   $2.9\pm1.1$  \\ 
SDSSJ1557+1540    & 18 &  $3.5\pm0.7$   &   $3.9\pm1.0$  \\ 
SDSSJ1342+1702    & 24 &  $1.1\pm0.3$   &   $2.1\pm1.0$  \\ 
Q-0115-30         & 29 &  $2.7\pm0.5$   &   $3.1\pm1.5$  \\ 
Q2355+0108        & 46 &  $3.4\pm0.8$   &   $3.7\pm2.6$  \\ 
SDSSJ0819+0823    & 50 &  $6.6\pm1.3$   &   $4.9\pm1.8$  \\ 
J015741-01062957  & -  &  $1.5\pm0.3$   &   $2.0\pm1.2$  \\ 
J020944+051713    & -  &  $1.0\pm0.3$   &   $1.8\pm1.5$  \\ 
J024401-013403    & -  &  $1.1\pm0.2$   &   $1.8\pm0.4$  \\ 
J033900-013318    & -  &  $3.4\pm0.7$   &   $2.8\pm1.0$  \\ 
J111113-080401    & -  &  $3.1\pm0.6$   &   $3.3\pm1.4$  \\ 
J193957-100241    & -  &  $3.0\pm0.6$   &   $3.7\pm1.1$  \\ 
J221527-161133    & -  &  $1.5\pm0.3$   &   $1.8\pm0.7$  \\ 
J230301-093930    & -  &  $4.5\pm0.9$   &   $3.8\pm1.5$  \\ 
J233446-090812    & 45 &  $3.4\pm0.8$   &   $3.3\pm0.8$  \\ 
\hline
\end{tabular}
\label{tab:ov_factors}
\tablefoot{
\tablefoottext{a}{ID in QSO MUSEUM.}
\tablefoottext{b}{Overdensity factor estimated as ratio of model normalizations with respect to the reference blank fields.}
\tablefoottext{c}{Overdensity factor estimated as ratio of the cumulative number of sources with respect to the reference blank fields.}
}
\end{table}

This first method is appropriate if the shape of the underlying model is consistent with that of the blank field model. However, for many of the fields there is evidence that the models are steeper, meaning that the overdensity could be flux dependent (Figs.~\ref{fig:counts}, \ref{fig:countsMUSEUM}, and \ref{fig:countsMAGG}), and higher than estimated with the first approach (\citealt{Nowotka2022}). To check this, we computed the cumulative number of sources for each field by integrating the models in the flux range defined by the minimum and maximum source fluxes, and compare it with the blank sky value within the same range.  
As expected, we find on average larger overdensity factors of $3.4\pm0.4$, $2.5\pm0.2$, $3.1\pm0.4$, and $2.3\pm0.3$, for the ELAN, all 17 quasars, and the QSO MUSEUM and MAGG fields, respectively.
The overdensity factors obtained with this second approach for all the fields are also listed in Table~\ref{tab:ov_factors}. Since this second approach does not suffer from any bias in the selection of a fiducial model shape, we base our discussion on these values (see Sect.~\ref{sec:overdensities}). 

We stress that the data at 450~$\mu$m do not provide enough statistics for a robust analysis of the counts as done for the 850~$\mu$m counts. However, for the ELAN fields in which we have better sensitivity, we checked the number of detected sources for fluxes above the 80\% and 50\% completeness level and compare it to the available blank field values at 450~$\mu$m (e.g., \citealt{TC2013b,Geach2013,Zavala2017}), taking into account the completeness and the deboosted flux for our sources. We found that the ELAN fields have on average $1^{+2.3}_{-1.0}$ ($6^{+5.8}_{-3.3}$) sources for fluxes $>30.4$~mJy ($>24.0$~mJy), which are consistent with the average blank field values from the aforementioned works, $0.6^{+1.3}_{-0.6}$ ($1.5^{+2.8}_{-1.4}$). 
The absence of a 450~$\mu$m excess provides further evidence that the 850~$\mu$m overdensities are likely at $z>2$.
Further, we note that most of the sources in our catalogs have a flux ratio $f_{850}/f_{450}$ consistent with expectations from the SMG template of the ALESS survey (\citealt{Swinbank2014,daCunha2015}) redshifted to our targeted fields. 
In particular, we found that only 1.5\% of all the sources in our catalogs have the flux ratio lower than the ALESS expectations. It is therefore likely that most of the detected sources are at a redshift similar to the targeted fields.

\FloatBarrier

\subsection{Source counts in radial apertures}
\label{sec:apert}

\begin{figure*}
\centering
\includegraphics[width=1.0\textwidth]{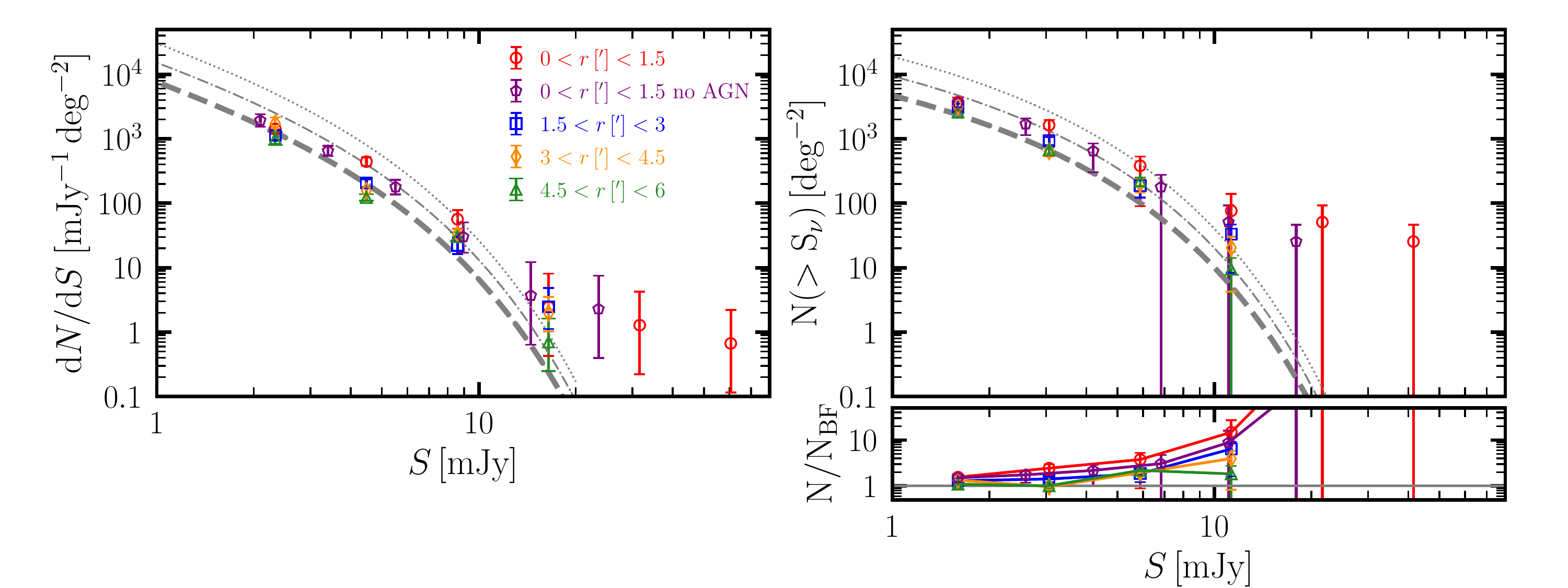}
\includegraphics[width=1.0\textwidth]{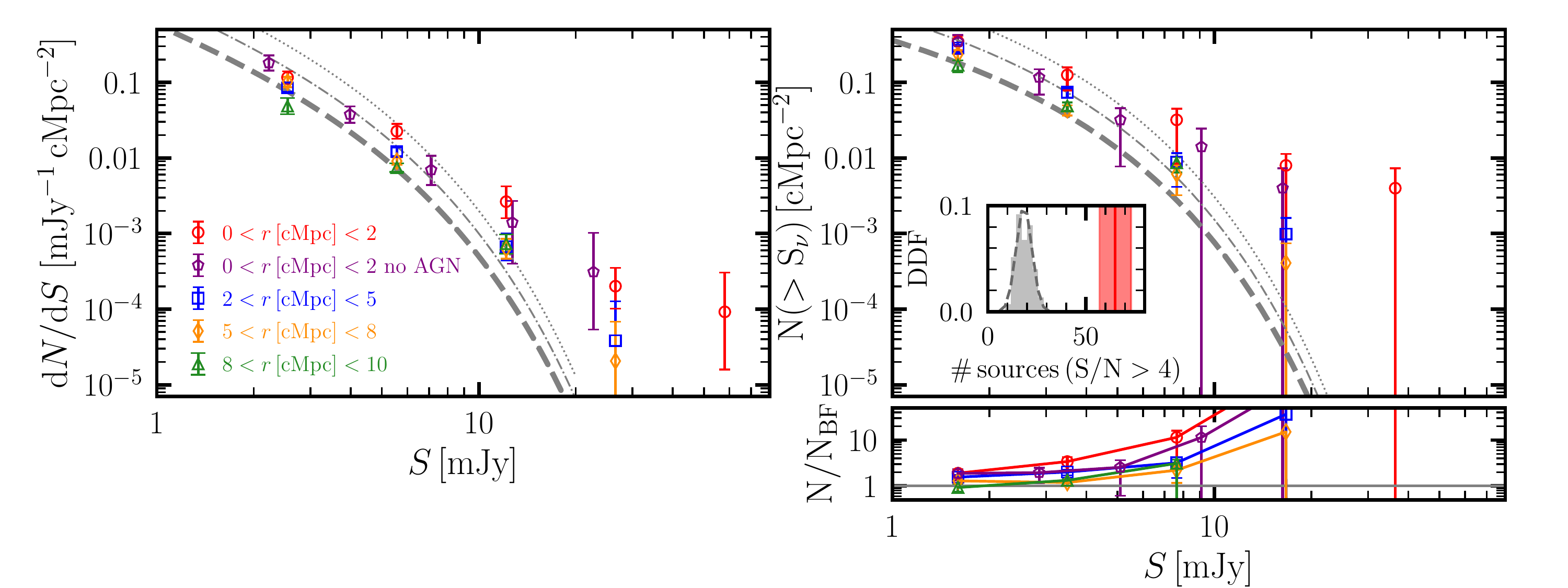}
\caption{Differential and cumulative number counts obtained in apertures. \emph{Top:} Comparison of the differential (left panel) and cumulative (right panel) number counts obtained in apertures (in arcminutes; see the legend and Sect.~\ref{sec:apert}) around the targeted systems with respect to an estimate for blank fields (dashed gray; \citealt{Geach2017}). The dot-dashed and dotted lines show models with the same $\gamma$ and $S_0$, but with $2\times$ and $4\times$ higher normalization ($N_0$), respectively. The panel in the bottom-right corner shows the ratio between the number of sources for each aperture and the reference blank field model. For a better visualization of the data points at low fluxes, this plot is cut at a ratio of 50, and therefore it does not show the uncertain data points at large fluxes, which by far exceed the blank field model. A  trend of decreasing overdensity from the inner to the outer aperture is evident. \emph{Bottom:} Similar to the top panel, but for the differential number counts obtained in apertures defined in comoving megaparsecs (see the legend and Sect.~\ref{sec:apert}) around the targeted systems with respect to an estimate for blank fields (dashed gray; \citealt{Geach2017}), but translated to comoving megaparsecs for the median redshift of our sample. The inset plot compares the number of detected sources with S/N$>4$ for the inner bin with the density distribution function obtained from 500 realizations of 20 mock blank fields, similar to what is done in Fig.~\ref{fig:sig_counts_inAp} with apertures in arcminutes.}
\label{fig:countsInAp}
\end{figure*}

The number of fields and the large number of sources detected at high significance (S/N$>4$) at 850~$\mu$m allow us to study the source counts in radial apertures around the targeted systems. A similar analysis has been performed by \citet{Zeballos2018}, who targeted a sample of 17 high-redshift radio galaxies (HzRGs) at $0.5<z<6.3$. They used radial apertures defined by radii every 1.5~arcmin, out to 4.5~arcmin, finding marginal evidence of an overdensity ($\sim3\times$ the blank fields) for the central aperture ($0<r<1.5$~arcmin). 
Given the large spread in redshift of their sources, this result could be biased as overdensities span different physical scales at different epochs.  

For this analysis we decided to follow two approaches: following the analysis in \citet{Zeballos2018} to allow a one-to-one comparison, but also repeating the analysis using apertures in comoving units. 
First, we compute the corrected source counts in apertures similar to \citet{Zeballos2018}: a circle of radius $r=1.5$~arcmin, and then three additional annuli with $1.5\arcmin<r<3\arcmin$, $3\arcmin<r<4.5\arcmin$, and $4.5\arcmin<r<6\arcmin$, basically covering most of the effective area of each map. 
We start from our catalogs and correct them for completeness and flux boosting using curves obtained as described in Sect.~\ref{sec:compl}. However, since the rms in our maps degrades toward their outskirts, we compute completeness and flux boosting for the different apertures in each map. Each source flux is then corrected by dividing by the obtained median boosting factor at its observed flux.  
The contribution of each source to the counts is computed by dividing its single occurrence by the completeness at its observed flux. We then computed the number counts in logarithmically spaced flux bins, covering the minimum and maximum flux in each aperture. 
The top panels in Fig.~\ref{fig:countsInAp} show the differential and cumulative number counts obtained with this method. For the inner aperture, we compute the source counts with two approaches: including (red circles) or excluding (purple pentagons) the possible counterparts to the targeted systems (i.e., sources at $r<10$~arcsec; see the tables in Appendix~\ref{app:catalogs}). The second approach is binned differently as the brightest sources are neglected. For both approaches, it can be seen that the inner aperture is consistently more overdense than the others, having counts $\sim3\times$ higher than blank fields at $S_{850\mu m}>4$~mJy.

\begin{table}
\scriptsize
\caption{Differential and cumulative source counts computed in areas centered on the targeted systems.}
\centering
\begin{tabular}{rrrr}
\hline
\hline
$S$     & d$N$/d$S$          & $S$            & $N(>S)$         \\      
(mJy)   & (mJy$^{-1}$)       & (mJy)          & (deg$^{-2}$)    \\
\hline
\multicolumn{4}{c}{$0<r<1.5$~arcmin} \Tstrut\Bstrut\\ 
\hline 
2.3  & 1427$^{+228}_{-267}$     &  1.6     &  3723$^{+665}_{-893}$   \Tstrut \\ 
4.5  &  442$^{+64}_{-74}$       &  3.1     &  1628$^{+331}_{-501}$   \Tstrut \\ 
8.6  &   57$^{+16}_{-21}$       &  5.9     &   382$^{+150}_{-292}$   \Tstrut \\ 
16.5 &    2.5$^{+5.7}_{-2.0}$   &  11.3    &    76$^{+63}_{-176}$  \Tstrut \\ 
31.6 &    1.3$^{+2.9}_{-1.1}$   &  21.6    &    51$^{+42}_{-117}$ \Tstrut \\ 
60.6 &    0.7$^{+1.5}_{-0.6}$   &  41.5    &    25$^{+21}_{-59}$ \Tstrut\Bstrut \\ 
\hline
\multicolumn{4}{c}{$1.5<r<3$~arcmin} \Tstrut\Bstrut\\
\hline
2.3  & 1139$^{+179}_{-210}$ &  1.6     &  3118$^{+382}_{-455}$  \Tstrut  \\ 
4.5  &  206$^{+26}_{-29}$   &  3.1     &   944$^{+118}_{-147}$  \Tstrut  \\ 
8.6  &   22$^{+6}_{-8}$     &  5.9     &   188$^{+45}_{-66}$    \Tstrut  \\ 
16.5 &  2.5$^{+2.4}_{-1.3}$ &  11.3    &    33$^{+14}_{-25}$    \Tstrut\Bstrut \\ 
\hline
\multicolumn{4}{c}{$3<r<4.5$~arcmin} \Tstrut\Bstrut\\
\hline
2.3  & 1748$^{+334}_{-405}$ &  1.6     &  3206$^{+582}_{-703}$  \Tstrut  \\ 
4.5  &  157$^{+18}_{-20}$   &  3.1     &   640$^{+91}_{-108}$   \Tstrut  \\ 
8.6  &   33$^{+6}_{-7}$     &  5.9     &   198$^{+40}_{-52}$    \Tstrut  \\ 
16.5 &  2.0$^{+0.9}_{-1.6}$ &  11.3    &    20$^{+10}_{-16}$    \Tstrut\Bstrut \\ 
\hline
\multicolumn{4}{c}{$4.5<r<6$~arcmin} \Tstrut\Bstrut\\
\hline
2.3  & 1018$^{+215}_{-266}$ &  1.6     &  2613$^{+383}_{-471}$   \Tstrut \\ 
4.5  &  123$^{+14}_{-15}$   &  3.1     &   671$^{+68}_{-80}$     \Tstrut \\ 
8.6  &   30$^{+5}_{-5}$     &  5.9     &   222$^{+29}_{-38}$     \Tstrut \\ 
16.5 &  0.7$^{+0.9}_{-0.5}$ &  11.3    &    9.5$^{+4.7}_{-9.6}$  \Tstrut\Bstrut \\ 
\hline
\end{tabular}
\label{tab:source_counts_inAp}
\end{table}

Interestingly, as can be better seen below the top-right panel of Fig.~\ref{fig:countsInAp}, there is a promising trend indicating a gradual decrease in the overdensity for $S_{850\mu}>4$~mJy from the inner to the outer aperture. The outer aperture seems still slightly overdense ($\sim2\times$) with respect to the field, probably meaning that the physical extent of the overdensities around these systems is larger than our maps. This finding would be in agreement with theoretical expectations by \citet{Chiang2013} that place the effective radius of protoclusters associated with halo masses $M_{\rm DM}>10^{12}$~M$_{\odot}$ at physical scales similar to the size of our maps ($\sim 10$~cMpc). 
Similar predictions have been provided by additional simulations (e.g., \citealt{Muldrew2015}), which qualitatively agree with the large extents for 
protoclusters reported in observations (e.g., \citealt{Casey2016}). Table~\ref{tab:source_counts_inAp} lists the differential and cumulative number counts for all apertures. For the inner aperture we only list the values including all sources, as the results of the two aforementioned approaches are consistent within the uncertainties.

\begin{figure}
\centering
\includegraphics[width=0.8\columnwidth]{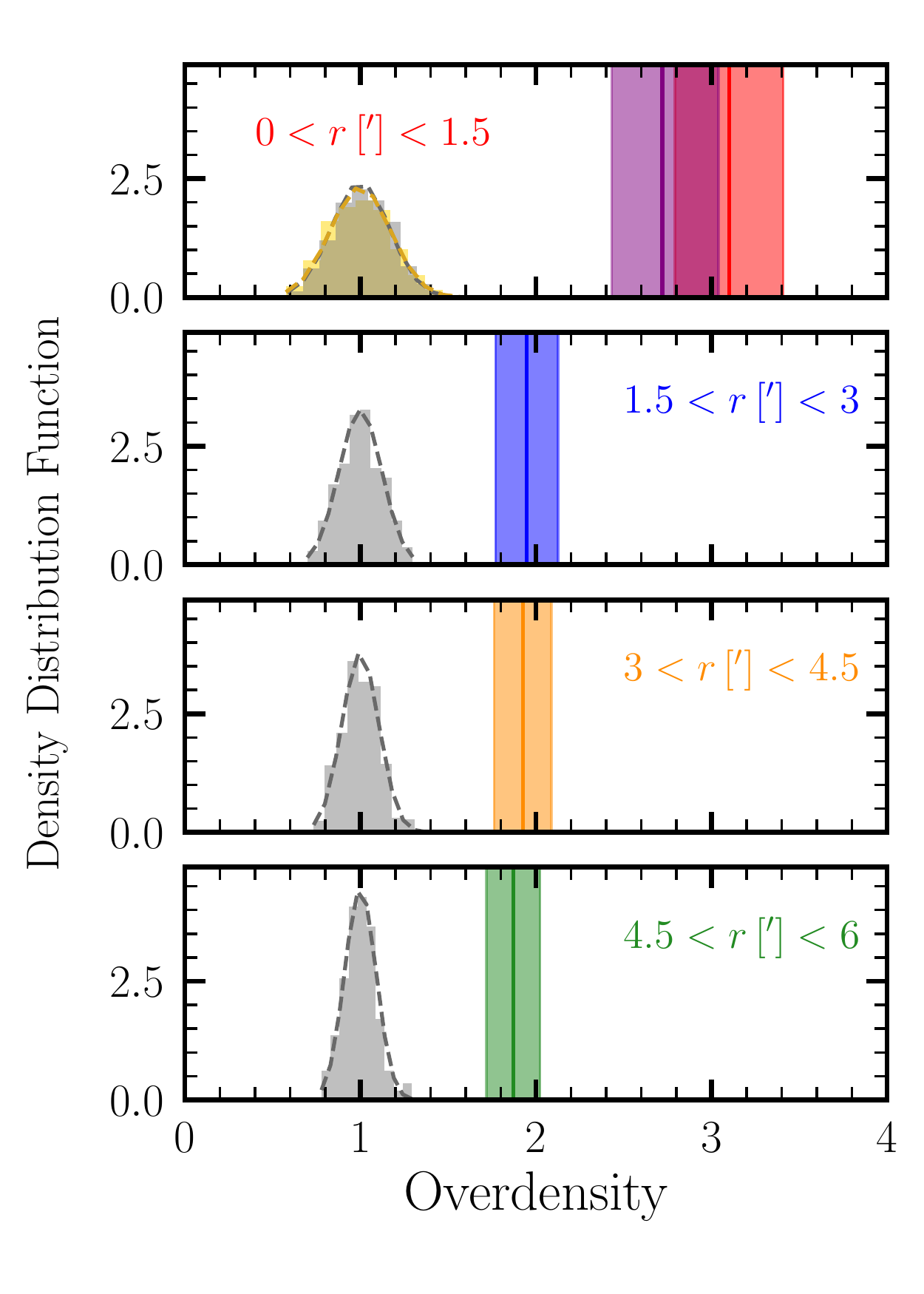}
\caption{Overdensities of detected sources with S/N$>4$ in different apertures in arcminutes (colored vertical lines) compared to the density distribution functions of overdensities from 500 realizations of 20 mock blank fields with similar noise properties as the 20 science maps, each divided by their mean value (gray distributions). The overdensity in each aperture is detected at a high significance, with the strength of the overdensity decreasing from the central to the outer aperture (see Table~\ref{tab:sig_inAp}). The errors on overdensities take into account the Poisson uncertainties on the counts. The significance of the overdensities also taking the Gaussian ``background'' into account is given in Table~\ref{tab:sig_inAp}.}
\label{fig:sig_counts_inAp}
\end{figure}

We quantified the significance of the obtained overdensities in different apertures by using the 10000 (500 per system) synthetic maps for blank fields obtained in Sect.~\ref{sec:NC} using the jackknife maps. 
Indeed, the comparison between real and mock maps allows us to estimate the significance of the radial trend because in the mock maps, sources are injected at random positions.
We focused on comparing the total number of sources detected in all 20 fields for S/N$>4$, which is the threshold for our catalogs. Specifically, we first computed the total number of sources per aperture for our catalogs, and then repeated the same experiment 500 times for samples of 20 mock blank fields, one for each targeted system (i.e., with the noise properties of that field). Comparing the number of sources obtained in the observed maps and in the mock maps allows us to obtain a significance without the need of deboosting or completeness corrections. Figure~\ref{fig:sig_counts_inAp} shows the density distribution functions obtained in each aperture for the 500 realizations of blank fields (histograms),  which can be well fit by a Gaussian, divided by their mean value. We then computed the overdensity in each aperture as the ratio between the total number of observed counts divided by the mean of the Gaussian distribution for the respective blank field realizations. The total number of observed counts (vertical line with $1\sigma$ Poisson uncertainty) in each aperture exceeds the blank field by a factor of at least $\sim2$ in the outer annuli and increases to $\sim3$ in the central aperture, in agreement with the previous analysis of number counts (top panels of Fig.~\ref{fig:countsInAp}). We estimated the significance of the observed overdensities in each aperture by assuming that the uncertainty on our measurement is described by Poisson statistics in the presence of a Gaussian background (the distributions for the blank-field). The significance can be then obtained following \citet{Vianello2018} (their Eq. 15). We find that the overdensities in each aperture are detected at $\geq 4.7\sigma$.
Table~\ref{tab:sig_inAp} lists the overdensities in each aperture, together with their significance and the parameters of the Gaussians. 

\begin{table}
\scriptsize
\caption{Data for the significance of the overdensities in different apertures around the targeted systems.}
\centering
\begin{tabular}{lcccc}
\hline
\hline
Aperture & Overdensity  & Significance & $\mu_{\rm Gauss}^{\rm BF}$ & $\sigma_{\rm Gauss}^{\rm BF}$
\Tstrut\Bstrut \\       
\hline
$0<r<1.5$~arcmin            & 3.1                 & 6.9  & 32.6   &   5.4 \Tstrut \\ 
$10$~arcsec~$<r<1.5$~arcmin & 2.7                 & 5.7  & 31.2   &   5.4  \\ 
$1.5<r<3$~arcmin            & 1.9                 & 4.7  & 61.6   &   7.5   \\ 
$3<r<4.5$~arcmin            & 1.9                 & 5.2  & 72.2   &   7.6  \\ 
$4.5<r<6$~arcmin            & 1.9                 & 5.4  & 79.6   &   7.1  \\ 
\hline
\end{tabular}
\label{tab:sig_inAp}
\end{table}

Further, we tested whether the use of apertures defined using comoving megaparsecs rather than arcminutes would result in an even stronger detection of overdensities as a function of aperture. We redid the previous analysis using a circle of radius $r=2$ cMpc and then three additional annuli with $2<r<5$, $5<r<8$, and $8<r<10$ cMpc, covering the portion of effective areas in common for different redshifts. The differential and cumulative number counts within these apertures are shown in the bottom panels of Fig.~\ref{fig:countsInAp}. As expected, given the relatively small redshift range, they confirm the results of the previous analysis. In particular, we find that the use of comoving megaparsecs slightly increase the overdensity value. For example, for the inner region we find $3.5\times$ higher counts at $S_{\rm 850\mu m}>4$~mJy. Once again, we quantify the significance of the overdensity, and find that the overdensity in the inner region is detected with a significance of 6.1 (see the inset in the bottom-right panel of Fig.~\ref{fig:countsInAp}). We further characterize the detected overdensities in Sect.~\ref{sec:spatialScale}.

\subsection{Overdensity factors versus system properties}
\label{sec:over_vs_prop}

\FloatBarrier

\begin{figure}
\centering
\includegraphics[width=1.0\columnwidth]{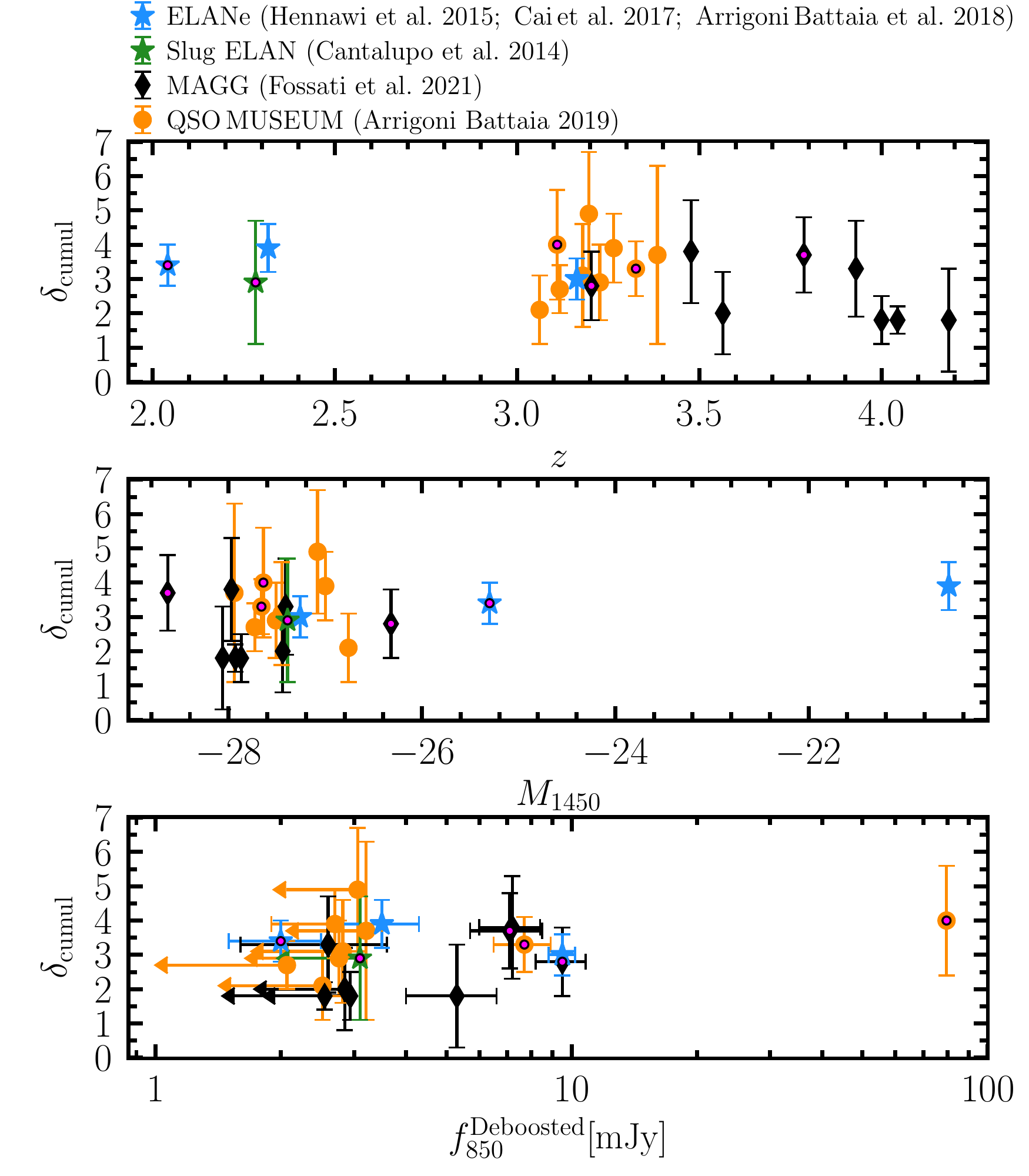}
\caption{Comparison of the obtained overdensity factors (Sect.~\ref{sec:overd}) versus some of the quasar's properties: redshift (top), $M_{1450}$ (middle), and $f_{850}^{\rm Deboosted}$ for
the quasar or ELAN counterpart (bottom). Magenta dots indicate systems with radio-loud sources.}
\label{fig:overFac_vs_QSOprop}
\end{figure}

In this section we investigate the presence of any trend between the obtained overdensity factors and the properties of the quasars and their associated extended Ly$\alpha$ emission. First we compare with the quasar properties. Figure~\ref{fig:overFac_vs_QSOprop} shows the overdensity factors plotted against redshift (top panel), quasar magnitude at rest-frame 1450~\AA, $M_{1450}$ (middle panel), and the deboosted flux at $850\ \mu$m, $f_{850}^{\rm Deboosted}$ for detections within 10~arcsec of the optical quasar position (bottom panel).
Different colors and symbols denote different samples: ELANe targeted in this work (blue stars), Slug ELAN (green star; from \citealt{Nowotka2022}), QSO MUSEUM fields (orange circles), and MAGG fields (black diamonds).
We note some interesting features in the data that need to be verified with larger samples and with spectroscopically confirmed overdensities. 

In the top panel of Fig.~\ref{fig:overFac_vs_QSOprop}, it is remarkable that the three fields at $z\geq4$ show the smallest source excess in the sample. This fact could be due to a weaker coevolution between SMGs and quasars at high redshifts, or to the known redshift distribution of SMGs in blank fields, which peaks at $z\sim2-3$ (e.g., \citealt{Dudzeviciute2020}). In other words, estimating overdensities with respect to blank fields at these high redshifts could result in a stronger dilution of the signal than at $z\sim2-3$. The middle panel is confirming that the luminosity of the central AGN is not linked to its large-scale environment. Quasars with similar absolute magnitudes can have a different excess of submillimeter sources. This is in agreement with clustering studies for quasars that show no dependence on the quasar luminosities in the range ($-28.7<M_i<-23.8$ or about $-27.4<M_{1450}<-22.5$ and in the redshift range $2.2\leq z \leq 3.4$ ; \citealt{Eftekharzadeh2015}).
The bottom panel shows that central radio loud quasars have counterparts at 850~$\mu$m with $f_{850}^{\rm Deboosted}\gtrsim 7$~mJy, and all have 850~$\mu$m source excesses $\gtrsim2.8\times$ the blank fields. Similarly high values are found also for the two ELANe with a radio-loud companion. Six out of 14 (or $43^{+26}_{-17}$\%) radio-quiet fields show lower cumulative overdensities than the radio-loud objects. Therefore, being a radio-loud quasar seems to guarantee a relatively high 850~$\mu$m source excess. This ubiquitous excess around radio-loud quasars is in contrast with the more heterogeneous source counts found around HzRGs (e.g., \citealt{Dannerbauer2014,Zeballos2018}). This fact could be related to the different inclination of the AGN jets with respect to the line-of-sight, and how they are expected to align with the large-scale structure. Indeed, jets should be aligned perpendicular to the large-scale filaments traced by the SMGs' positions (e.g., \citealt{Zeballos2018} and references therein). If this is indeed the case, for some HzRGs (as they have the jet on the plane of the sky), the filaments could be aligned along the line-of-sight, making the overdensity of galaxies more compact and difficult to detect, on average. Instead, radio-loud quasars with unresolved radio emission (as in our sample) are expected to have their jet (if any) almost aligned with our line-of-sight (e.g., \citealt{Sbarrato2022}), implying a filament direction roughly on the plane of the sky. We find a first piece of evidence of this effect when looking at the different results for the overdensities obtained in apertures between our study and \citet{Zeballos2018}. On average, our fields are overdense out to larger distances than those around HzRGs, which only show a clear excess for $<1.5$~arcmin. This hypothesis could be tested by spectroscopically confirming the association of galaxies and reconstructing in 3D the large-scale structure around these systems.

\begin{figure}
\centering
\includegraphics[width=1.0\columnwidth]{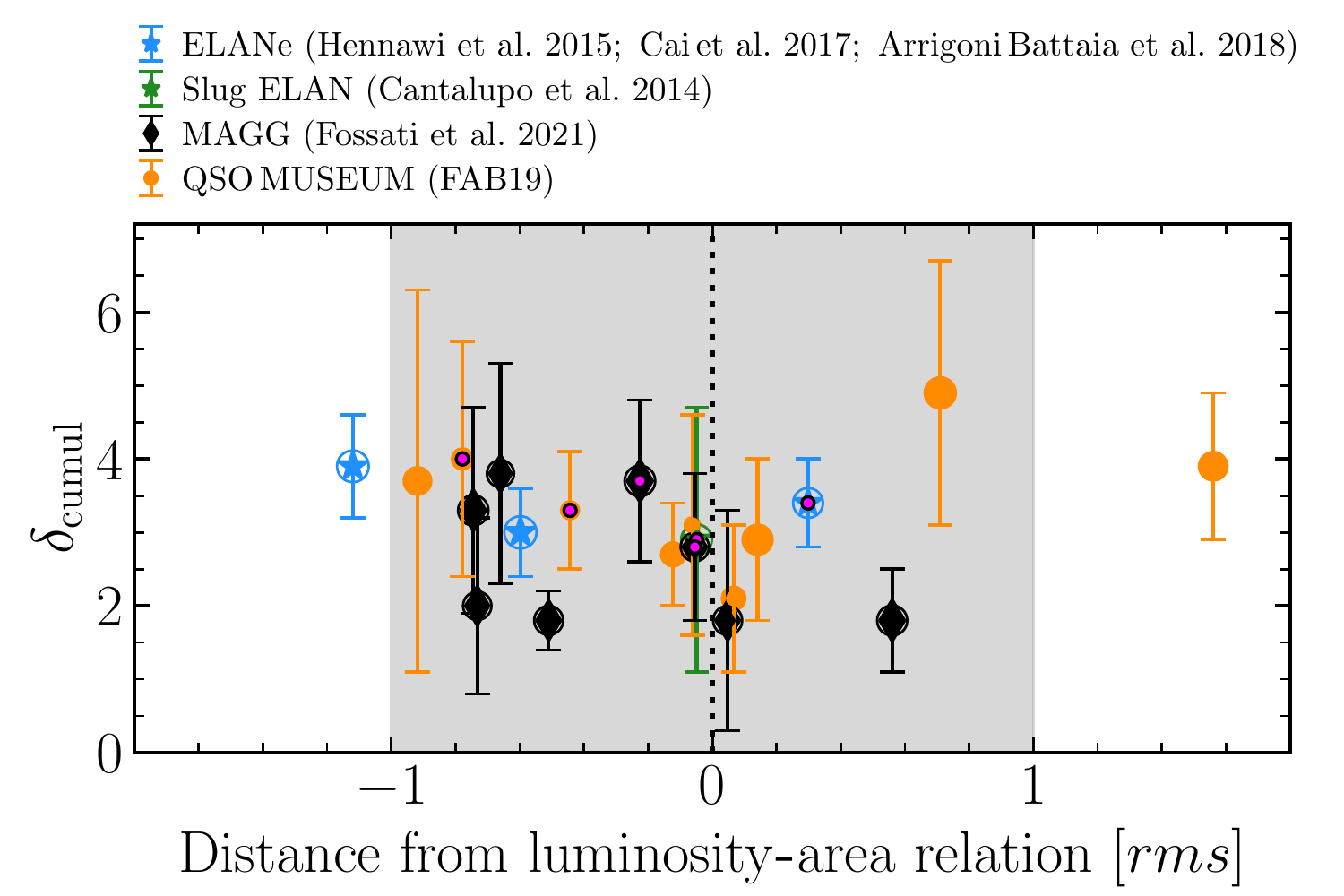}
\caption{Cumulative overdensity factor $\delta_{\rm cumul}$ obtained in Sect.~\ref{sec:overd} versus distance from the luminosity-comoving area relation for quasars' Ly$\alpha$ nebulae, in units of its rms. 
For the luminosity-area relation we use the reference inliers fit in \citet{FAB2023b}. 
Each data point has a circle that is logarithmically scaled following the comoving area of the Ly$\alpha$ nebulae. Magenta dots indicate systems with radio-loud sources. The dotted line with shaded area indicates the relation and one $rms=0.17$~dex.}
\label{fig:LumArea_overFac}
\end{figure}

Second, we looked for relations between the overdensity factors and the properties of the Ly$\alpha$ nebulae.
Specifically, as we are comparing extended structures at different redshifts, we computed luminosities and comoving areas for the Ly$\alpha$ nebulae at a fixed
redshift-corrected common threshold in surface brightness (SB) using the maps from the works by \citet{cantalupo14}, \citet{hennawi+15}, \citet{Cai2016}, \citet{FAB2019}, and \citet{Fossati2021}. As
done in \citet{FAB2023b}, 
we chose the SB threshold $1.23\times10^{-17}$~erg~s$^{-1}$~cm$^{-2}$~arcsec$^{-2}$ at $z=2.0412$ (Jackpot), and obtained the threshold at different redshifts by considering the cosmological SB dimming $(1+z)^4$. At each redshift these values are well above the $2\sigma$ threshold usually used to select extended Ly$\alpha$ emission. 

We then compare the obtained values against the reference inliers fit of the luminosity-area relation in \citet{FAB2023b}, 
and look for any trend with respect to the cumulative overdensity factors.
Figure~\ref{fig:LumArea_overFac} shows the results of this exercise. Specifically, the plot shows the overdensity as a function of the distance from the luminosity-area relation (without taking the absolute value). Also, the size of the data points indicates the comoving area of the nebulae.
Importantly, the MAGG quasar's nebulae lie on the luminosity-area relation discovered in that work, which  
uses only $2<z<3.5$ quasars (dotted line with rms region). Therefore, this fact confirms that the relation is in place for higher redshifts (up to $z=4.2$). Regarding the overdensity factors, given their uncertainties, there is no clear trend with respect to the distance from the relation.

\section{Discussion}
\label{sec:discussion}

\subsection{Do ELANe sit in similarly overdense fields as other quasars?}
\label{sec:overdensities}

\begin{figure}
\centering
\includegraphics[width=0.8\columnwidth]{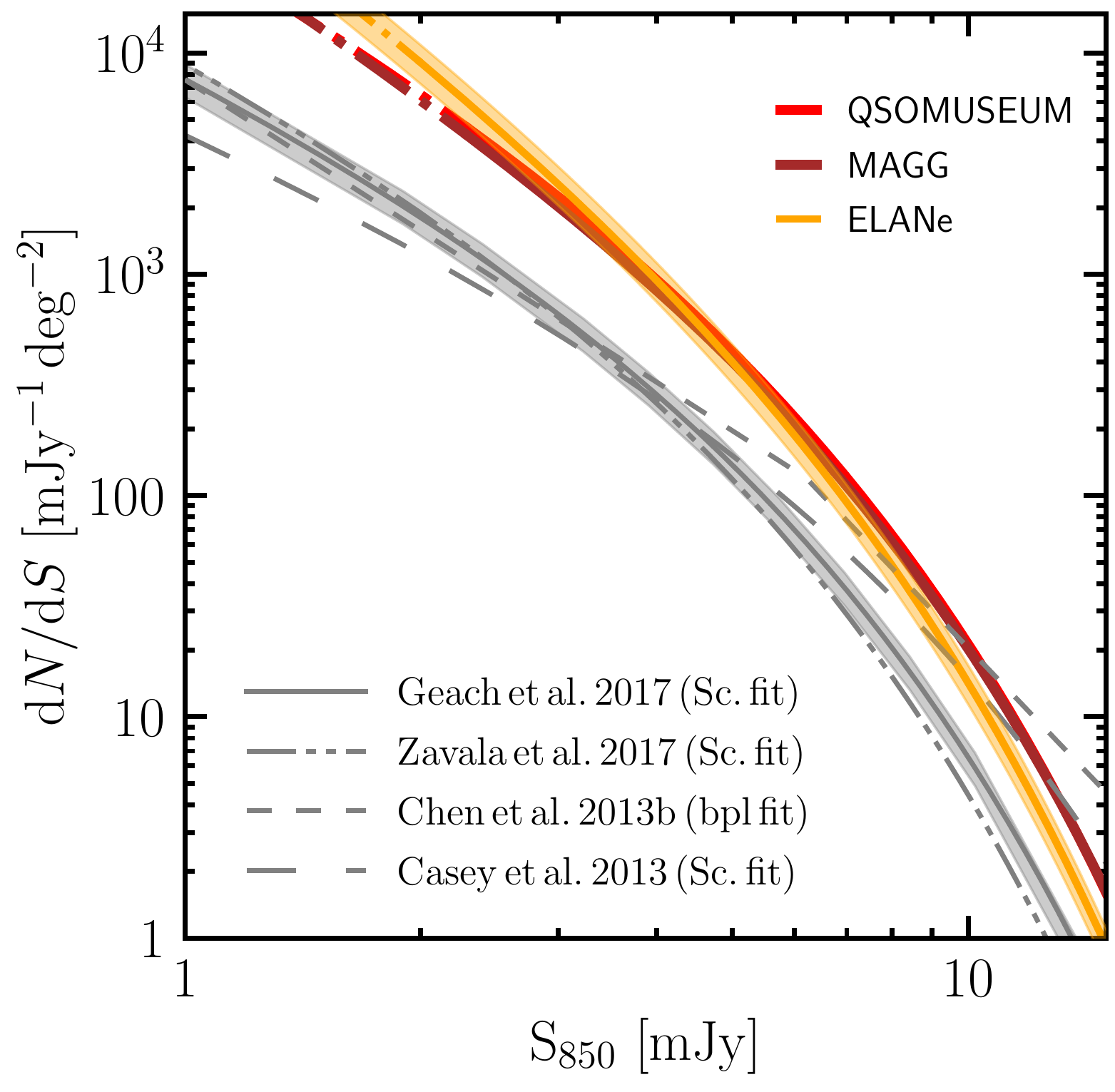}
\caption{Average differential number counts for ELANe (yellow), MAGG (brown), and QSO MUSEUM (red) quasars shown in comparison with those from blank fields (gray). We indicate the representative uncertainties on these average estimates for quasars and blank fields around the ELANe and Geach et al. counts, respectively.}
\label{fig:comp_diff_counts_all}
\end{figure}

\citet{Nowotka2022} obtained 850~$\mu m$ number counts for three ELAN fields, finding on average overdensities of 
$3.6\pm0.6$. As we show in Sect.~\ref{sec:overd}, the overdensities around our sample are similar to those around ELANe. We further emphasize this by obtaining average differential number counts for the ELAN and the other quasars, from QSO MUSEUM and the MAGG sample. Figure~\ref{fig:comp_diff_counts_all} shows the comparison between these average curves and the blank field counts. All three curves show similar overdensities. Given the rarity of the ELANe, this fact could mean that they represent only a short transitory phenomenon in a massive system evolution and/or a peculiar illumination or activity. To verify this picture and the ubiquity of overdensities around the current sample and in general around high-z quasars, we need to spectroscopically confirm the association of the detected sources. Indeed, literature studies start to provide evidence for CO-emitting companions and/or overdensities of CO emitters associated with high-z quasars, but on smaller scales than here probed (e.g., \citealt{Decarli2017,TC2021,JianruiLi2021,GarciaVergara2022}). A follow-up ALMA survey targeting the brightest 850~$\mu$m detections in this work is already on-going (Wang et al. in prep.).

\FloatBarrier

\subsection{A characteristic spatial scale and orientation of the overdensities}
\label{sec:spatialScale}

\begin{figure*}
\centering
\includegraphics[width=0.35\textwidth]{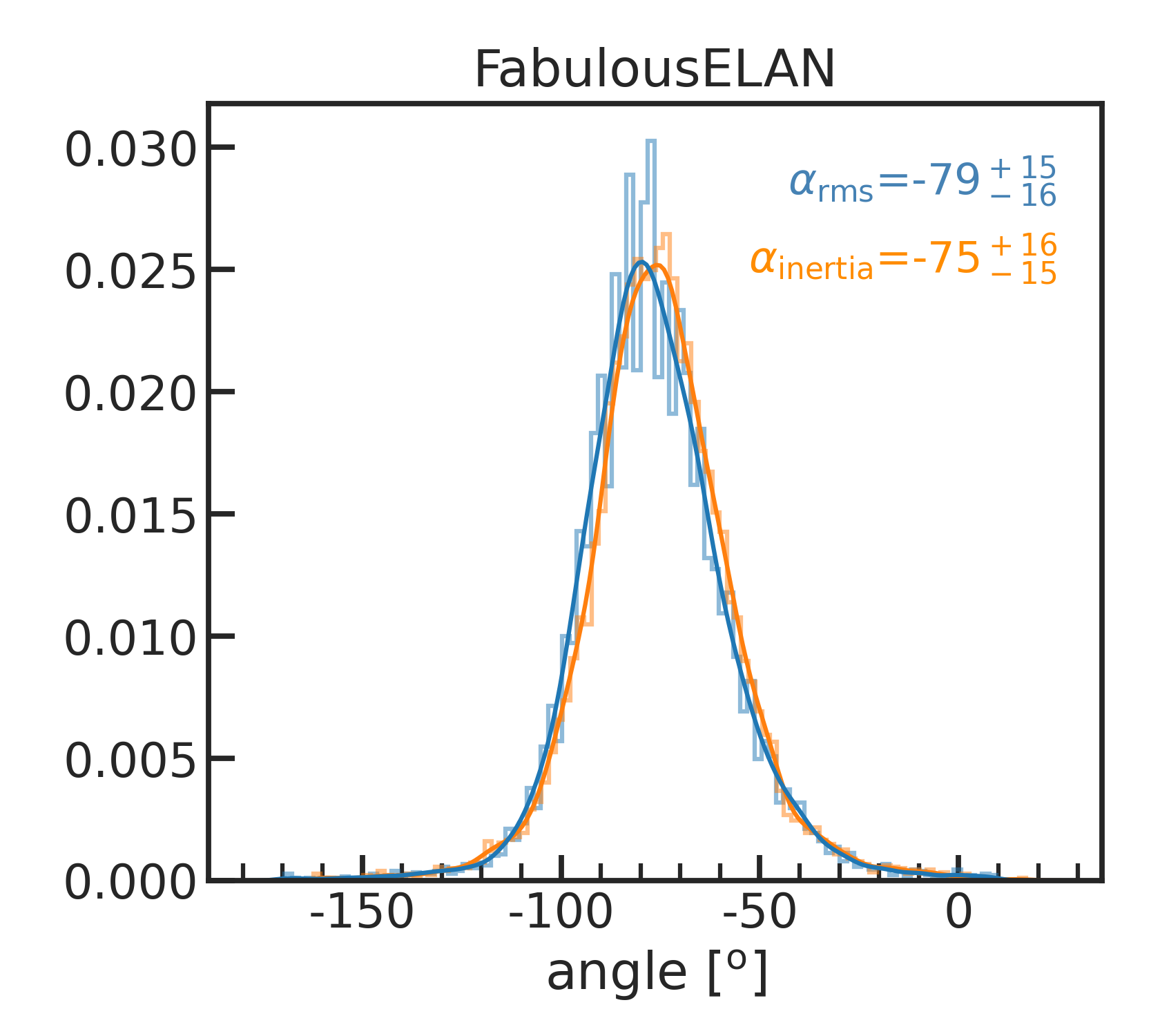}
\includegraphics[width=0.35\textwidth]{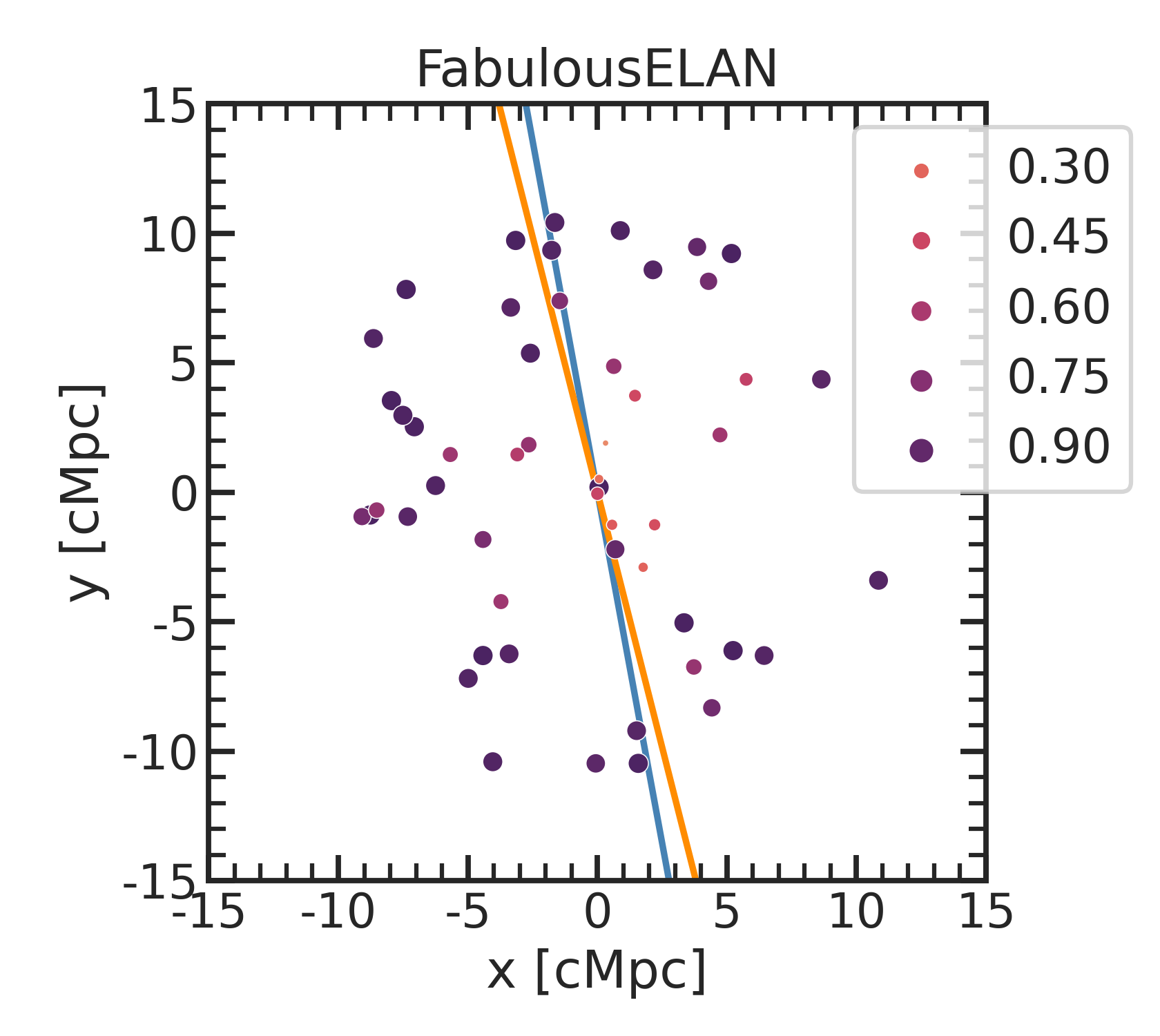}
\includegraphics[width=0.24\textwidth, trim=0 -3.2cm 0 0]{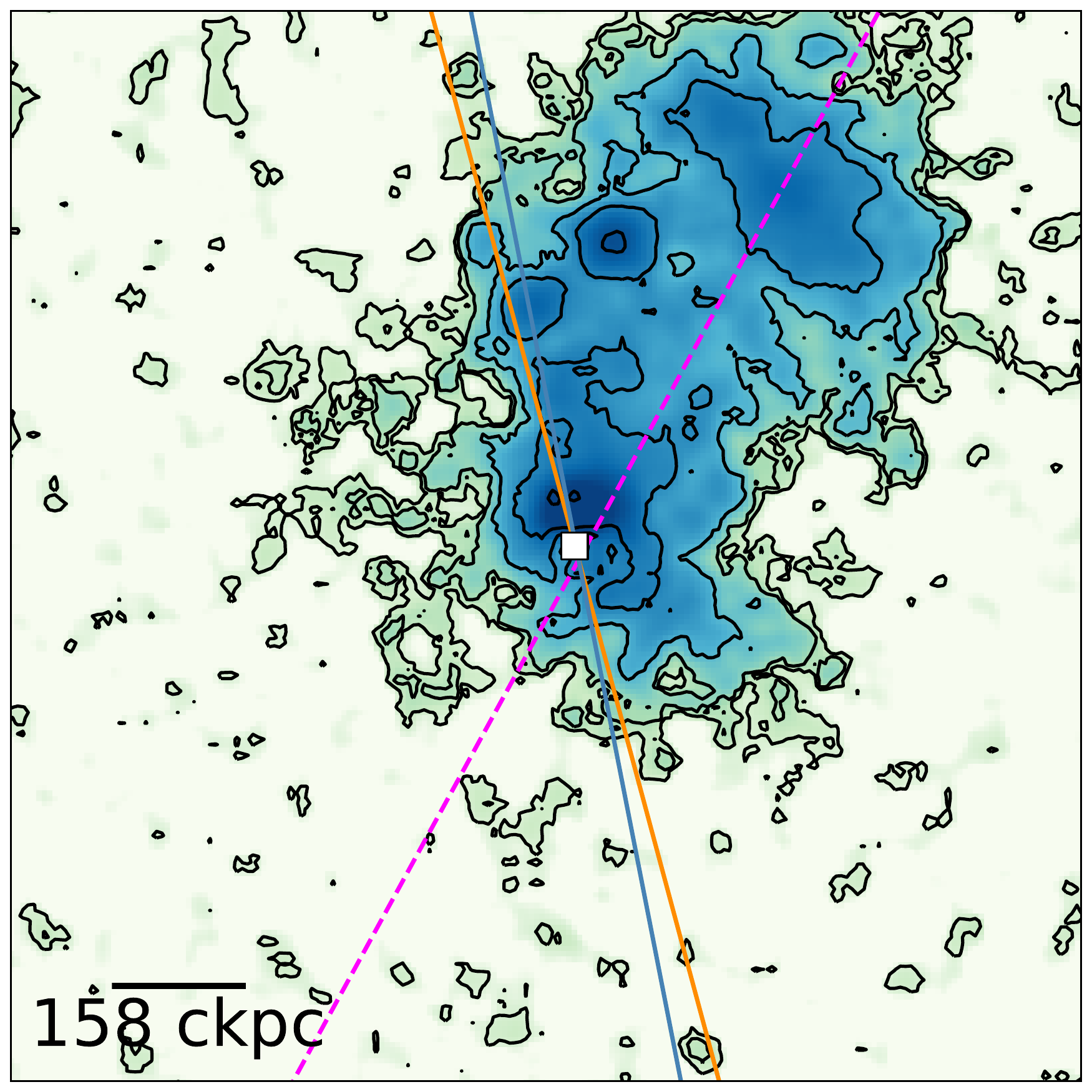}

\includegraphics[width=0.35\textwidth]{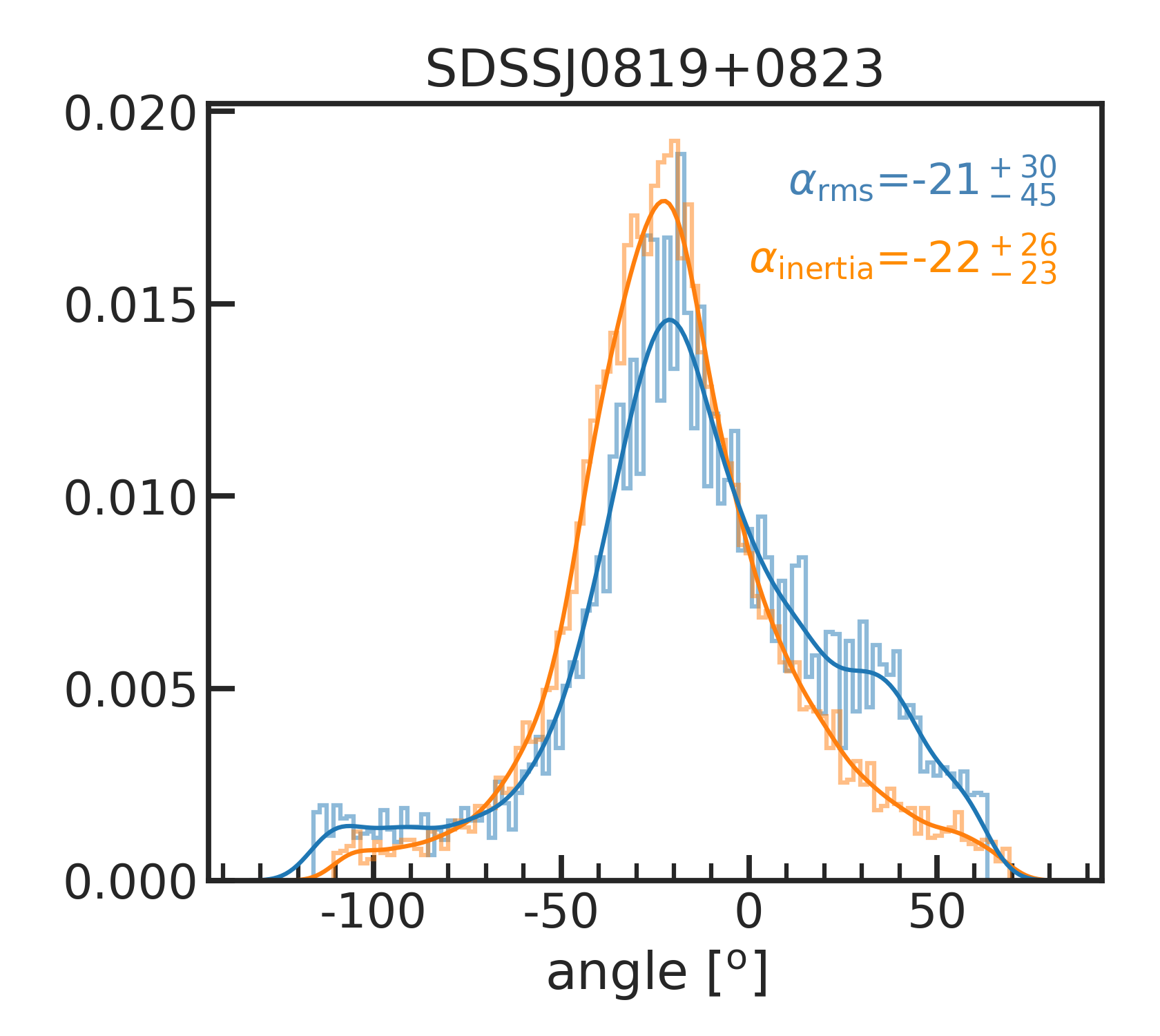}
\includegraphics[width=0.35\textwidth]{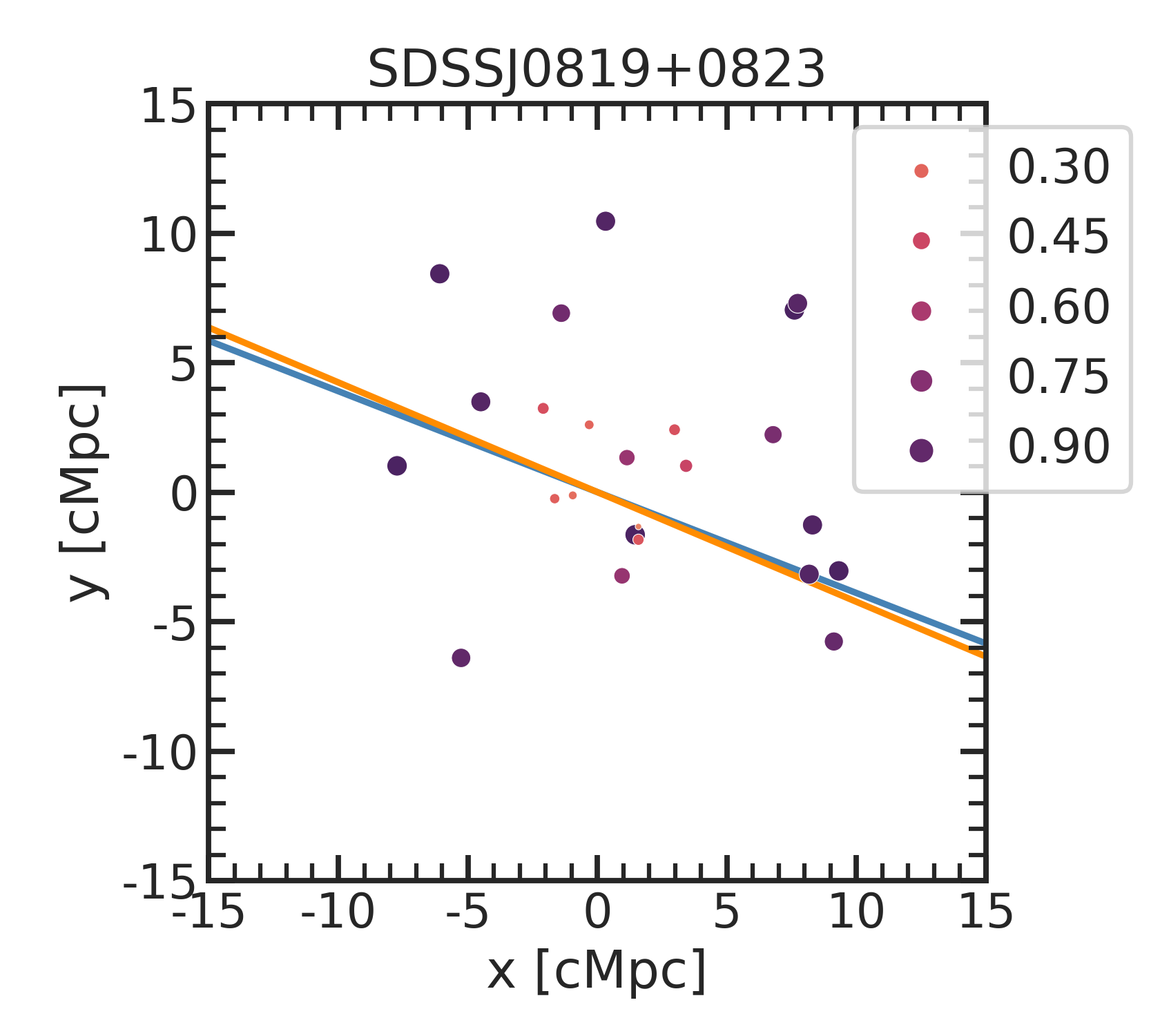}
\includegraphics[width=0.24\textwidth, trim=0 -3.2cm 0 0]{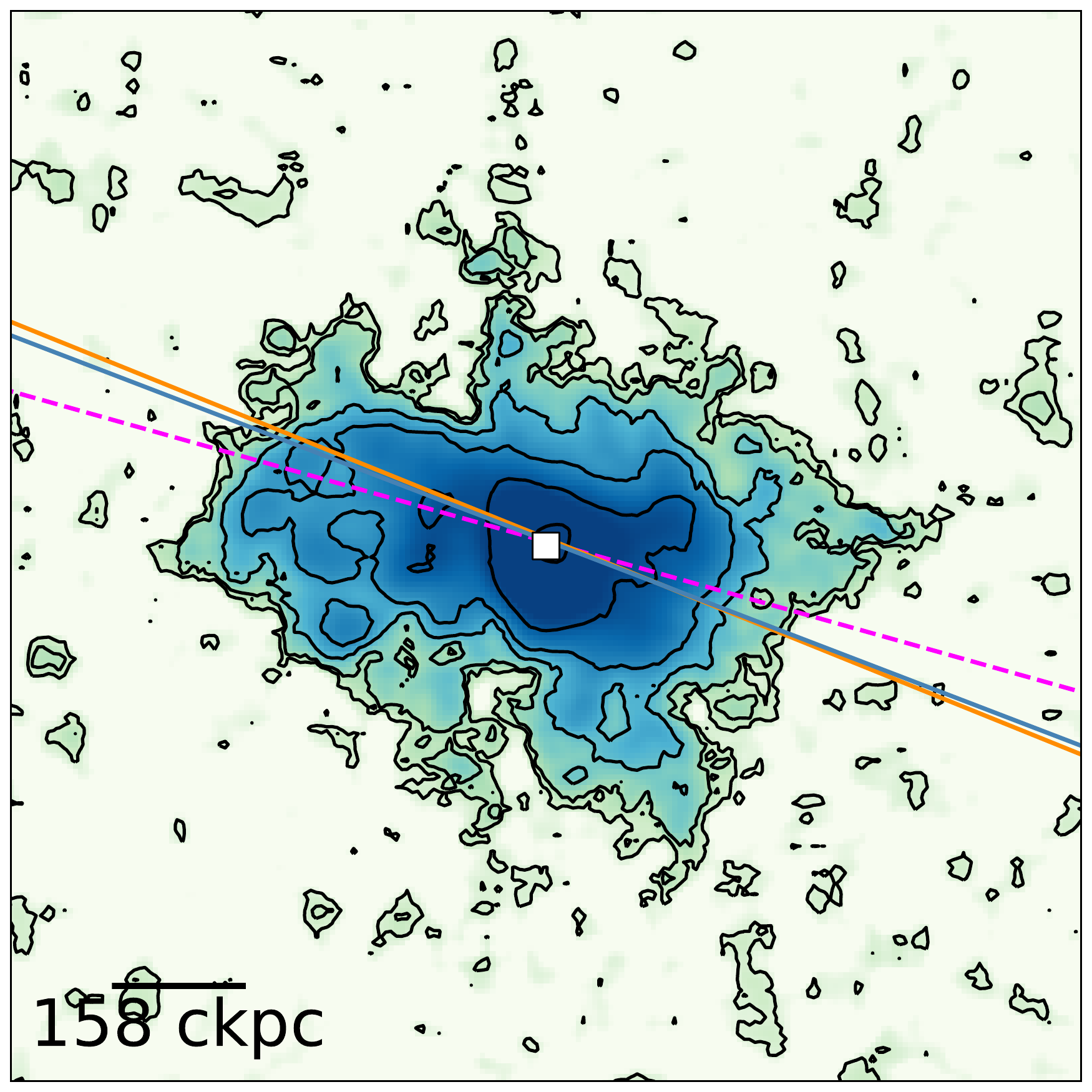}

\includegraphics[width=0.35\textwidth]{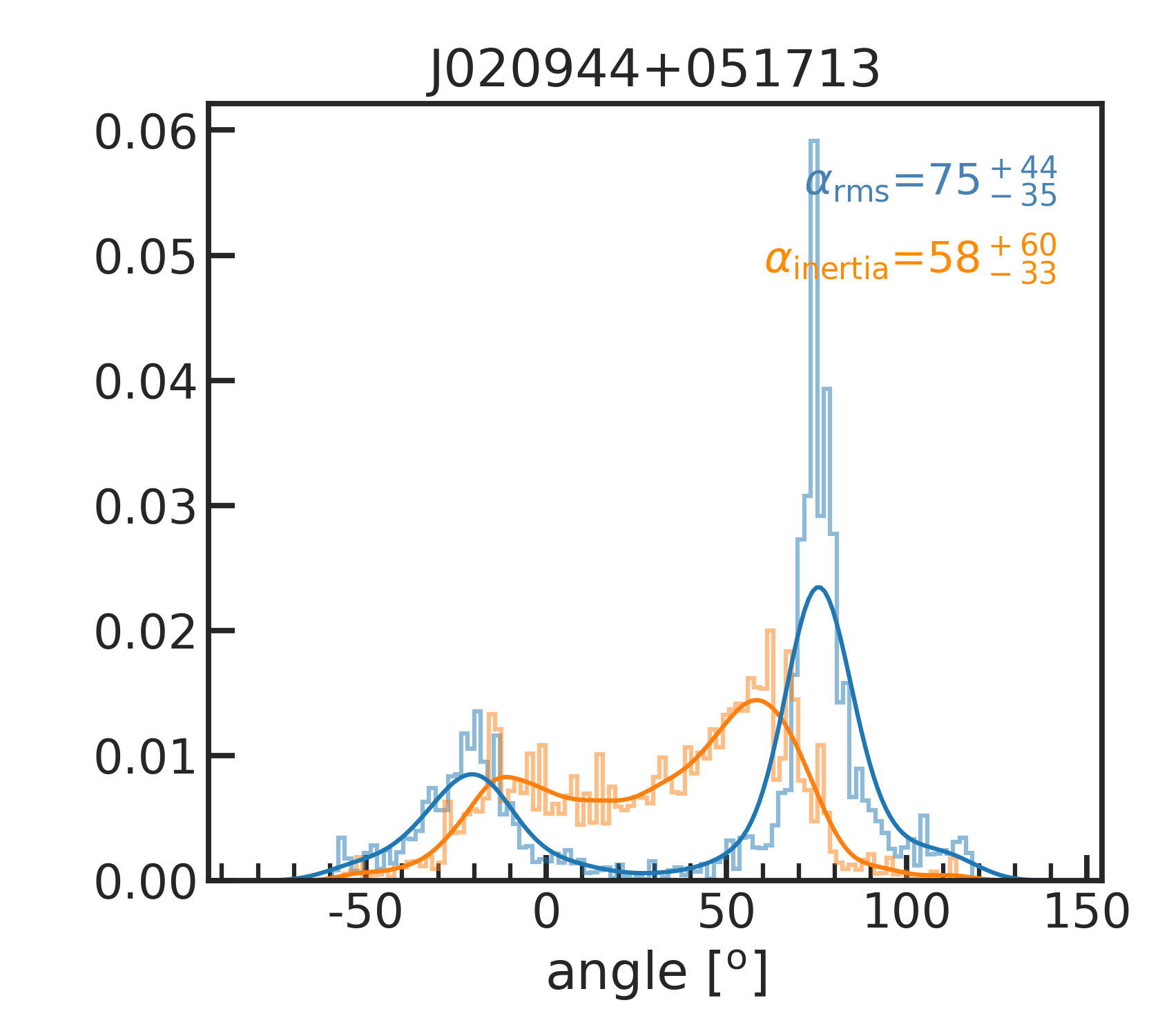}
\includegraphics[width=0.35\textwidth]{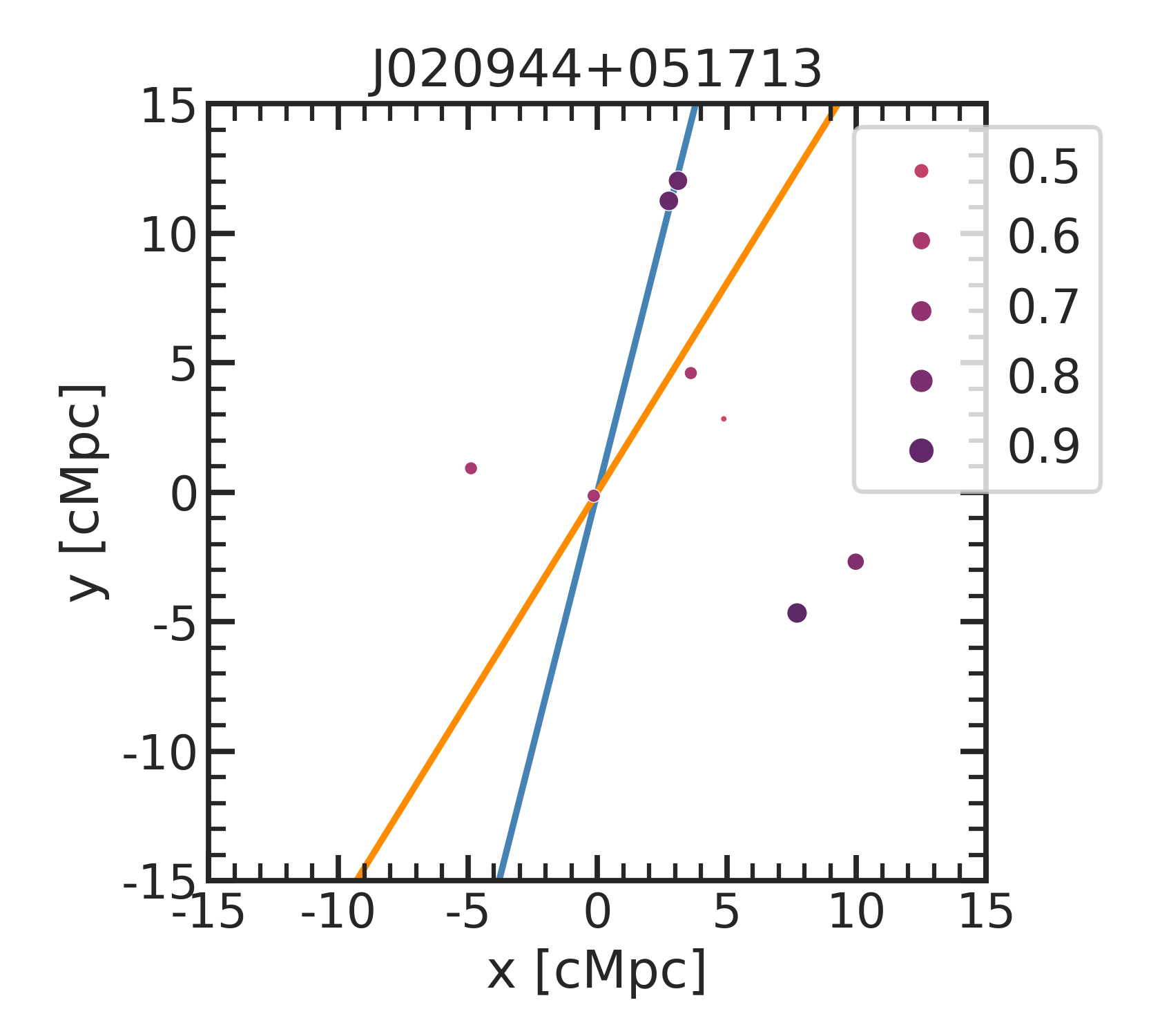}
\includegraphics[width=0.24\textwidth, trim=0 -3.2cm 0 0]{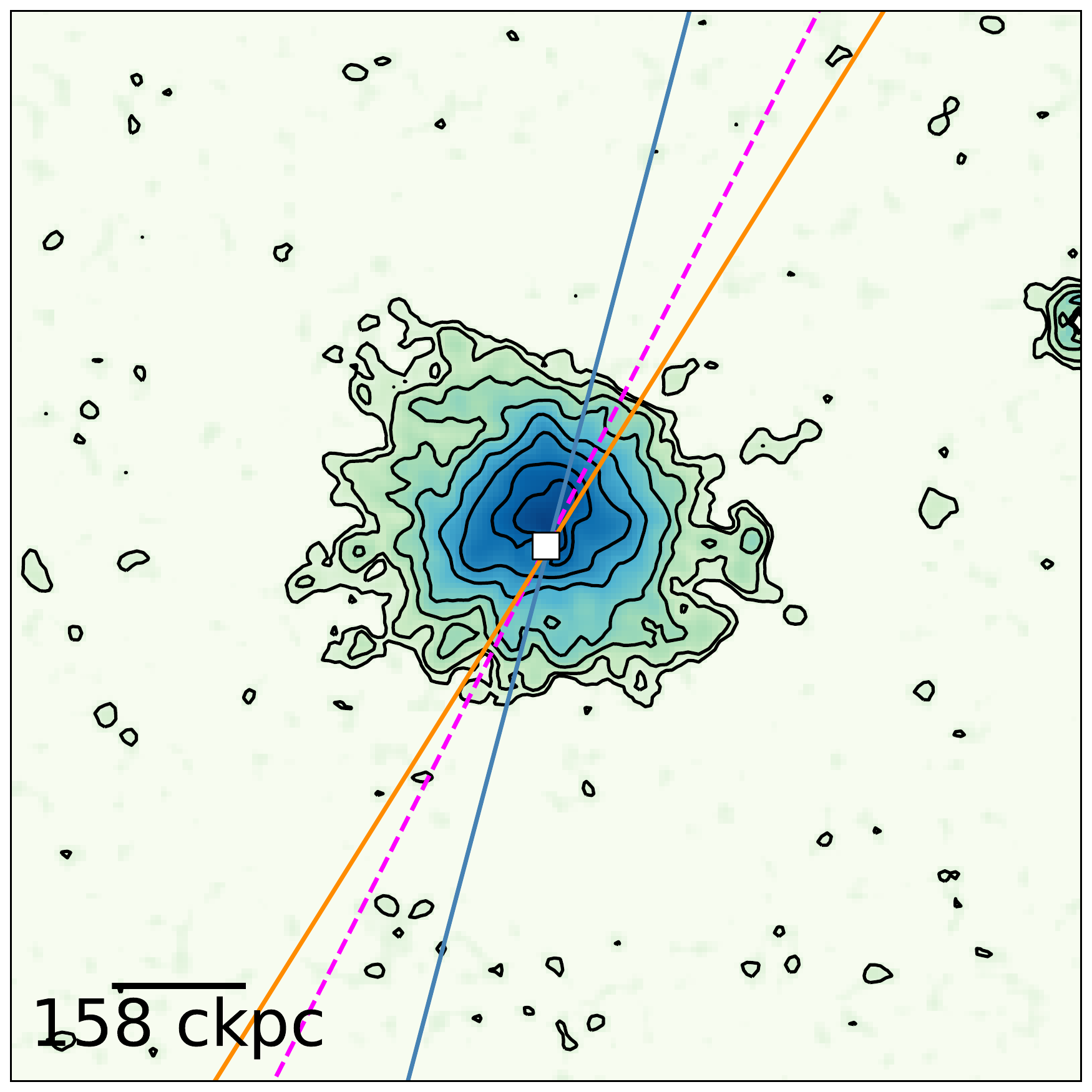}
\caption{Density distribution functions for the direction of the overdensity (left), source distributions with the direction of the overdensity overlaid (center), and Ly$\alpha$ SB maps (right) with the direction of the overdensity and the Ly$\alpha$ major axis overlaid (magenta). In each plot, the direction of the overdensity is color-coded in blue or orange depending on the method used, rms or inertia tensor, respectively (see Sect.~\ref{sec:spatialScale}). We show these plots for one ELAN (Fabulous) and for the two quasar fields with the largest (SDSSJ0819+0823) and smallest (J020944+051713) overdensity factor in our sample. In the central panels, the source sizes and colors indicate their completeness, which is used as a weight in the calculation of the direction of the overdensity. The Ly$\alpha$ SB maps are shown with a logarithmic color scheme stretching from $10^{-19}$ to $5\times10^{-17}$~erg~s~cm$^{-2}$~arcsec$^{-2}$, and with contours at S/N$=2$, $4$, $10$, $20$, and $50$. We stress that the Ly$\alpha$ maps are on much smaller scales (see their scale bars) than the 850~$\mu$m detections in the central panels.} 
\label{fig:angle_over}
\end{figure*}

Section~\ref{sec:apert} shows that the overdensities around the targeted systems are more significant within a radius of 2~cMpc, and then smoothly drops for larger circular apertures, without reaching the blank field counts for the explored area (out to $10$~cMpc).
To go beyond the simple circular apertures, we used two methods to define preferred directions for each of the 20 fields, assuming that the detected overdensities are associated with the central sources.
In the first method, we sampled the directions on the plane of the sky that pass through the main quasar position in steps of 0.5$^{\rm o}$ (angle $\alpha$ measured from east to north), and computed the rms distance of the sources to each direction\footnote{Each source in the field is weighted by its completeness.}.
Then, we took the direction $\alpha_{\rm rms}$ as the preferred orientation, which minimizes the rms. Straightforwardly, if there exists a well defined angle for which the rms distance is very small, it means that the sources are distributed along a projected large-scale structure or filament defined by that angle. To constrain well $\alpha_{\rm rms}$, we draw 10000 random subsamples (90\%) of sources for each field (bootstrap with replacement), and for each one of them find the angle minimizing the rms distance. The distributions of these angles for three fields (one ELAN, the most overdense quasar field, and the least overdense quasar field) are given by the blue histograms in the left panels of Fig.~\ref{fig:angle_over} (top, central and bottom). The corresponding distribution for the remaining 17 fields are shown in Figs.~\ref{fig:angle_over_app1},~\ref{fig:angle_over_app2},~\ref{fig:angle_over_app3}, and ~\ref{fig:angle_over_app4}. As the histograms can be quite noisy, we smoothed them using the Gaussian kernel density estimate to find the value of $\alpha_{\rm rms}$ where the distribution peaks. We associated errors with this peak value by computing the 34 percentiles left and right of peak. 

The second method involves computing the 2D inertia tensor from the position of the sources and  
taking as preferred direction $\alpha_{\rm inertia}$ the eigenvector corresponding to the smallest eigenvalue \citep{Zeballos2018}. As for the rms method, when computing the inertia tensor, we weight the sources by their completeness. We draw 10000 random subsamples (90\%) of sources for each field to construct the distributions of $\alpha_{\rm inertia}$, from which we find the values corresponding to the peak, and the 34 percentiles left and right of peak. The distributions of 
$\alpha_{\rm inertia}$ are given by the orange histograms in the left panels of Figs.~\ref{fig:angle_over},~\ref{fig:angle_over_app1},~\ref{fig:angle_over_app2},~\ref{fig:angle_over_app3}, and ~\ref{fig:angle_over_app4}. The angles $\alpha_{\rm inertia}$ are also measured north of east. 

\begin{figure*}
\centering
\includegraphics[width=0.70\textwidth]{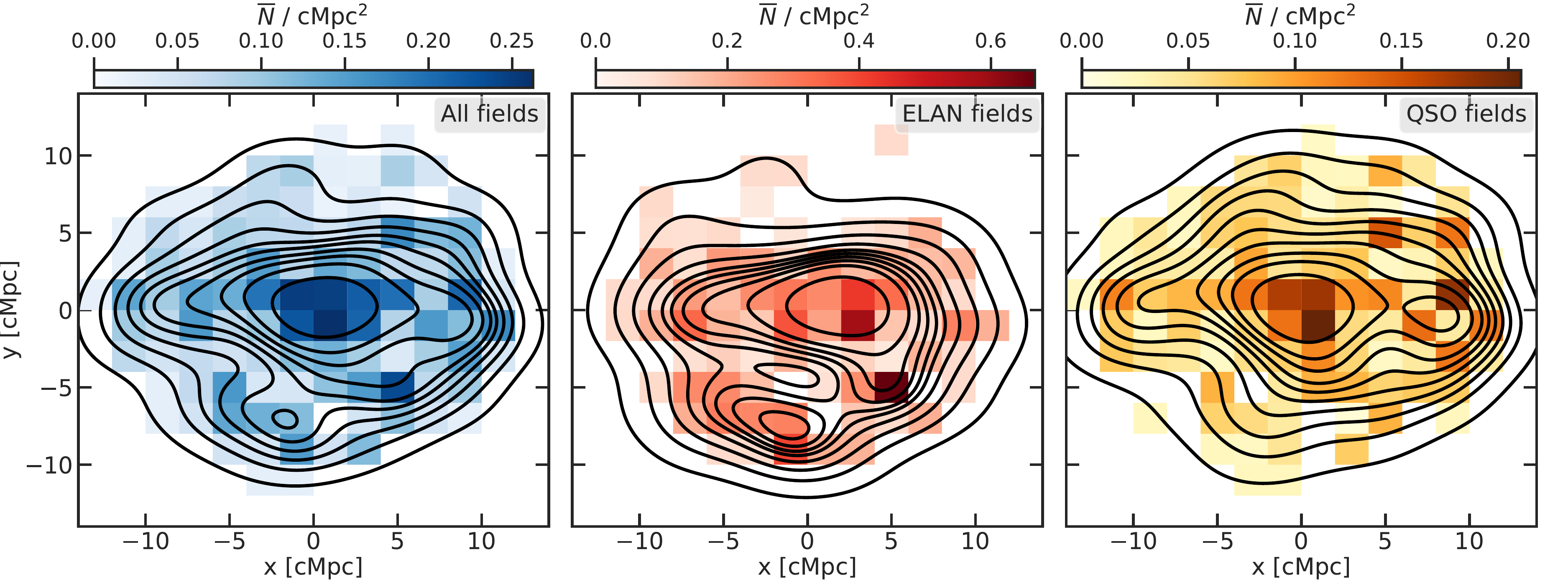}
\includegraphics[width=0.27\textwidth]{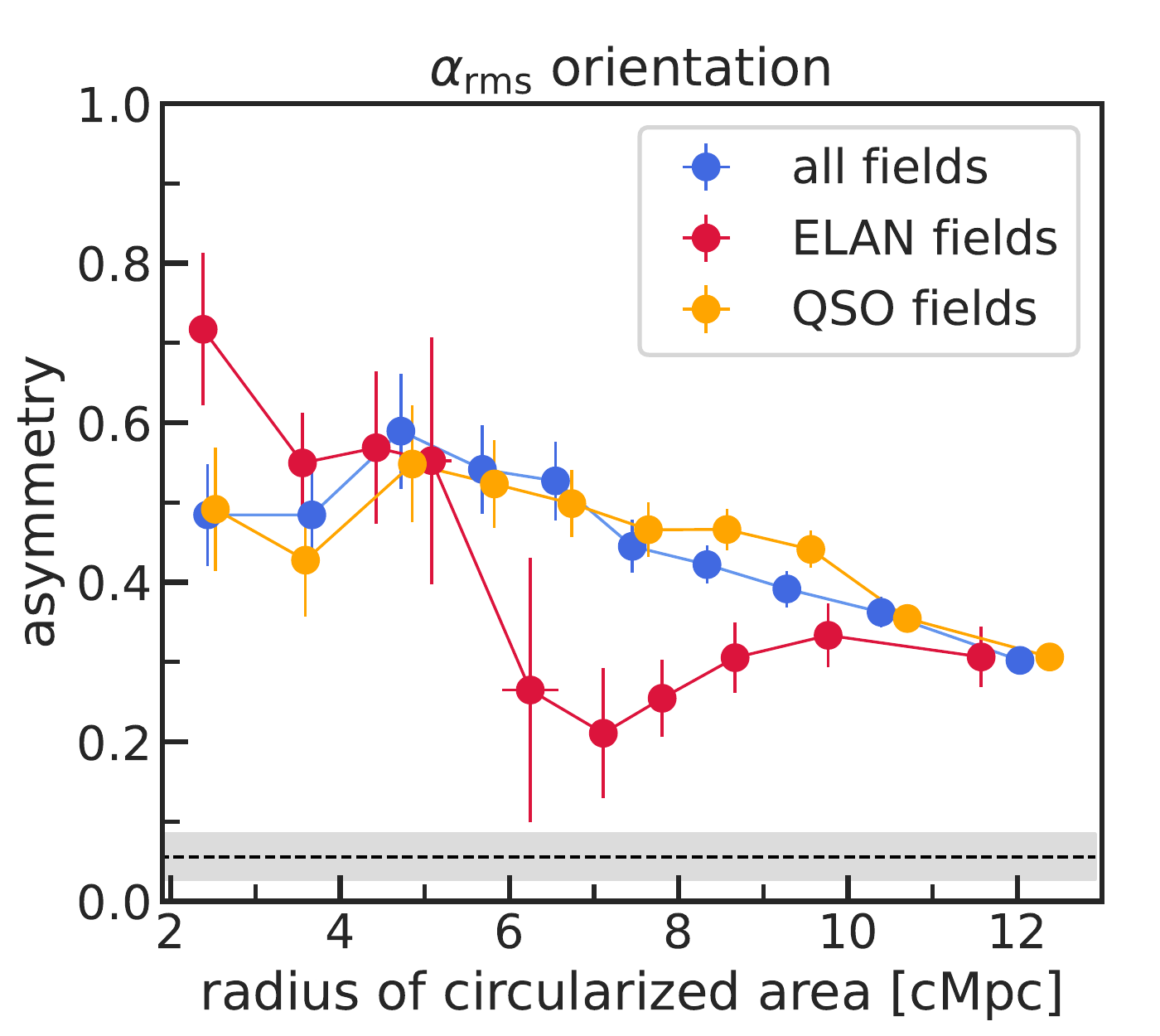}

\includegraphics[width=0.70\textwidth]{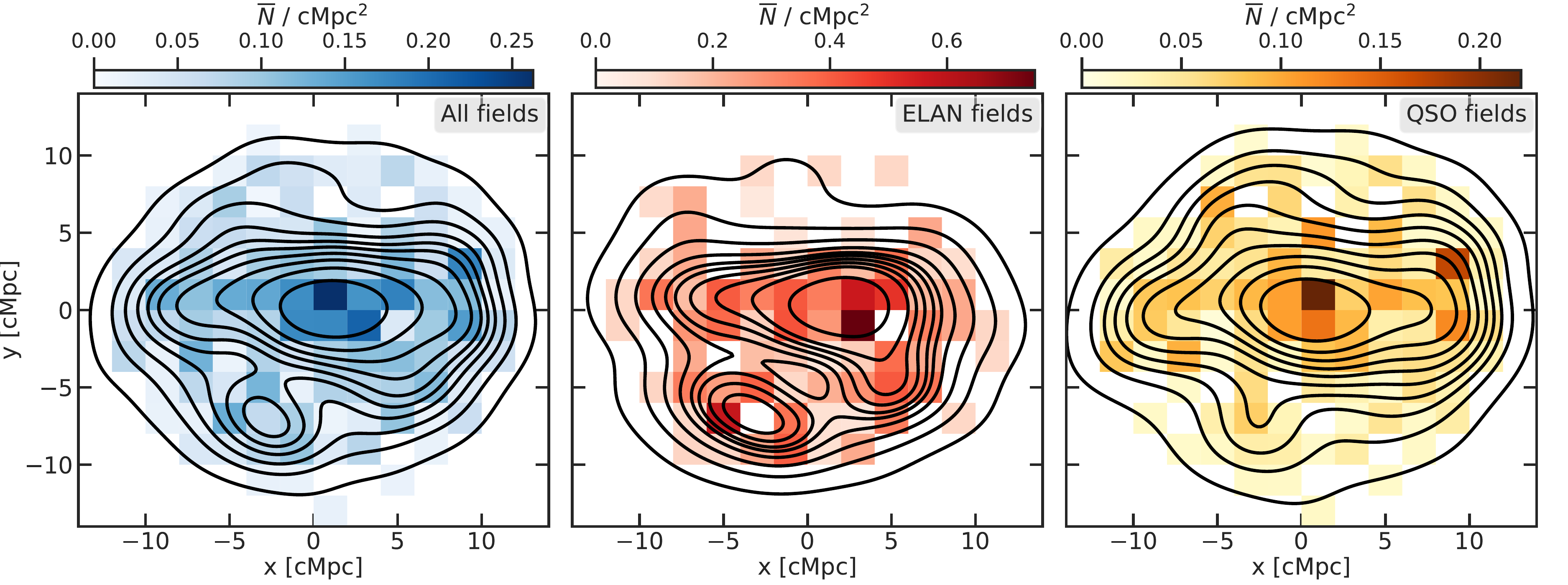}
\includegraphics[width=0.27\textwidth]{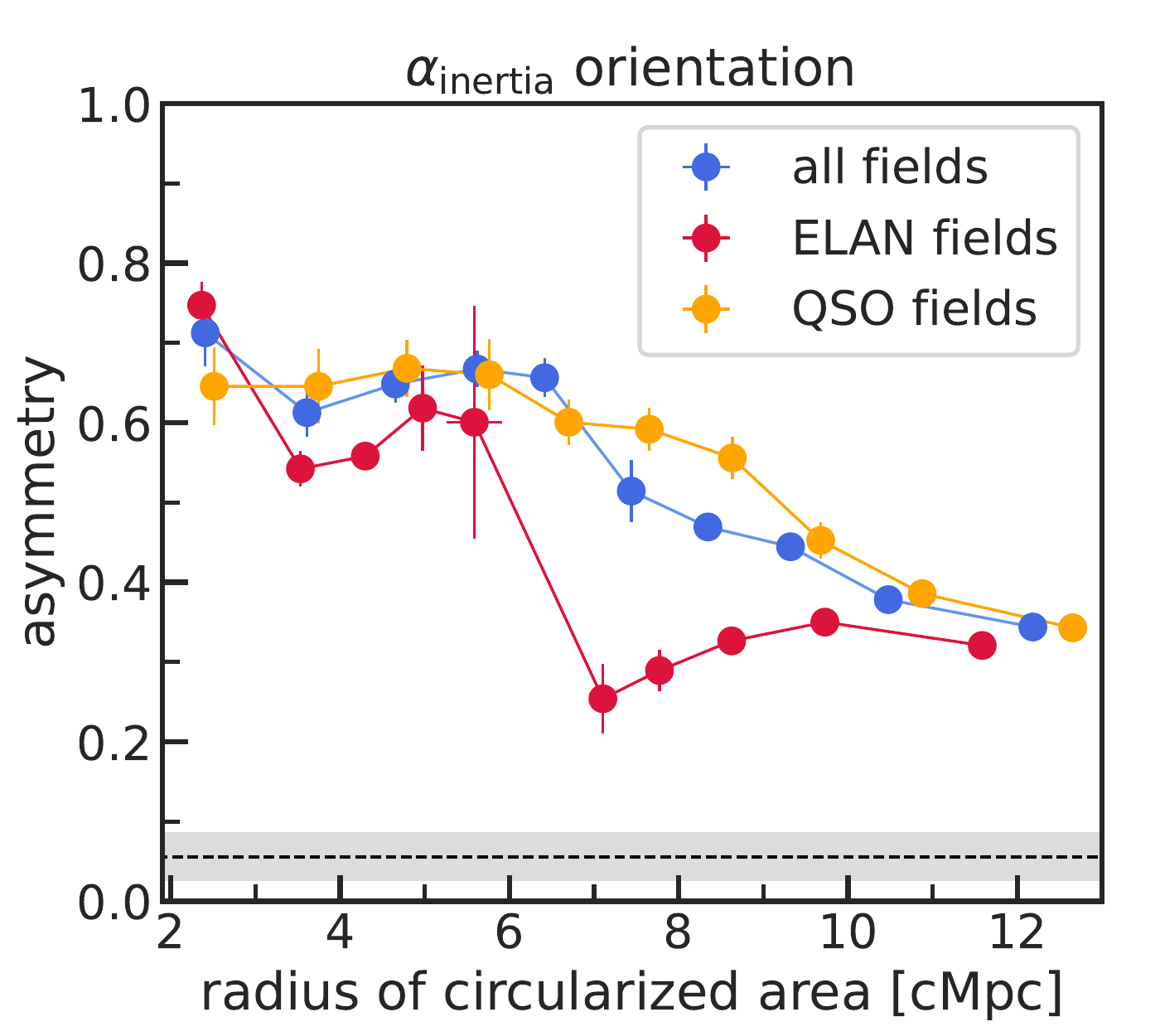}
\caption{Stacked source distribution maps and their asymmetry profiles. \emph{Left:} Stacked maps of mean number of sources per cMpc$^{\rm 2}$ for all the fields (left), only the ELANe (center), and only the quasar fields (right). For the maps on the top row the preferred direction is given by $\alpha_{\rm rms}$ and the ones on the bottom by $\alpha_{\rm inertia}$. The black contours in each map enclose increasing fractions of sources from 0.1 for the innermost contour to 1.0 for the outermost one in steps of 0.1. 
\emph{Right:} Asymmetry profiles of the stacked source distribution maps. The x-axis gives the radius of the circularized area of the contours, $\sqrt{area/\pi}$, in comoving megaparsecs. From left to right, the points correspond to the asymmetry within the contours enclosing increasing fractions of sources, from 0.1 to 1.0 in steps of 0.1. For the profiles in the top panel, the stacking has been done after rotating each field by its $\alpha_{\rm rms}$, while for the ones in the bottom panel the fields have been rotated by $\alpha_{\rm inertia}$. The horizontal dashed black line together with its gray band (in both panels) marks the level of asymmetry of 0.06$\rm\pm$0.03 corresponding to stacks of all 20 fields where the positions of the sources have been randomized.} 
\label{fig:2Dmaps_andprofiles}
\end{figure*}

At this point, we can rotate each field such that its preferred direction, given either by $\alpha_{\rm rms}$ or by $\alpha_{\rm inertia}$, becomes the new x-axis, and stack them together. The maps of Fig.~\ref{fig:2Dmaps_andprofiles} show the results of this exercise, where all the 20 fields have been used to construct the maps on the left panels, while the central and the right panels show the stacked maps of mean number of sources (corrected for completeness) per cMpc$^{\rm 2}$ for the ELAN and quasar fields, respectively.     
For the maps on the top row the preferred direction is given by $\alpha_{\rm rms}$, while for the ones on the bottom by $\alpha_{\rm inertia}$. The black contours in each map enclose increasing fractions of sources from 0.1 for the innermost contour to 1.0 for the outermost one in steps of 0.1. From these stacks of surface number density maps, we can see that the distribution of sources is more asymmetric (elongated along the x-direction) for the ELAN fields than for the quasar ones when looking at the central part of the maps. Also, it is clear that for all three cases, contours become increasingly more circular when moving outward from the center. 

We quantify the asymmetry in the distribution of sources within these contours in the right most panels of Fig.~\ref{fig:2Dmaps_andprofiles}. In practice, we use all the sources within each main contour (we do not consider the small secondary ones) to compute the Stokes parameters $Q$ and $U$, again weighting each source by its completeness $w_i$\footnote{$Q=M_{xx}-M_{yy}$ and $U=2M_{xy}$, where $M_{xy}=\sum_i{\frac{w_i x_i y_i}{x_i^2+y_i^2}}/\sum_i{w_i}$, and ($x_i$,$y_i$) is the position of source $i$.}. From the Stokes parameters we can estimate the asymmetry as $(1-\sqrt{Q^2+U^2})/(1+\sqrt{Q^2+U^2})$. If all the sources are collapsed along a single direction, the asymmetry reaches its maximum value of 1, while for randomly distributed sources the asymmetry would be 0. The colored points in the profiles of  Fig.~\ref{fig:2Dmaps_andprofiles} correspond to the stacks obtained after rotating each field by its peak $\alpha$ angle ($\alpha_{\rm rms}$ and $\alpha_{\rm inertia}$ for the top and bottom panel, respectively). The errors on the points have been obtained as the standard deviation of the asymmetry at each level over 1000 stacks, where each stack has been constructed using rotation angles drawn (independent for each field) from the angle distributions around the peaks. The horizontal dashed black line gives the prediction for the average stack of all 20 fields when the positions of the sources are randomized (distributed uniformly in the field-of-view). 

Irrespective of the method used to define a preferential angle (rms distance and inertia tensor for the top and bottom panels, respectively), there is a marked difference between the asymmetry profile of ELAN and quasar fields. The ELAN fields show a broken profile, with large asymmetries (from 0.5 to 0.8) for the region enclosing $\sim$50\% of the sources (within $\sim$6~cMpc of the central object), and lower asymmetries ($\sim$0.3) for regions enclosing more than half of the sources. In contrast, quasar fields have a slowly decreasing profile, from asymmetries of $\sim$0.5-0.6 in the inner regions to $\sim$0.3 for the full field-of-view. Given the low number statistics, it is important to judge these results in comparison with the asymmetry 
level corresponding to positions of the sources randomly distributed within the field-of-view (horizontal dashed black line in both right most panels of Fig.~\ref{fig:2Dmaps_andprofiles}). For all three combinations of stacks (only ELAN, only quasars, or all fields), the profiles of asymmetry are clearly above the level of 0.06$\rm\pm$0.03 predicted for randomly distributed sources. An additional interesting result is that the stack of only ELAN fields reveals clearly the presence of companion overdensities to the overdensity centered on the main quasar position, at (0,0) in the maps. The presence of these likely merging large-scale structures explains the broken asymmetry profile in ELAN fields. Indeed, the companion structure circularizes the overdensity on smaller scales in comparison to quasar fields. In Appendix~\ref{app:second_sub} we show that (i) the secondary peak in the ELAN stacked map is due to chance superposition of overdensities in the individual fields, and (ii) the asymmetry profile is robust against 180 degree individual rotations of each of the three ELAN fields.

\begin{figure}
    \centering
    \includegraphics[width=0.9\columnwidth]{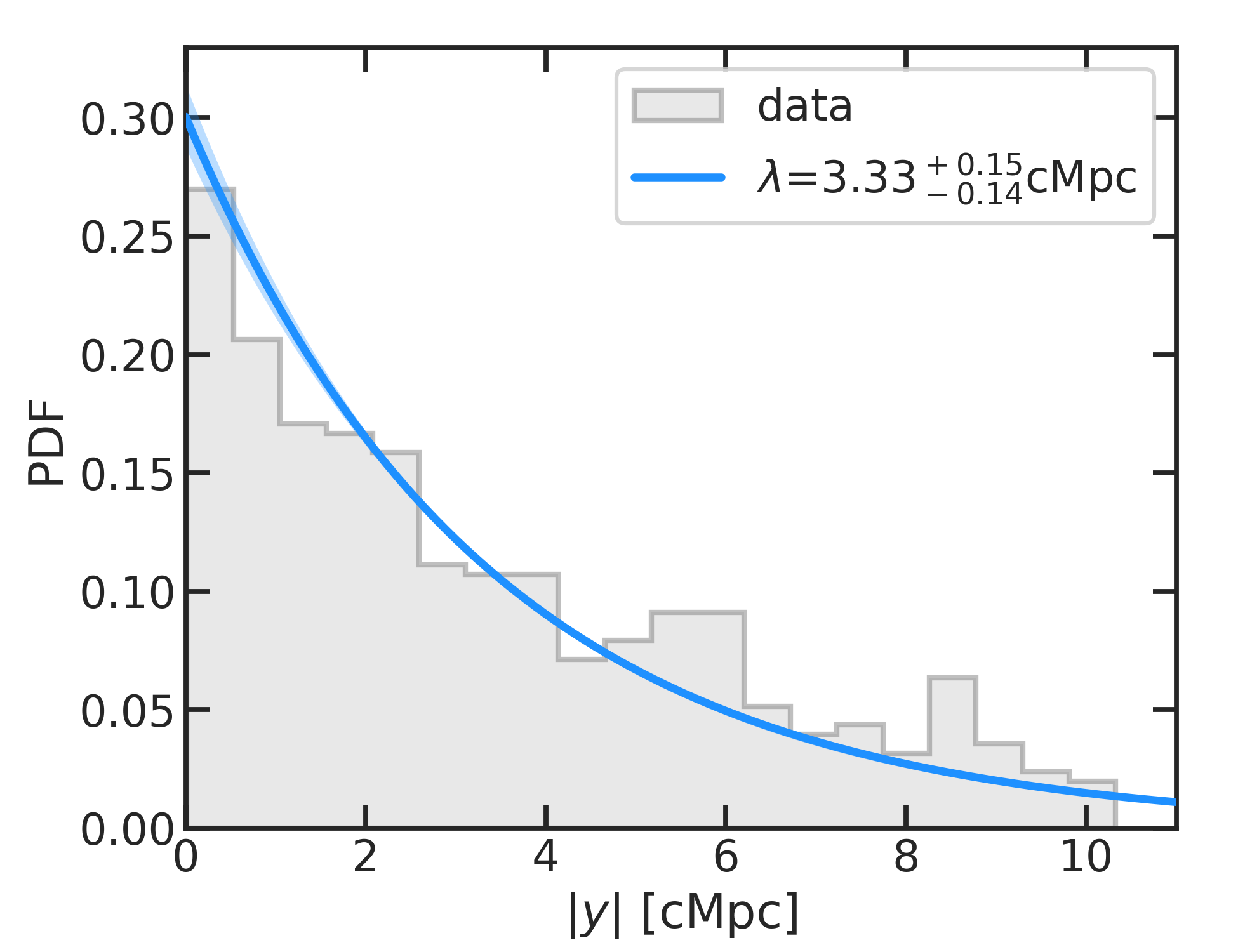}
    \caption{Empirical distribution function for the distances to the x-axis of the sources in the stack of fields rotated according to $\alpha_{\rm rms}$ (gray histogram). The blue curve indicates the 
    most likely exponential distribution that fits the data, while the shaded area represents the 1$\sigma$ uncertainty on the scale parameter $\lambda$.}
    \label{fig:PDF_scale}
\end{figure}

Further, we quantify a scale width for the elongated structure evident in the stacked maps and reminiscent of a large-scale filament. We focus on the stacked map obtained by using all fields rotated according to their $\alpha_{\rm rms}$, and compute the distribution of distances ($|y|$) from the x-axis, which we consider the major axis of the structure. From this analysis we exclude the sources within a radius of 0.5~cMpc from the center, which roughly corresponds to the virial radius of quasar host halos, resulting in a sample of $n=489$ sources. The obtained empirical probability distribution function (ePDF) is then fitted with an exponential $ {\rm PDF}(|y|) = e^{-|y|/\lambda}/\lambda$ using EMCEE (\citealt{emcee}) to find the scale parameter $\lambda=3.33_{-0.14}^{+0.15}$~cMpc  
(Fig.~\ref{fig:PDF_scale}). 
We assume a typical log-likelihood ${\rm log}(L)=-n\, {\rm log}(\lambda) - (\sum_{i=1}^{n}\, |y|_i)/(n \lambda)$  
and a flat prior on the scale parameter $\lambda$ ($0.1<\lambda<10$). Further, we use the Kolmogorov-Smirnov (KS) test (e.g., \citealt{Kolmogorov1933}) to determine whether our ePDF is indeed consistent with an exponential. We find that the ePDF is drawn from an exponential function with 95\% confidence ($KS=0.0598$, p-value$=0.0605$)\footnote{
Initially we fitted our data with the more generic gamma function $\Gamma(k,\lambda)=(|y|^{k-1}e^{-x/\lambda})/(\lambda^k \Gamma(k))$, which reduces to the exponential when the shape parameter $k=1$. We found the shape and scale parameters to be $k=1.14^{+0.07}_{-0.06}$ and $\lambda=2.92^{+0.22}_{-0.20}$, respectively. The KS test confirms that a gamma function is consistent with the data (KS$=0.0573$ and p-value$=0.0806$), but given the marginal difference with an exponential we prefer to rely on the fit with only one parameter. }.  

The obtained scale width is comparable with the diameter of thick filaments at high redshift measured in cosmological simulations from their DM and gas (e.g., \citealt{Cautun2014,Zhu2021}). 
Intriguingly, thick filaments are expected to live in overdense areas (see, e.g., Fig. 40 in \citealt{Cautun2014}), which is reassuring given that on average the targeted fields host overdensities.
However, we stress that the filament thickness is computed in different ways in cosmological simulation (e.g., using the beta model; \citealt{Zhu2021}) and frequently only focus at the smooth DM and gas mass distribution along the direction perpendicular to the filament spine, while we use halo/galaxy counts. A more robust comparison with cosmological simulations will be the focus of future works.

\FloatBarrier

\subsection{The alignment  of Ly$\alpha$ nebulae and overdensities}
\label{sec:comp_angles}

\FloatBarrier

\begin{table}[!hb]
\scriptsize
\caption{Preferred directions of the overdensities and major axis of the Ly$\alpha$ nebulae.} 
\centering
\begin{tabular}{lcrrr}
\hline
\hline
Quasar  & ID\tablefootmark{a} & $\alpha_{\rm rms}$ [$^{\rm o}$]\tablefootmark{b} & $\alpha_{\rm inertia}$ [$^{\rm o}$]\tablefootmark{c} & $\beta$ [$^{\rm o}$]\tablefootmark{d} \\        
\hline\\[-1ex]
Jackpot           & -  &  $25^{+42}_{-55}$    &   $40^{+22}_{-25}$   & $-77$ \\[1ex] 
MAMMOTH-I         & -  &  $80^{+26}_{-27}$    &   $75^{+12}_{-12}$   & $-77$ \\[1ex]
Fabulous          & 13 &  $-79^{+15}_{-16}$   &   $-75^{+15}_{-16}$  & $61$  \\[1ex] 
J0525-233         &  3 &  $100^{+7}_{-6}$     &   $89^{+29}_{-18}$   & $88$  \\[1ex] 
SDSSJ1209+1138    &  7 &  $-13^{+10}_{-10}$   &   $-31^{+26}_{-26}$  & $-45$ \\[1ex] 
SDSSJ1025+0452    & 10 &  $22^{+24}_{-20}$    &   $21^{+14}_{-12}$   & $-71$ \\[1ex] 
SDSSJ1557+1540    & 18 &  $-33^{+13}_{-15}$   &   $-42^{+11}_{-12}$  & $-21$ \\[1ex] 
SDSSJ1342+1702    & 24 &  $-57^{+26}_{-37}$   &   $-62^{+20}_{-20}$  & $46$  \\[1ex] 
Q-0115-30         & 29 &  $-2^{+42}_{-54}$    &   $-94^{+52}_{-55}$  & $-11$ \\[1ex] 
Q2355+0108        & 46 &  $37^{+51}_{-44}$    &   $25^{+65}_{-31}$   & $41$  \\[1ex] 
SDSSJ0819+0823    & 50 &  $-21^{+30}_{-45}$   &   $-22^{+26}_{-23}$  & $-16$ \\[1ex] 
J015741-01062957  & -  &  $51^{+14}_{-15}$    &   $57^{+12}_{-14}$   & $-68$ \\[1ex] 
J020944+051713    & -  &  $75^{+44}_{-35}$    &   $58^{+60}_{-33}$   & $63$  \\[1ex] 
J024401-013403    & -  &  $-60^{+14}_{-17}$   &   $-51^{+19}_{-21}$  & $-16$ \\[1ex] 
J033900-013318    & -  &  $61^{+12}_{-12}$    &   $56^{+27}_{-25}$   & $80$  \\[1ex] 
J111113-080401    & -  &  $48^{+16}_{-26}$    &   $56^{+18}_{-19}$   & $18$  \\[1ex] 
J193957-100241    & -  &  $-51^{+29}_{-39}$   &   $8^{+55}_{-61}$    & $69$  \\[1ex] 
J221527-161133    & -  &  $90^{+49}_{-30}$    &   $11^{+59}_{-51}$   & $-87$ \\[1ex] 
J230301-093930    & -  &  $7^{+31}_{-32}$     &   $43^{+38}_{-27}$   & $30$  \\[1ex] 
J233446-090812    & 45 &  $-25^{+24}_{-41}$   &   $-100^{+44}_{-49}$ & $7$   \\[1ex] 
\hline
\end{tabular}
\tablefoot{
\tablefoottext{a}{ID in QSO MUSEUM.}
\tablefoottext{b}{Angle north of east for the preferred direction of the overdensity using the rms method.}
\tablefoottext{c}{Angle north of east for the preferred direction of the overdensity using the 2D inertia tensor.}
\tablefoottext{d}{Major axis of the Ly$\alpha$ nebulae.}
}
\label{Tab:angles_over_Lya}
\end{table}

In this section we investigate whether the large-scale Ly$\alpha$ emission on hundreds of kiloparsecs around the targeted systems has a preferred direction with respect to the orientation of the 850~$\mu$m overdensities found on megaparsec scales. 
As the reference angle for the Ly$\alpha$ nebulae, we used their major axis direction, $\beta$, obtained from the Stokes parameters weighted by the Ly$\alpha$ SB within the $2\sigma$ isophote of each nebulae, as already done  in \citet{FAB2019} for the QSO MUSEUM sample. 
On the other hand, for the overdensities, we used the preferred directions $\alpha_{\rm rms}$ and $\alpha_{\rm inertia}$ obtained in Sect.~\ref{sec:spatialScale}.
All these angles are listed in Table~\ref{Tab:angles_over_Lya} and the obtained directions are shown in the Figs.~\ref{fig:angle_over},~\ref{fig:angle_over_app1},~\ref{fig:angle_over_app2},~\ref{fig:angle_over_app3}, and ~\ref{fig:angle_over_app4}, together with the SB map of each system.

\begin{figure*}
    \centering
    \includegraphics[width=0.28\textwidth]{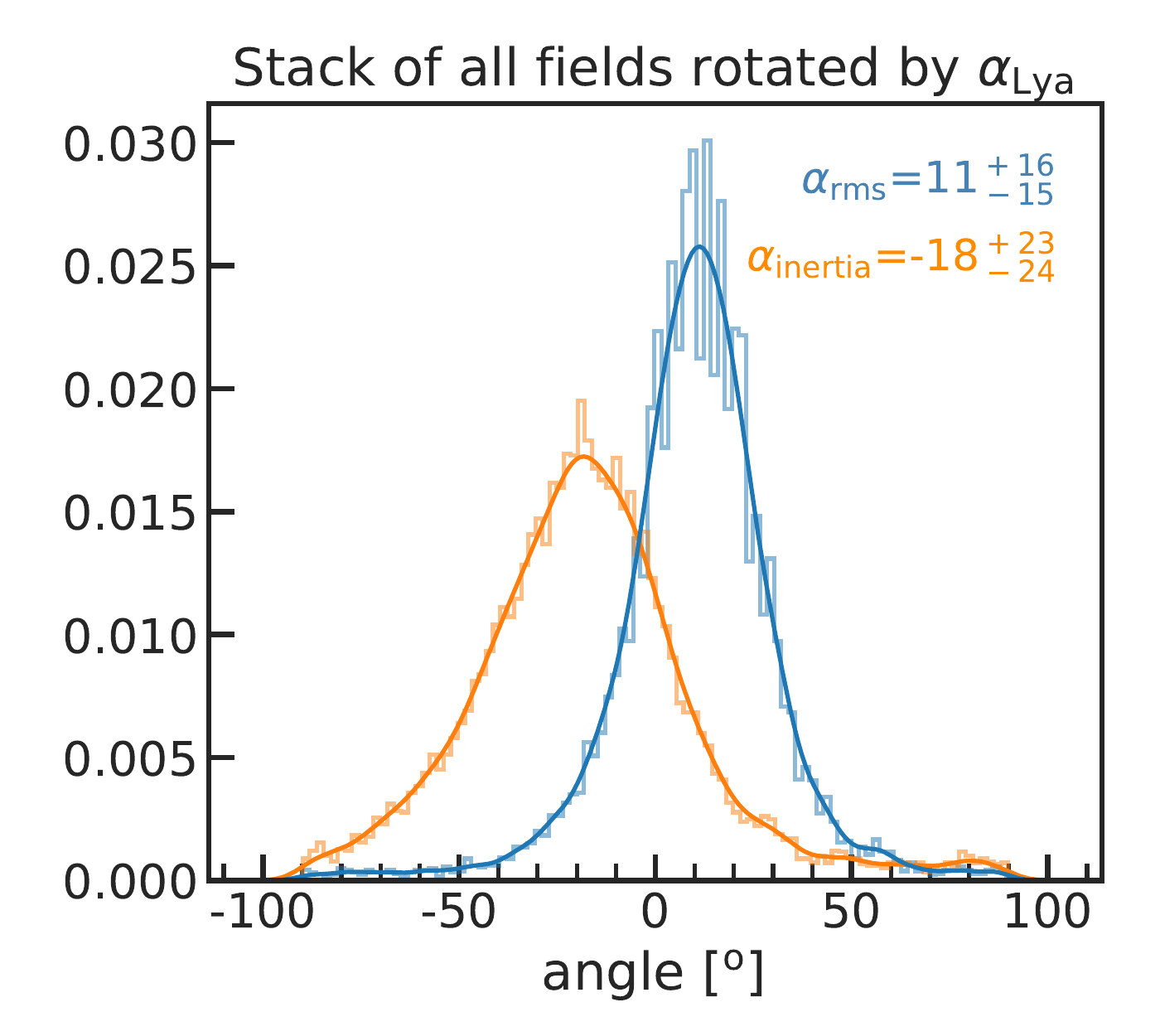}
    \includegraphics[width=0.28\textwidth]{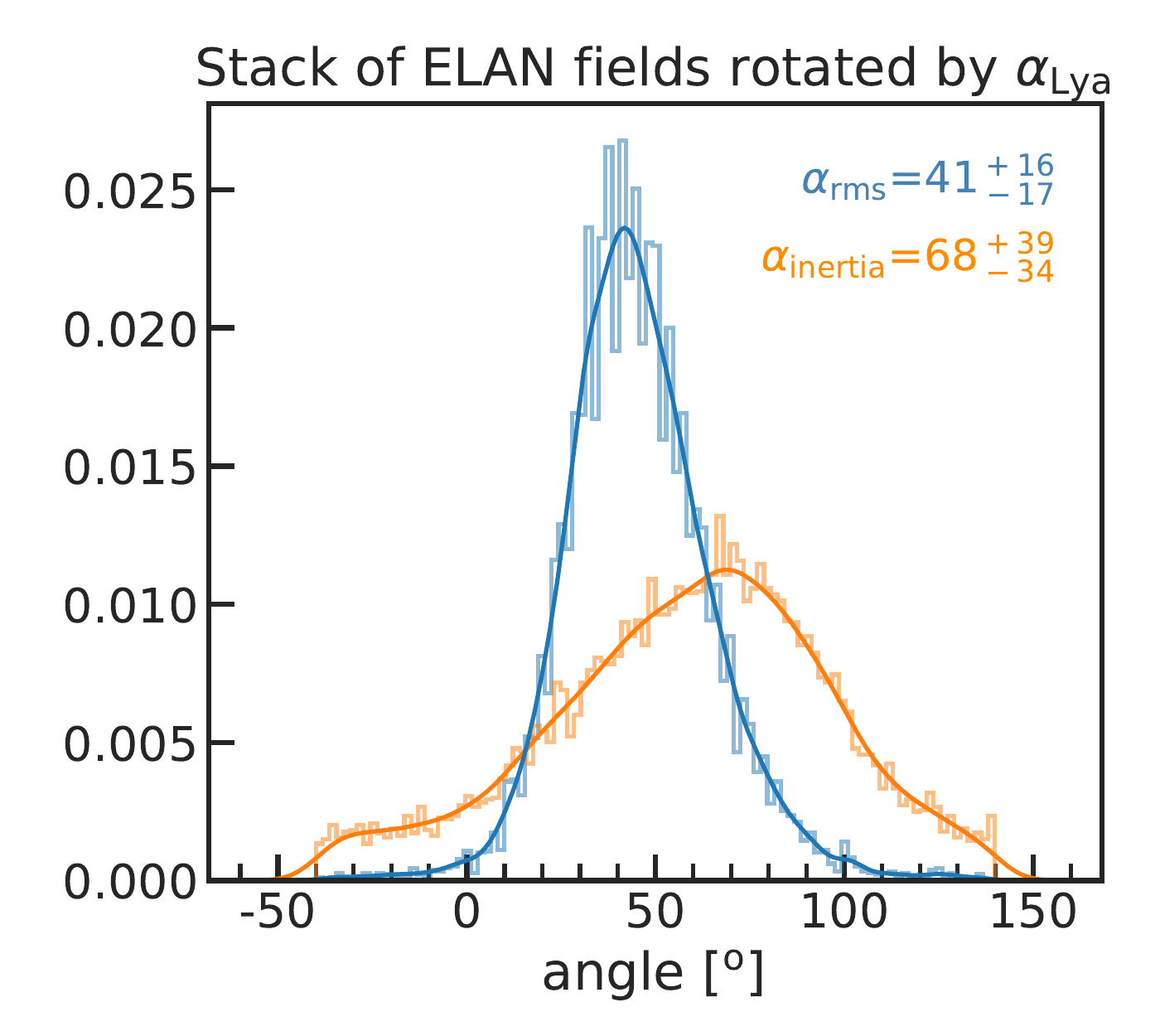}
    \includegraphics[width=0.28\textwidth]{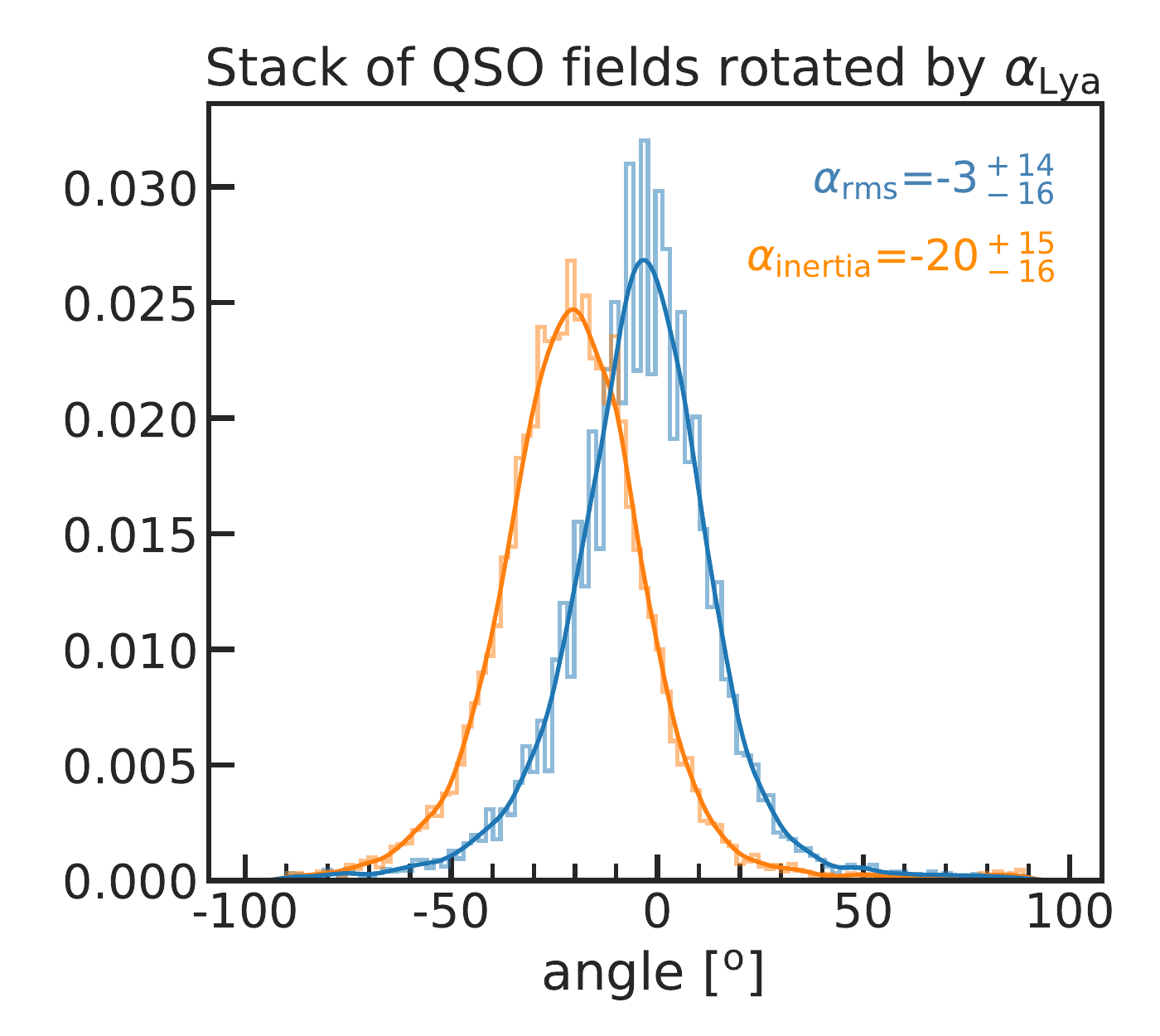}
    \caption{Distributions of preferred angles (measured with respect to the x-axis) for the stack of all fields (left), ELAN only (central), and quasars only (right), after each individual field has been first oriented with the x-axis along the direction of its Ly$\alpha$ nebula. Both the minimum rms distance and inertia methods result in angles compatible with 0, within their respective errors, for the case of all the field stacks and the case of only the quasar stack. The stack of rotated ELAN fields results in angles that are not compatible with 0.}
    \label{fig:angle_stack_after_Lyarot}
\end{figure*}

Comparing the angles, we find that the majority of the Ly$\alpha$ nebulae have major axis at less than $45$ degrees from the preferred directions of the overdensities. Specifically, 75\% and 60\% of the sample when comparing to $\alpha_{\rm rms}$ and $\alpha_{\rm inertia}$, respectively. If we focus only on the fields with $| \alpha_{\rm rms} - \alpha_{\rm inertia} |<30$~degrees, we find that $73$\% of them (or 11 out of 15) have their Ly$\alpha$ nebula major axis at less than 45 degrees from the direction of their overdensity determined by $\alpha_{\rm rms}$ or $\alpha_{\rm inertia}$. The alignment between Ly$\alpha$ emission and overdensities suggests that the extended Ly$\alpha$ emission traces, for most of the systems, the large-scale structure in which the quasars are embedded. This finding is in agreement with the work by \citet{Costa2022}, which studied the extended Ly$\alpha$ emission around high-redshift quasars using cosmological zoom-in simulation post processed with a Ly$\alpha$ radiative transfer code. Indeed, they showed that for quasars host galaxies close to face-on (as it should be more probable for quasars), the Ly$\alpha$ emission is more extended and traces the surrounding large-scale structure (see, e.g., their Fig.~1 for a visual example). On the other hand, more inclined galaxies would hinder the propagation of Ly$\alpha$ photons, resulting in more lopsided nebulae that do not trace the large-scale structure  well.

To further test the discovered alignment and measure an average angle between Ly$\alpha$ major axis and the overdensity, we combined the information from all the fields. Specifically, we first consider the catalog detections for each field and rotate them so that the Ly$\alpha$ major axis align with the x-axis. Then, we use all the sources from all fields simultaneously to compute the preferred direction of the overdensity $\alpha_{\rm rms}^{\rm all}$ and $\alpha_{\rm inertia}^{\rm all}$, as previously done for the individual fields.
The left panel of Fig.~\ref{fig:angle_stack_after_Lyarot} shows that $\alpha_{\rm rms}^{\rm all}=$11$_{\rm -15}^{\rm +16}$ and $\alpha_{\rm inertia}^{\rm all}=$-18$_{\rm -24}^{\rm +23}$, ultimately proving that, on average, the Ly$\alpha$ nebulae are oriented along the same direction as the overdensities they inhabit. 

We also did the same exercise for the separate stacks of rotated ELAN and quasar fields (central and right panels of Fig.~\ref{fig:angle_stack_after_Lyarot}).\ This revealed that the ELANe are actually fields that do not have the nebulae orientated along their host large-scale 850$\mu$m source distributions but show a preferred relative angle ($\alpha_{\rm rms}^{\rm ELAN}=$41$_{\rm -17}^{\rm +16}$ and $\alpha_{\rm inertia}^{\rm ELAN}=$68$_{\rm -34}^{\rm +39}$). This is in contrast to the stack of rotated quasar fields, which clearly proves that these Ly$\alpha$ nebulae are orientated along the large-scale structure ($\alpha_{\rm rms}^{\rm quasar}=$-3$_{\rm -16}^{\rm +14}$ and $\alpha_{\rm inertia}^{\rm quasar}=$-20$_{\rm -16}^{\rm +15}$). These results can be understood in terms of the particular evolutionary stage and environment that ELANe showcase. Indeed, several works have reported evidence that ELANe pinpoint the location of protocluster cores given the richness of their environment on small scales and on hundreds of kiloparsecs (e.g., \citealt{hennawi+15,FAB2018b,Nowotka2022}). The misalignment between the main large-scale structure and the extended Ly$\alpha$ could therefore be due to the latter tracing inspiraling halo accretion (a.k.a. mergers, interactions) of active galaxies within these protocluster cores (e.g., \citealt{FAB2018,TC2021,LiQ2021}). On the other hand, single quasars powering Ly$\alpha$ nebulae likely live in relatively less turbulent and active small-scale environment, and therefore the central quasar is probably better at illuminating their host large-scale structure without contamination of other substructures: a node at the intersection of various filaments. This hypothesis is further strengthen by the difference in the asymmetry profiles and the presence of the secondary overdensity peak for the stacked map for ELAN fields reported in Sect.~\ref{sec:spatialScale} (Fig.~\ref{fig:2Dmaps_andprofiles}). This likely companion overdensity would be further evidence in favor of more disturbed environments around ELANe.

Further confirmation of our findings requires spectroscopic redshifts for the 850~$\mu$m detections. As previously said, this effort is ongoing using sensitive interferometers (Wang et al. in prep.). Also, it is necessary to enlarge the submillimeter surveyed area around these sources to pinpoint the 3D location of the large-scale filaments hosting them, and verify on which scales the overdensity reaches field values.

\section{Summary}
\label{sec:summary}

We searched for submillimeter sources in the fields of 3 ELANe and 17 $z\sim3-4$ quasars (9 from QSO MUSEUM and 8 from MAGG) using the JCMT/SCUBA-2 instrument and surveying an area of $\sim 125$~arcmin$^{2}$ (or $\sim 450$~cMpc$^2$) for each field, with an average central noise of $0.9$~mJy~beam$^{-1}$ at 850~$\mu$m and $11.7$~mJy~beam$^{-1}$ at 450~$\mu$m.
This work aimed to confirm and extend  the effort by \citet{FAB2018b} and \citet{Nowotka2022}, who show that ELANe pinpoint the location of excess 850~$\mu$m counts, further indicating that they likely inhabit protoclusters. Our analysis of this large data set includes:

\begin{itemize}
    \item The discovery of 523 and 102 sources  (S/N$>4$ or detected in both bands at S/N$>3$)  at 850~$\mu$m and 450~$\mu$m, respectively. For each field, we include catalogs in Appendix~\ref{app:catalogs}.
    \item The calculation of differential number counts at 850~$\mu$m and of their underlying count models with self-consistent simulations, which allowed us to compute overdensity factors with respect to blank fields. We find that submillimeter sources with $S_{\rm 850\mu m}\gtrsim 3$~mJy are overabundant in 75\% of the targeted fields, by a factor of $2-7$ within a radius of $\sim10$~cMpc (Sect.~\ref{sec:overd}). The significance of the overdensity in each field varies between $1\sigma$ and $6\sigma$ depending on the method used in computing the overdensity factors, as shown by \citet{Nowotka2022}. When normalized to the fiducial blank-field counts, the weighted average overdensity factors are $ 1.9\pm0.2$, $1.8\pm0.1$, $2.3\pm0.2$, and $1.6\pm0.1$ for the ELANe, all quasars together, the QSO MUSEUM objects, and the MAGG sample, respectively. When using a cumulative method over the flux density range probed, we find larger values of $3.4\pm0.4$, $2.5\pm0.2$, $3.1\pm0.4$, and $2.3\pm0.3$ for the ELANe, all quasars together, the QSO MUSEUM objects, and the MAGG sample, respectively. This difference is due to underlying models that are steeper than those of blank fields. ELANe therefore show similar overdensities as other quasar fields (see also Sect.~\ref{sec:overdensities}).
    \item The finding, at high significance ($\gtrsim5\sigma$), of a radial decrease in the overdensity through the calculation of differential and cumulative number counts in radial apertures using the whole sample (Sect.~\ref{sec:apert}). Specifically, the overdensity decreases from $>3$ within 1.5 arcmin (or about 2~cMpc) to $\sim2$ at the edge of the field. This result suggests that the physical extent of the overdensities around these systems is larger than our maps, in agreement with theoretical expectations that place the effective radius of protoclusters associated with halo masses $M_{\rm DM}>10^{12}$~M$_{\odot}$ at scales similar to the size of our maps ($\sim10$~cMpc; \citealt{Chiang2013,Muldrew2015}).
    \item The comparison of the computed overdensity factors with quasars and Ly$\alpha$ nebula properties, from which we find interesting features in the data that should be explored further (Sect.~\ref{sec:over_vs_prop}): (i) $z\geq 4$ quasar fields show the smallest source excess, (ii) quasars with similar absolute magnitudes can have a different excess of submillimeter sources, and (iii) being a radio-loud quasar seems to guarantee a 850~$\mu$m source excess $\gtrsim 2.8$ times that of the blank field.   
    \item The discovery of preferred directions for the overdensities of 850~$\mu$m sources, which allowed us to stack the fields to unveil an elongated structure reminiscent of a large-scale filament with a scale width of $\approx 4$~cMpc (Sect.~\ref{sec:spatialScale}). 
    We also determine an asymmetry profile for the overdensities. A significant difference is found between ELAN and quasar fields. Overdensities around ELANe show a broken asymmetry profile, with large asymmetries ($0.5$ to $0.8$) for the region enclosing $\sim50\%$ of the sources (or a radius of $\lesssim6$~cMpc), and lower asymmetries ($\sim0.3$) for outer regions. On the other hand, quasar fields are characterized by a slowly decreasing profile, from large asymmetries ($\sim 0.6$) in the inner regions to $\sim 0.3$ in the outer scales probed. This finding is connected to the presence of a secondary overdensity peak or likely merging large-scale structures in ELAN fields.
       \item The discovery of a trend for the major axis of the Ly$\alpha$ nebulae surrounding the central quasars to align along the direction of the 850~$\mu$m overdensities (Sect.~\ref{sec:comp_angles}). This result suggests that the Ly$\alpha$ emission on scales of hundreds of kiloparsecs  traces the central portion of the projected large-scale structure on megaparsec scales. While this is true for quasar fields, ELANe do not have the Ly$\alpha$ nebulae oriented along their host large-scale 850~$\mu$m source distributions, likely indicating that the small-scale (hundreds of kiloparsecs) turbulent and active environment of ELANe contaminates the signal from more diffuse gas with a signal associated with active merging substructures.
\end{itemize}

Overall, our work showcases the richness of ELAN and $z\sim3-4$ quasar megaparsec environments at submillimeter wavelengths and paves the way to study the connection between the extended Ly$\alpha$ emission and the surrounding large-scale structure for these massive nodes of the cosmic web. Confirming member associations of the current detections with follow-up observations is ongoing (ALMA and NOEMA). These datasets will allow us to further understand the spatial and kinematic distribution of dusty star-forming galaxies around ELANe and quasars and characterize their cosmic environment.

\begin{acknowledgements}
We thank the anonymous referee for their careful reading of the manuscript and useful suggestions.
We thank Elisabeta Lusso and Ian Smail for providing comments to an earlier draft of this work, and Daniela Gal\'arraga-Espinosa for fruitful discussions on filament sizes estimations from cosmological simulations.
The James Clerk Maxwell Telescope is operated by the East Asian Observatory on behalf 
of The National Astronomical Observatory of Japan; Academia Sinica Institute of Astronomy and Astrophysics; 
the Korea Astronomy and Space Science Institute; the Operation, Maintenance, and Upgrading Fund for Astronomical Telescopes 
and Facility Instruments, budgeted from the Ministry of Finance (MOF) of China and administrated by the Chinese Academy of Sciences (CAS), 
as well as the National Key R\&D Program of China (No. 2017YFA0402700). 
Additional funding support is provided by the Science and Technology Facilities Council 
of the United Kingdom and participating universities in the United Kingdom and Canada.
A.O. is funded by the Deutsche Forschungsgemeinschaft (DFG, German Research Foundation) –- 443044596.
C.-C.C. acknowledges support from the National Science and Technology Council of Taiwan (NSTC 109-2112-M-001-016-MY3 and 111-2112-M-001-045-MY3), 
as well as Academia Sinica through the Career Development Award (AS-CDA-112-M02).
This project has received funding from the European Research Council (ERC) under the European Union's Horizon 2020 research and 
innovation programme (grant agreement No 757535) and by Fondazione Cariplo, grant No 2018-2329.

\end{acknowledgements}

\bibliographystyle{aa} 
\bibliography{allrefs}

\begin{appendix} 

\section{Flux conversion factors}
\label{app:FCFs}

\FloatBarrier

In this work we adopt the recommended FCFs listed on the 
JCMT web pages\footnote{\url{https://www.eaobservatory.org/jcmt/instrumentation/continuum/scuba-2/calibration/}}. For completeness, we list them here in Table~\ref{tab:FCFs}. Our conclusions are unchanged when using the old FCFs values from \citet{Dempsey2013}.

\begin{table}[htp]
\footnotesize
\caption{FCFs used in this work.}
\centering
\begin{tabular}{lcc}
\hline
\hline
Date    & FCF$_{450\mu m}$                & FCF$_{850\mu m}$           \\ 
        &     (Jy beam$^{-1}$ pW$^{-1}$)  & (Jy beam$^{-1}$ pW$^{-1}$) \\
\hline
Before June 30, 2018  &    $531\pm93$                &       $516\pm42$                     \\
After June 30, 2018   &    $472\pm76$                &       $495\pm32$                     \\

\hline
\end{tabular}
\label{tab:FCFs}
\end{table}

\FloatBarrier
\clearpage

\section{S/N histograms}
\label{app:histos}

In this appendix we show for completeness the normalized histograms of the S/N values for the pixels within the effective area of the 850 and 450~$\mu$m maps for each field.
For the 850~$\mu$m histograms we indicate the S/N threshold used for the counts analysis (dashed vertical line).

\begin{center} 
\begin{minipage}{\textwidth}
\includegraphics[width=1.0\textwidth]{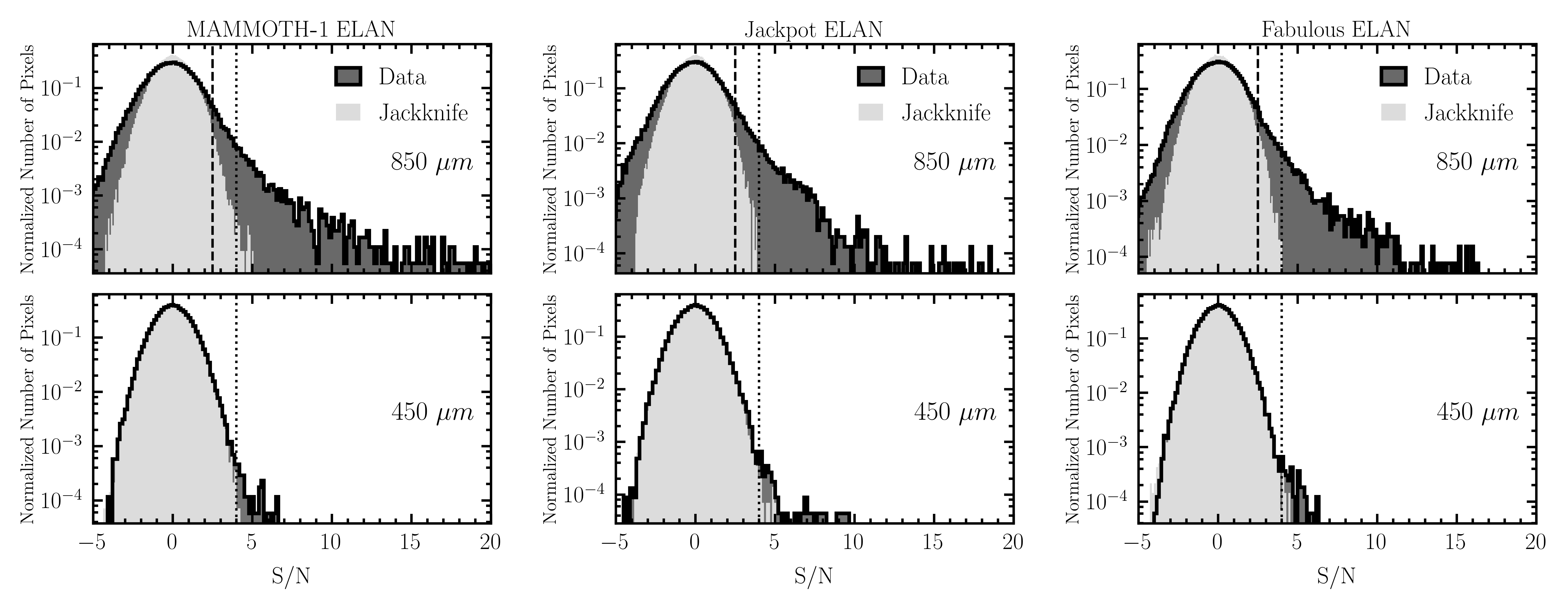}
\captionof{figure}{Normalized histograms of the S/N values for the pixels within the effective area of the 850 and 450~$\mu$m maps. The light and dark gray histograms indicate the distributions of the jackknife maps and of the data, respectively. The data, especially at 850~$\mu$m, show excesses at high S/N where sources contribute to the distribution. 
The dashed vertical line represents the cut used in the simulations of the number counts (see Table~\ref{tab:und_models}). The dotted line represents the threshold (4$\sigma$) above which we catalog sources.}
\label{fig:SNhistos1}
\end{minipage}
\end{center}

\FloatBarrier

\begin{figure*}
\centering
\includegraphics[width=1.0\textwidth]{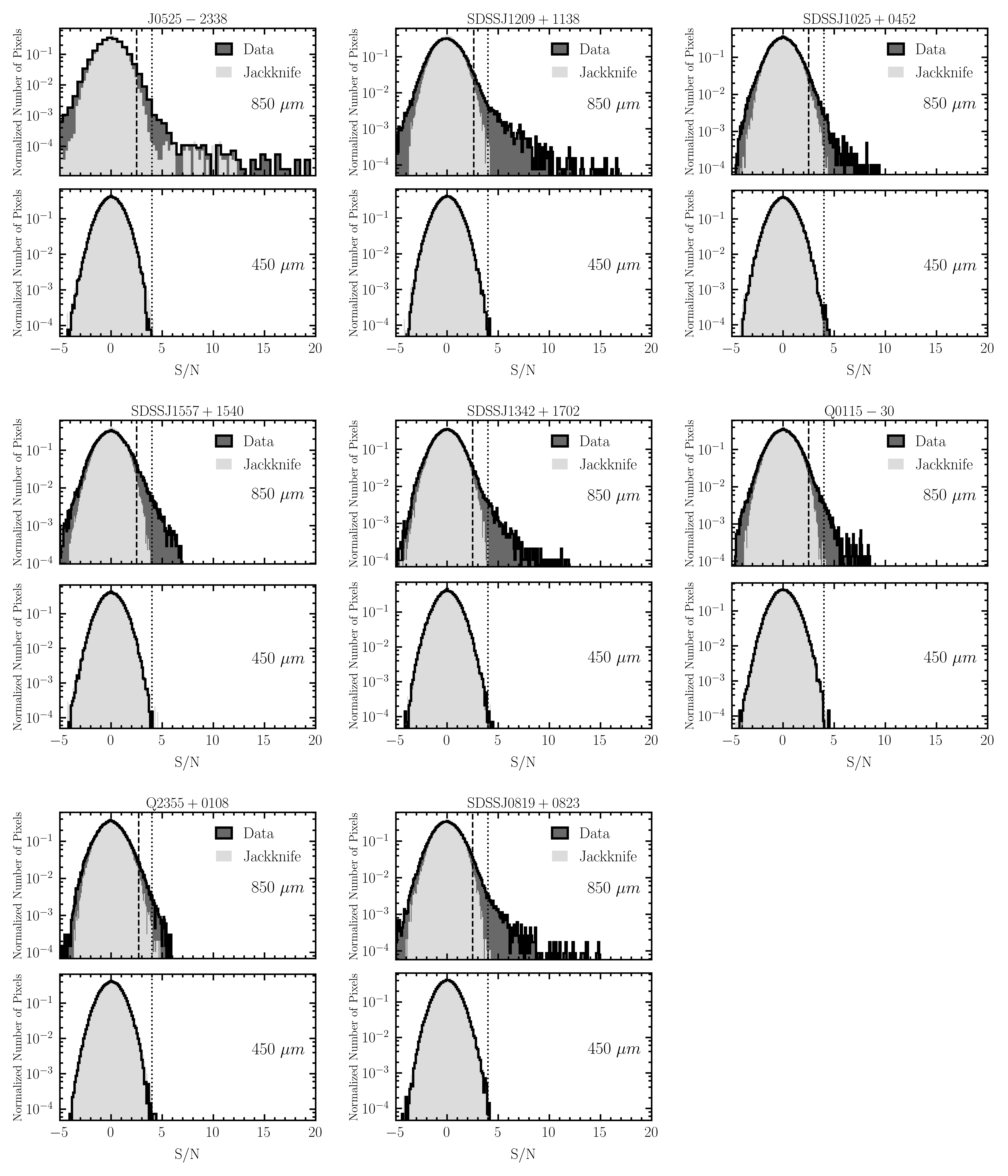}
\caption{Same as Fig.~\ref{fig:SNhistos1}, but for eight quasar fields from the QSO MUSEUM survey. Note that for J0525-2338 the jackknife map is affected by the very bright detection associated with this quasar. This should capture the number of spurious sources at high S/N present in the real data and close to this detection.}
\label{fig:SNhistos2}
\end{figure*}

\begin{figure*}
\centering
\includegraphics[width=1.0\textwidth]{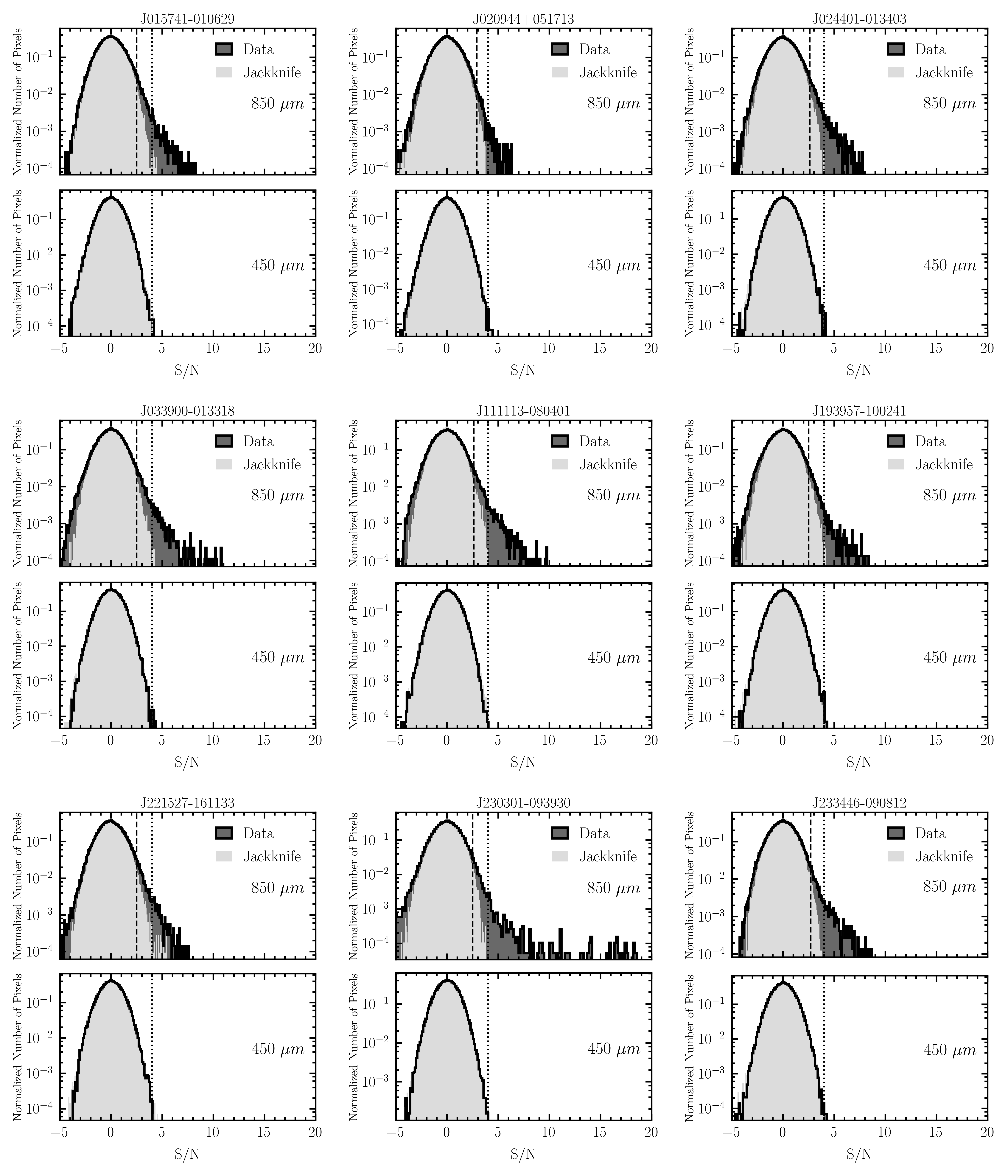}
\caption{Same as Fig.~\ref{fig:SNhistos1}, but for nine quasar fields from the MAGG survey (the last target J233446-090812 is also part of the QSO MUSEUM survey).}
\label{fig:SNhistos3}
\end{figure*}

\FloatBarrier

\section{S/N maps at 850~$\mu$m and 450~$\mu$m}
\label{app:maps}
In this appendix we give the obtained maps for the remaining 17 fields not shown in Fig.~\ref{fig:Maps1}.
\FloatBarrier

\begin{center} 
\begin{minipage}{\textwidth}
\centering
\includegraphics[width=0.81\textwidth]{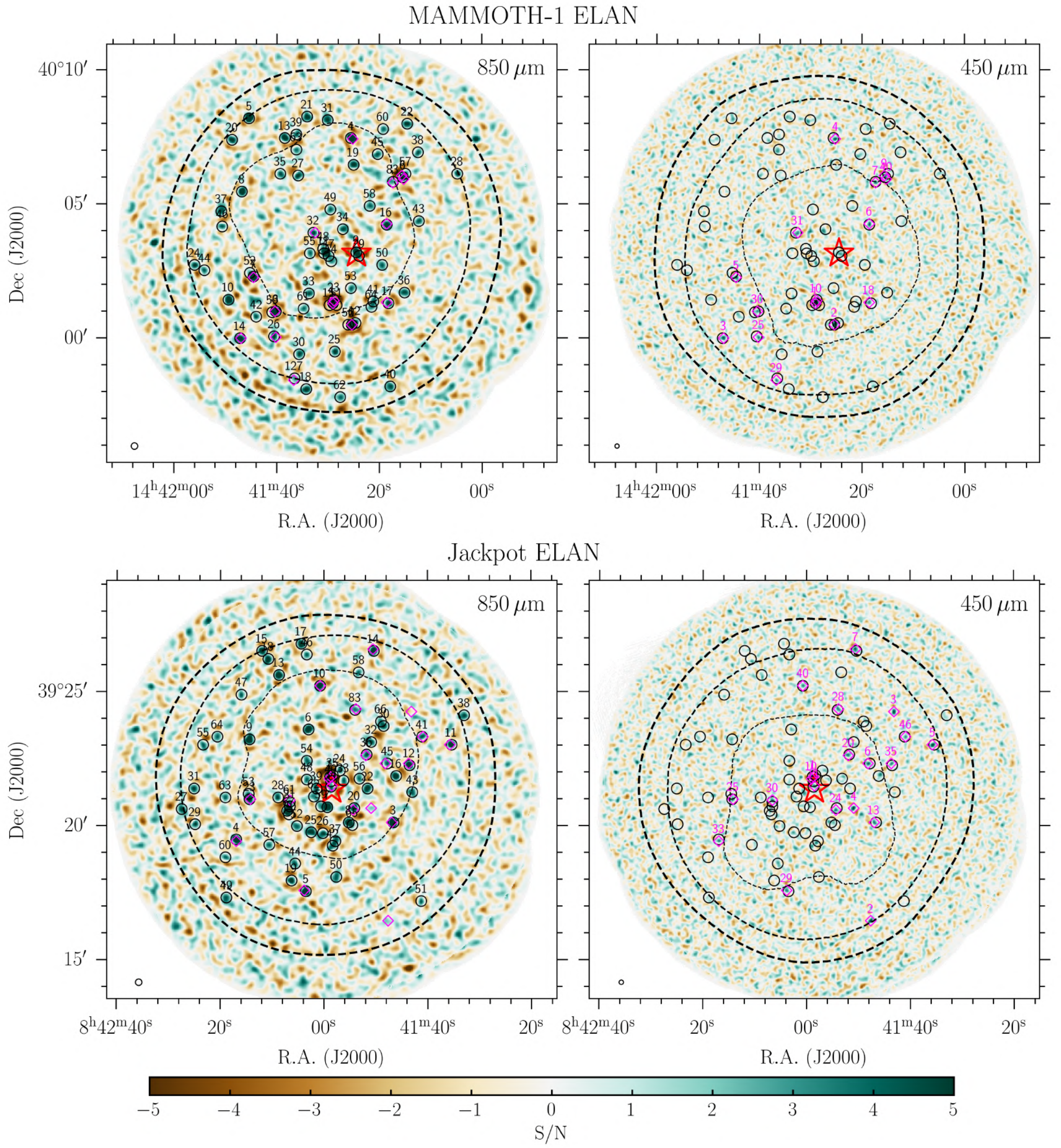}
\captionof{figure}{Same as for Fig.~\ref{fig:Maps1}, but for two ELAN fields, Jackpot and MAMMOTH-1.
}
\label{fig:MapsELAN}
\end{minipage}
\end{center}

\begin{figure*}
\centering
\includegraphics[width=0.81\textwidth]{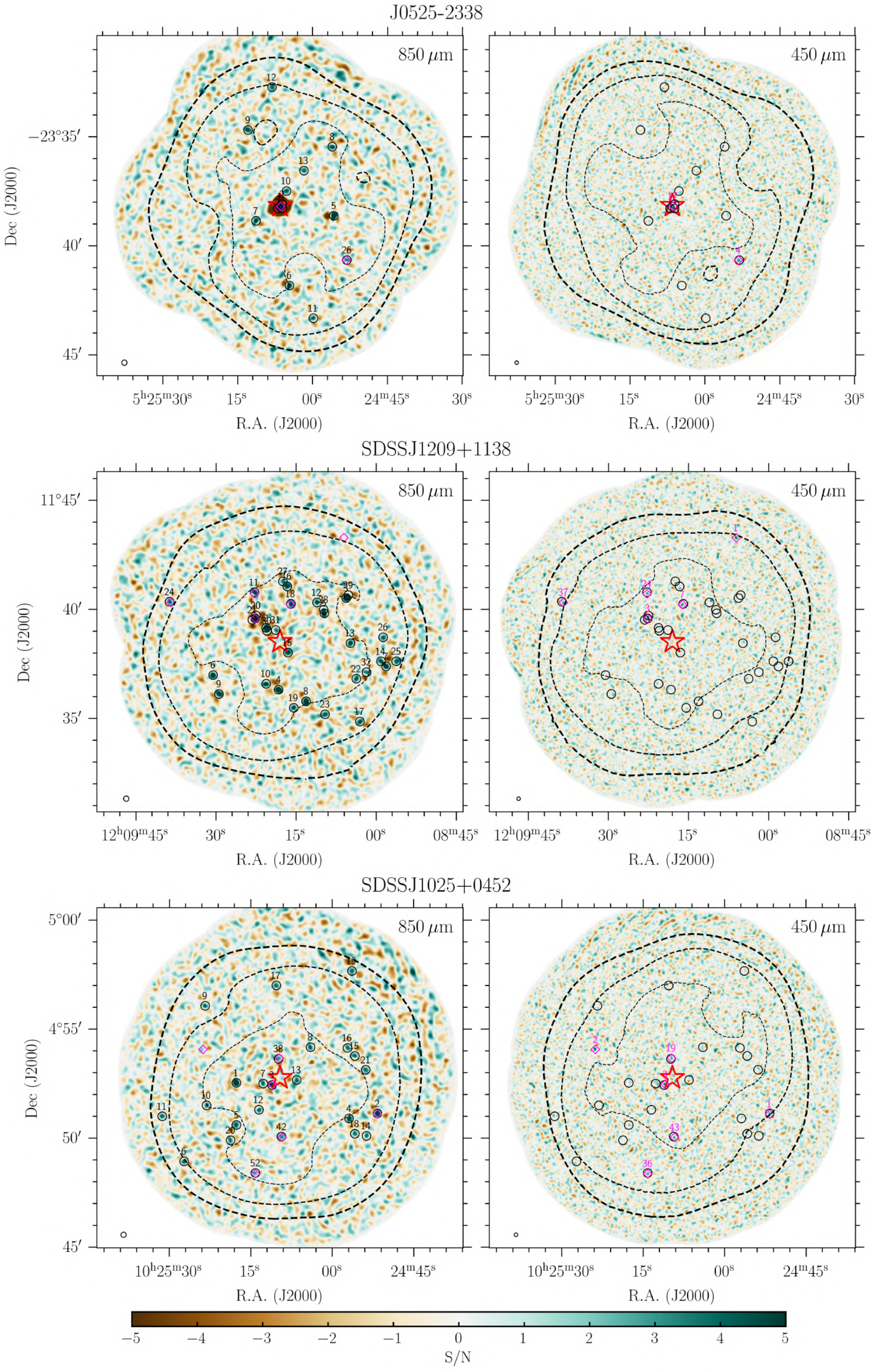}
\caption{Same as for Fig.~\ref{fig:Maps1}, but for three quasar fields from the QSO MUSEUM survey: J0525-2338, SDSSJ1209+1138, and SDSSJ1025+0452.
}
\label{fig:Maps2}
\end{figure*}

\begin{figure*}
\centering
\includegraphics[width=0.81\textwidth]{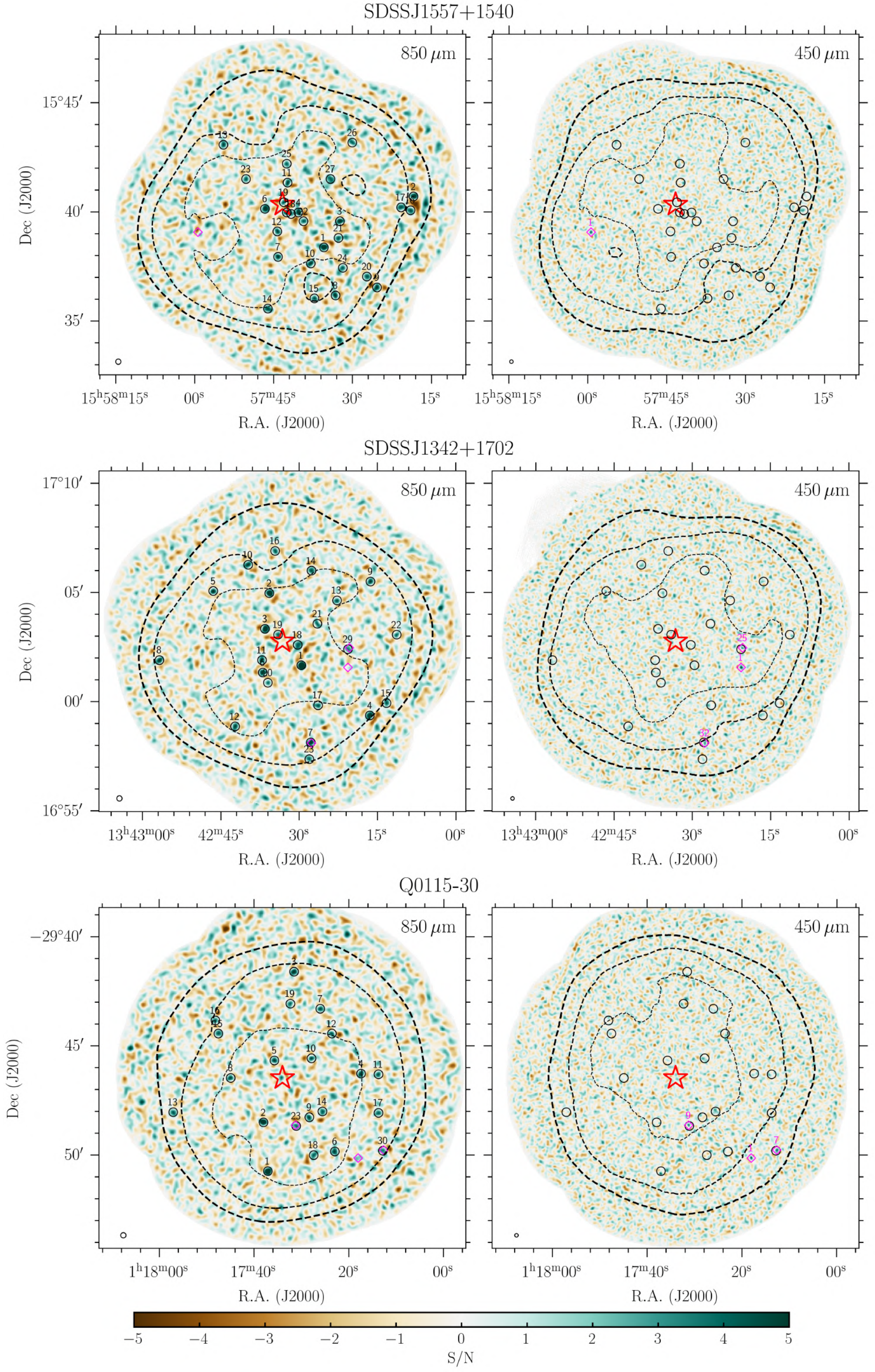}
\caption{Same as for Fig.~\ref{fig:Maps1}, but for three quasar fields from the QSO MUSEUM survey: SDSSJ1557+1540, SDSSJ1342+1702, and Q0115-30.
}
\label{fig:Maps3}
\end{figure*}

\begin{figure*}
\centering
\includegraphics[width=0.81\textwidth]{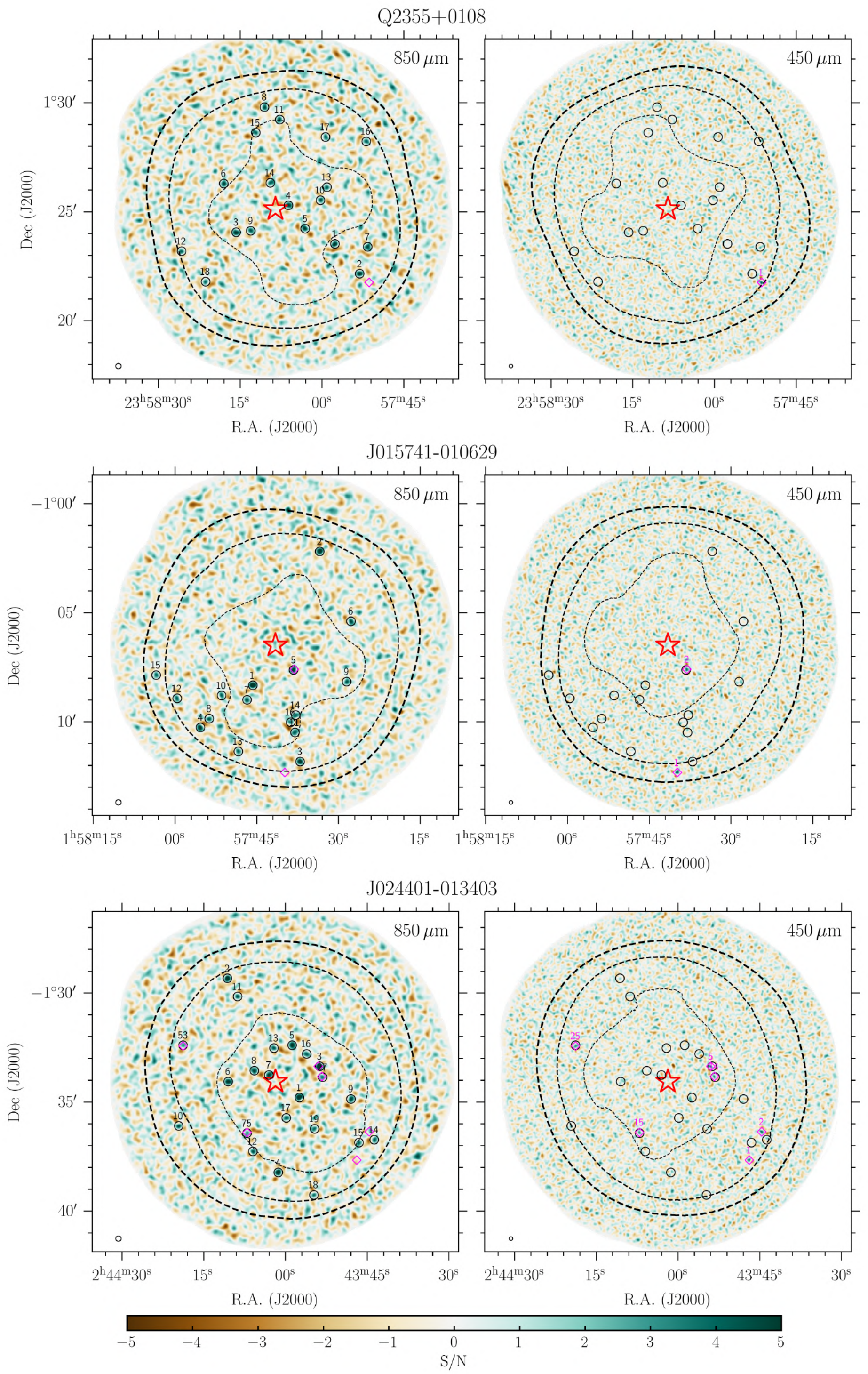}
\caption{Same as for Fig.~\ref{fig:Maps1}, but for one quasar field from the QSO MUSEUM survey, Q2355+0108, and two quasar fields from the MAGG survey, J015741-010629 and J024401-013403.}
\label{fig:Map4}
\end{figure*}

\begin{figure*}
\centering
\includegraphics[width=0.81\textwidth]{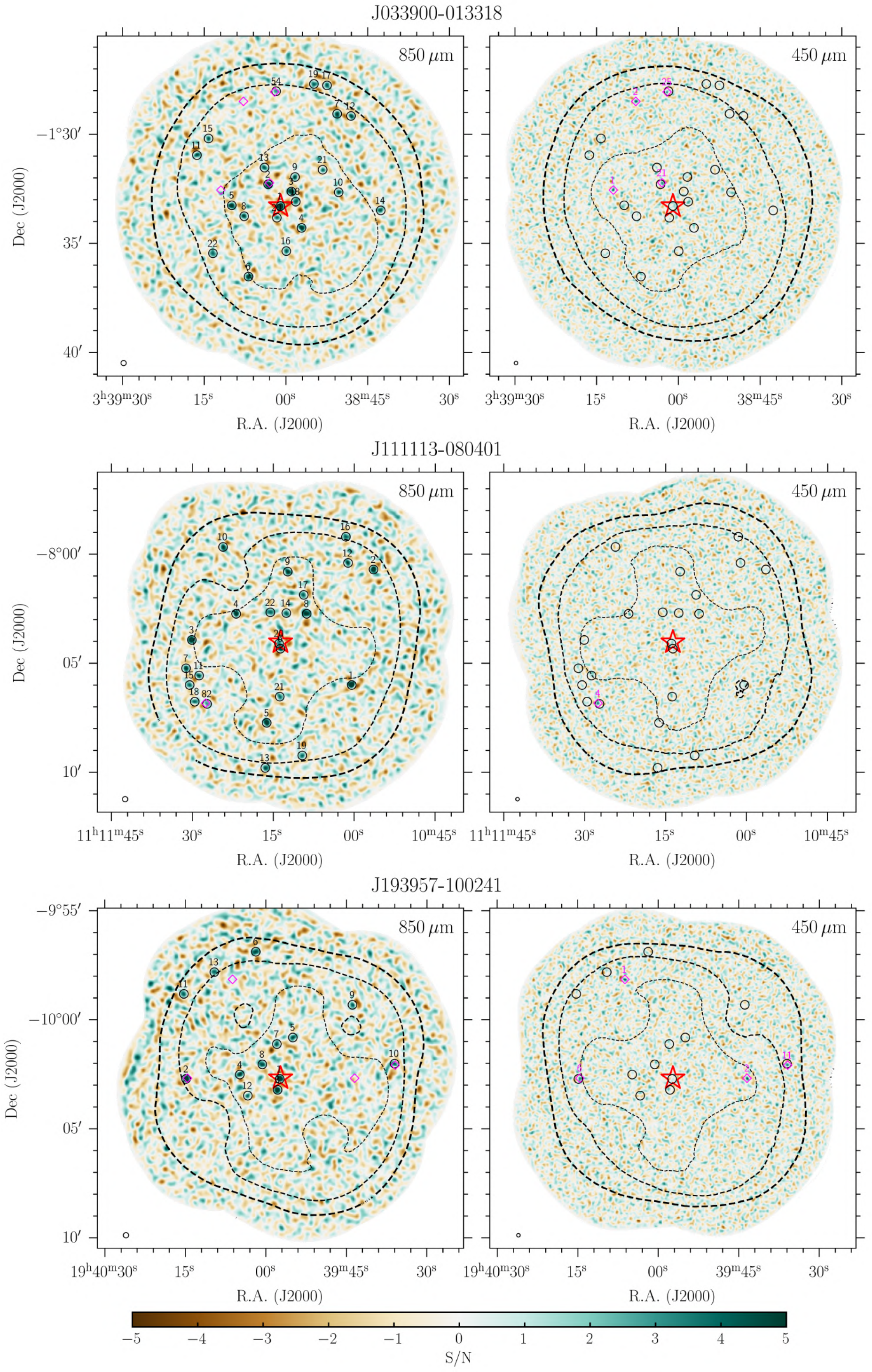}
\caption{Same as for Fig.~\ref{fig:Maps1}, but for three quasar fields from the MAGG survey: J033900-013318, J111113-080401, and J193957-100241.
}
\label{fig:Maps6}
\end{figure*}

\begin{figure*}
\centering
\includegraphics[width=0.81\textwidth]{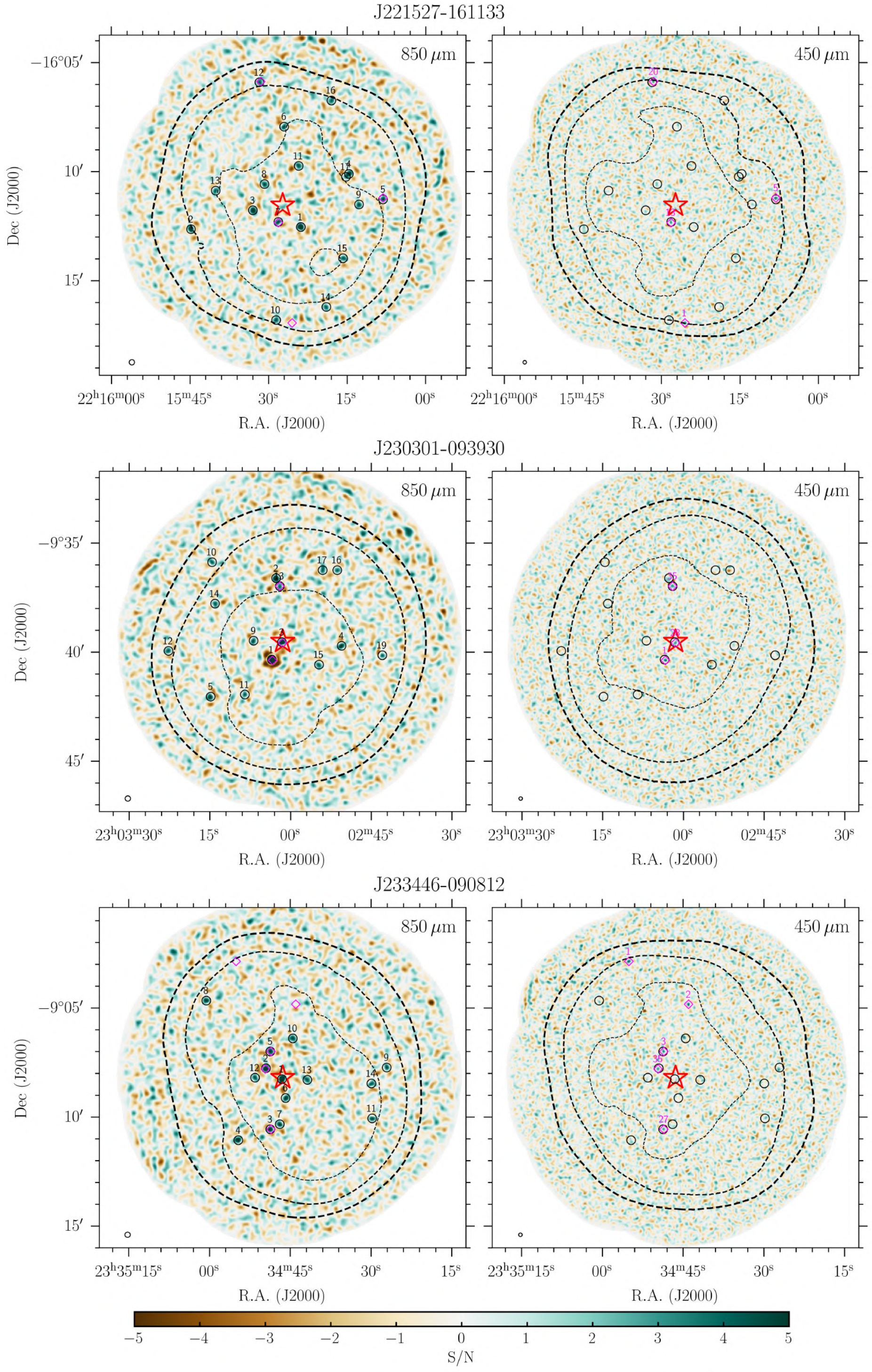}
\caption{Same as for Fig.~\ref{fig:Maps1}, but for three quasar fields from the MAGG survey: J221527-161133, J230301-093930, and J233446-090812 (this last field is also in the QSO MUSEUM survey).
}
\label{fig:Maps7}
\end{figure*}

\FloatBarrier

\section{Spurious sources}
\label{app:spurious}
\FloatBarrier

For completeness, we report the detailed results for each field and band on the number of spurious sources that could affect our catalogs, as discussed in Sect.~\ref{sec:rel}.

\begin{center} 
\begin{minipage}{\textwidth}
\scriptsize
\captionof{table}{Estimates of numbers of spurious sources in our catalogs.}
\centering
\begin{tabular}{lcccccc}
\hline
\hline
Quasar  & Neg$_{850}$\tablefootmark{a} & Neg$_{450}$\tablefootmark{a} & jackknife$_{850}$\tablefootmark{b} & jackknife$_{450}$\tablefootmark{b} & $3<$S/N$_{850}<4$\tablefootmark{c} & $3<$S/N$_{450}<4$\tablefootmark{c} \\ 
\hline
Jackpot           & 36 & 2 & 0 & 2 & 0(0.0) & 3(0.5) \\
MAMMOTH-I         & 31 & 1 & 3 & 2 & 0(0.0) & 1(0.5) \\
Fabulous          & 22 & 1 & 1 & 5 & 0(0.1) & 3(0.3) \\
J0525-233         & 41 & 1 & 1 & 1 & 1(0.0) & 1(0.2) \\
SDSSJ1209+1138    & 14 & 0 & 3 & 1 & 1(0.1) & 0(0.3) \\
SDSSJ1025+0452    & 5  & 0 & 1 & 0 & 0(0.1) & 1(0.2) \\
SDSSJ1557+1540    & 13 & 1 & 0 & 2 & 0(0.0) & 1(0.2) \\
SDSSJ1342+1702    & 5  & 1 & 1 & 1 & 0(0.1) & 2(0.2) \\
Q-0115-30         & 12 & 1 & 1 & 3 & 1(0.0) & 0(0.2) \\
Q2355+0108        & 1  & 1 & 3 & 2 & 0(0.0) & 0(0.2) \\
SDSSJ0819+0823    & 7  & 1 & 0 & 2 & 0(0.1) & 1(0.3) \\
J015741-01062957  & 2  & 2 & 1 & 2 & 0(0.0) & 0(0.2) \\
J020944+051713    & 5  & 3 & 0 & 2 & 0(0.0) & 0(0.2) \\
J024401-013403    & 5  & 1 & 0 & 0 & 0(0.0) & 1(0.3) \\
J033900-013318    & 5  & 0 & 1 & 2 & 0(0.0) & 1(0.2) \\
J111113-080401    & 8  & 1 & 0 & 1 & 0(0.0) & 3(0.2) \\
J193957-100241    & 4  & 2 & 3 & 1 & 0(0.1) & 0(0.2) \\
J221527-161133    & 4  & 0 & 2 & 3 & 0(0.0) & 0(0.2) \\
J230301-093930    & 10 & 2 & 0 & 1 & 0(0.0) & 2(0.2) \\
J233446-090812    & 2  & 2 & 0 & 0 & 0(0.0) & 0(0.2) \\
\hline
\end{tabular}
\label{tab:spurious}
\tablefoot{
\tablefoottext{a}{Estimate of the number of spurious sources at ${\rm S/N}>4$ obtained from the inverted maps.}
\tablefoottext{b}{More reliable estimate of the number of spurious sources at ${\rm S/N}>4$ obtained from the jackknife maps.}
\tablefoottext{c}{Estimate of the number of spurious sources for the range $3<{\rm S/N}<4$ obtained from the jackknife maps. In brackets we list the estimate for the 1000 repetitions (see
Sect.~\ref{sec:rel}).}
}
\end{minipage}
\end{center}

\FloatBarrier
\clearpage

\section{Completeness tests}
\label{app:completeness}

In this appendix we report the figure for the completeness tests of each field and band.

\begin{figure}[!htb]
\centering
\includegraphics[width=1.0\columnwidth]{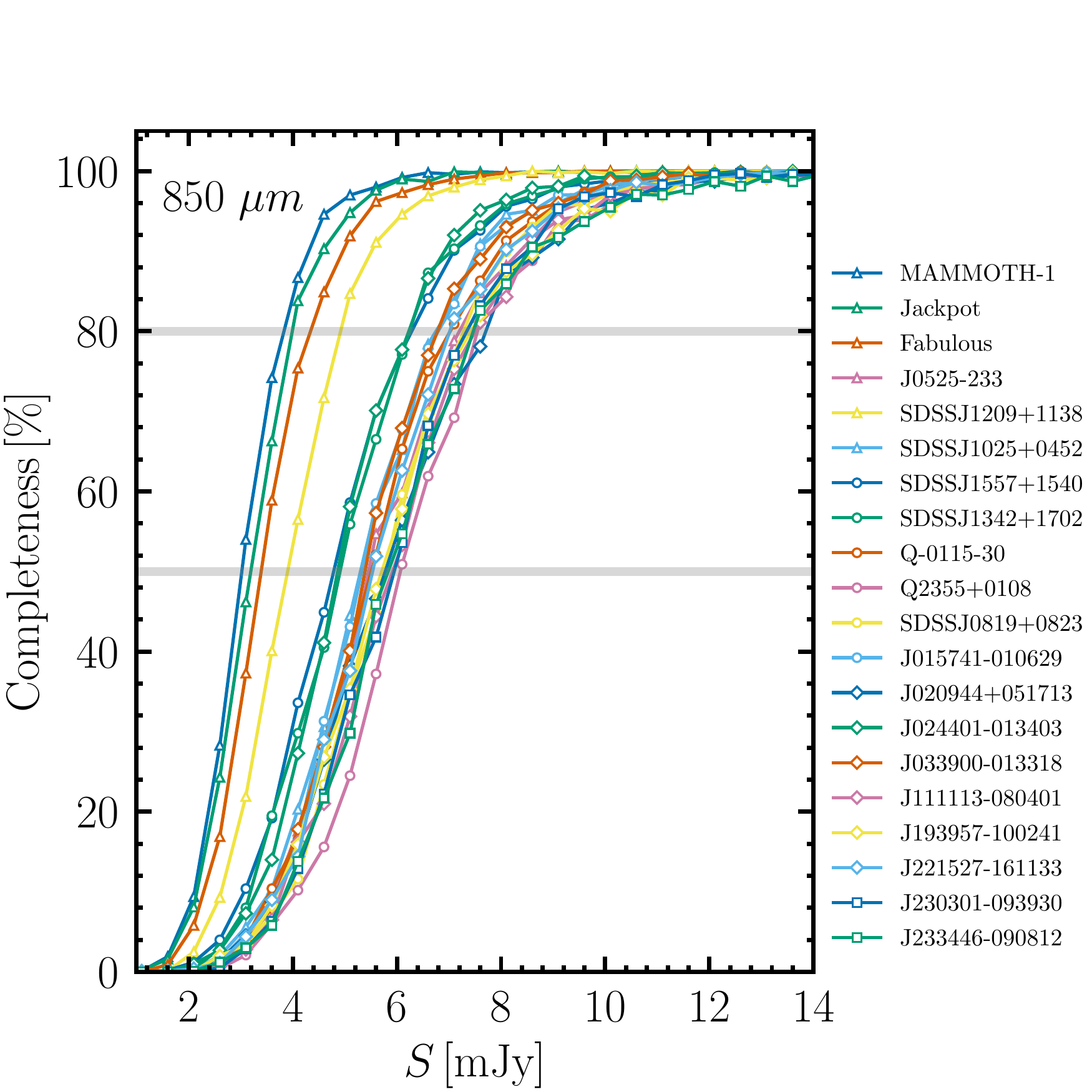}
\includegraphics[width=1.0\columnwidth]{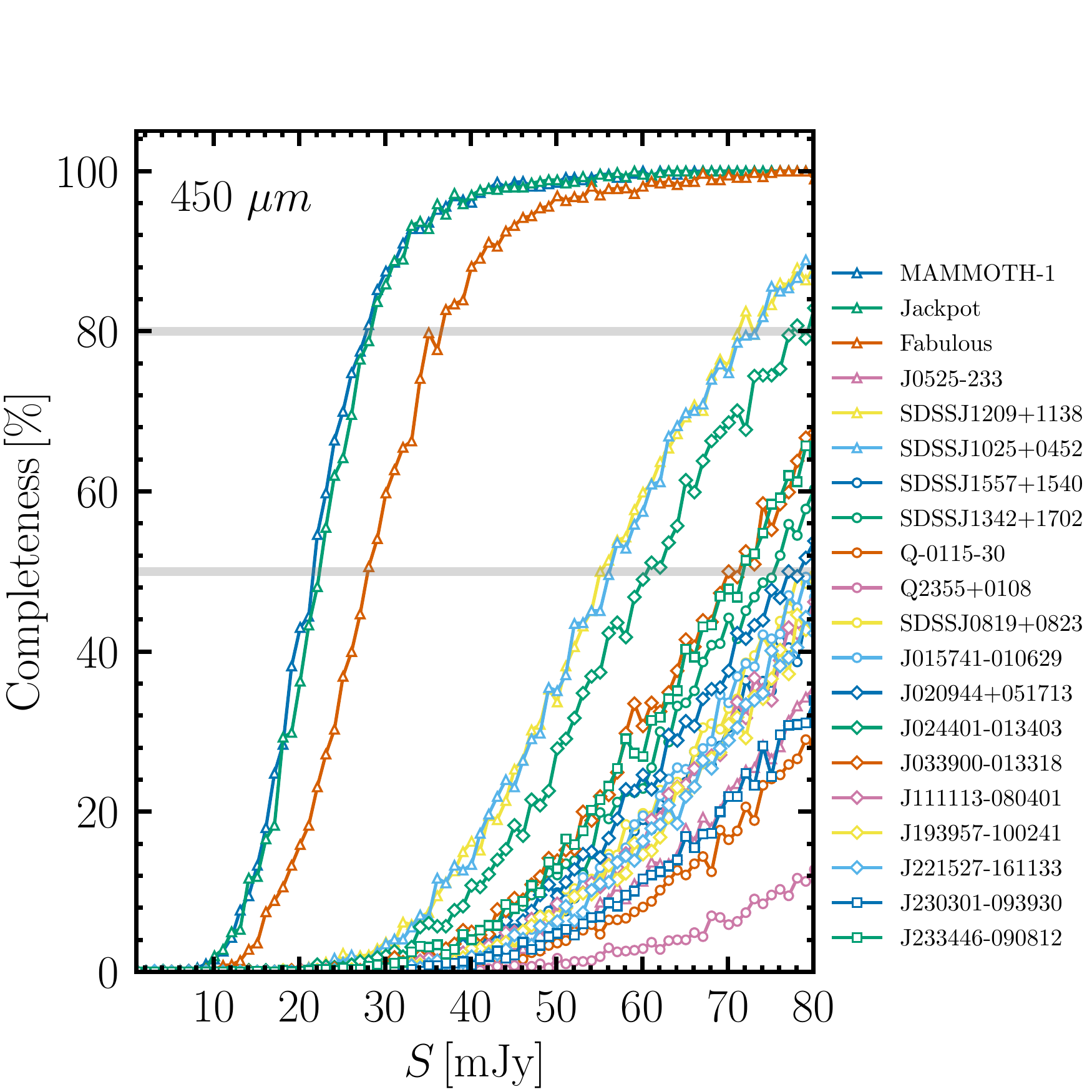}
\caption{Completeness for sources above $4\sigma$ as a function of flux for each field and at 850~$\mu$m (top) and at 450~$\mu$m (bottom).}
\label{fig:compl}
\end{figure}

\section{Source catalogs}
\label{app:catalogs}

\FloatBarrier

In this appendix we report all the source catalogs for the twenty fields in this study, built following the procedure
discussed in Sect.~\ref{sec:extr_cat}. Flux boosting and positional uncertainty are computed as done in \citet{Nowotka2022}. 
All catalogs are ordered following the S/N of the sources. For the three ELAN fields, we kept the naming from \citet{Nowotka2022} for the sources in common.
Detections at $<10\arcsec$ from the central system (quasar or ELAN) are indicated by an asterisk next to their name. These sources are likely counterparts to the ELANe and quasars.
We stress that caution has to be used when working with these catalogs for follow-up observations as some of these sources could be spurious, especially at 450~$\mu$m (Sect.~\ref{sec:extr_cat}).
The tables of the catalogs F1 to F40 are only available in electronic form at the CDS via anonymous ftp to cdsarc.cds.unistra.fr (130.79.128.5)
or via this link\footnote{\url{https://cdsarc.cds.unistra.fr/cgi-bin/qcat?J/A+A/}}.

\FloatBarrier

\clearpage
\section{Differential number counts and Monte Carlo simulations for all fields}
\label{app:diff_counts}
In this appendix we show the differential number counts and the results of the Monte Carlo simulations for all the fields not shown in Fig.~\ref{fig:counts}.

\FloatBarrier

\begin{center} 
\begin{minipage}{\textwidth}
\centering
\includegraphics[width=0.67\textwidth]{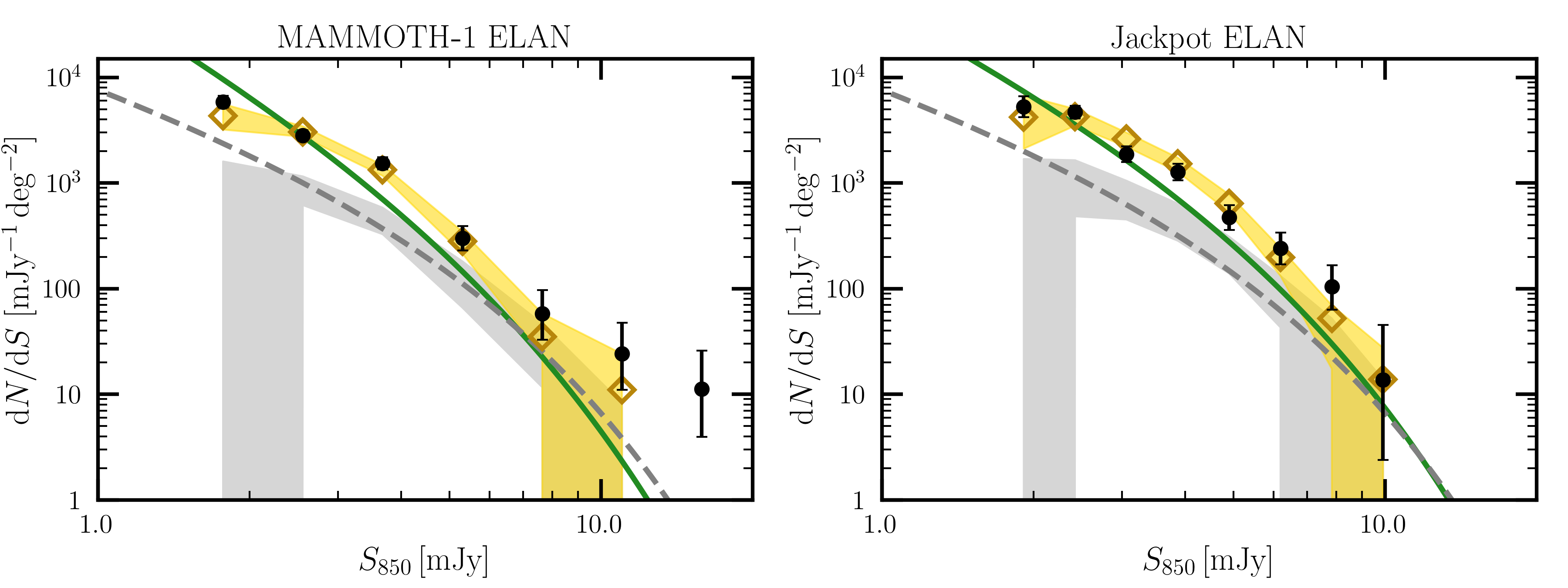}
\captionof{figure}{Same as Fig.~\ref{fig:counts}, but for the two ELAN fields, MAMMOTH-1 and Jackpot.}
\label{fig:countsELAN}
\end{minipage}
\end{center}

\begin{center} 
\begin{minipage}{\textwidth}
\centering
\includegraphics[width=1.0\textwidth]{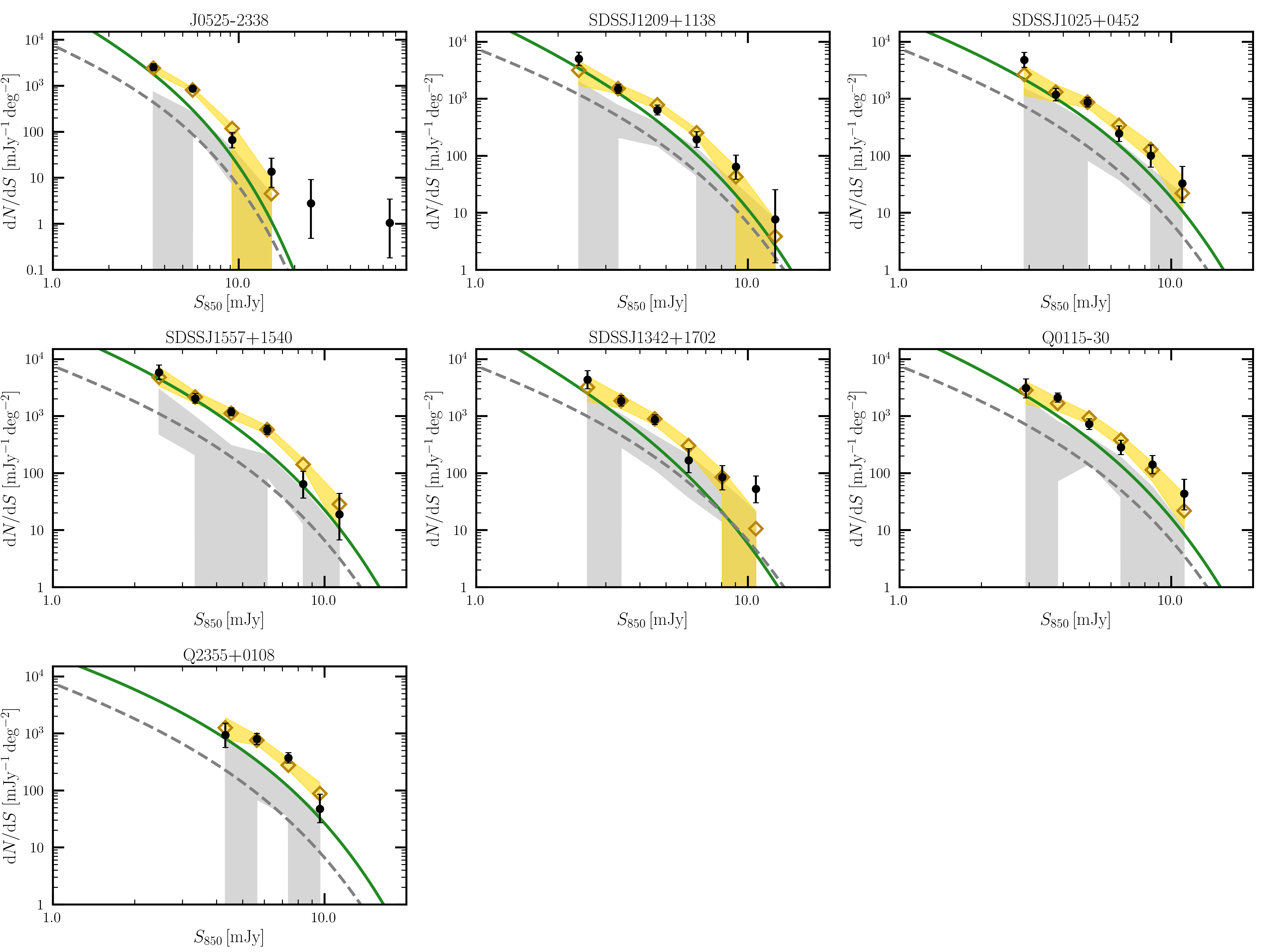}
\captionof{figure}{Same as Fig.~\ref{fig:counts}, but for the eight fields from the QSO MUSEUM (see Table~\ref{tab:sample}).}
\label{fig:countsMUSEUM}
\end{minipage}
\end{center}

\begin{figure*}
\centering
\includegraphics[width=1.0\textwidth]{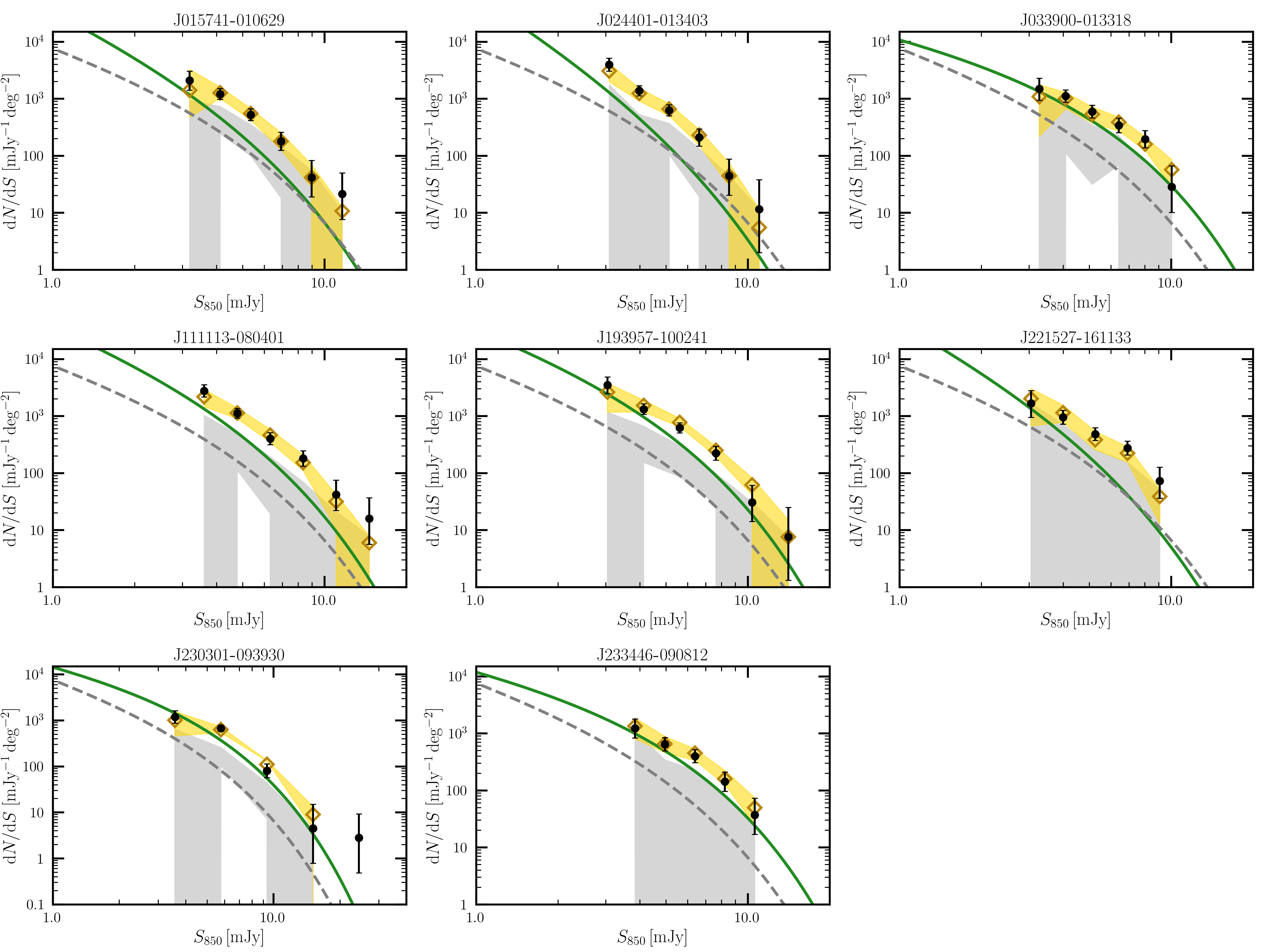}
\caption{Same as Fig.~\ref{fig:counts}, but for the nine fields from the MAGG survey (the last field J233446-090812 is also from the QSO MUSEUM survey; see Table~\ref{tab:sample}).}
\label{fig:countsMAGG}
\end{figure*}

\FloatBarrier

\section{Different stacked maps for the ELAN fields}
\label{app:second_sub}

\FloatBarrier

\begin{figure*}
\centering
\includegraphics[width=0.3\textwidth]{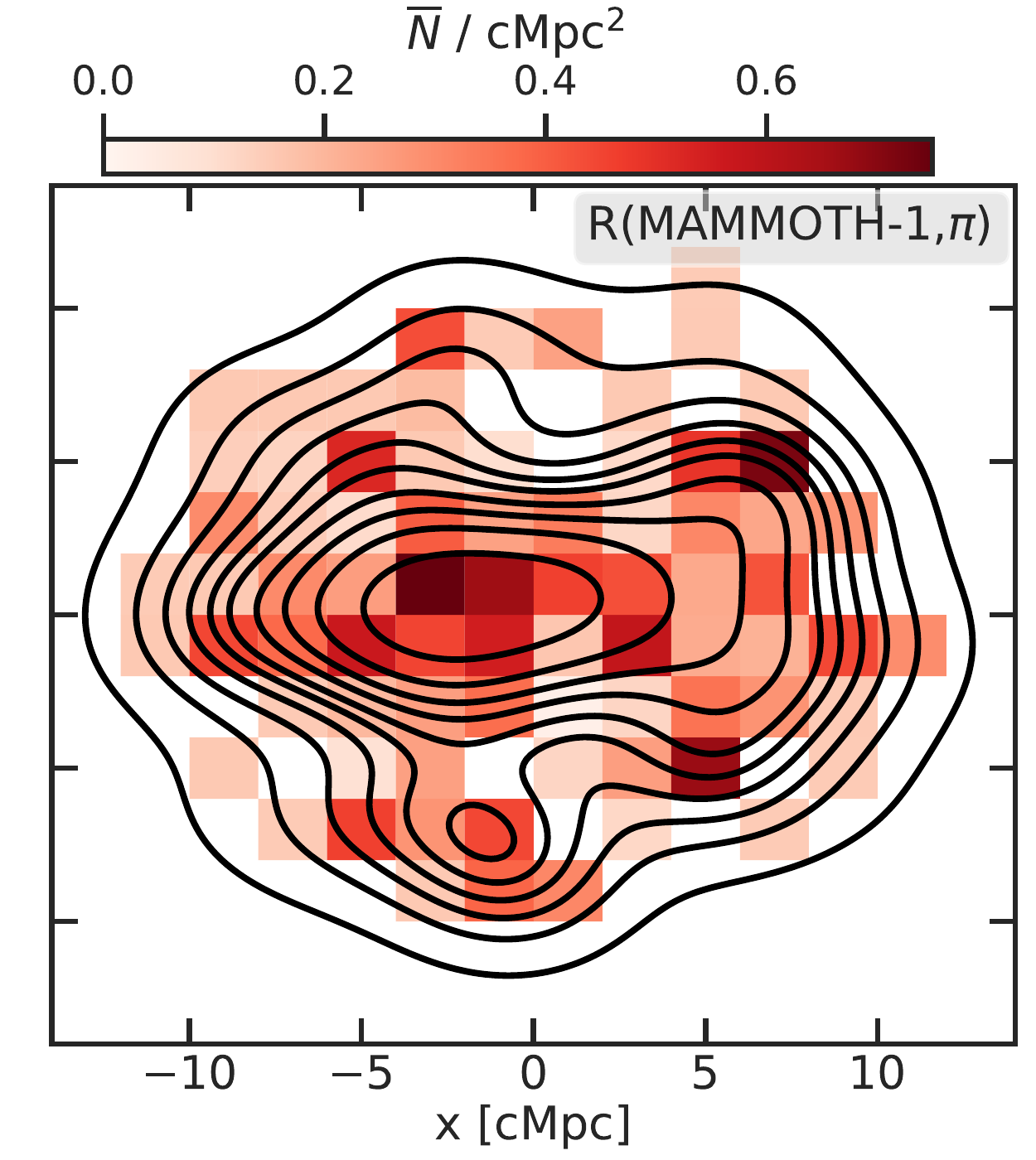}
\includegraphics[width=0.3\textwidth]{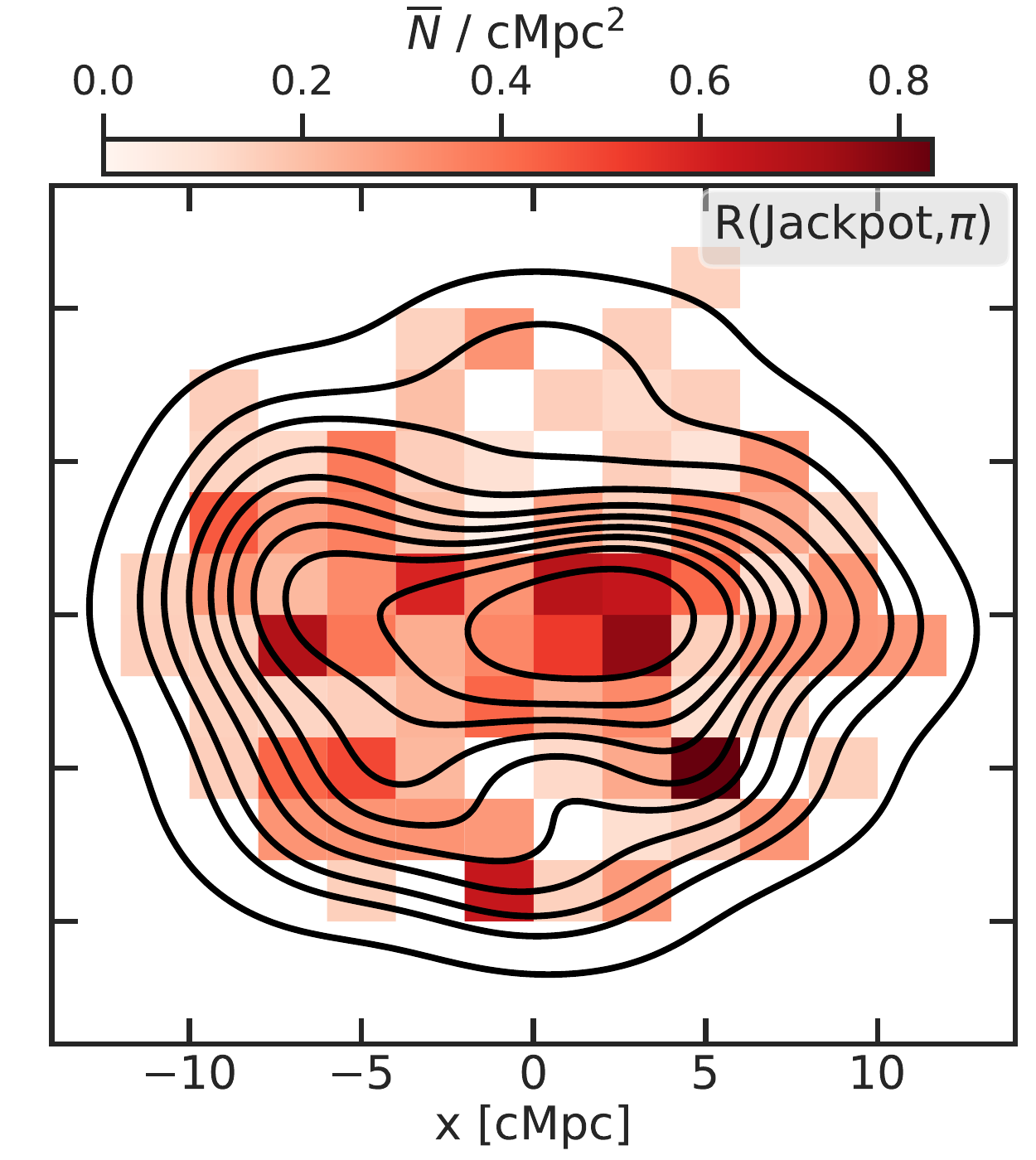}
\includegraphics[width=0.3\textwidth]{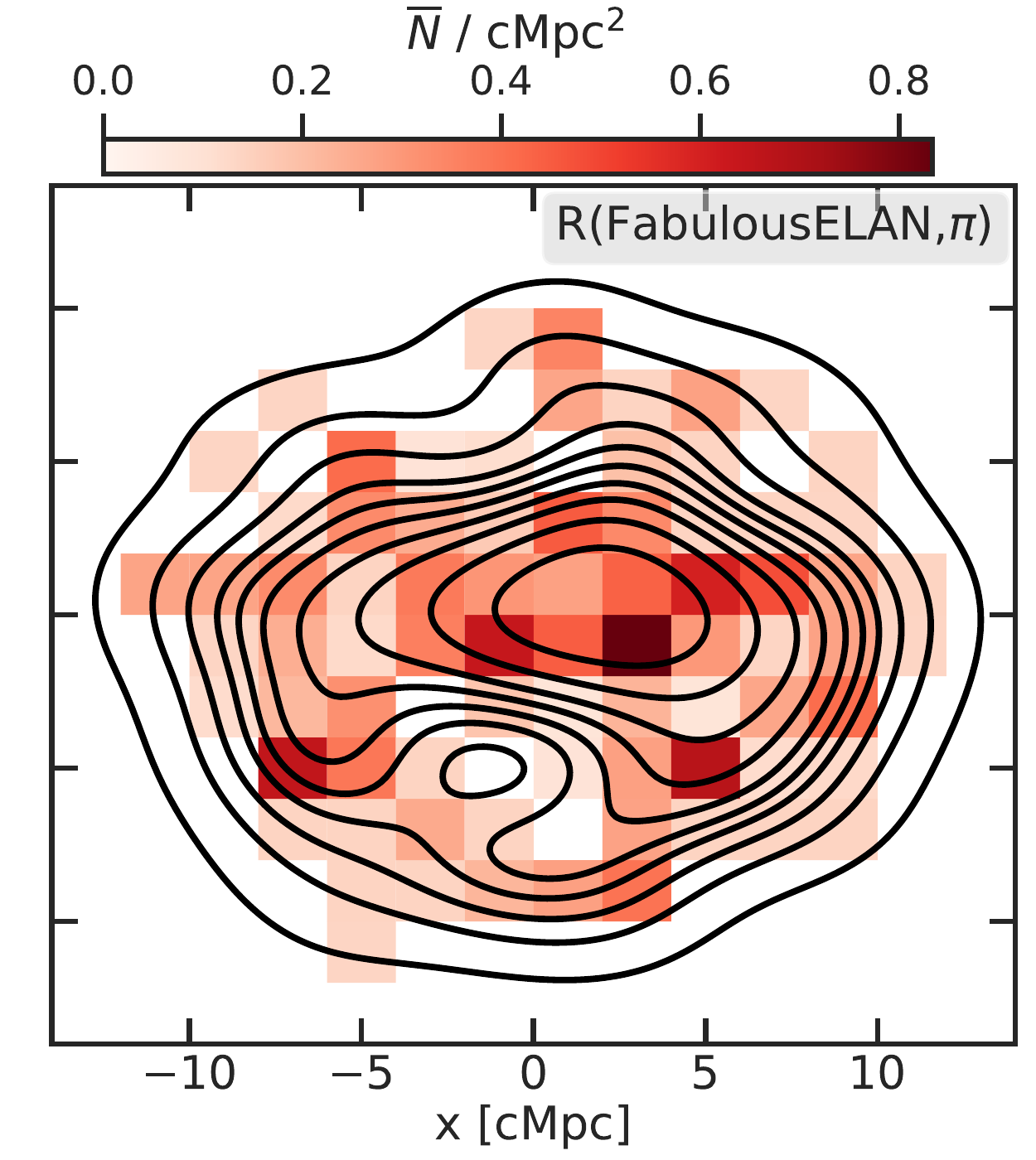}
\caption{Versions of the stacked ELAN fields maps in Fig.~\ref{fig:2Dmaps_andprofiles}, where one of the three fields (indicated in each panel) is rotated by an extra 180 degrees after its first rotation by $\alpha_{\rm rms}$.  These maps demonstrate that the secondary contours in the central panels of Fig.~\ref{fig:2Dmaps_andprofiles} are due to the chance alignment of major substructures in the individual ELAN fields.}
\label{fig:ELAN_maps}
\end{figure*}

\begin{figure}
\centering
\includegraphics[width=0.8\columnwidth]{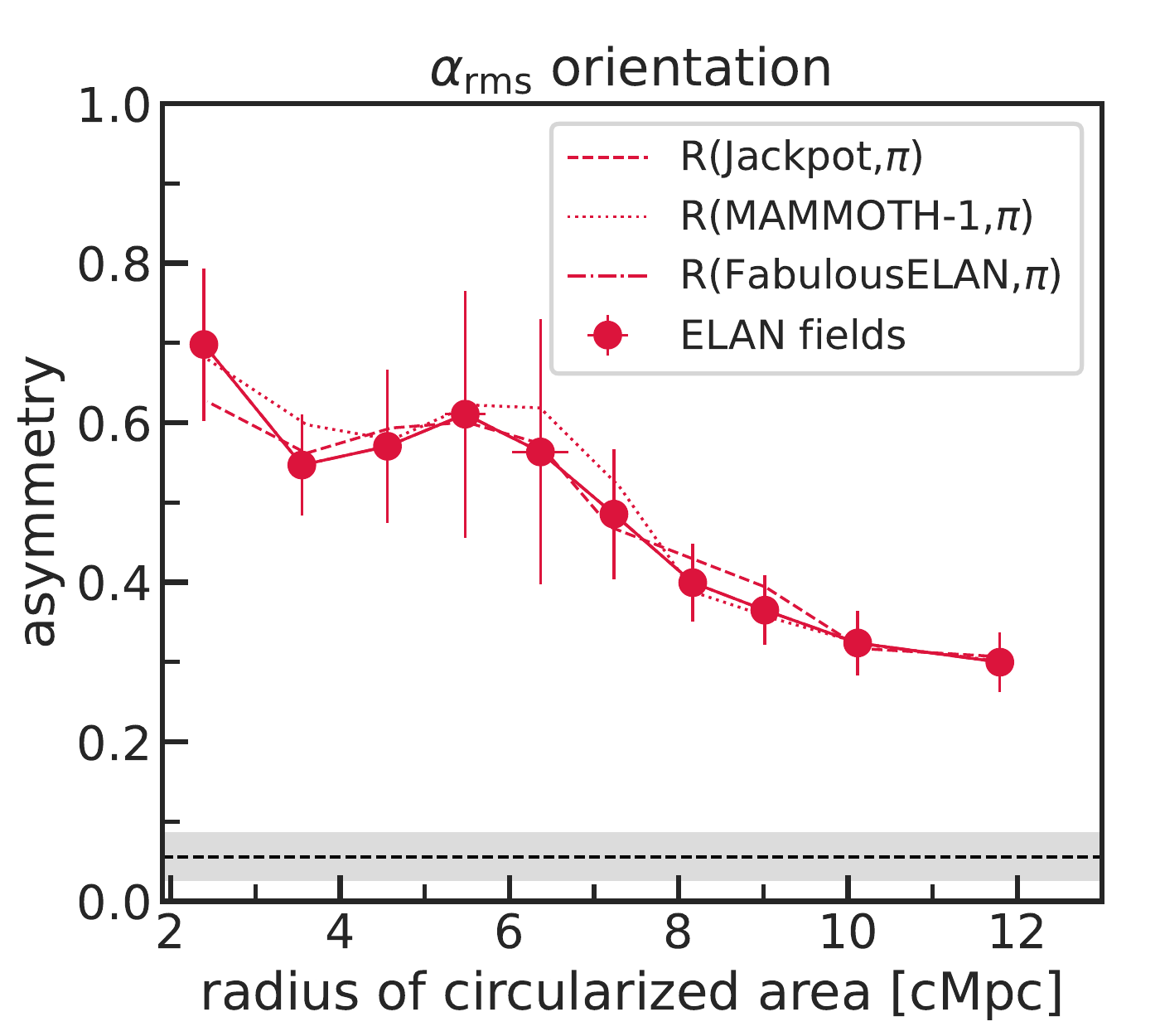}
\caption{Asymmetry profiles (dashed, dotted, and dashed-dotted curves) extracted from the contours in Fig.~\ref{fig:ELAN_maps} compared with the corresponding profile for the stacked ELAN fields after rotating the individual fields by 
$\alpha_{\rm rms}$. The extra rotation by 180 degrees for one of the three fields results in very similar asymmetry profiles. This means that the shape of these profiles is stable against the rotation of the three ELAN fields.}
\label{fig:test_asymmetry}
\end{figure}

In Sect.~\ref{sec:spatialScale} we show that the ELAN fields have a stacked map showing a secondary substructure and that their asymmetry profile decreases abruptly with respect to those of quasars (Fig.~\ref{fig:2Dmaps_andprofiles}). This finding is evident when obtaining the stacked maps by rotating the fields using the direction of the overdensities estimated with both the rms distances ($\alpha_{\rm rms}$) or the 2D inertia tensor ($\alpha_{\rm inertia}$) because the two angles are consistent for these fields. 
In this appendix we show that the second substructure is due to chance superposition of structures in each field and that the decrease in the asymmetry profile does not change significantly depending on the 
alignment of the three available fields. In particular, we generated three additional stacked maps by rotating one of the ELAN fields by additional 180 degrees, after its first rotation with $\alpha_{\rm rms}$. 
Figure~\ref{fig:ELAN_maps} shows that the maps are all characterized by secondary peaks, while Fig.~\ref{fig:test_asymmetry} shows that the asymmetry profile for these different maps and the one in the main text are consistent. Therefore, ELAN fields are characterized by dense substructures in the surrounding of the central Ly$\alpha$ nebula, which remain evident in the stacked maps of a few fields.

\FloatBarrier

\section{Preferred direction of the overdensities}
\label{app:over_direction}

For completeness, in this appendix we show the histograms for the calculation of the preferred direction of the overdensities $\alpha_{\rm rms}$ and $\alpha_{\rm inertia}$, together with a visualization of the discovered directions overlaid on the source distribution and Ly$\alpha$ nebula for all the remaining fields not shown in the main text (see Sect.~\ref{sec:spatialScale}). 

\begin{figure*}
\centering
\includegraphics[width=0.35\textwidth]{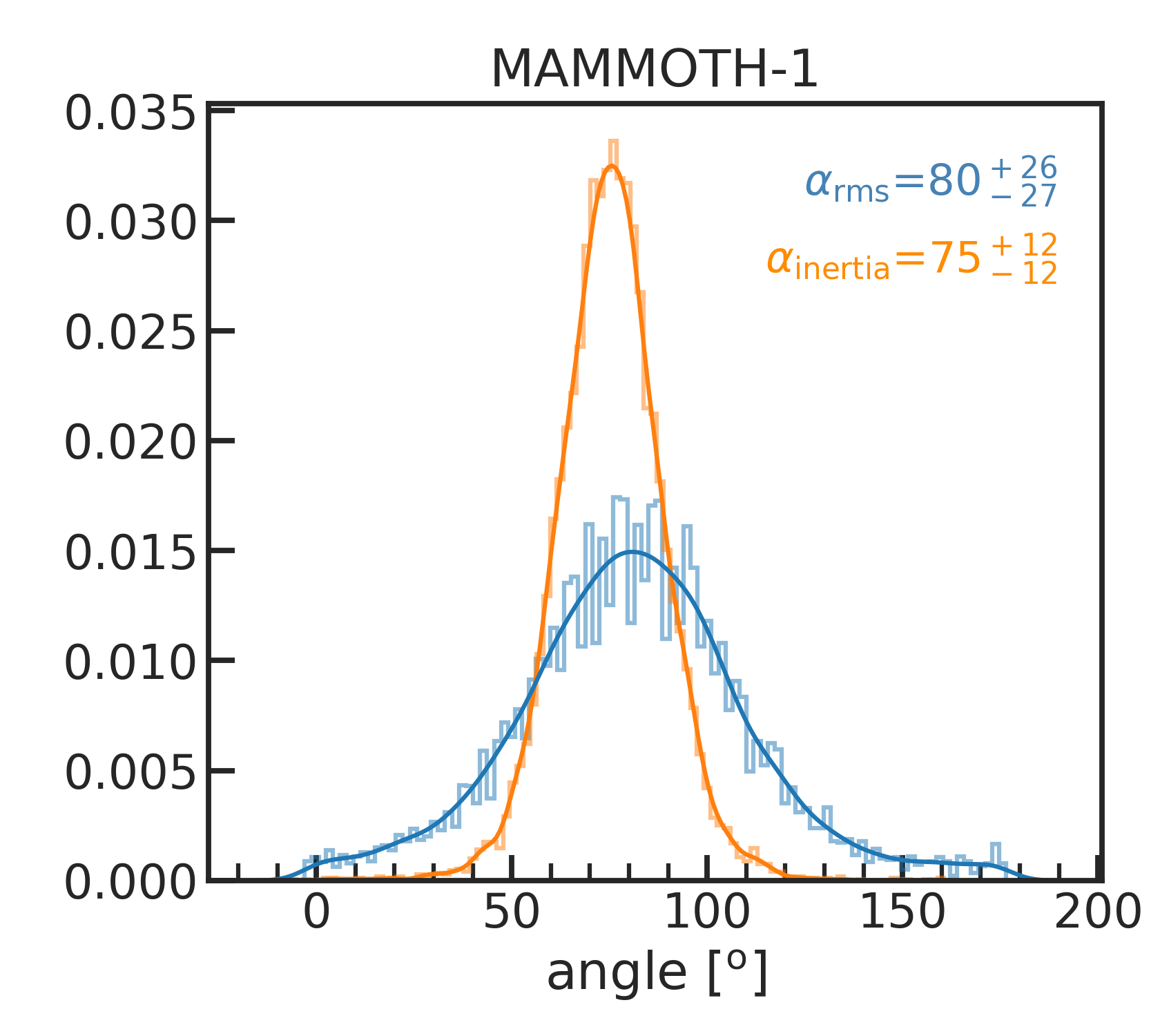}
\includegraphics[width=0.35\textwidth]{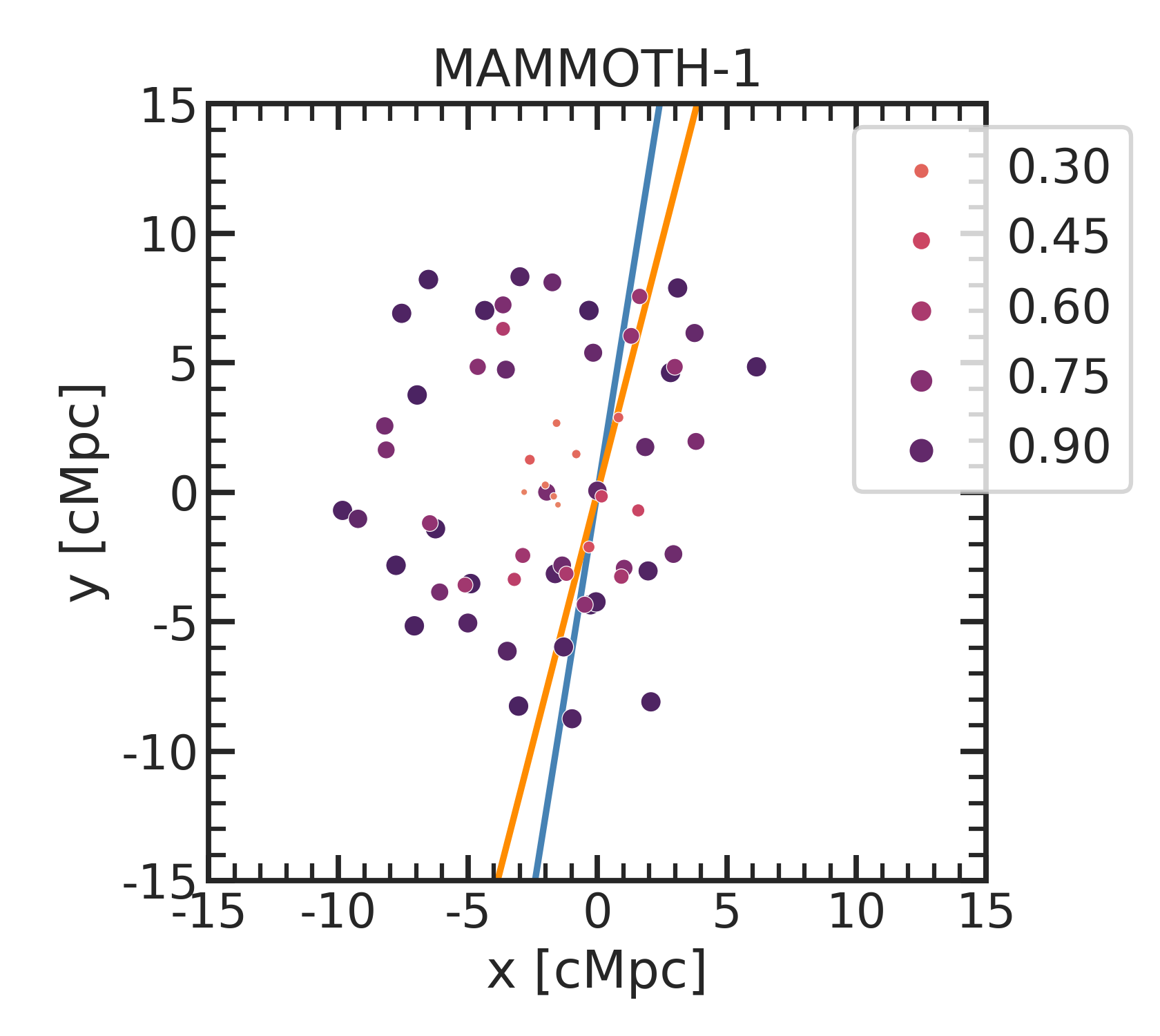}
\includegraphics[width=0.24\textwidth, trim=0 -3.2cm 0 0]{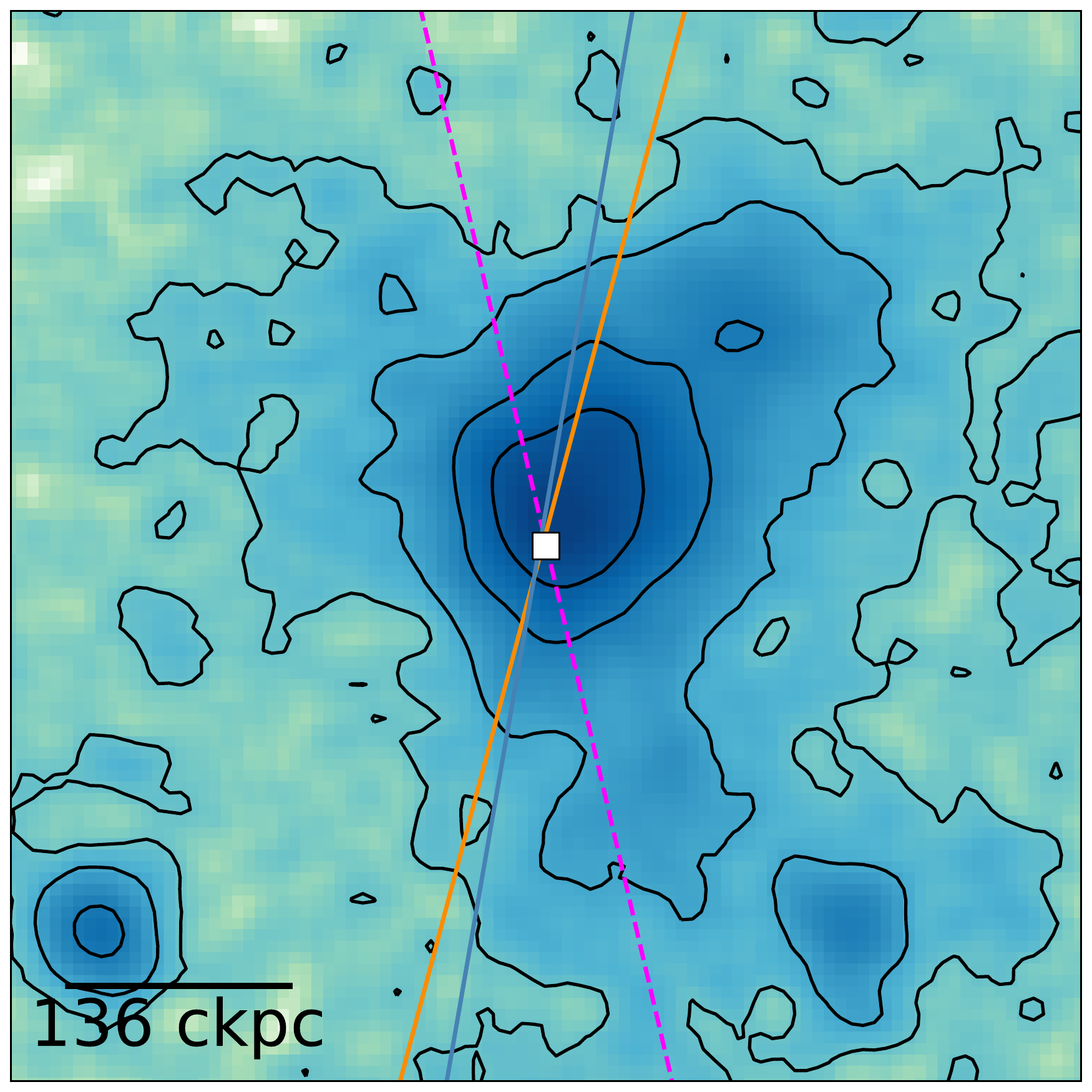}

\includegraphics[width=0.35\textwidth]{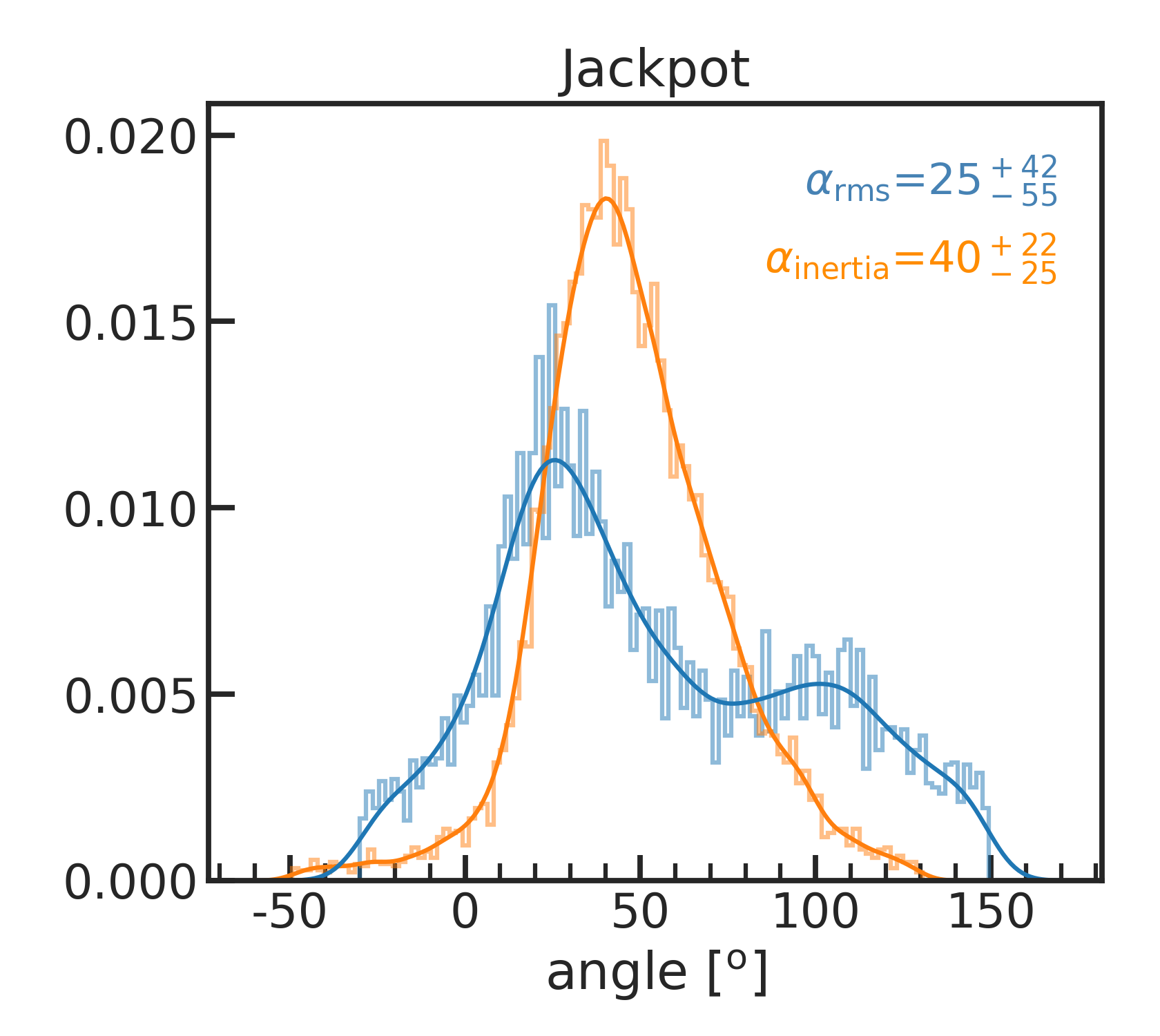}
\includegraphics[width=0.35\textwidth]{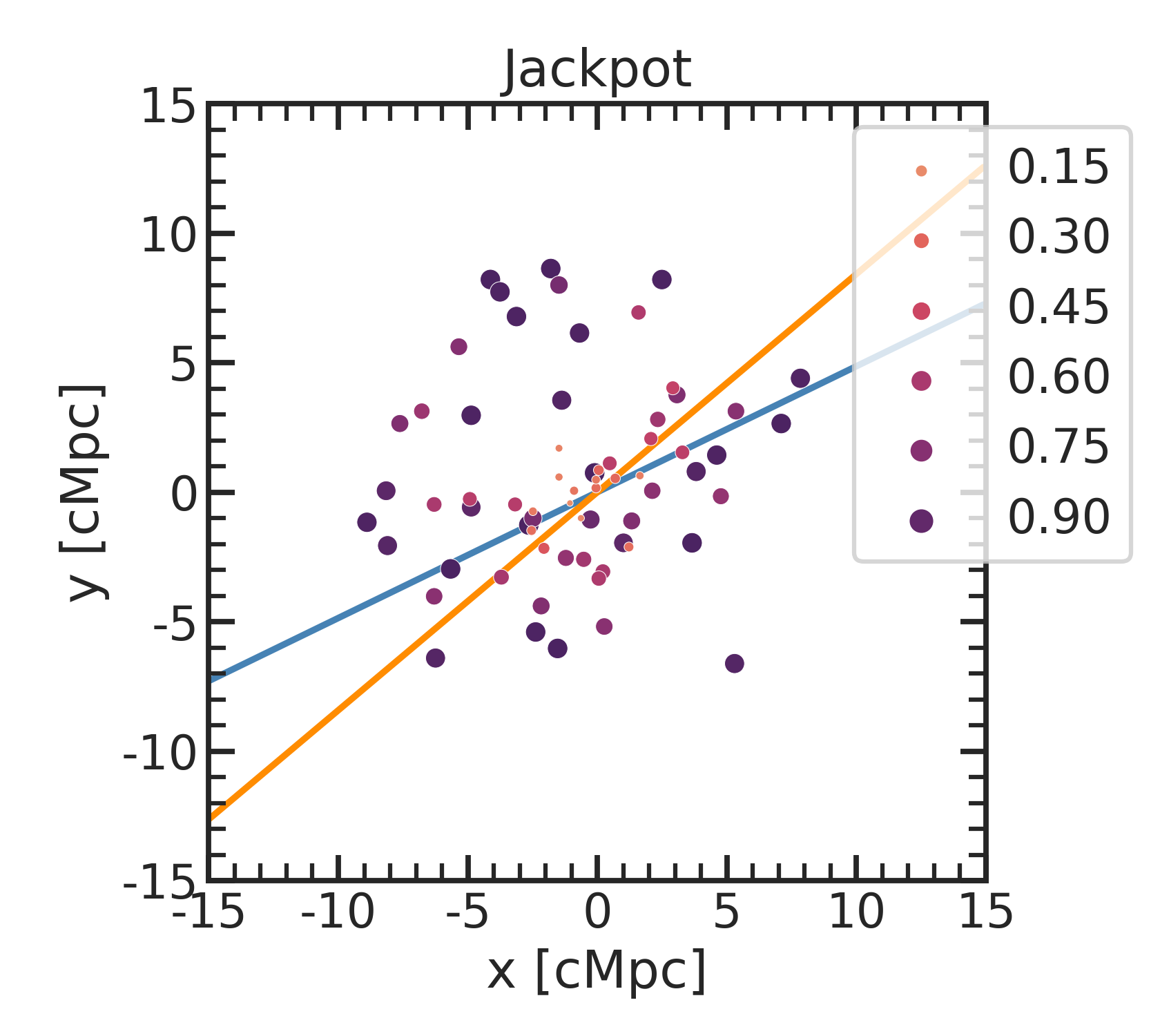}
\includegraphics[width=0.24\textwidth, trim=0 -3.2cm 0 0]{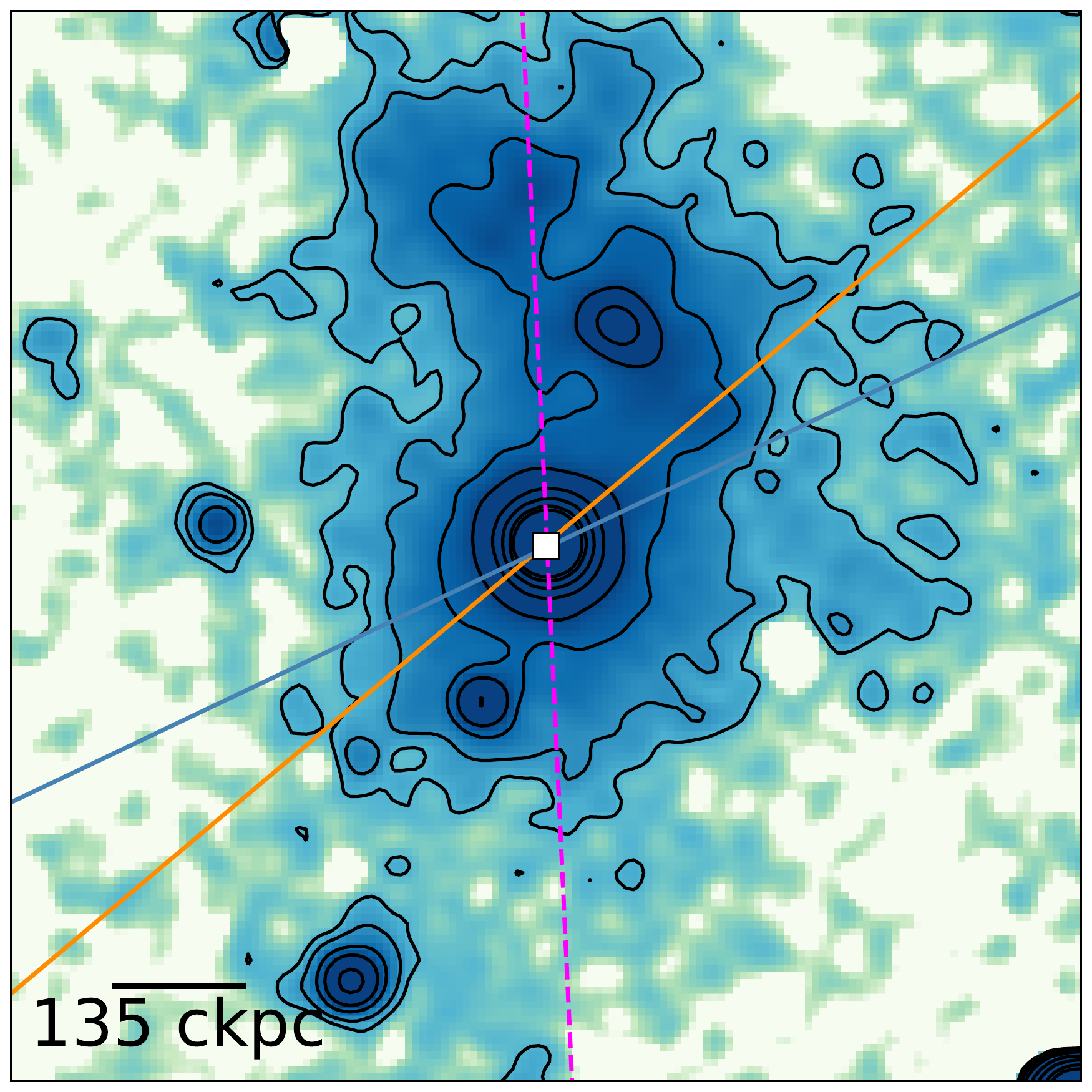}
\caption{Same as Fig.~\ref{fig:angle_over}, but for the MAMMOTH-1 and Jackpot fields.}
\label{fig:angle_over_app5}
\end{figure*}

\begin{figure*}
\centering
\includegraphics[width=0.35\textwidth]{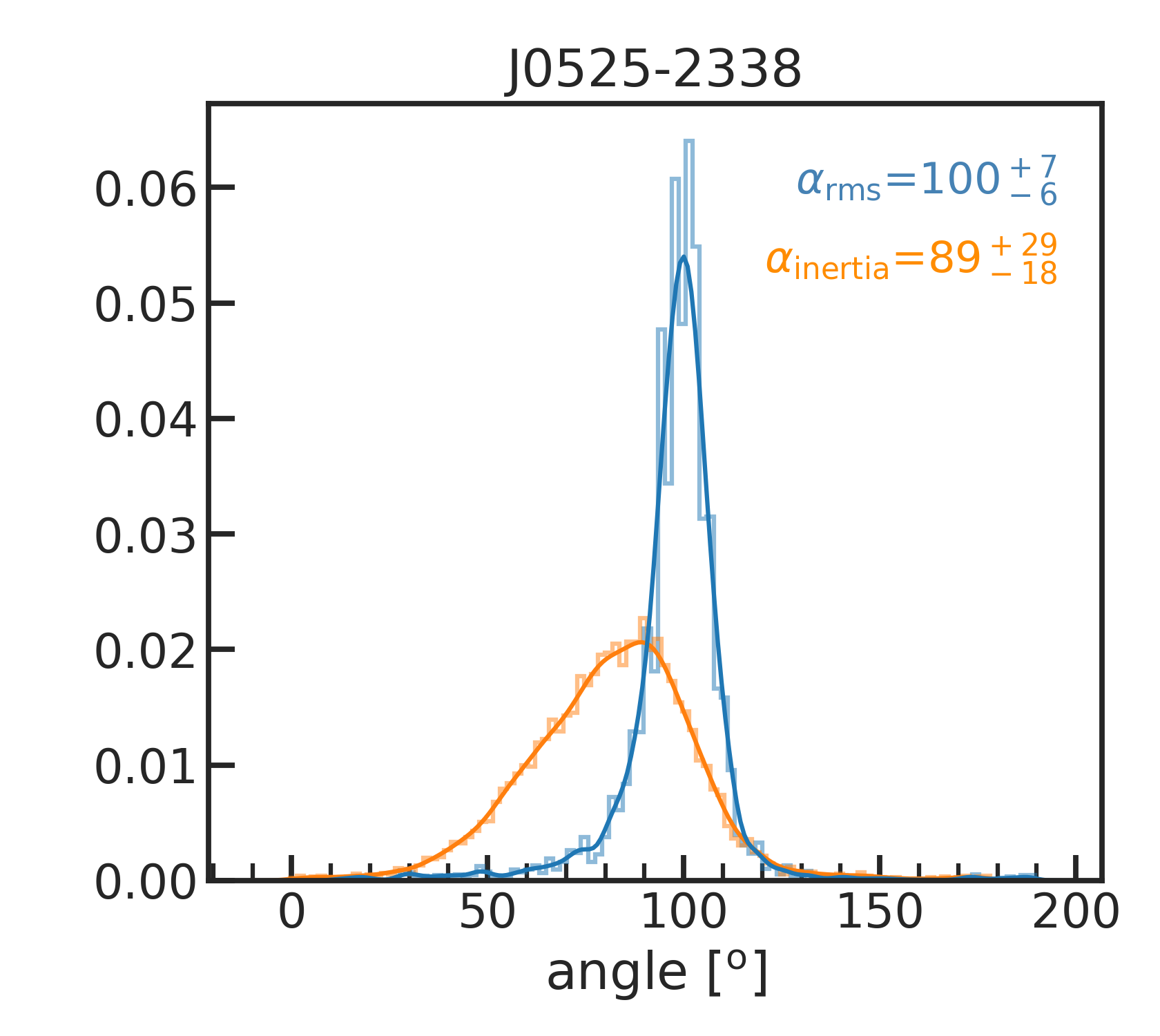}
\includegraphics[width=0.35\textwidth]{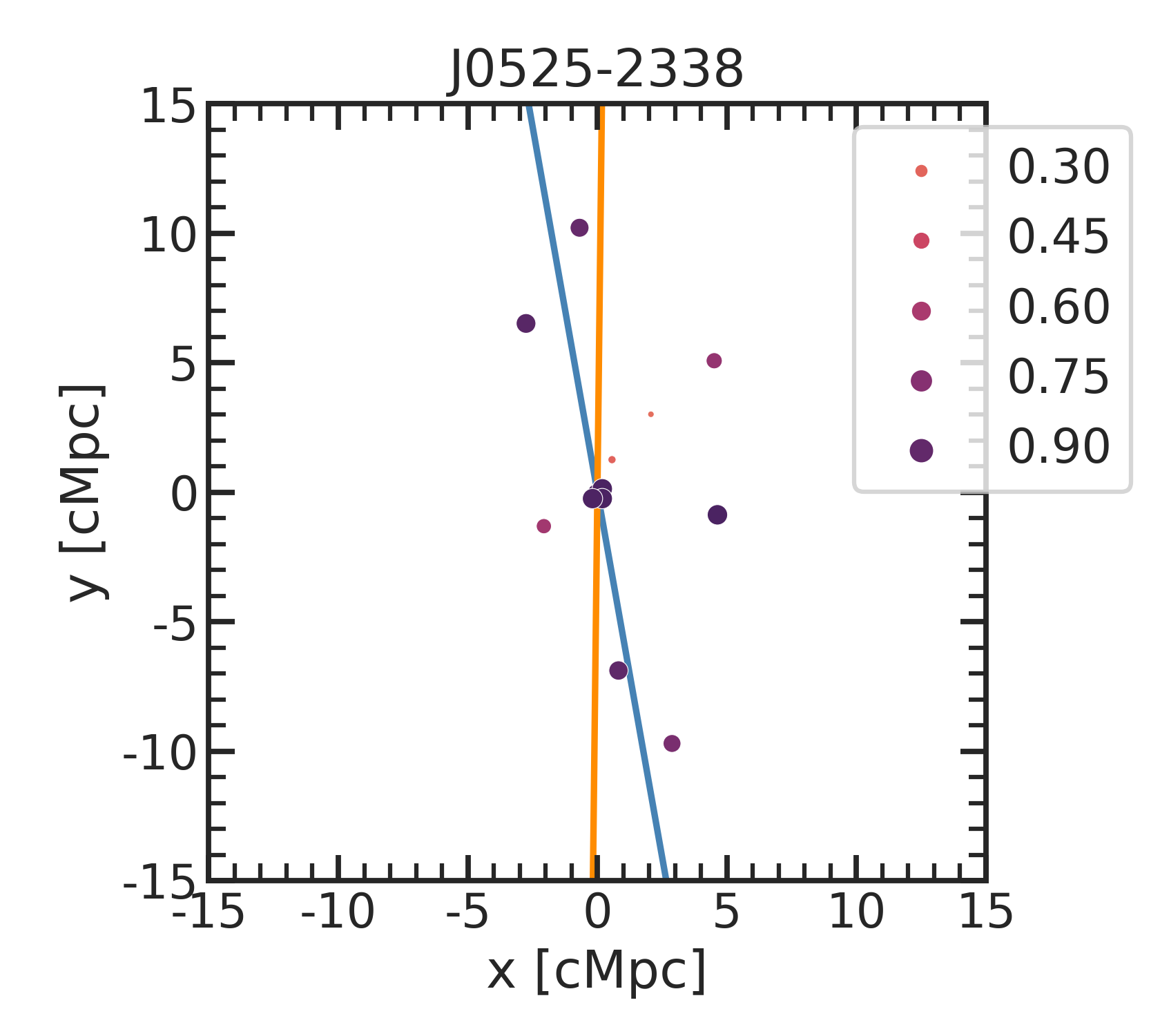}
\includegraphics[width=0.24\textwidth, trim=0 -3.2cm 0 0]{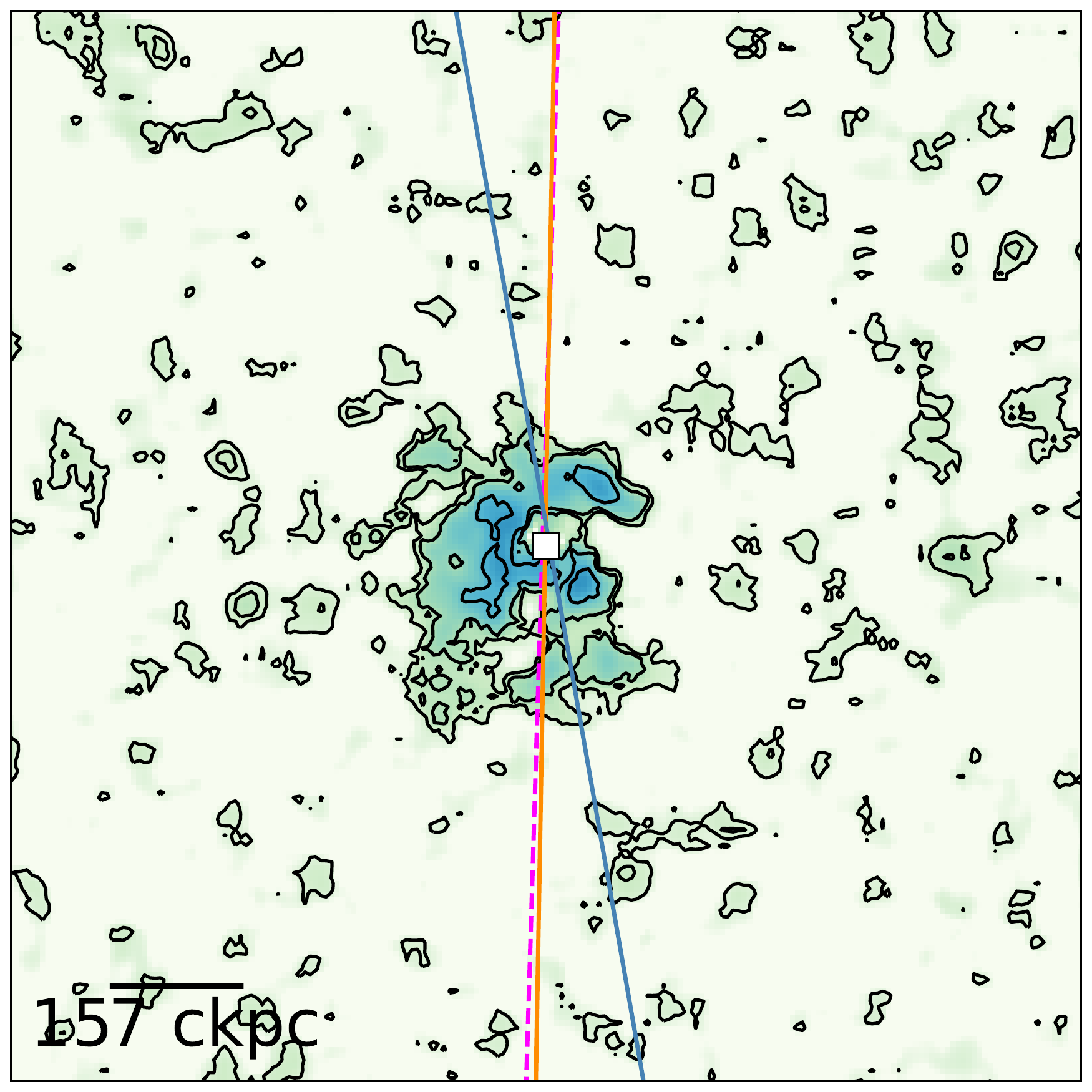}

\includegraphics[width=0.35\textwidth]{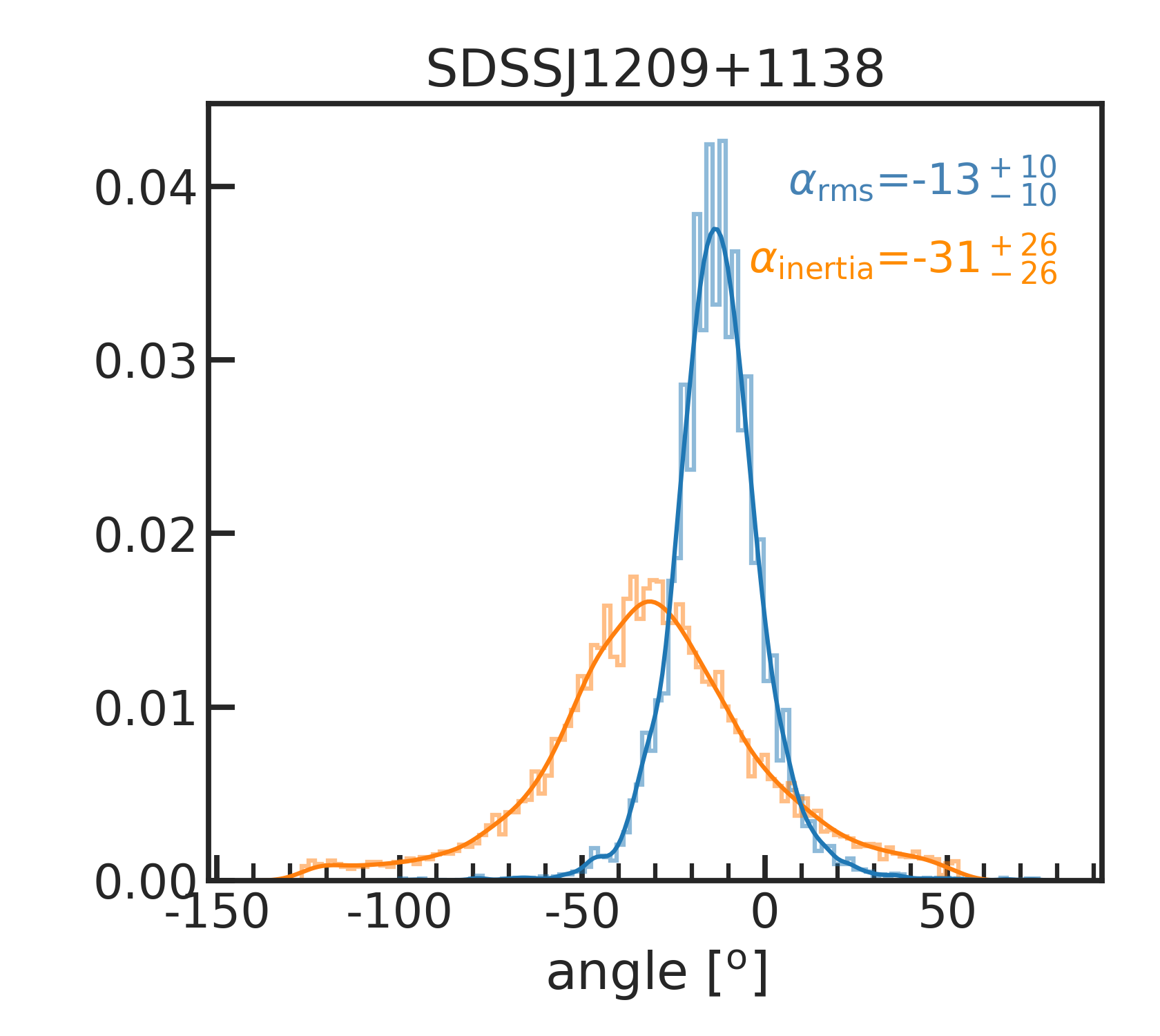}
\includegraphics[width=0.35\textwidth]{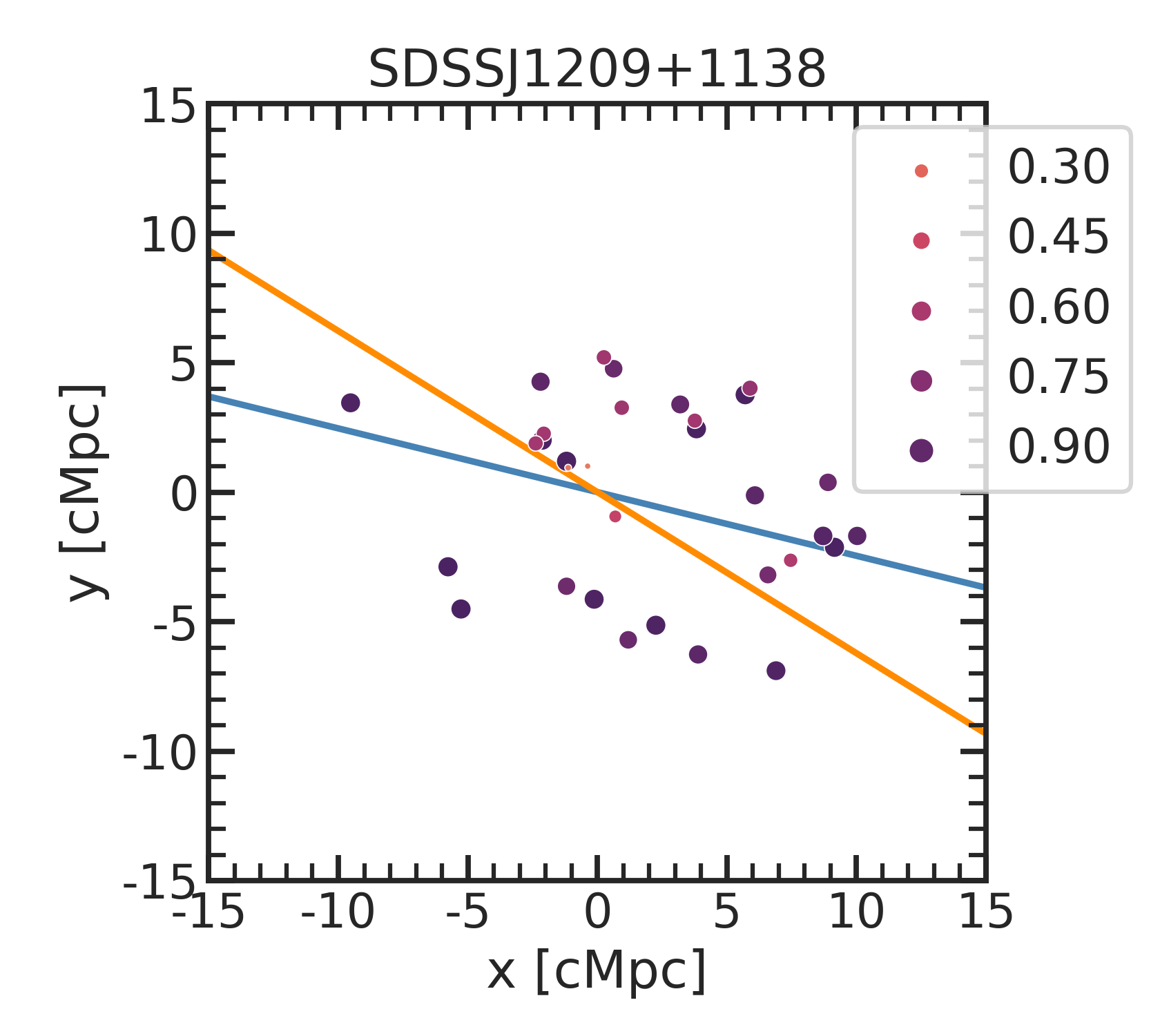}
\includegraphics[width=0.24\textwidth, trim=0 -3.2cm 0 0]{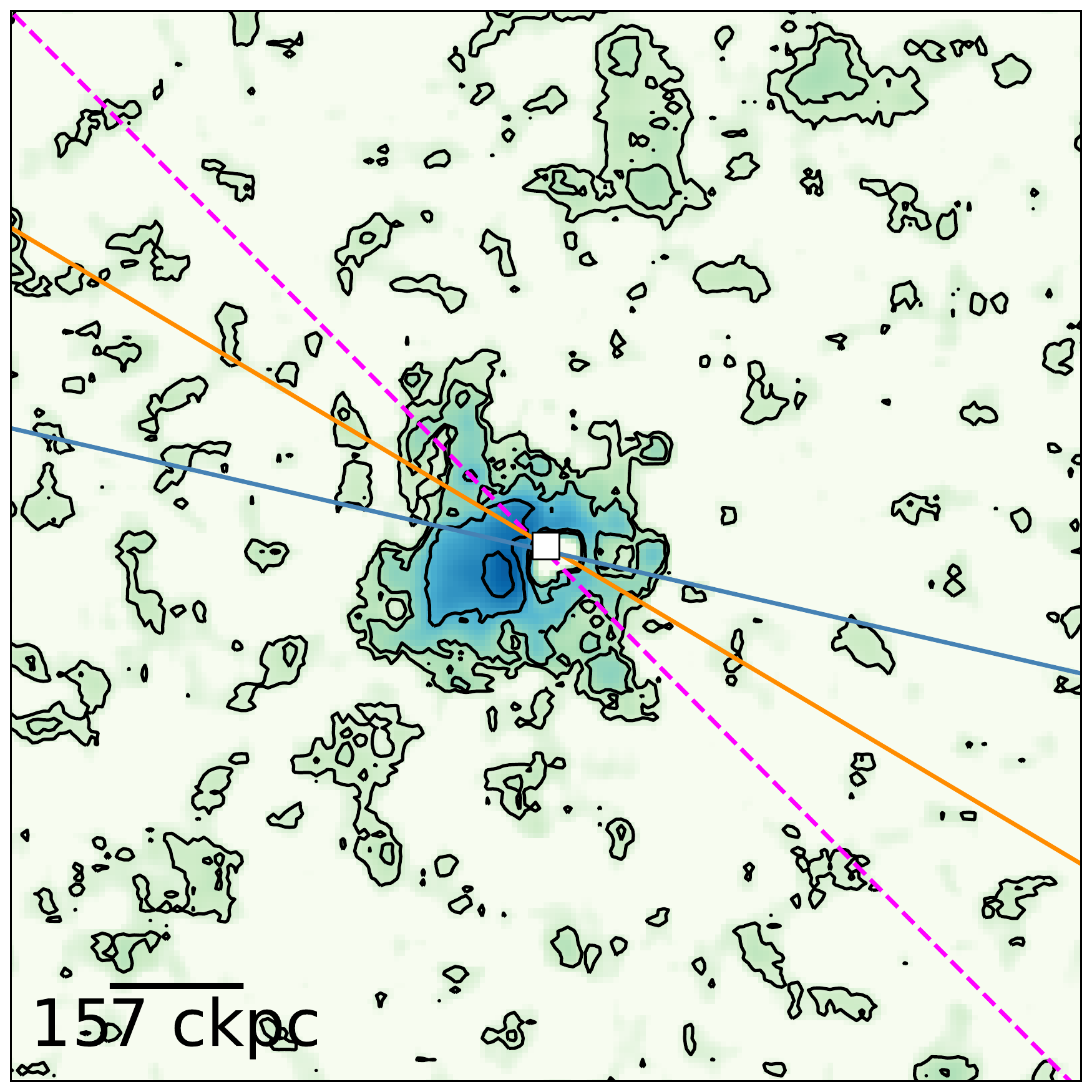}

\includegraphics[width=0.35\textwidth]{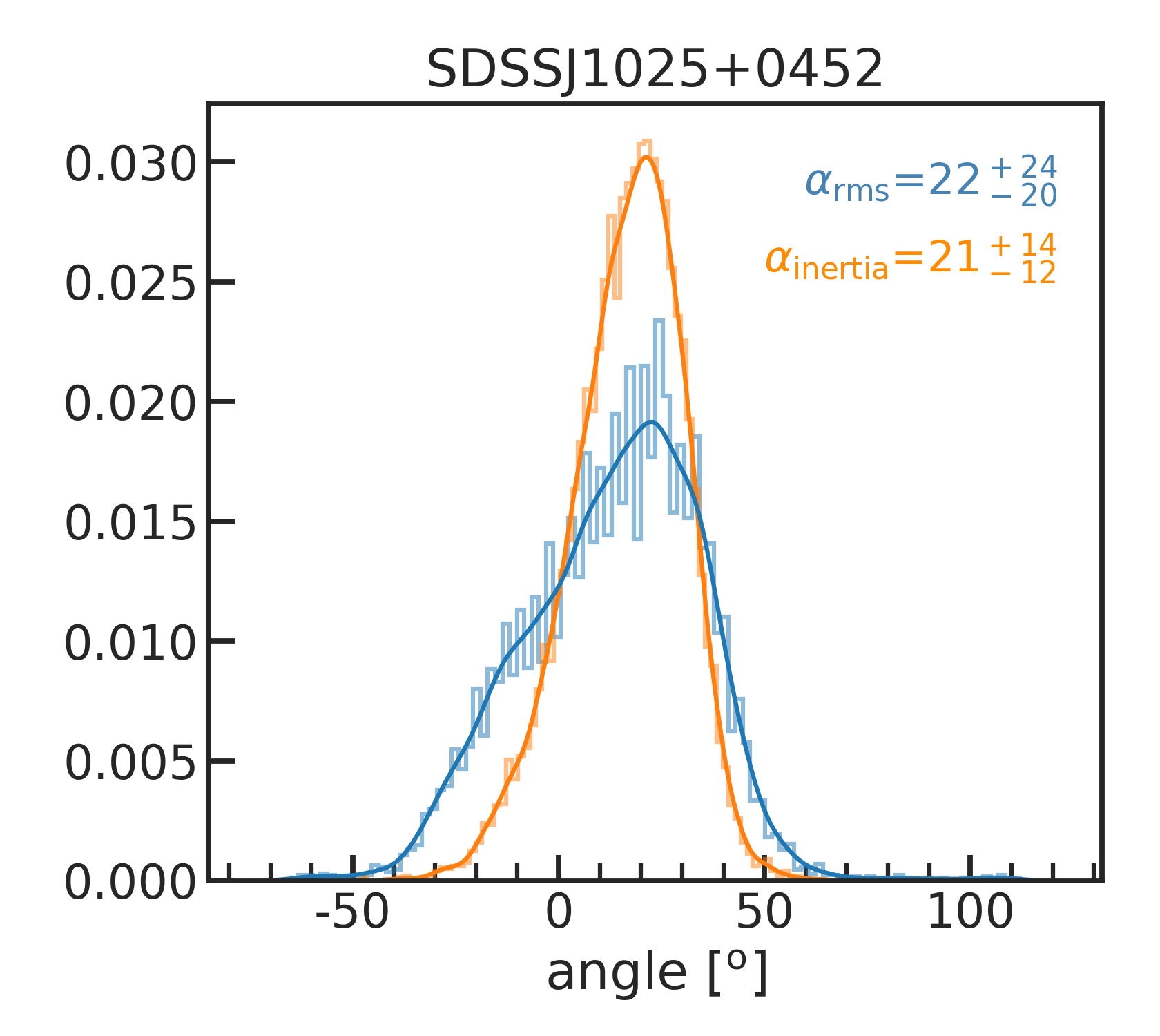}
\includegraphics[width=0.35\textwidth]{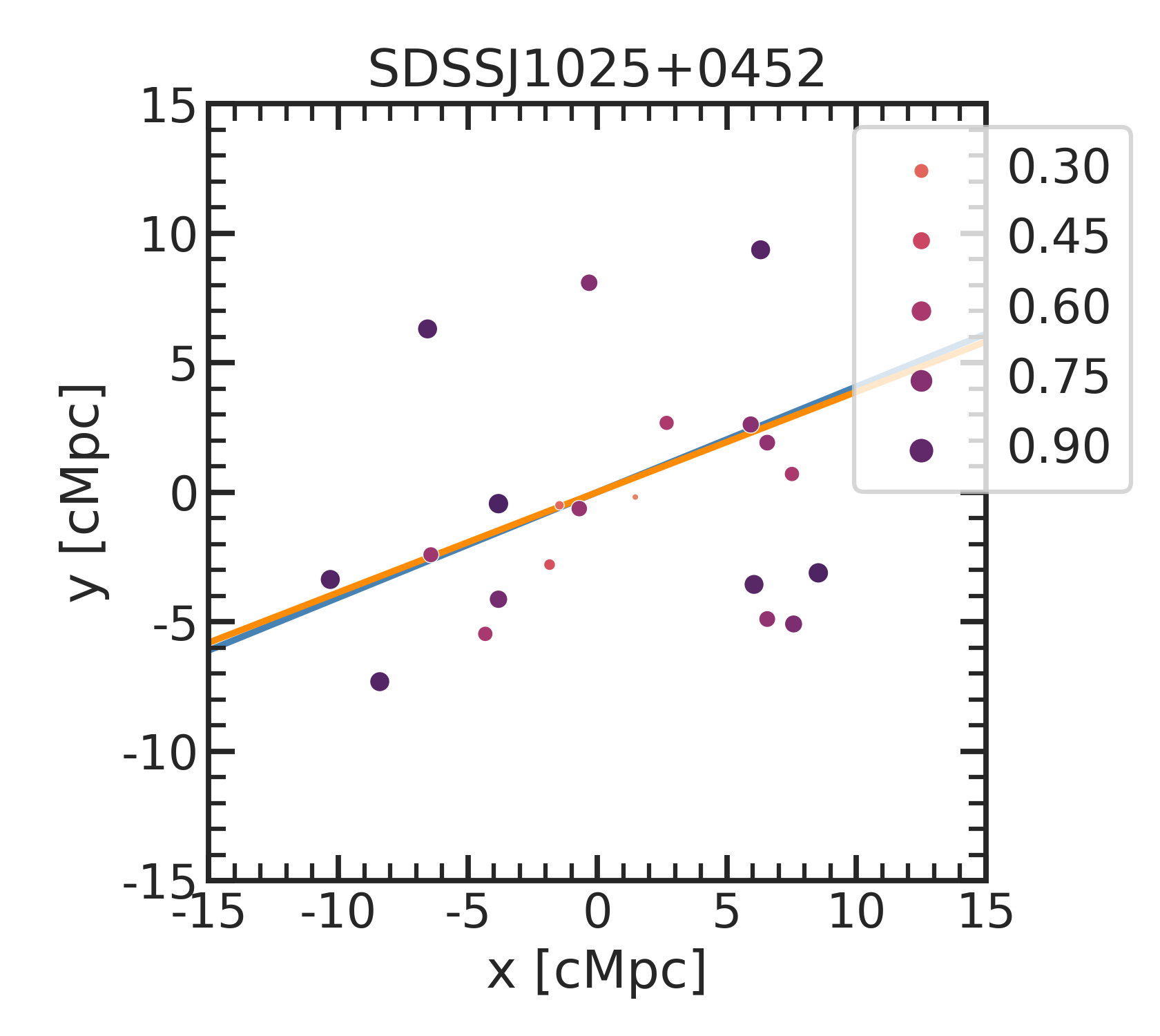}
\includegraphics[width=0.24\textwidth, trim=0 -3.2cm 0 0]{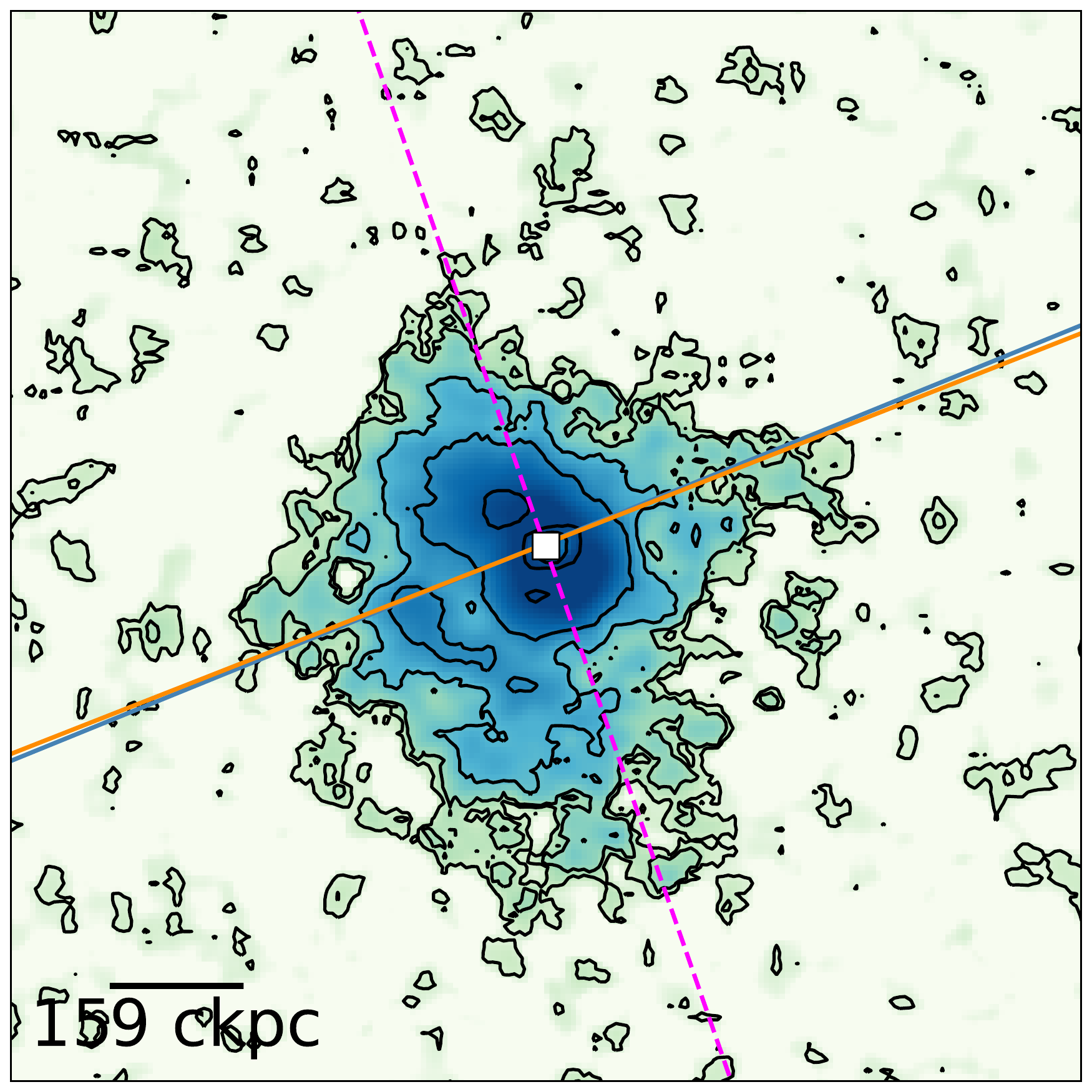}

\includegraphics[width=0.35\textwidth]{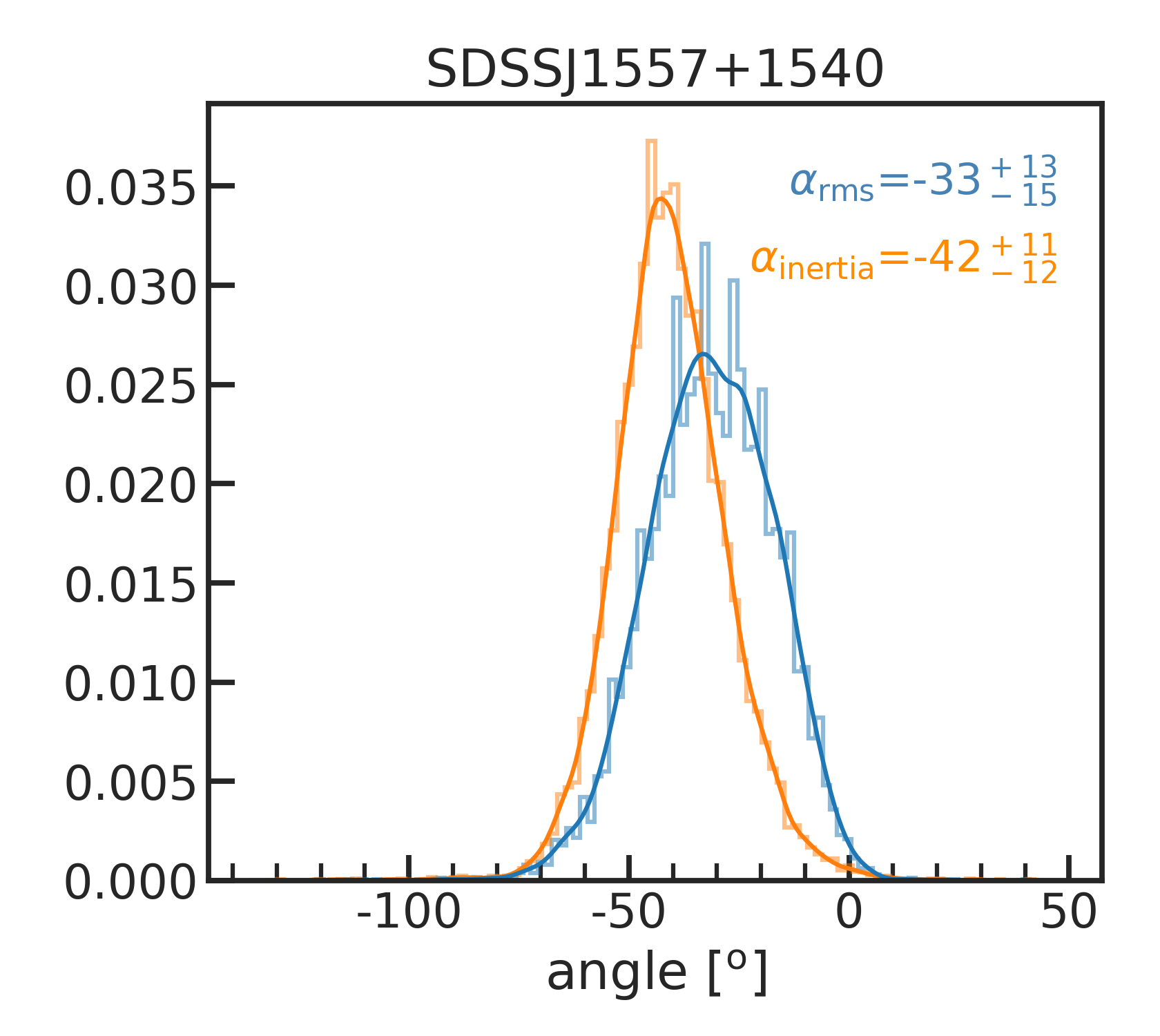}
\includegraphics[width=0.35\textwidth]{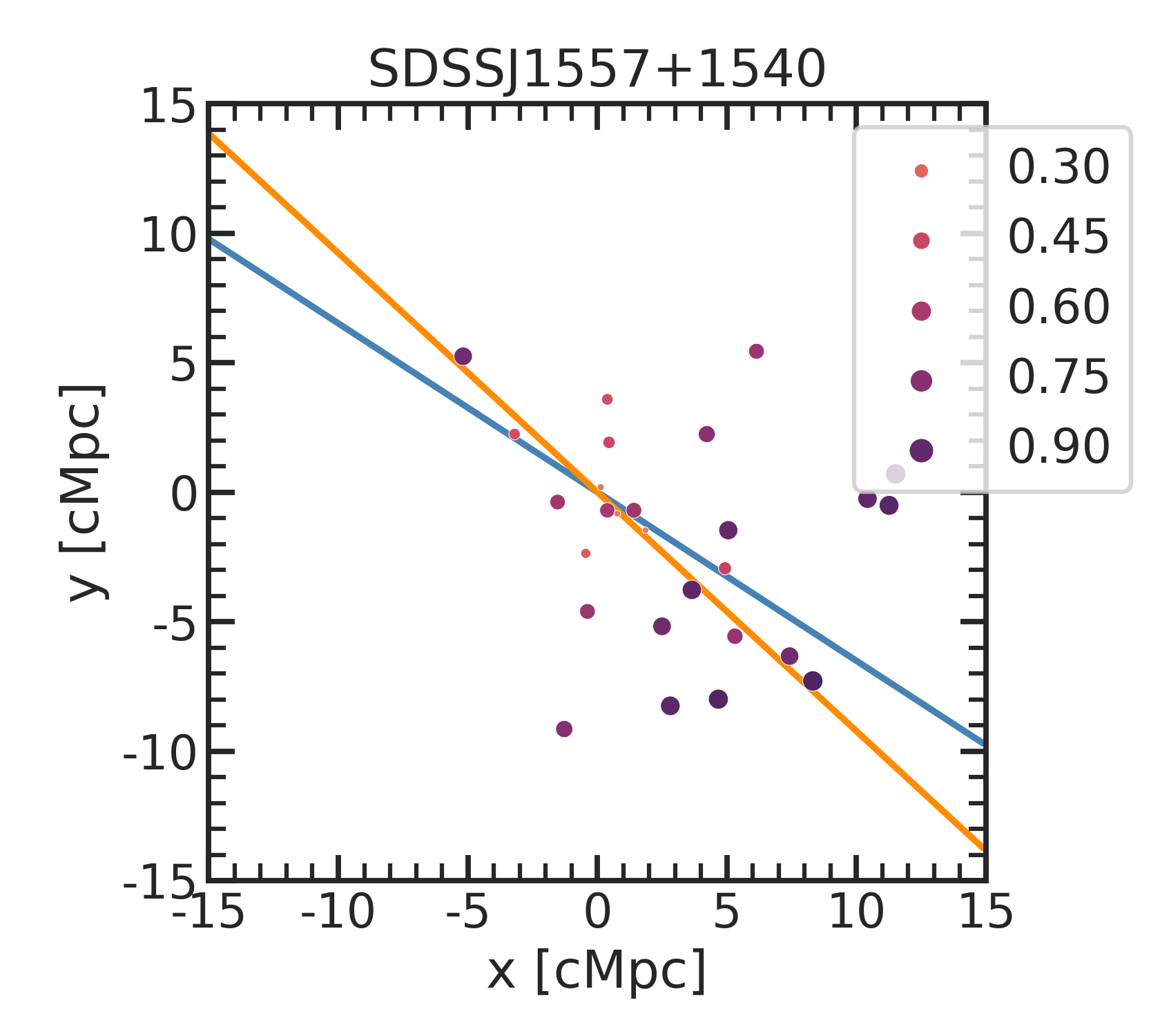}
\includegraphics[width=0.24\textwidth, trim=0 -3.2cm 0 0]{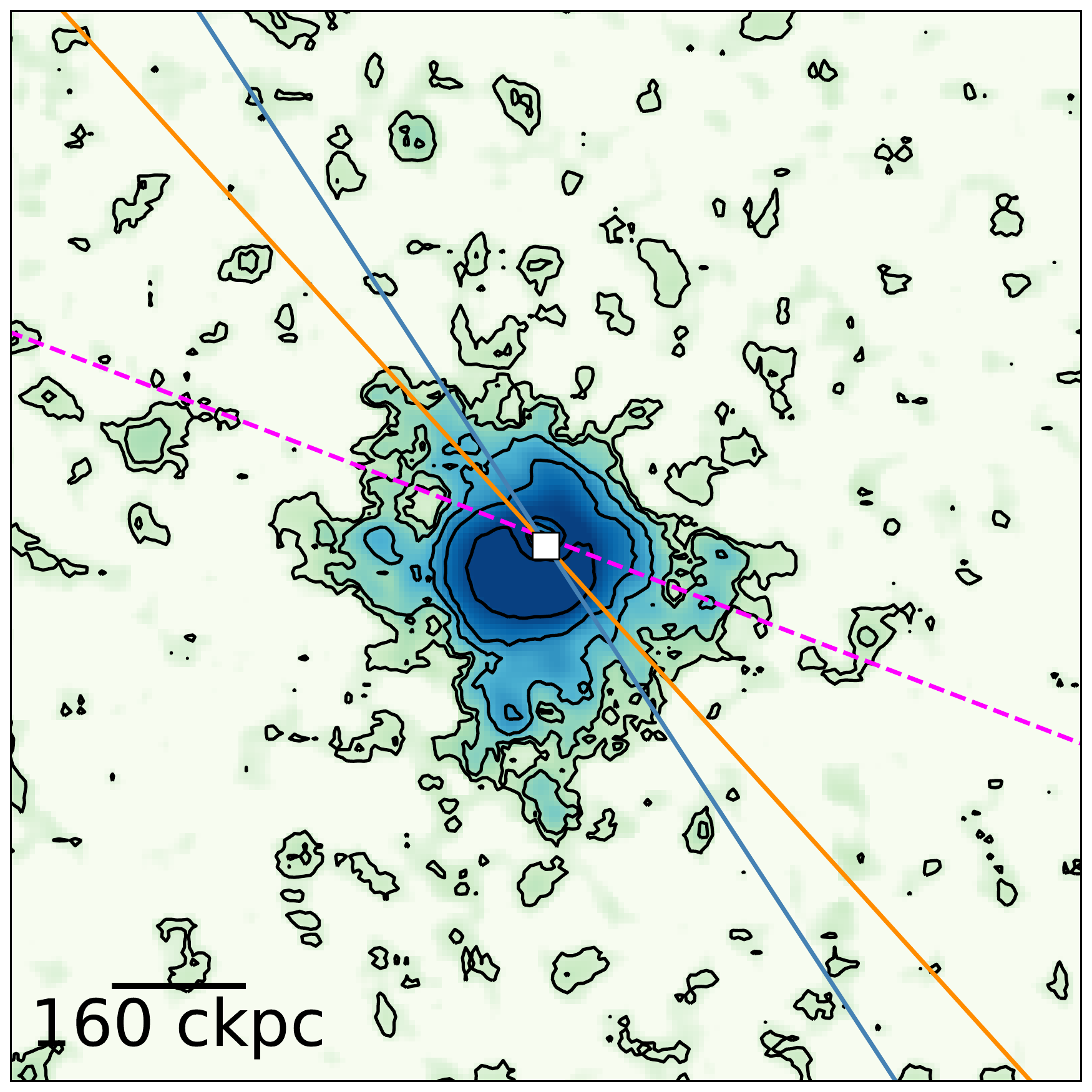}
\caption{Same as Figs.~\ref{fig:angle_over} and \ref{fig:angle_over_app5}, but for another four sources (see the labels).}
\label{fig:angle_over_app1}
\end{figure*}

\begin{figure*}
\centering
\includegraphics[width=0.35\textwidth]{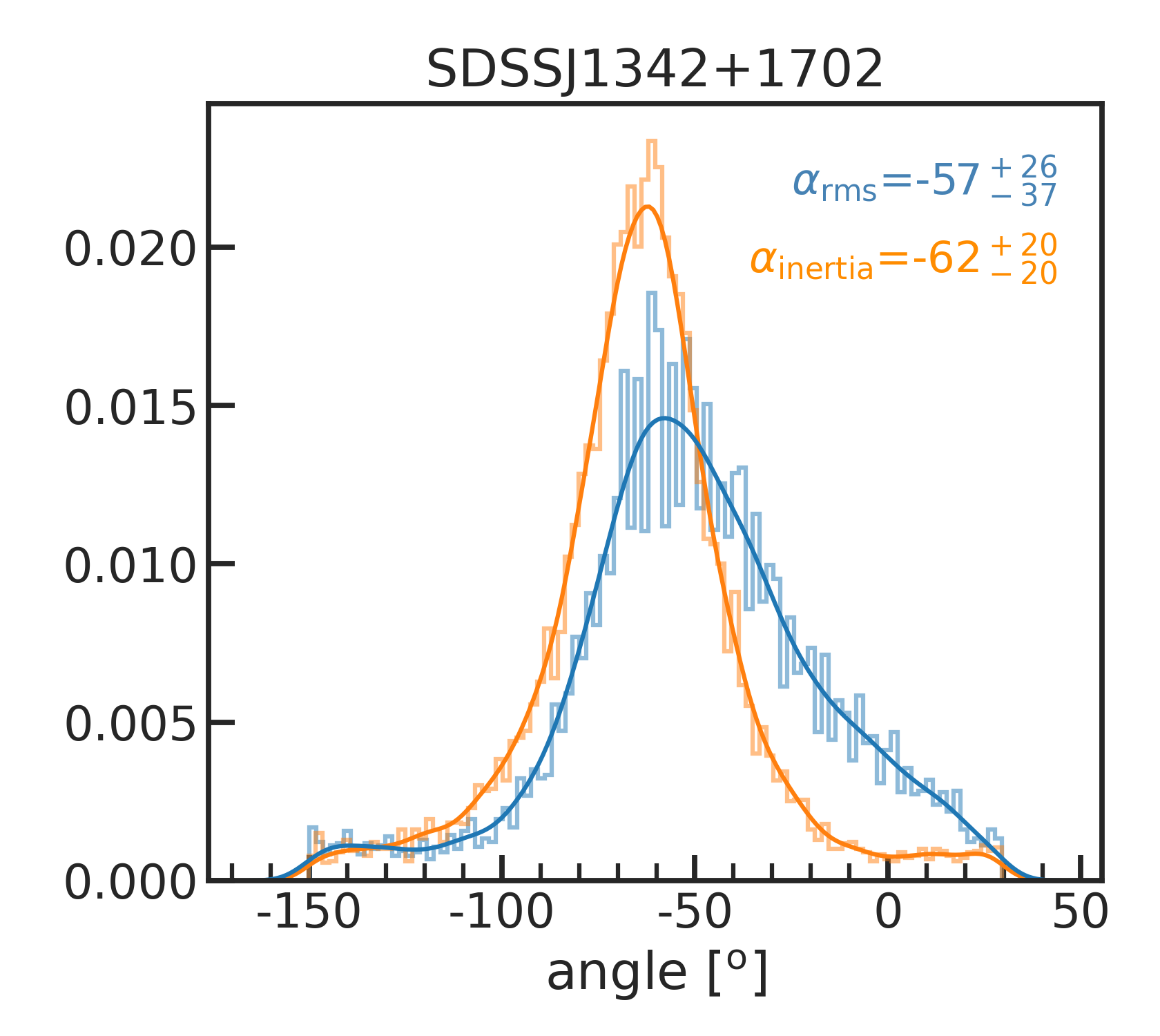}
\includegraphics[width=0.35\textwidth]{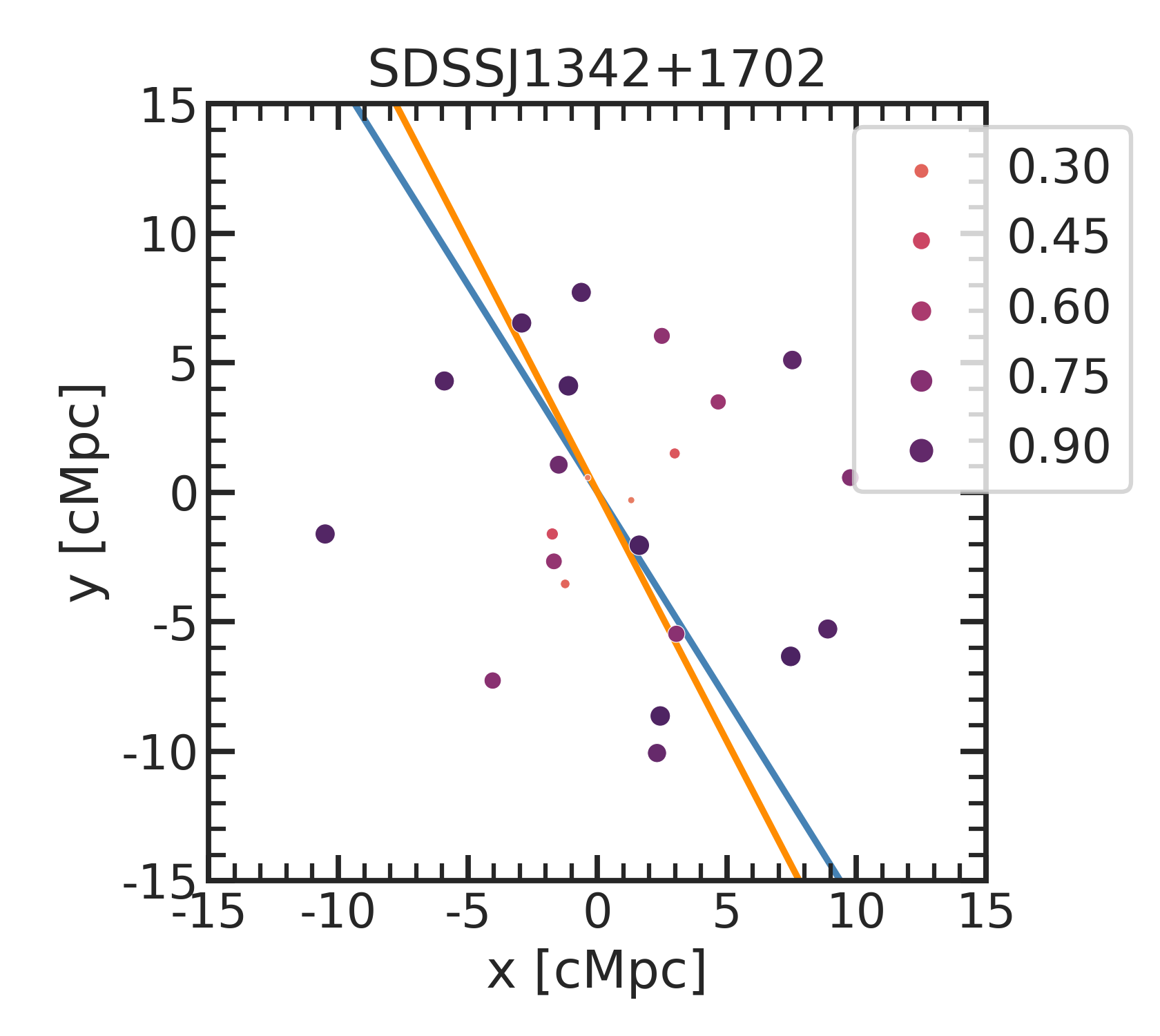}
\includegraphics[width=0.24\textwidth, trim=0 -3.2cm 0 0]{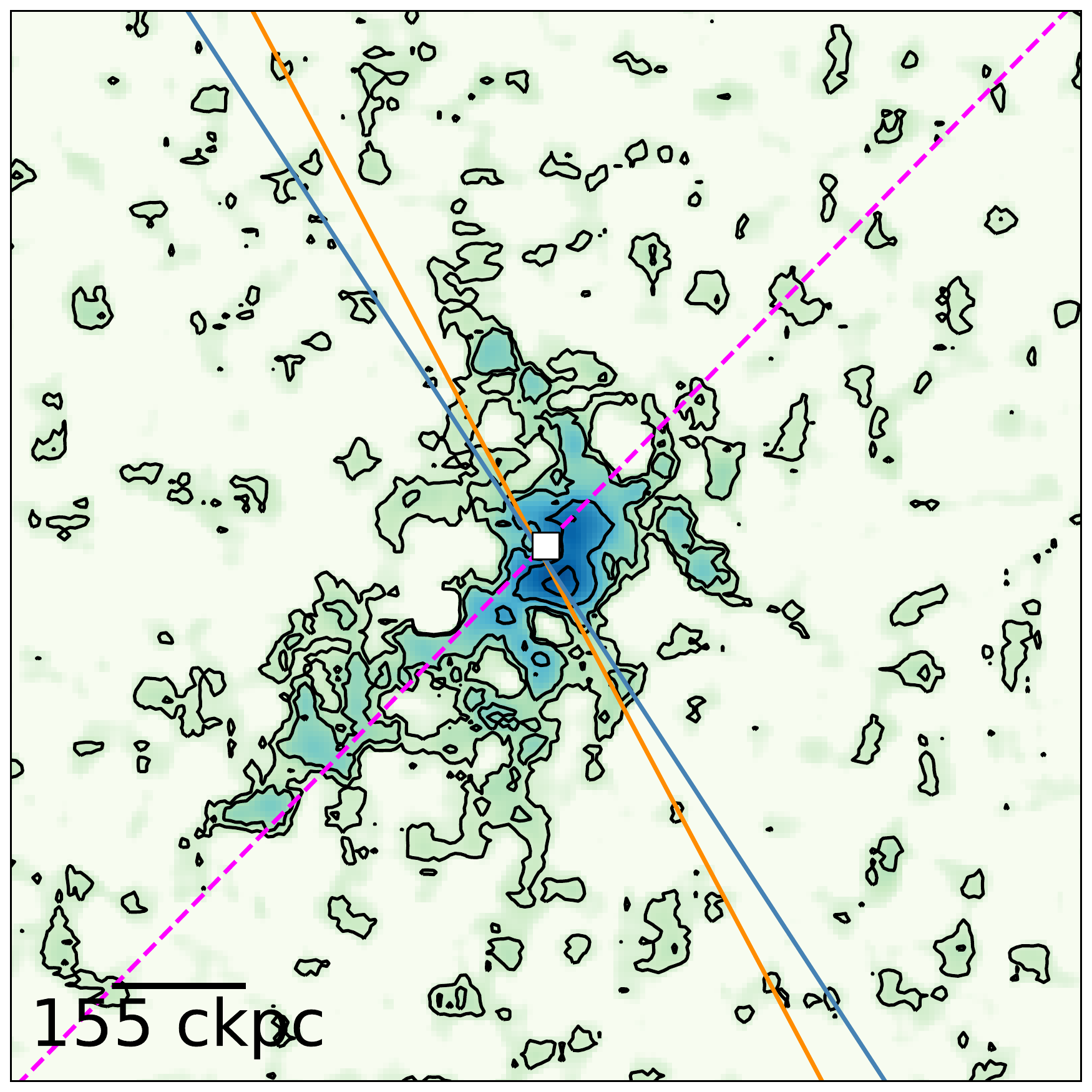}

\includegraphics[width=0.35\textwidth]{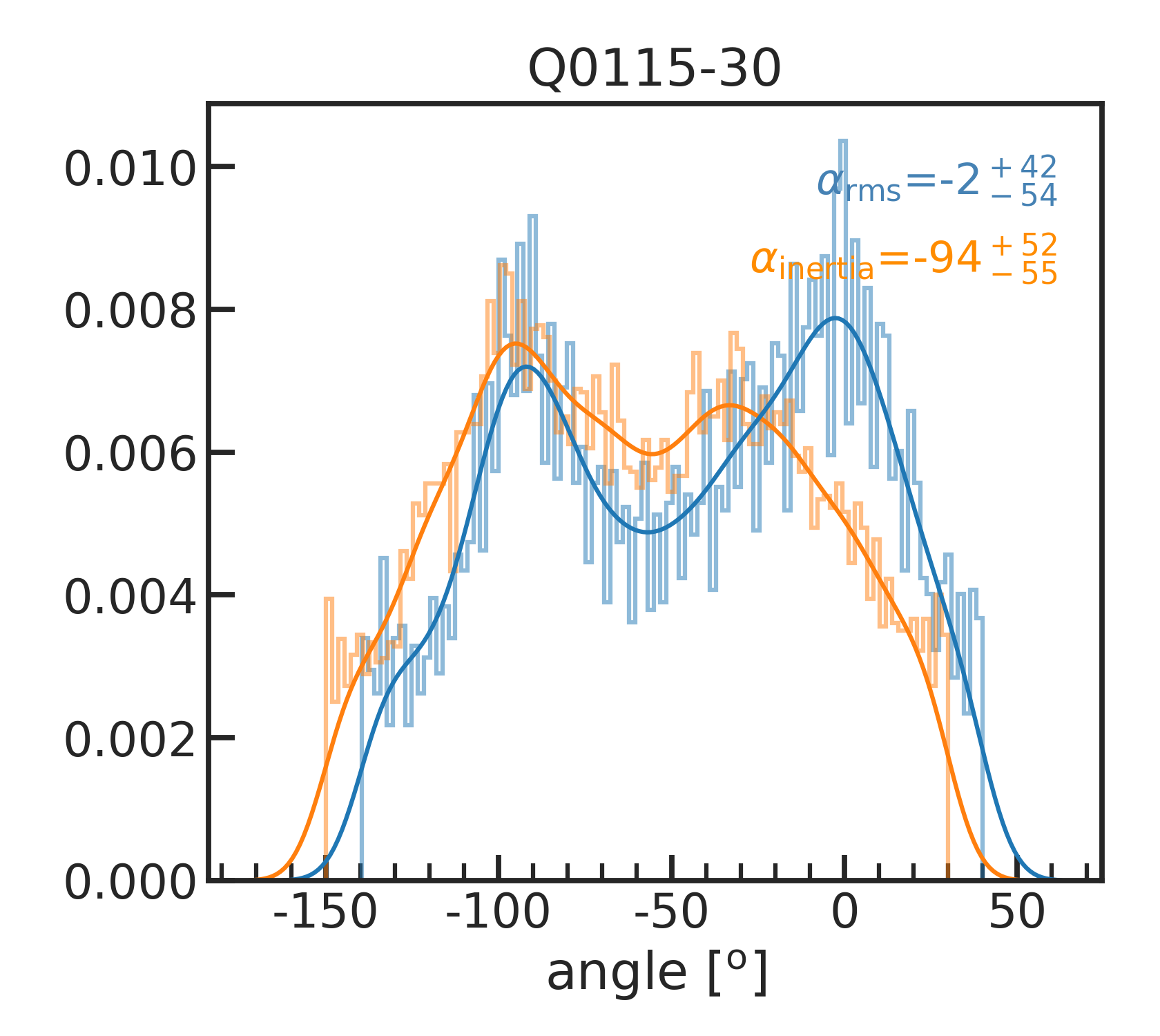}
\includegraphics[width=0.35\textwidth]{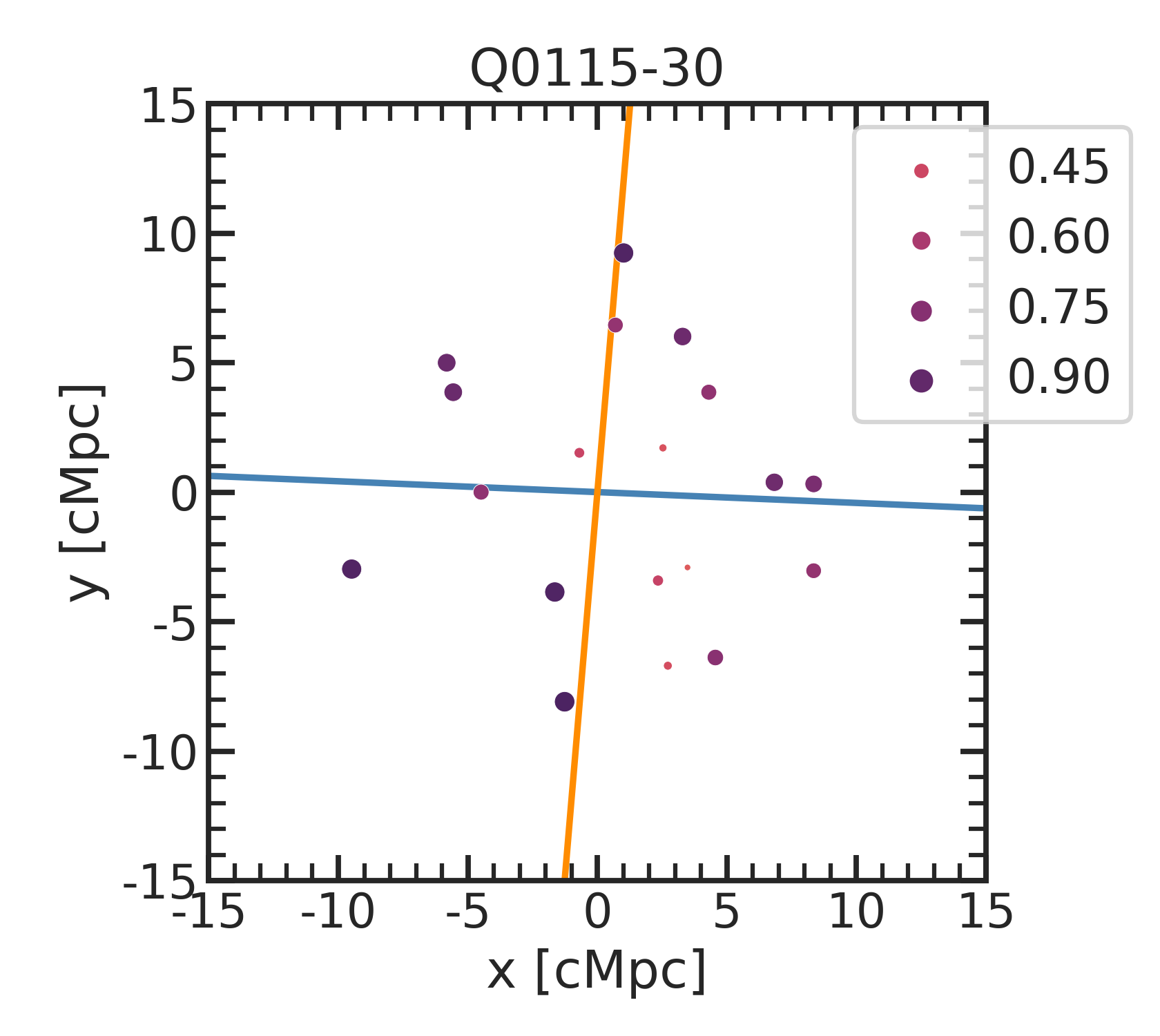}
\includegraphics[width=0.24\textwidth, trim=0 -3.2cm 0 0]{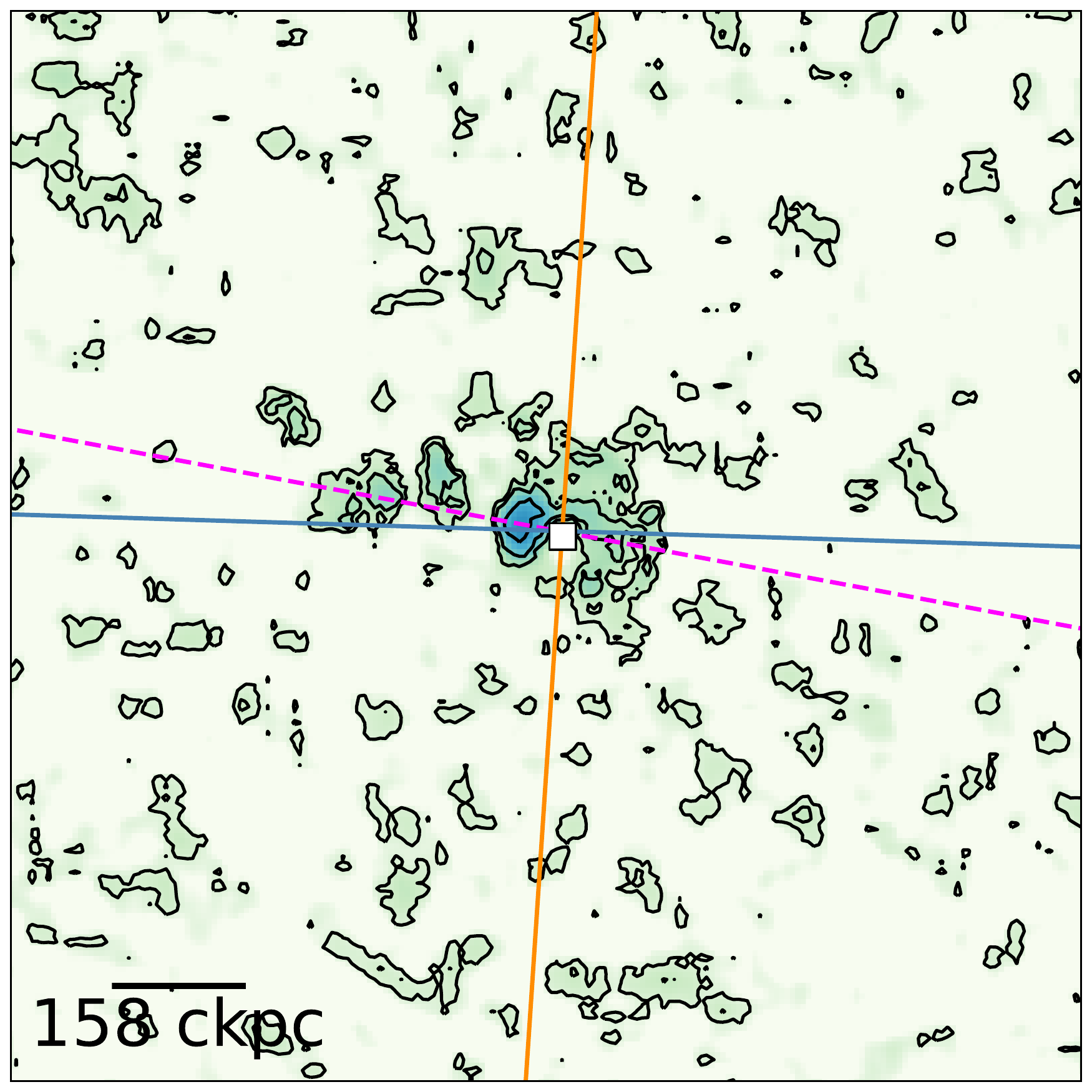}

\includegraphics[width=0.35\textwidth]{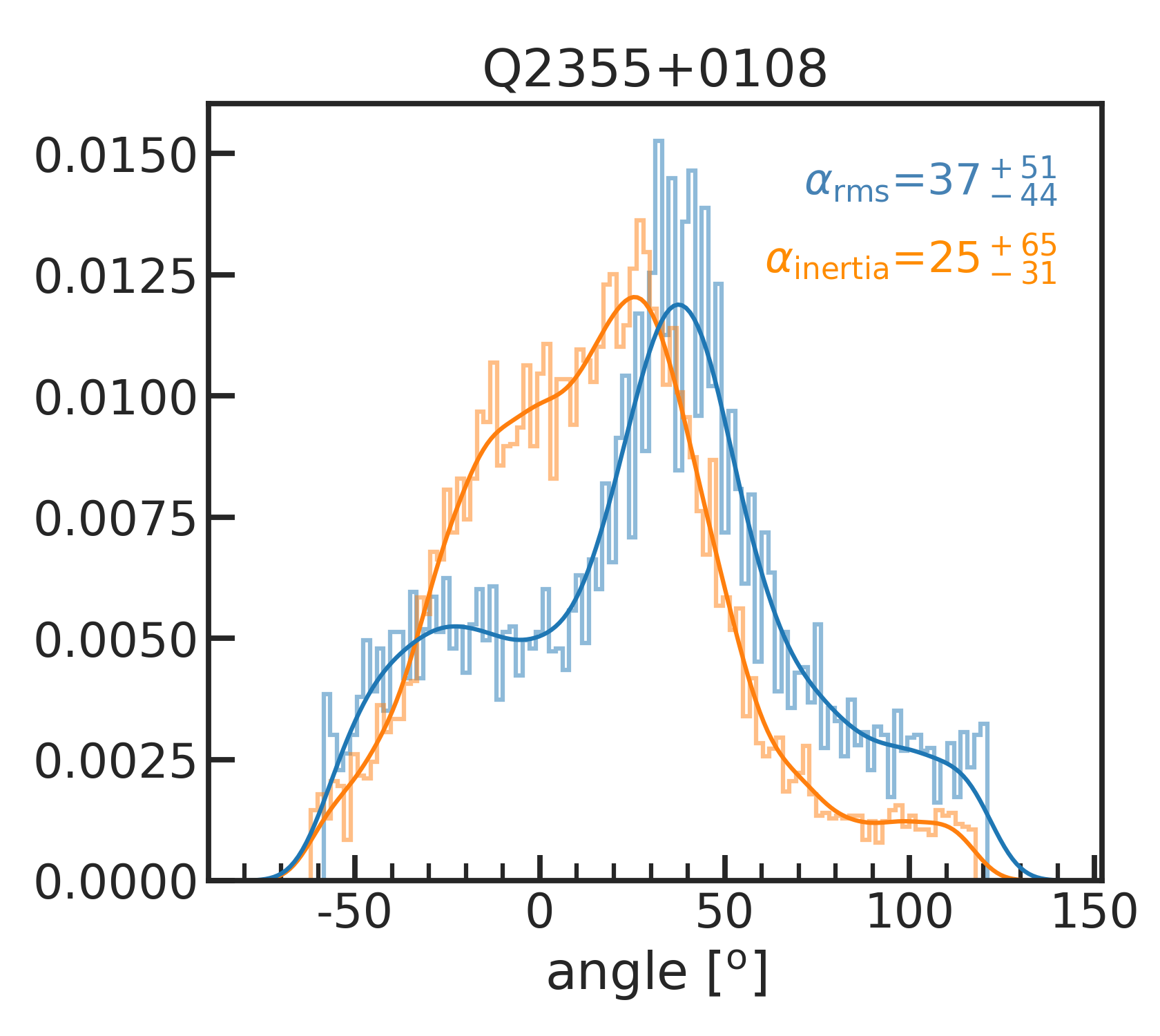}
\includegraphics[width=0.35\textwidth]{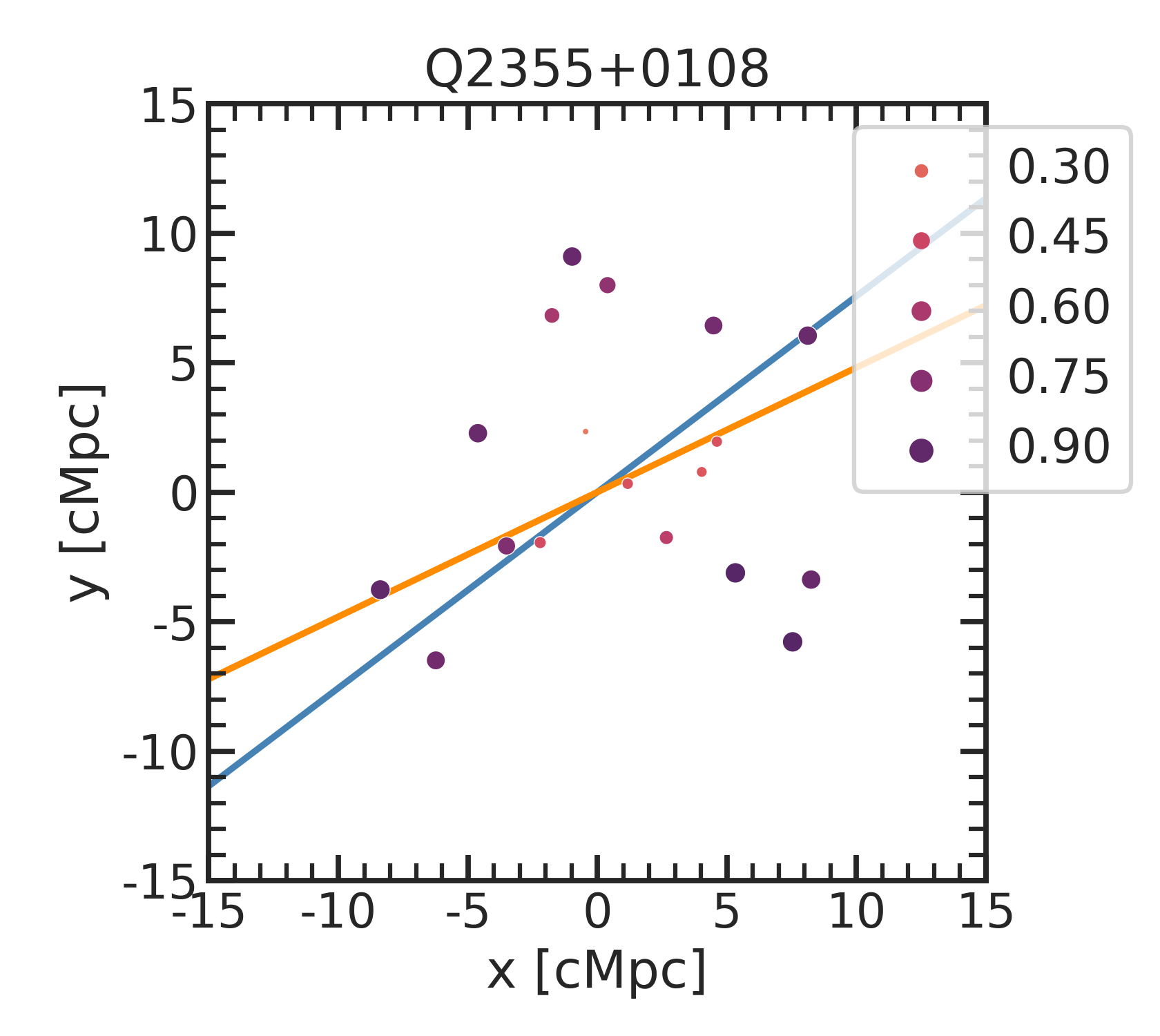}
\includegraphics[width=0.24\textwidth, trim=0 -3.2cm 0 0]{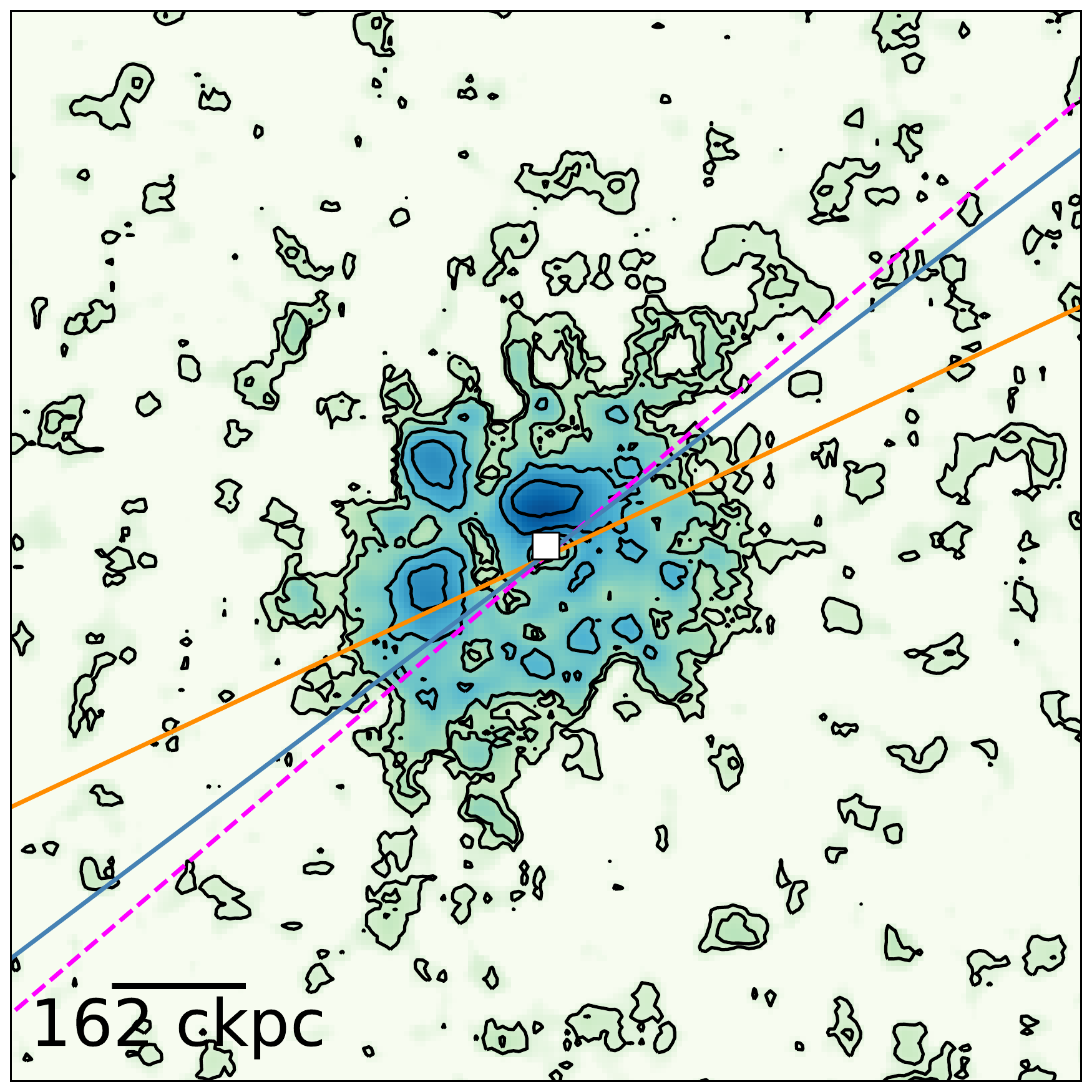}

\includegraphics[width=0.35\textwidth]{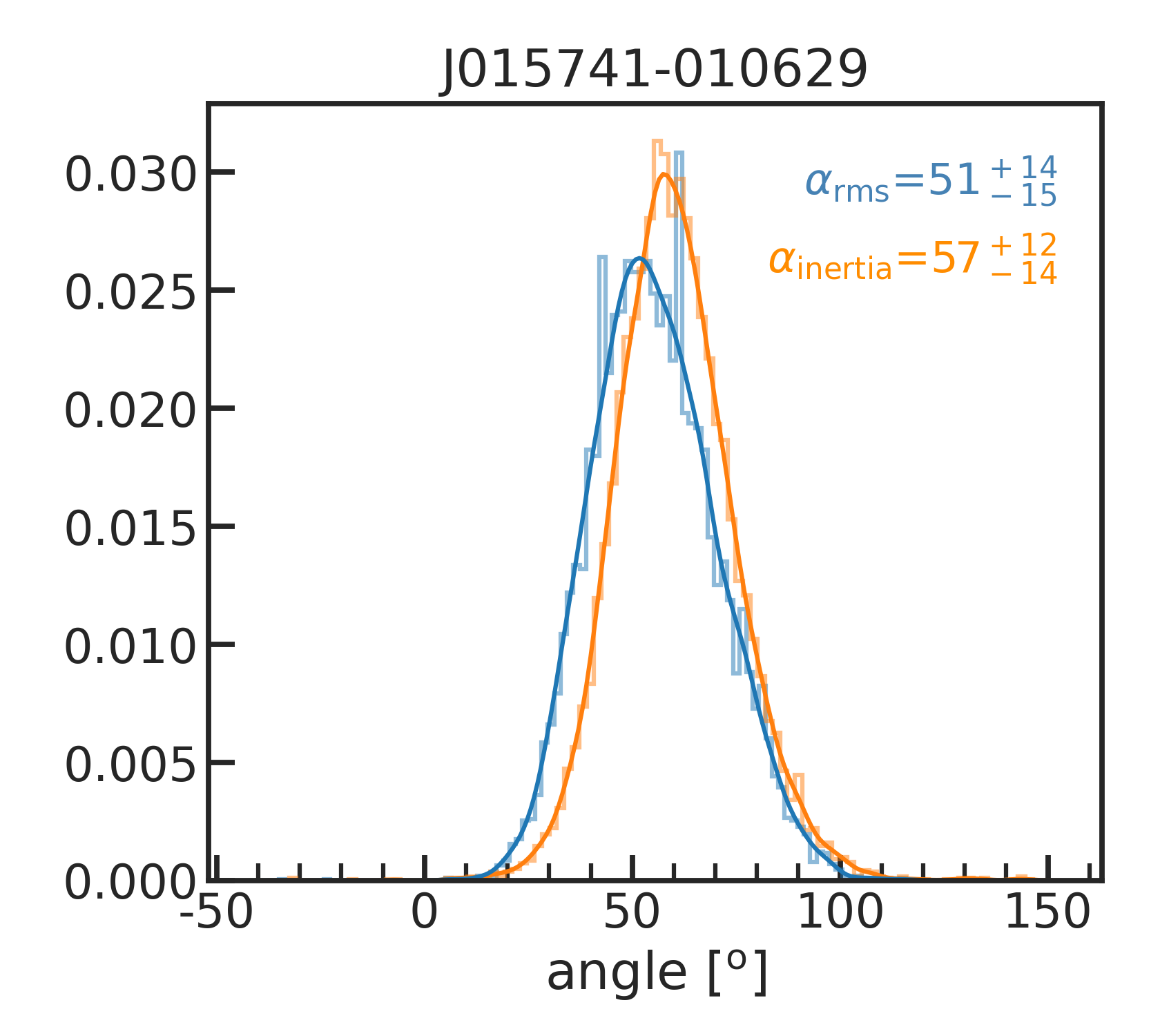}
\includegraphics[width=0.35\textwidth]{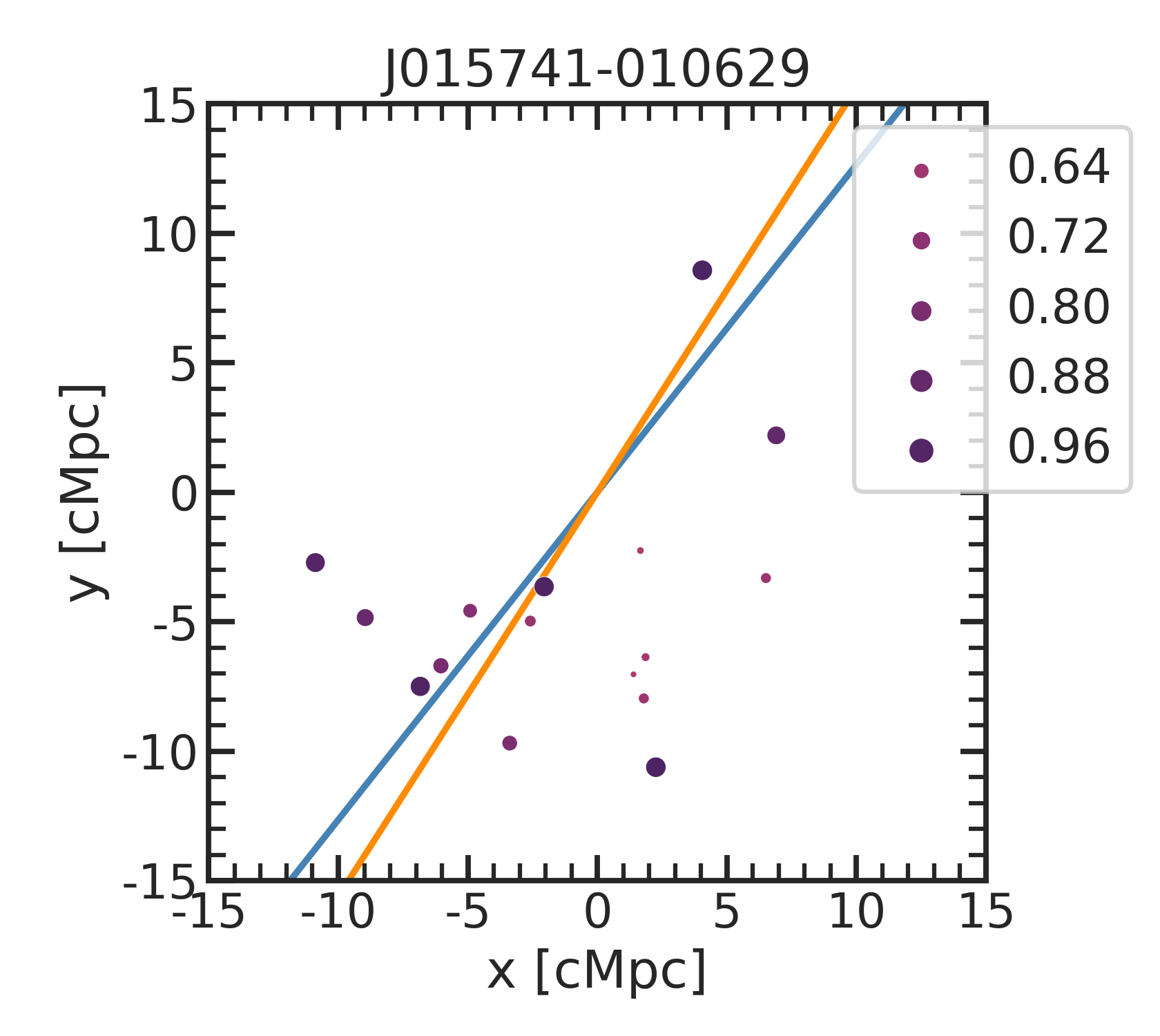}
\includegraphics[width=0.24\textwidth, trim=0 -3.2cm 0 0]{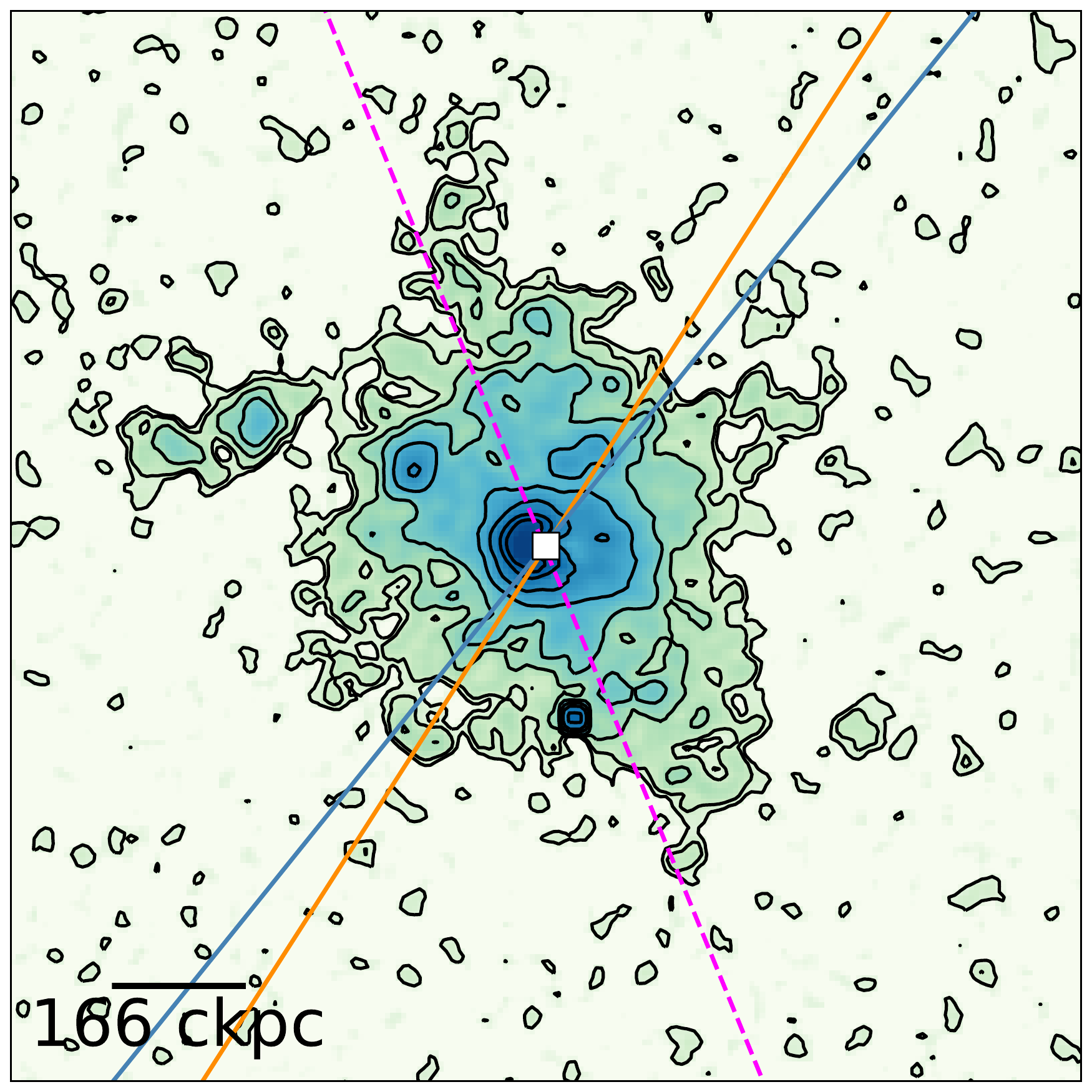}
\caption{Same as Figs.~\ref{fig:angle_over}, \ref{fig:angle_over_app5}, and \ref{fig:angle_over_app1}, but for another four sources (see the labels).}
\label{fig:angle_over_app2}
\end{figure*}

\begin{figure*}
\centering
\includegraphics[width=0.35\textwidth]{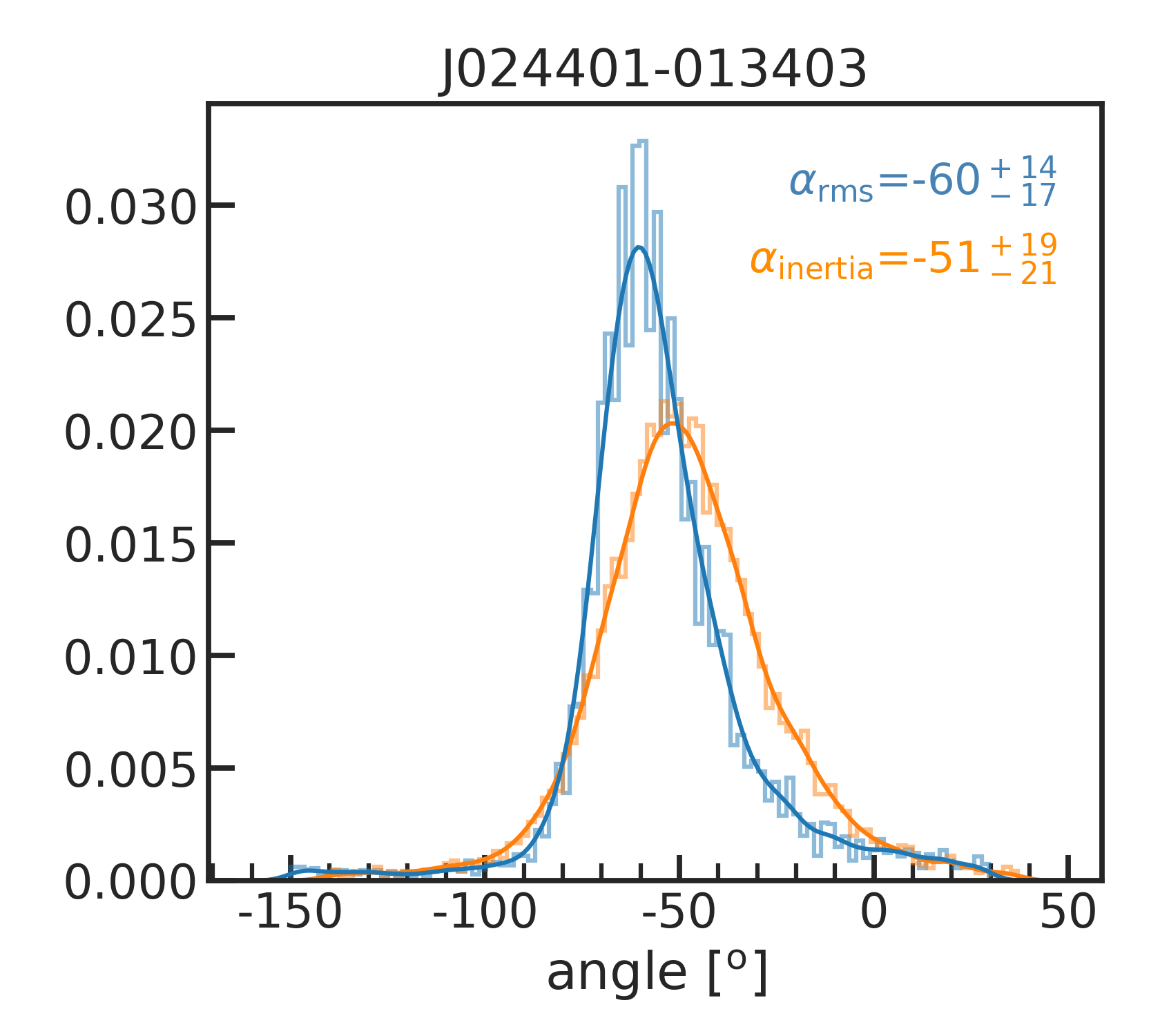}
\includegraphics[width=0.35\textwidth]{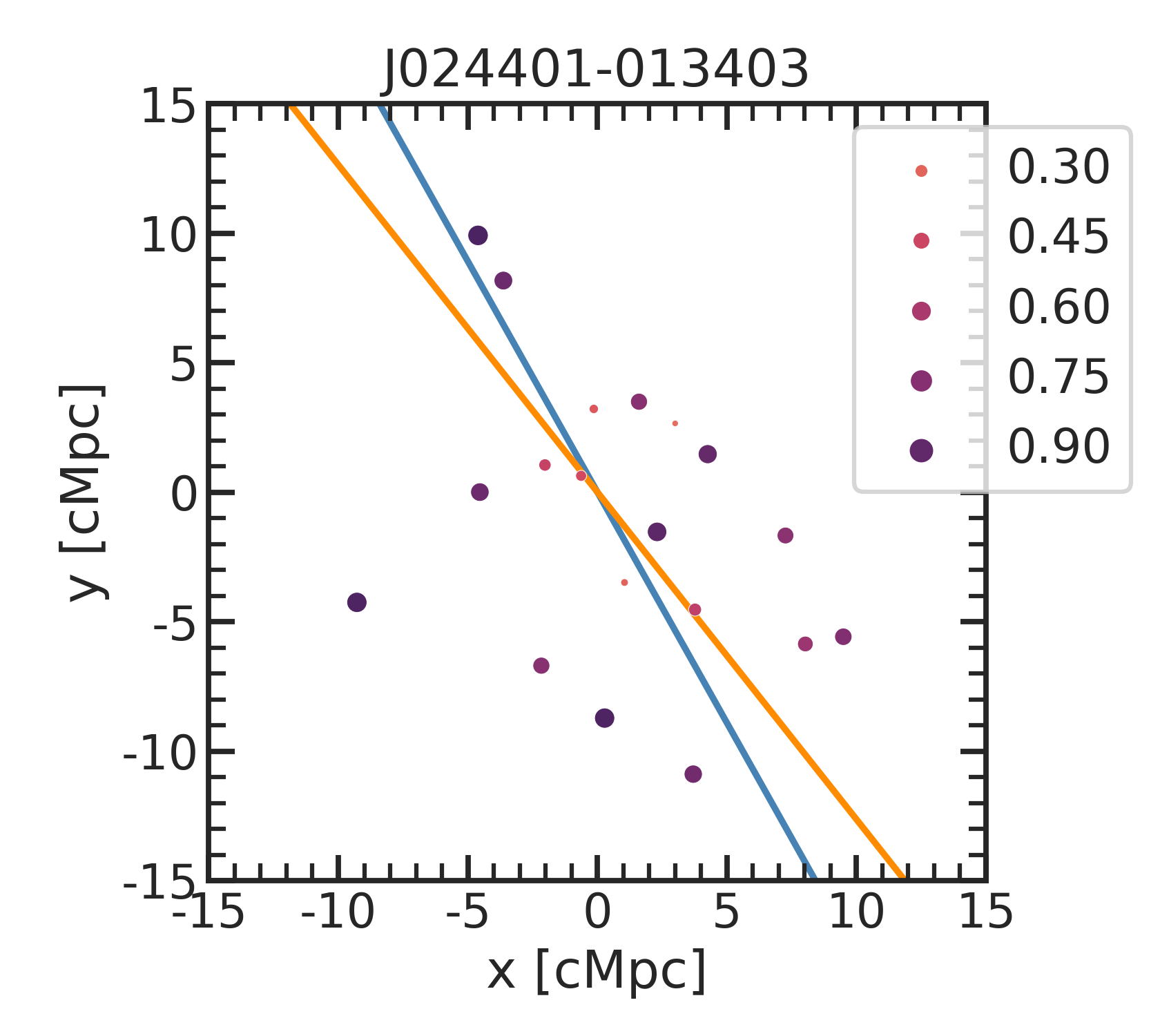}
\includegraphics[width=0.24\textwidth, trim=0 -3.2cm 0 0]{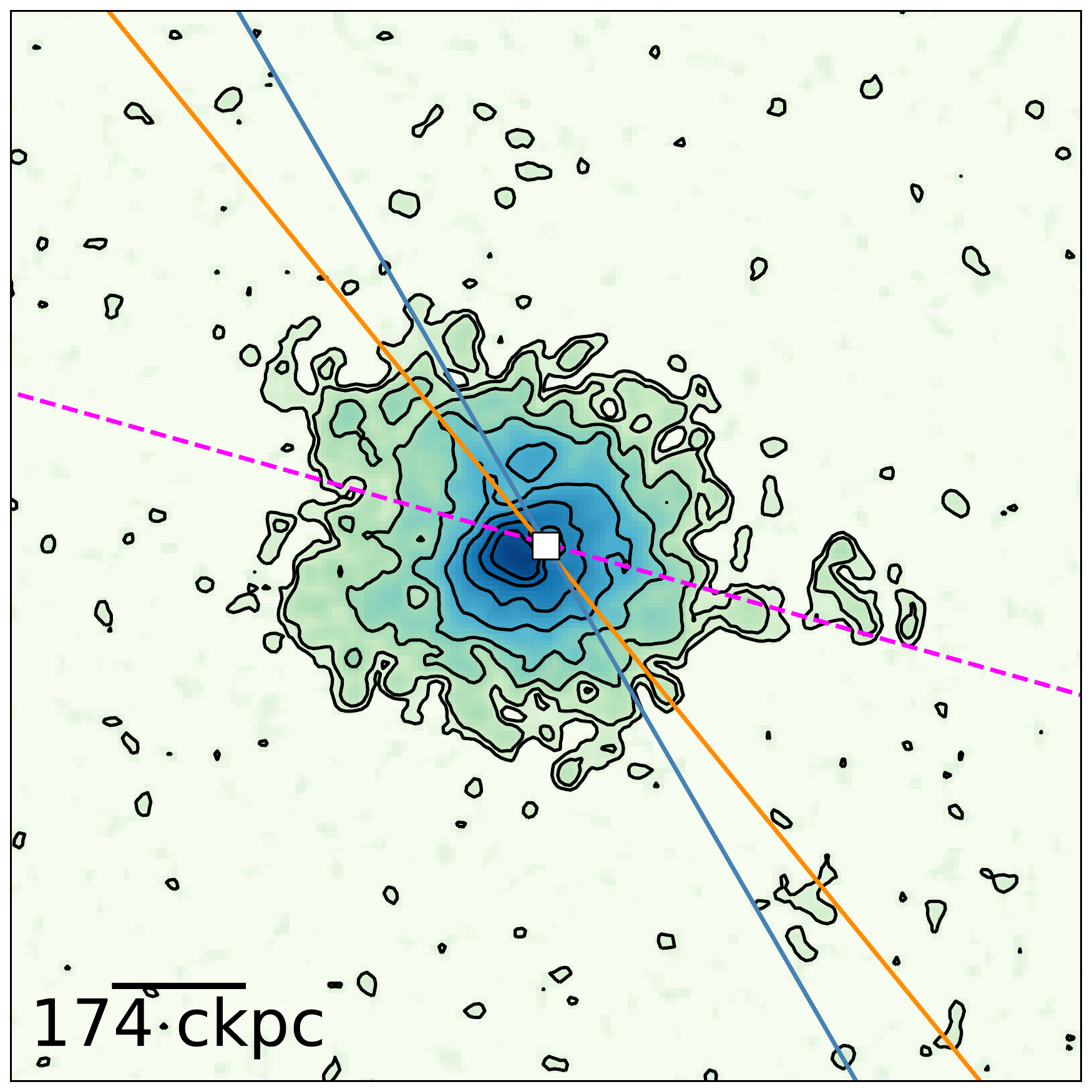}

\includegraphics[width=0.35\textwidth]{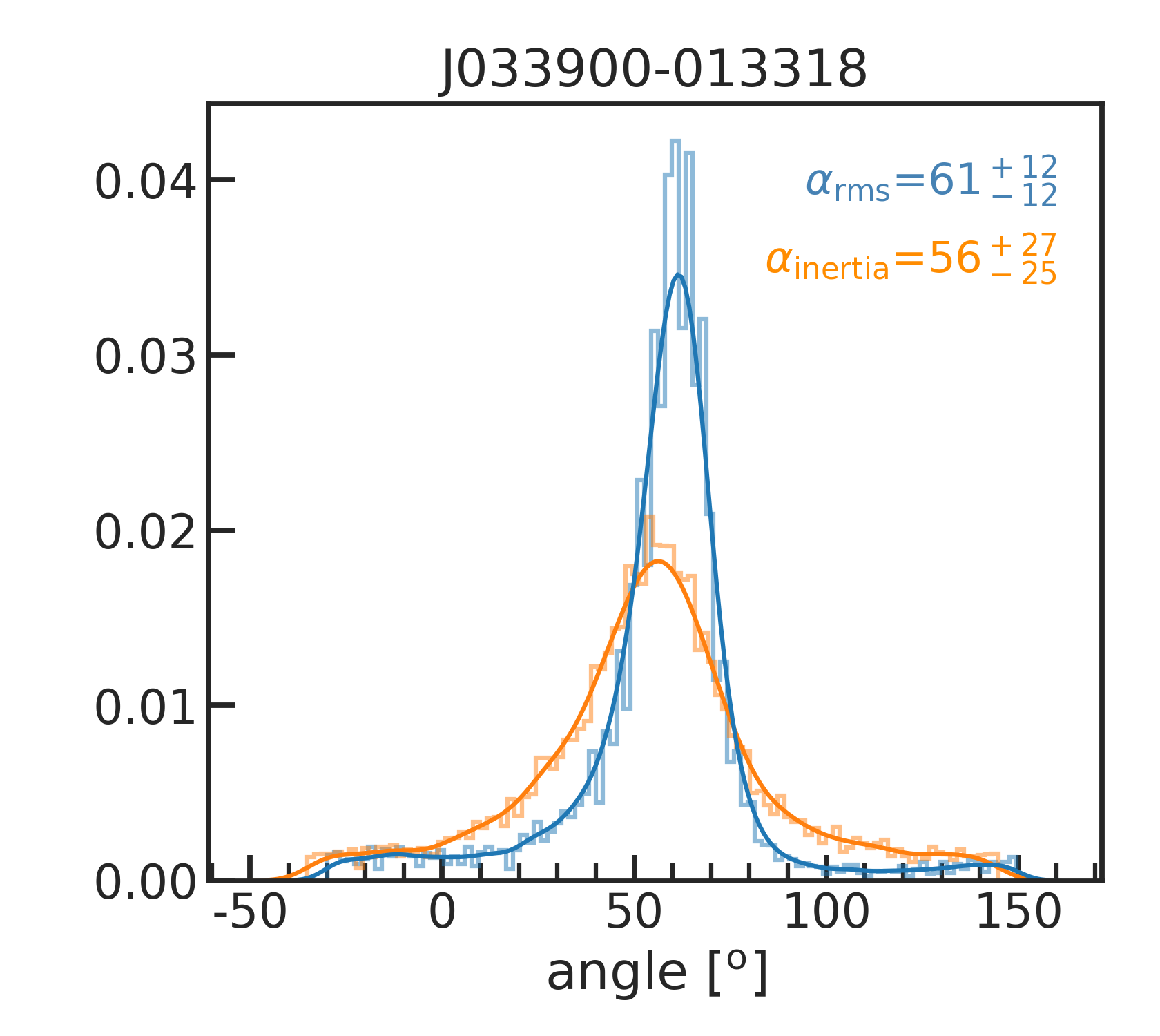}
\includegraphics[width=0.35\textwidth]{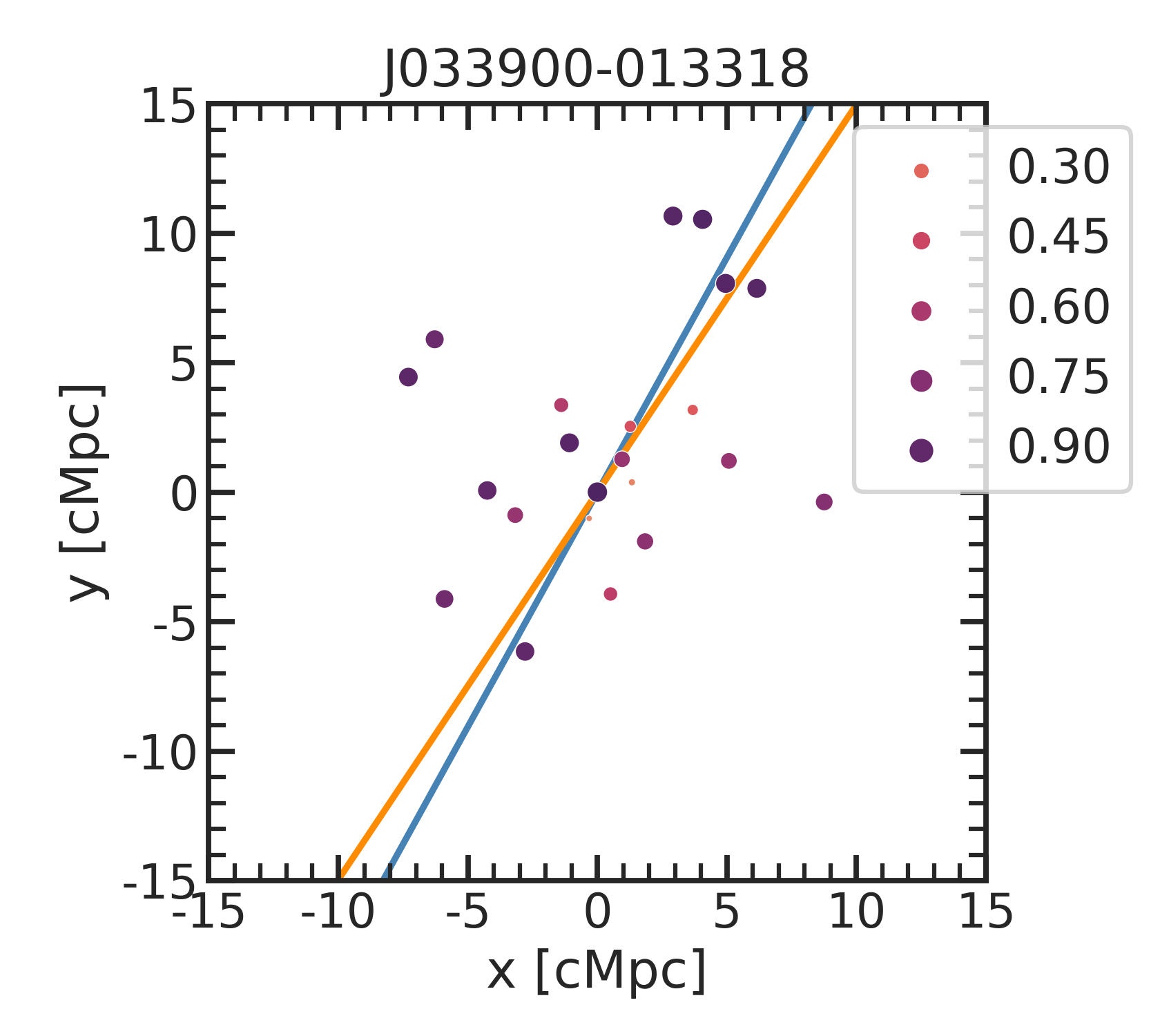}
\includegraphics[width=0.24\textwidth, trim=0 -3.2cm 0 0]{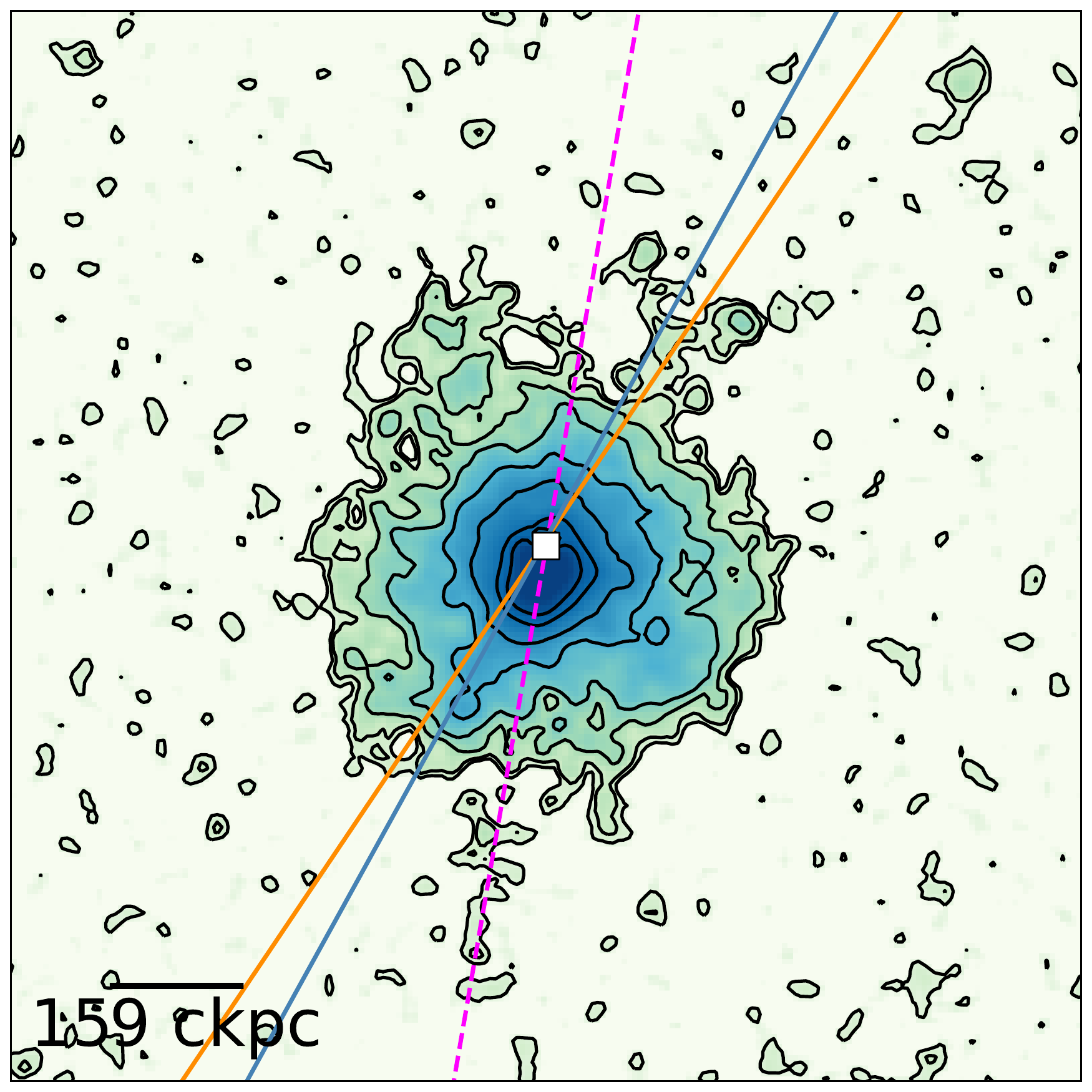}

\includegraphics[width=0.35\textwidth]{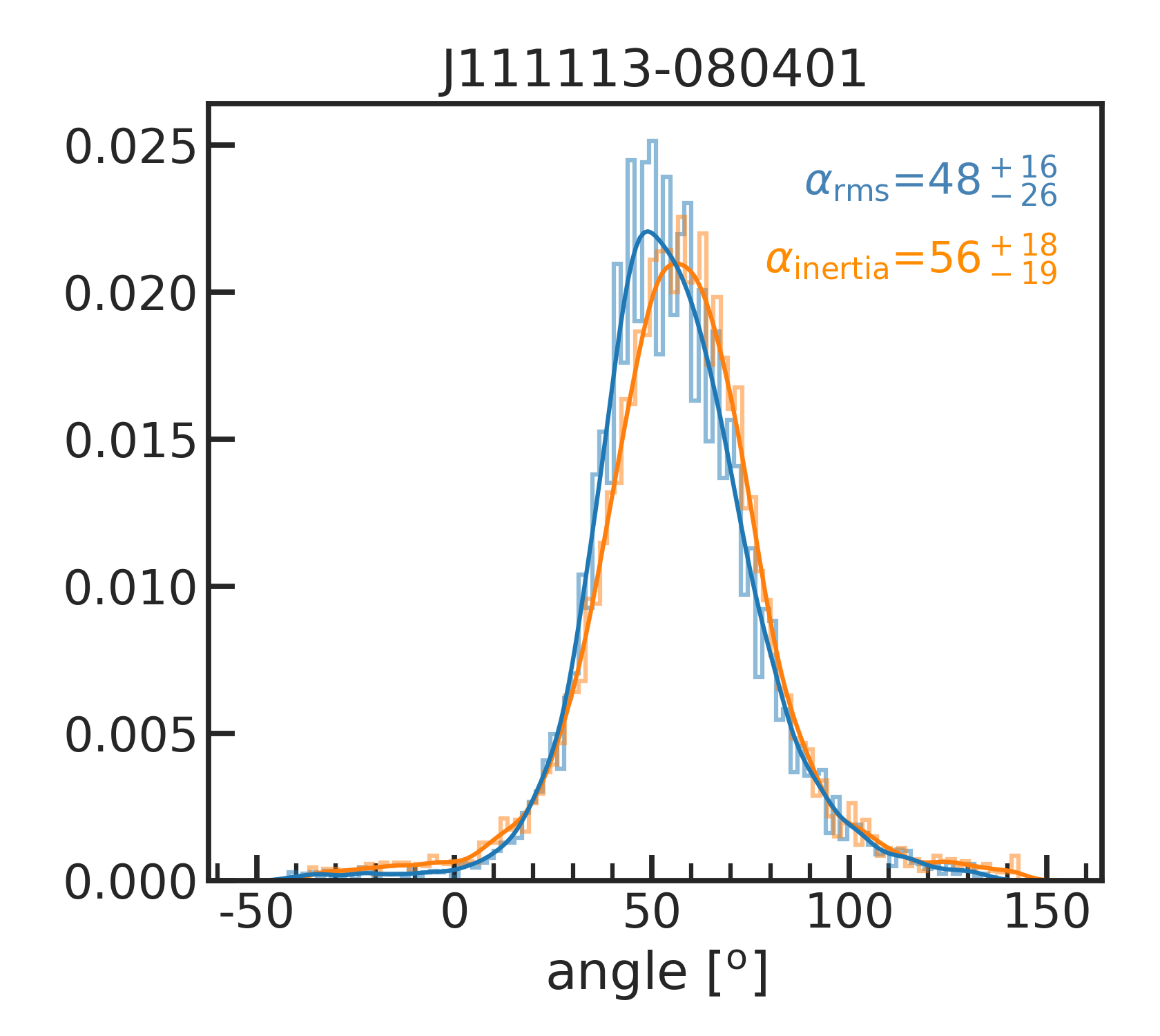}
\includegraphics[width=0.35\textwidth]{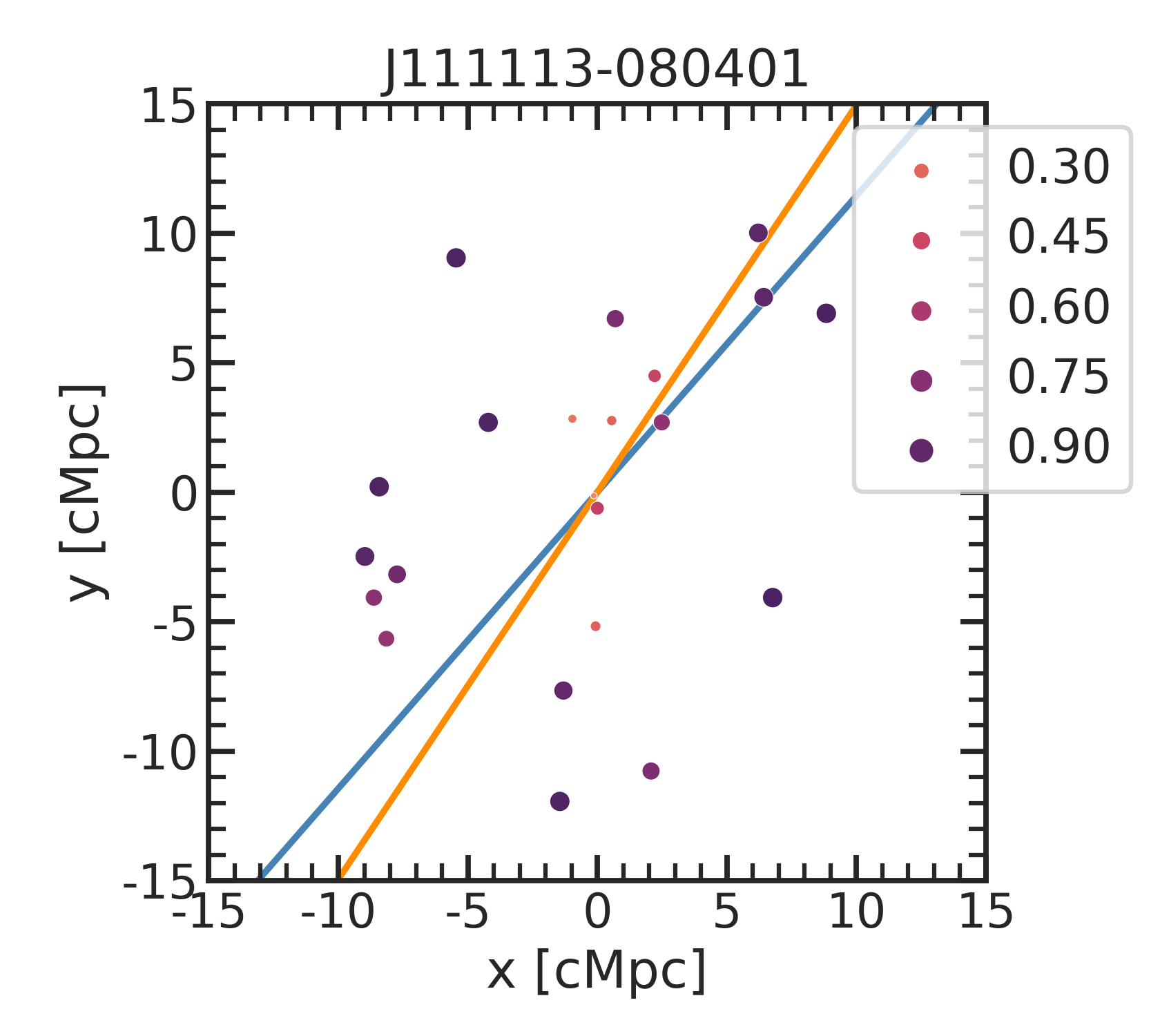}
\includegraphics[width=0.24\textwidth, trim=0 -3.2cm 0 0]{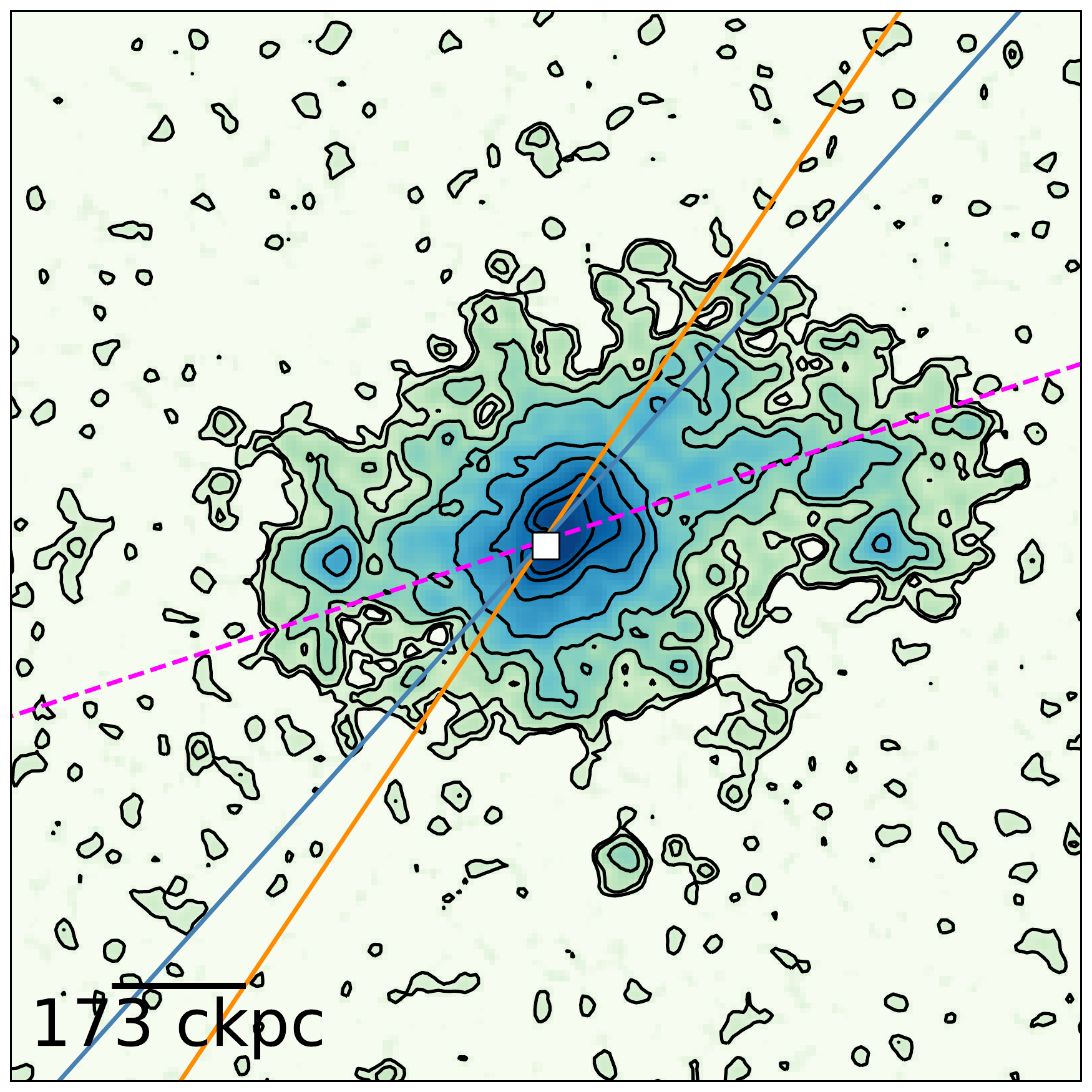}

\includegraphics[width=0.35\textwidth]{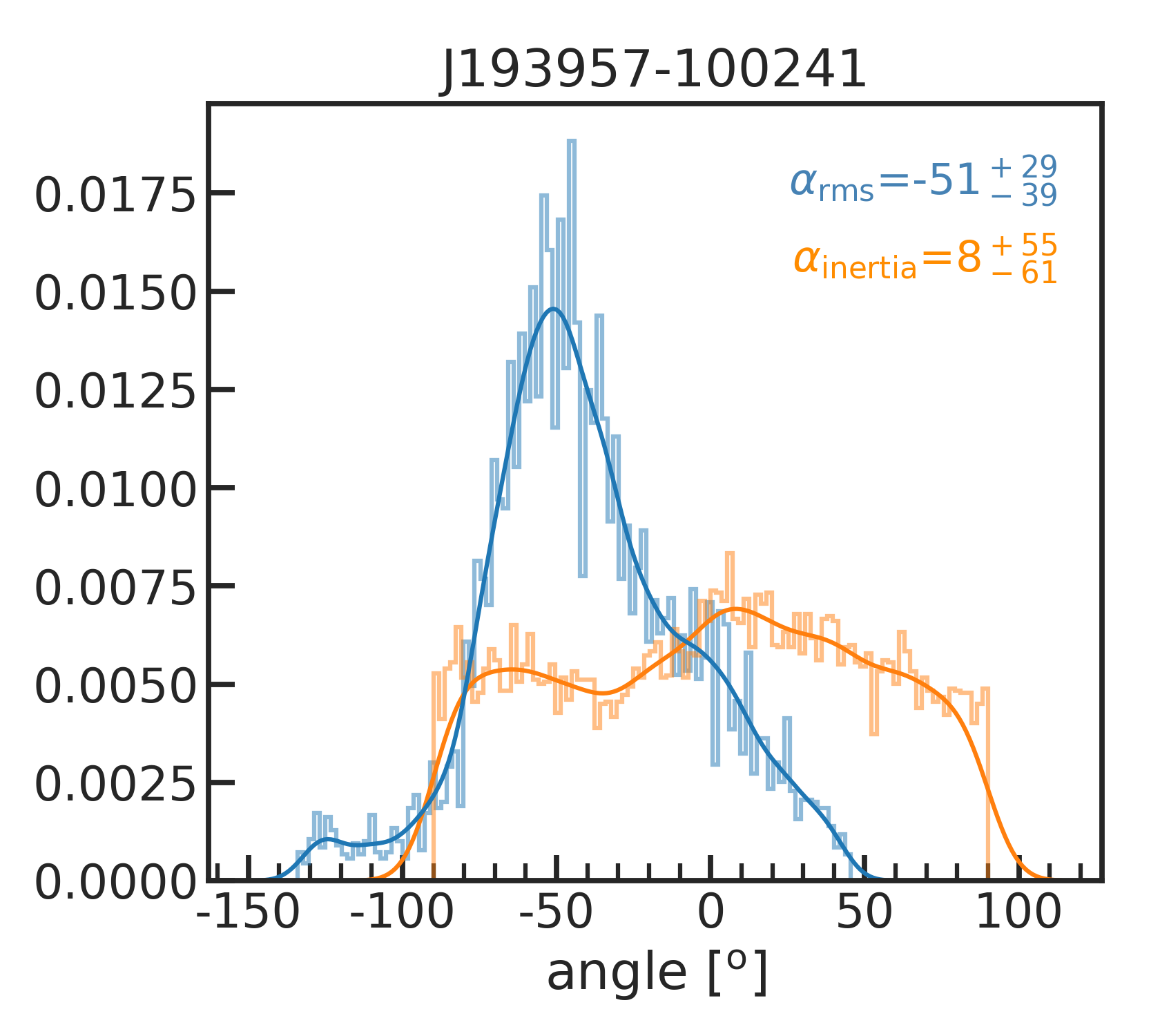}
\includegraphics[width=0.35\textwidth]{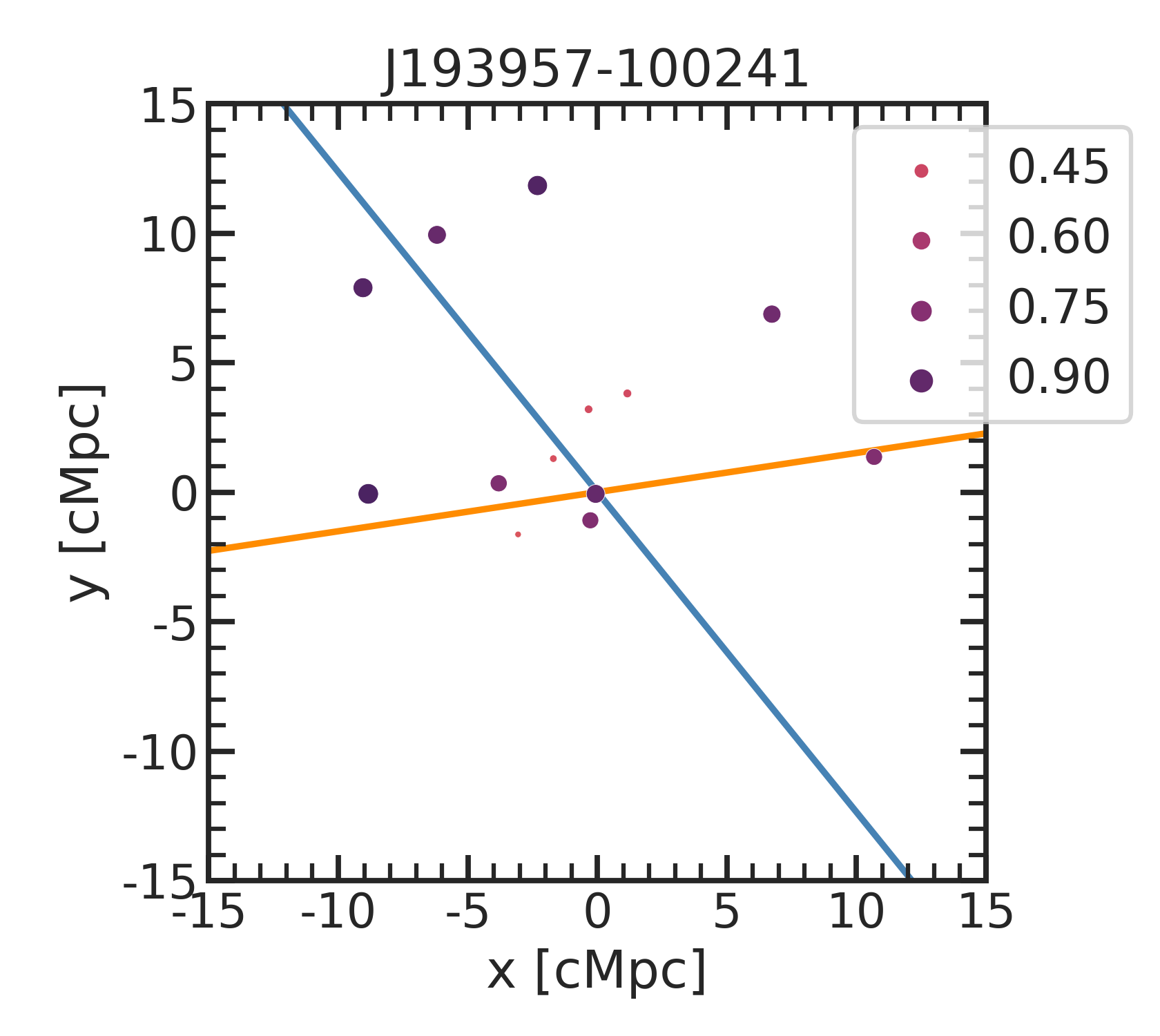}
\includegraphics[width=0.24\textwidth, trim=0 -3.2cm 0 0]{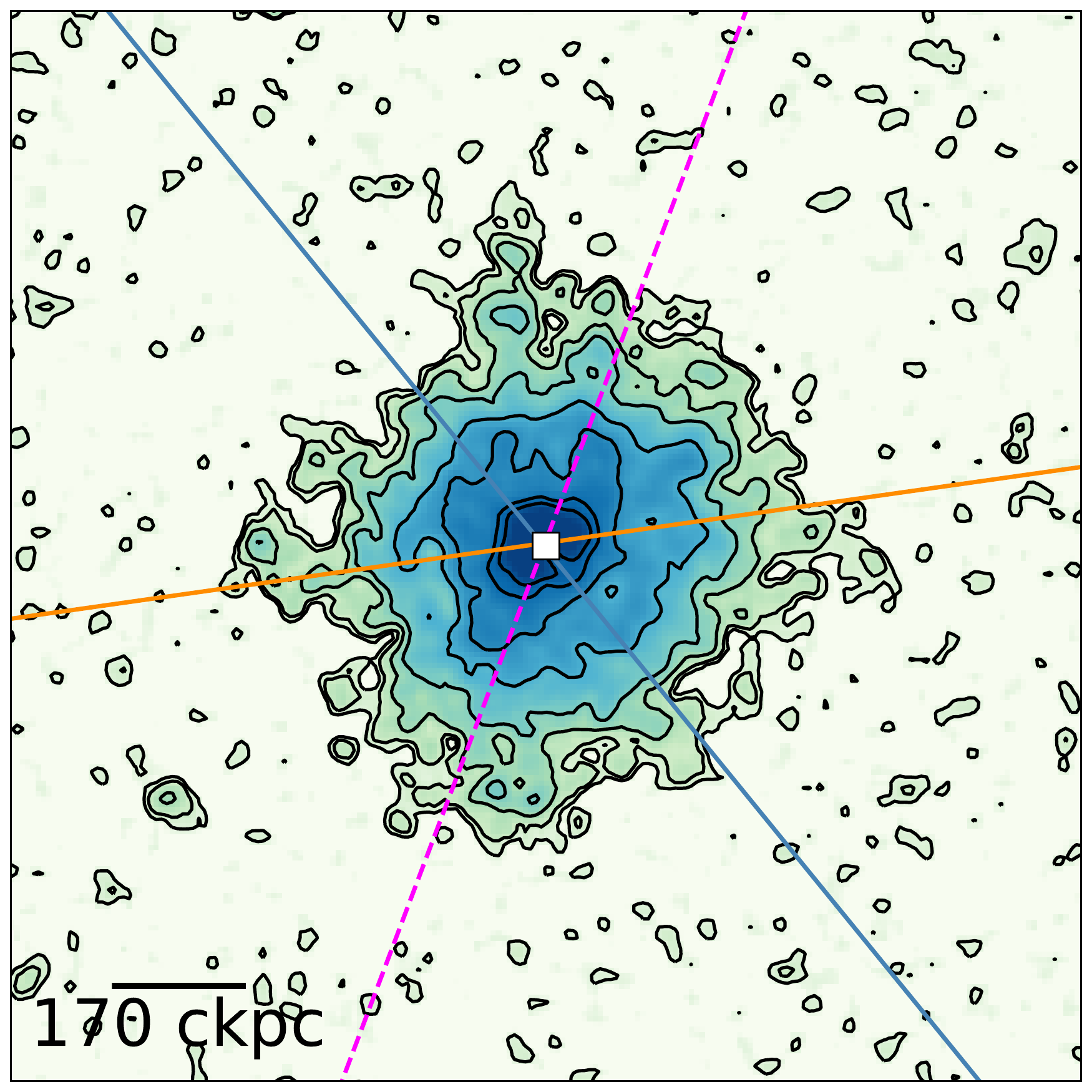}
\caption{Same as Figs.~\ref{fig:angle_over}, \ref{fig:angle_over_app5}, \ref{fig:angle_over_app1}, and \ref{fig:angle_over_app2}, but for another four sources (see the labels).}
\label{fig:angle_over_app3}
\end{figure*}

\begin{figure*}
\centering
\includegraphics[width=0.35\textwidth]{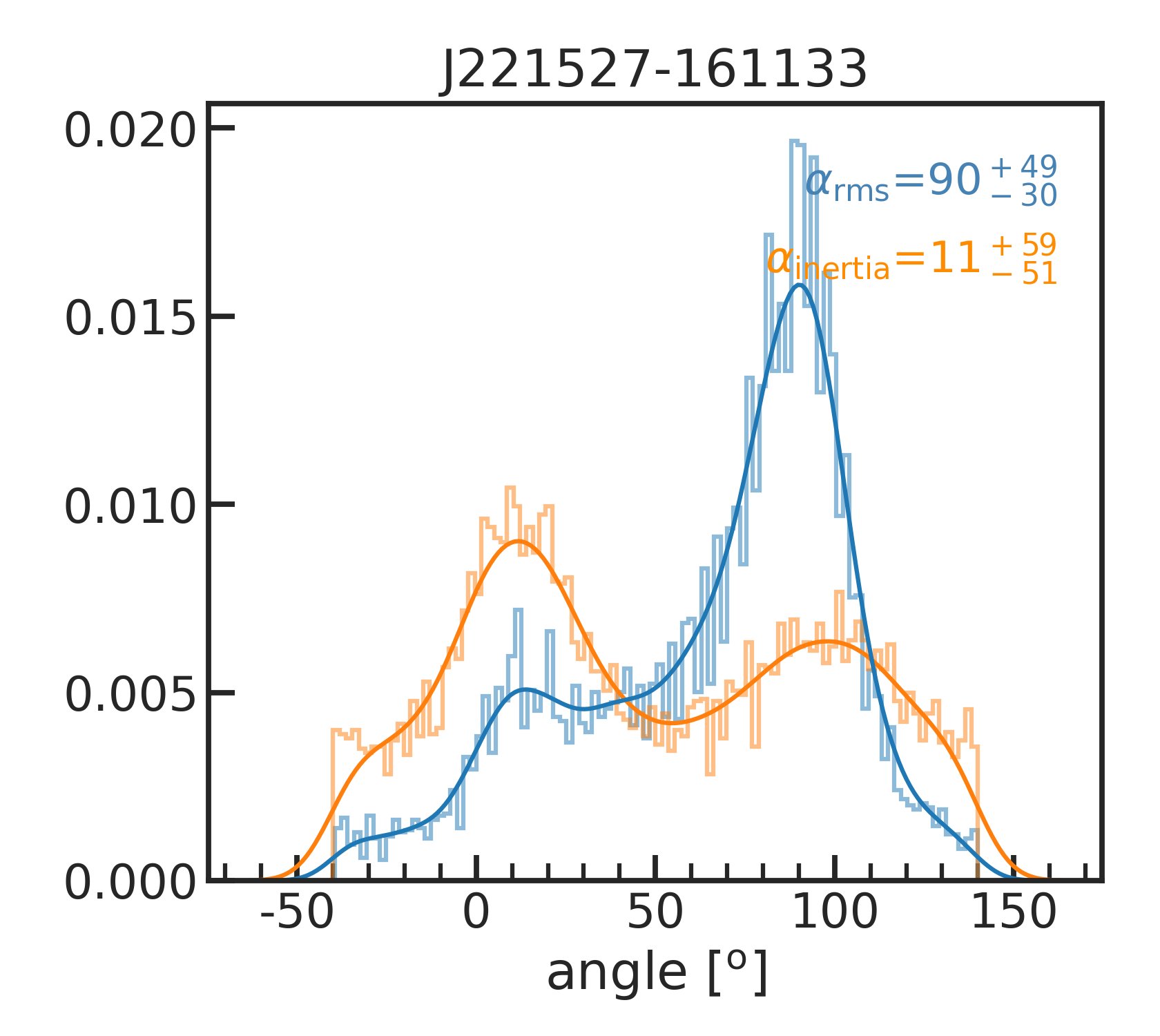}
\includegraphics[width=0.35\textwidth]{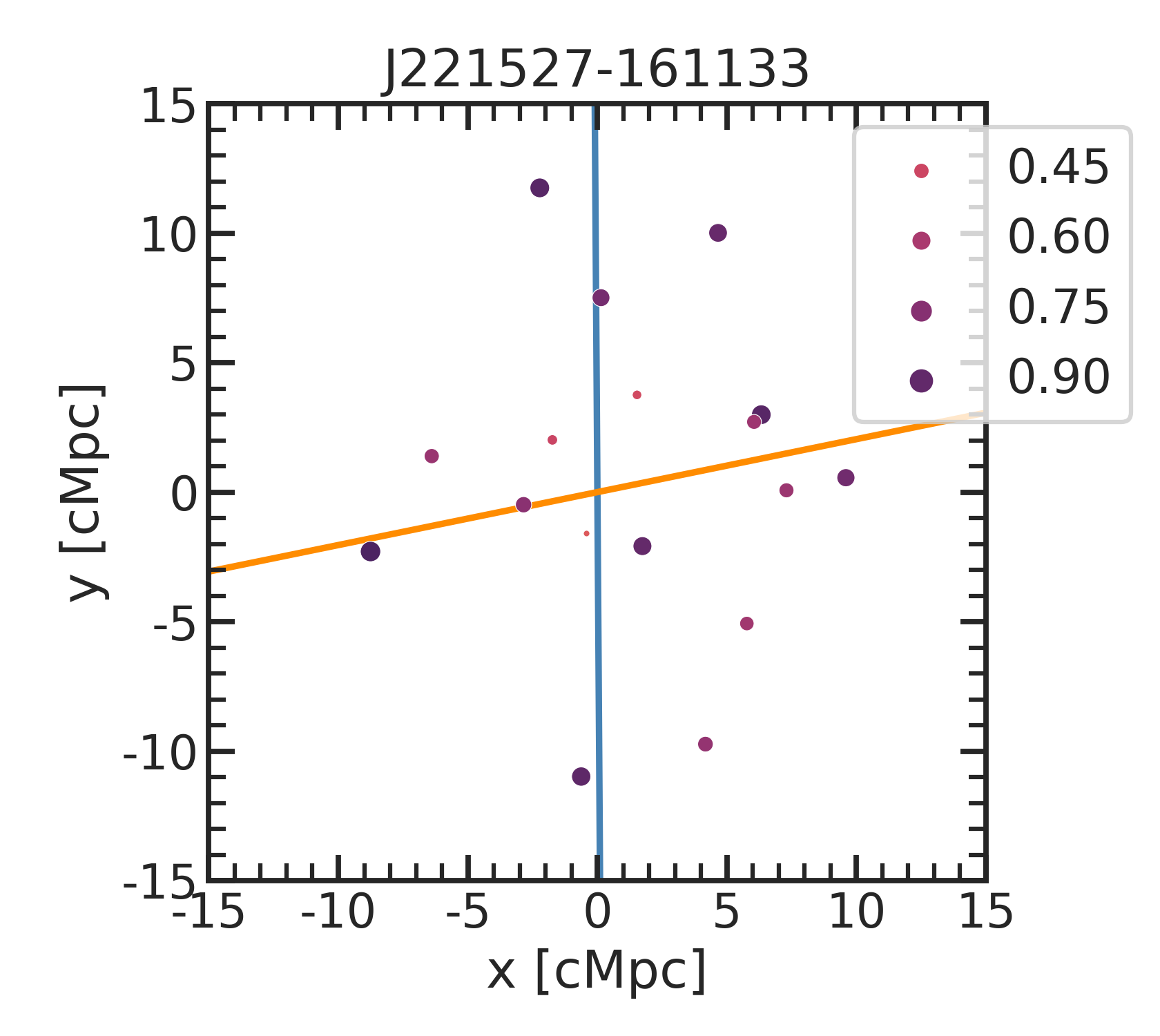}
\includegraphics[width=0.24\textwidth, trim=0 -3.2cm 0 0]{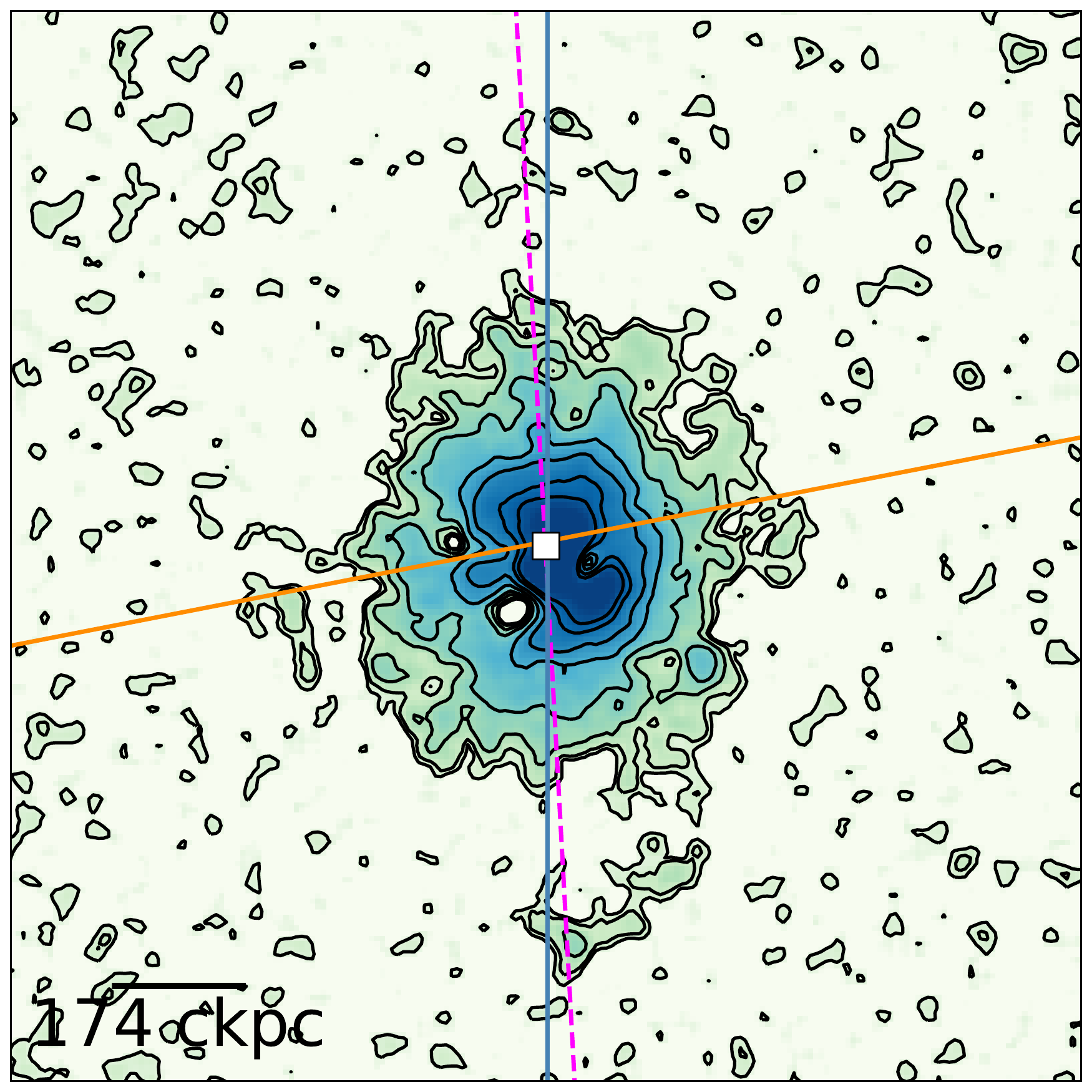}

\includegraphics[width=0.35\textwidth]{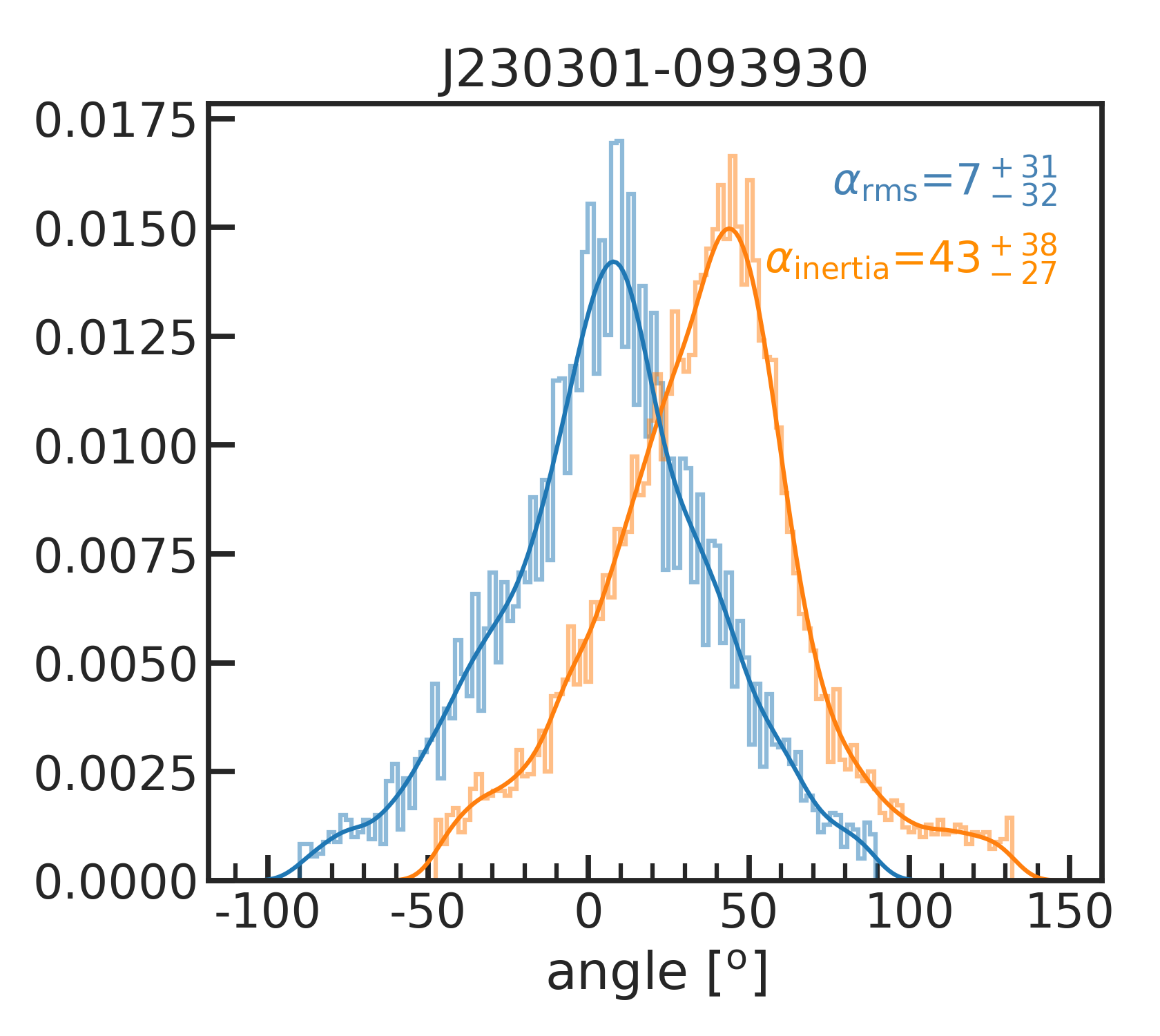}
\includegraphics[width=0.35\textwidth]{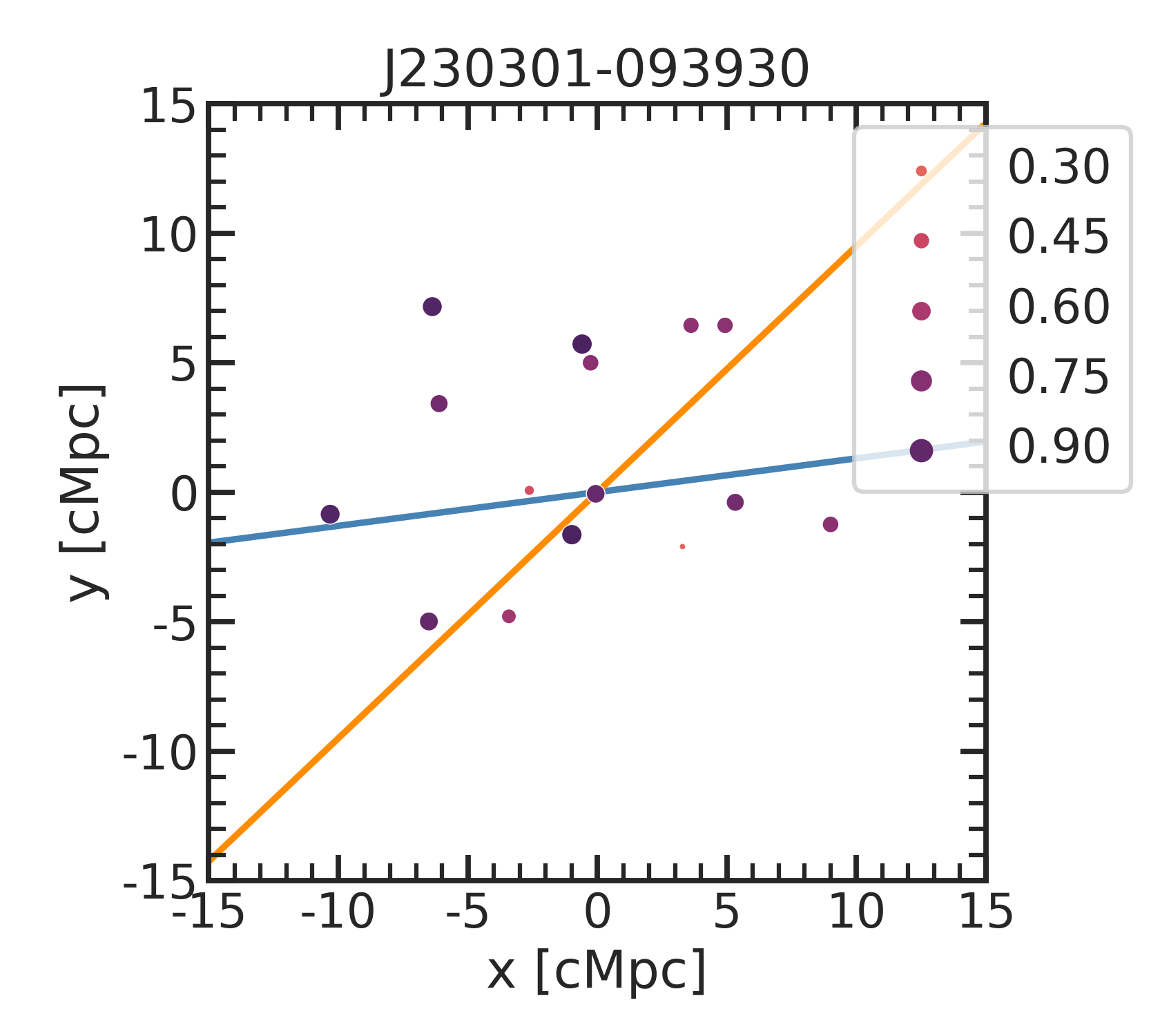}
\includegraphics[width=0.24\textwidth, trim=0 -3.2cm 0 0]{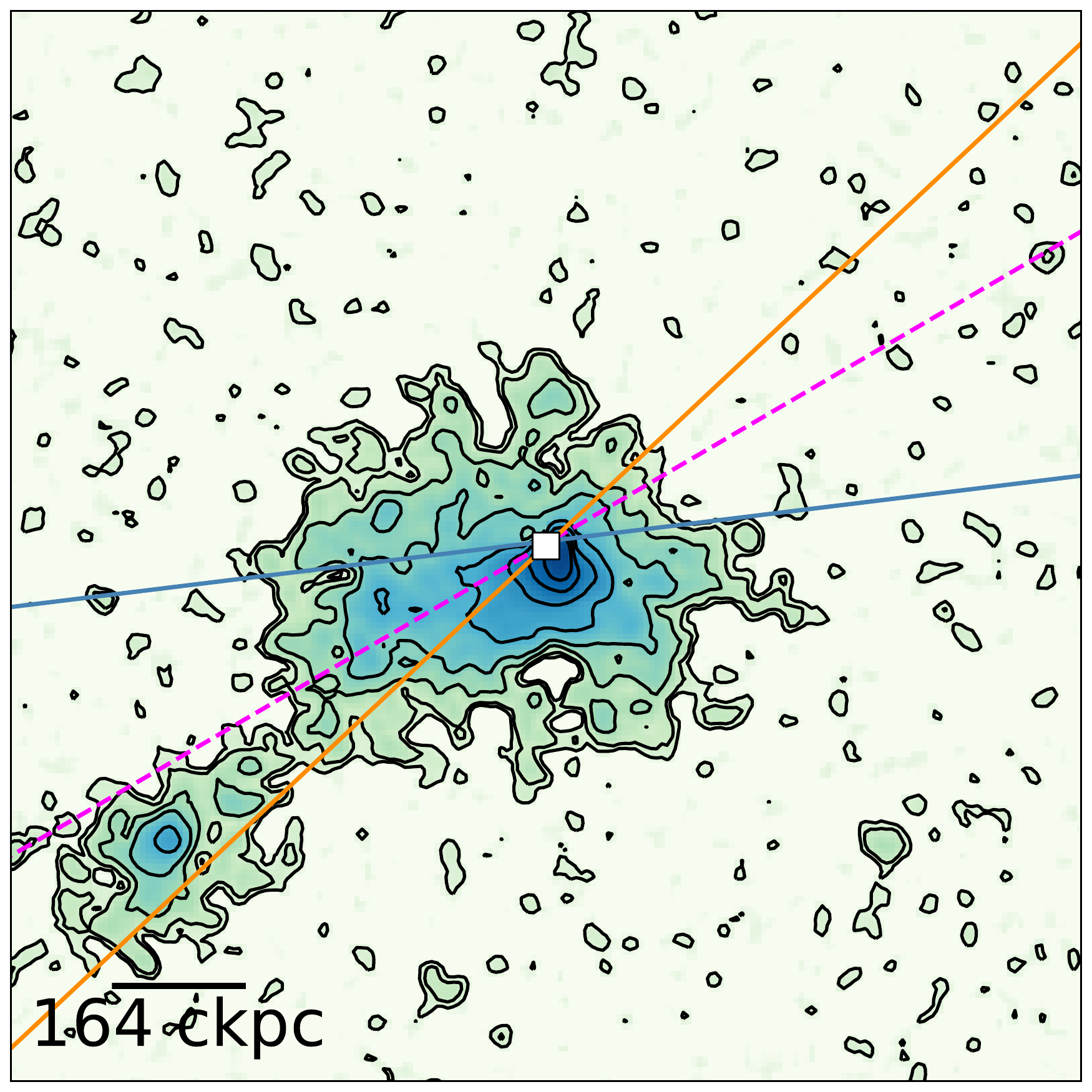}

\includegraphics[width=0.35\textwidth]{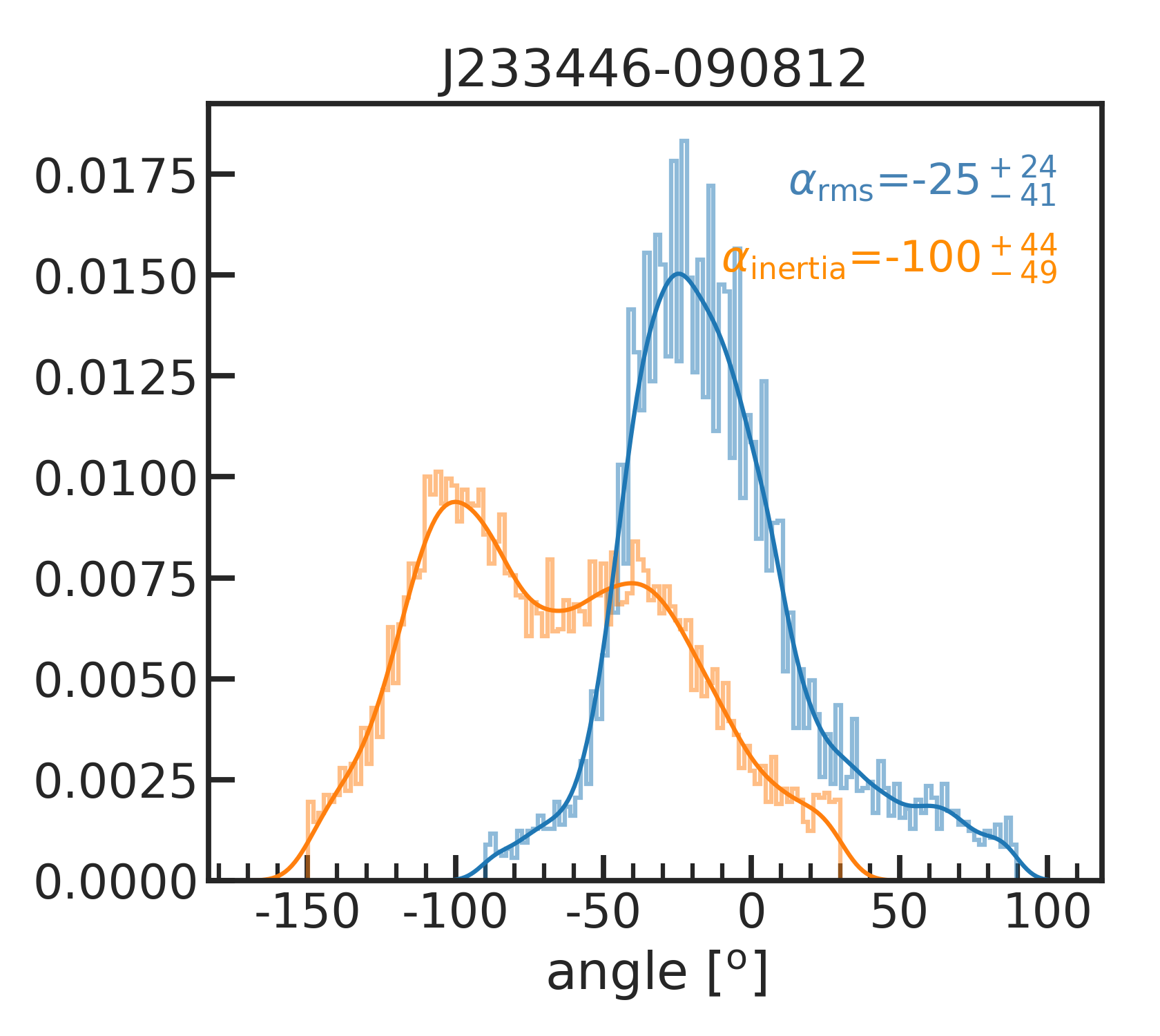}
\includegraphics[width=0.35\textwidth]{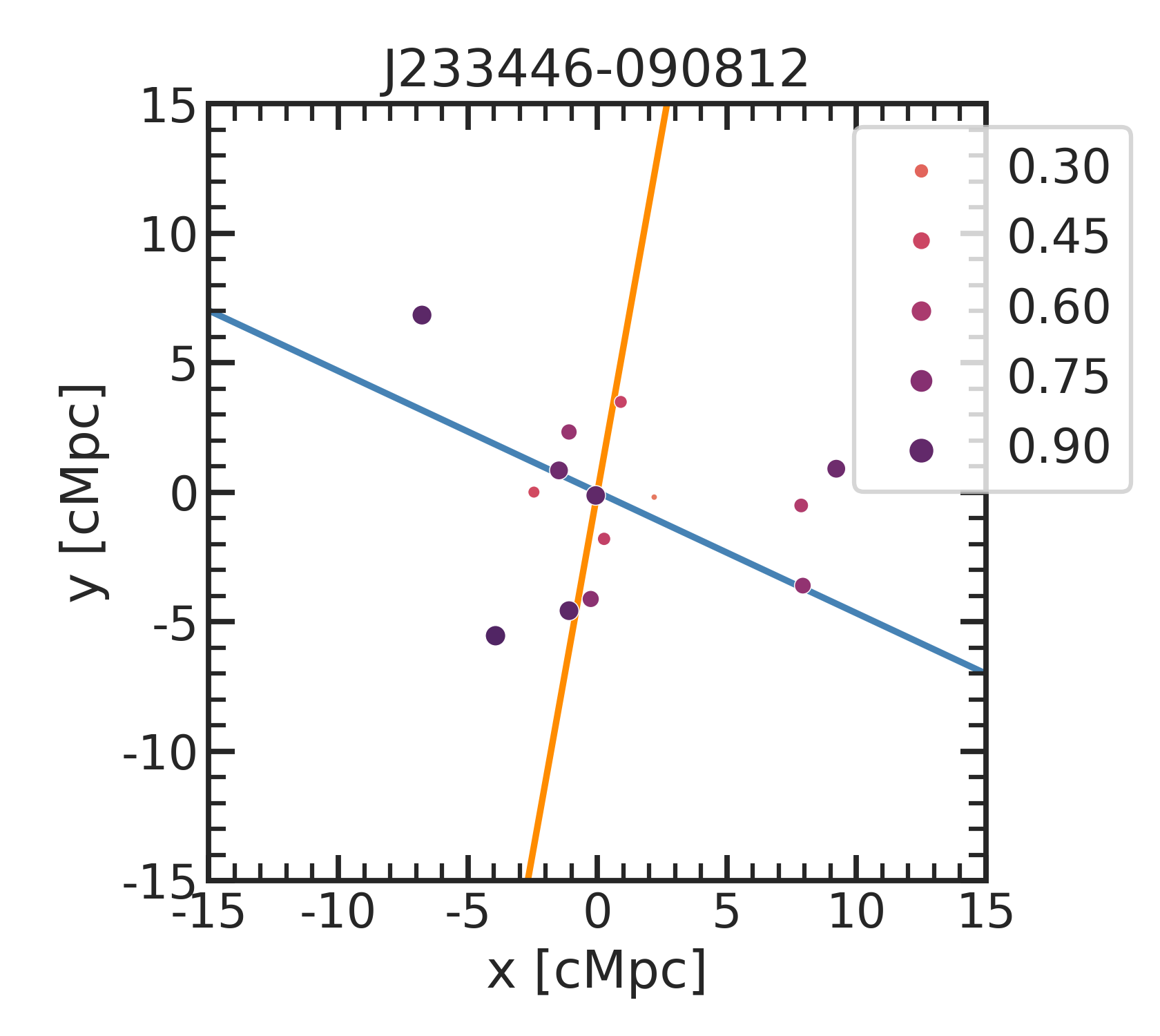}
\includegraphics[width=0.24\textwidth, trim=0 -3.2cm 0 0]{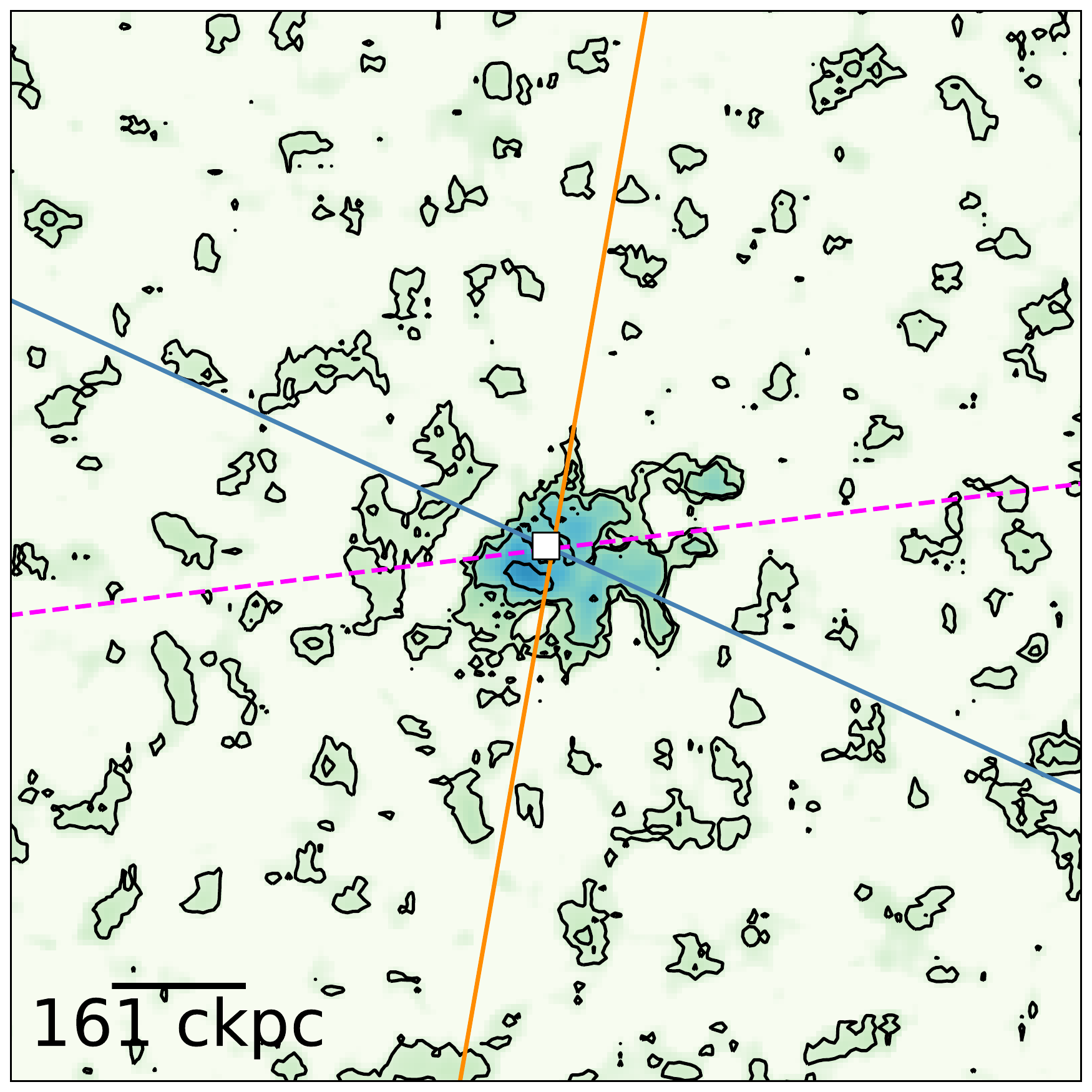}
\caption{Same as Figs.~\ref{fig:angle_over}, \ref{fig:angle_over_app5}, \ref{fig:angle_over_app1}, \ref{fig:angle_over_app2}, and \ref{fig:angle_over_app3}, but for another three sources (see the labels).}
\label{fig:angle_over_app4}
\end{figure*}

\FloatBarrier

\end{appendix}

\end{document}